\documentclass{aa}  
\usepackage[flushleft]{threeparttable}
\usepackage{float}
\usepackage{graphics}
\usepackage{graphicx}
\usepackage{amssymb, amsmath}
\usepackage{tikz}
\usetikzlibrary{graphs, positioning, quotes, shapes.geometric}
\usepackage{amstext}
\usepackage[T1]{fontenc}
\usepackage{mathrsfs}
\usepackage[normalem]{ulem}
\usepackage{natbib,twoopt}
\usepackage{hyperref} 
\hypersetup{
     colorlinks=true,
     linkcolor=blue,
     filecolor=blue,
     citecolor = black,      
     }
\usepackage{soul,xcolor}
\usepackage{booktabs}
\usepackage{txfonts}
\makeatletter
\def\@to{to}
\def\as     {\ifmmode {\rlap.}$\,$''$\,$\! \else ${\rlap.}$\,$''$\,$\!$\fi}
     \def\decsec  {\ifmmode {\rlap.}$\,$^{\rm s}$\,$\! \else ${\rlap.}$\,$^{\rm s}$\,$\!$\fi}\def\decs  {\ifmmode {\rlap.}$\,$^{\rm s}$\,$\! \else ${\rlap.}$\,$^{\rm s}$\,$\!$\fi}
\makeatother
\makeatletter
 \newcommandtwoopt{\citetads}[3][][]{%
  \nonstopmode%
  \href{http://adsabs.harvard.edu/abs/#3}{\def\hyper@linkstart##1##2{}%
    \let\hyper@linkend\@empty\citet[#1][#2]{#3}}%
  \errorstopmode}
\makeatother

\newcommand{\rah}{$^{\mbox{\scriptsize h}}$}

\newcommand{\ras}{$^{\mbox{\scriptsize s}}$}
\newcommand{\decd}{$^{\circ}$}
\newcommand{\decm}{$'$}

\begin{document}

\title{The evolution of temperature and density structures of OB cluster-forming molecular clumps}

 \author{Y. Lin
          \inst{1}\thanks{Member of the International Max-Planck Research School (IMPRS)
for Astronomy and Astrophysics at the Universities of Bonn and
Cologne.}  \and F. Wyrowski \inst{1} \and H. B. Liu \inst{2} \and A. Izquierdo \inst{3} \and T. Csengeri \inst{4} \and S. Leurini \inst{5}  \and K. M. Menten \inst{1}}

   \institute{Max Planck Institute for Radio Astronomy, Auf dem H\"{u}gel 69, 53121 Bonn\\
              \email{ylin@mpifr-bonn.mpg.de}
              \and
              Institute of Astronomy and Astrophysics, Academia Sinica, 11F of Astronomy-Mathematics Building, AS/NTU No.1, Sec. 4, Roosevelt Rd, Taipei 10617, Taiwan, ROC
              \and 
               European Southern Observatory, Karl-Schwarzschild-Str. 2, 85748 Garching bei München, Germany
         \and
        OASU/LAB-UMR5804, CNRS, Universit\'e Bordeaux, all\'ee Geoffroy Saint-Hilaire, 33615 Pessac, France
        \and 
        INAF – Osservatorio Astronomico di Cagliari, Via della Scienza 5, I-09047 Selargius (CA), Italy
       }


\abstract
{OB star clusters originate from parsec-scale massive molecular clumps, while individual stars may form out of $\lesssim$0.1 pc scales dense cores. The thermal properties of the clump gas are key factors governing the fragmentation process, and are closely affected by gas dynamics and feedback of forming stars.}
{We aim to understand the evolution of temperature and density structures on the intermediate-scale ($\lesssim$0.1-1 pc) extended gas of massive clumps. This gas mass reservoir is critical for the formation of OB clusters due to their extended inflow activities and intense thermal feedback during and after the formation.}
{We performed $\sim$0.1 pc resolution observations of multiple molecular line tracers (e.g., CH$_{3}$CCH, H$_{2}$CS, CH$_{3}$CN, CH$_{3}$OH) which cover a wide range of excitation conditions, towards a sample of eight massive clumps. The sample covers different stages of evolutions, and includes infrared-weak clumps and sources that are already hosting an H\textsc{ii} region, spanning a wide luminosity-to-mass ratio ($L/M$) range from $\sim$1 to $\sim$100 ($L_{\odot}$/$M_{\odot}$).
Based on various radiative transfer models, we constrain the gas temperature and density structures and establish an evolutionary picture, aided by a spatially-dependent virial analysis and abundance ratios of multiple species. } 
{We determine temperature profiles varying between 30-200 K over a continuous scale, from the center of the clumps out to 0.3-0.4 pc radii. The clumps' radial gas density profiles, described by radial power-laws with slopes between -0.6 and $\sim$-1.5, are steeper for more evolved sources, as suggested by results based on both dust continuum, representing the bulk of the gas ($\sim$10$^{4}$ cm$^{-3}$), and CH$_{3}$OH lines probing the dense gas ($\gtrsim$10$^{6}$-10$^{8}$ cm$^{-3}$) regime.
The density contrast between the dense gas and the bulk gas increases with evolution, and may be indicative of spatially and temporally varying star formation efficiencies.  
The radial profiles of the virial parameter show a global variation towards a sub-virial state as the clump evolves. The line-widths probed by multiple tracers decline with increasing radius around the central core region and increase in the outer envelope, with a slope shallower than the case of the supersonic turbulence ($\sigma_{\mathrm v}$$\,\propto\,$$r^{0.5}$) and the subsonic Kolmogorov scaling ($\sigma_{\mathrm v}$$\,\propto\,$$r^{0.33}$). In the context of evolutionary indicators for massive clumps, we also find that the abundance ratios of [CCH]/[CH$_{3}$OH] and [CH$_{3}$CN]/[CH$_{3}$OH] show correlations with clump $L/M$.} 
{}

\maketitle

\section{Introduction}\label{sec:intro}
Massive star-forming clumps are progenitors of OB clusters (\citealt{Williams00}, \citealt{Motte18}). They have masses of typically $\gtrsim$1000 $M_{\odot}$ over a spatial scale of $\sim$1 pc. Fragmentation and accretion processes of OB star clusters are strongly influenced by the evolution of the kinematics, density and temperature structure of pc-scale gas clumps (\citealt{Girichidis11}, \citealt{Lee19}, \citealt{Padoan19}), and vice versa  (\citealt{Krumholz12}, \citealt{Offner09}, \citealt{Hennebelle20}). Particularly, the stellar initial mass function (IMF) appears to be universal that varies weakly from one environment to another in the Milky Way, indicating that the formation of the most massive stars is deterministic, favoring particular physical environments instead of randomly occurring in molecular clouds (\citealt{Kroupa13}). This, together with dominant feedback caused by massive stars, may determine the evolutionary track of massive clumps. Accordingly, observational evidence can be collected by sampling a wide range of clumps at different evolutionary stages. 

The process of gas collapse resulting in protostars has been studied for decades, among the first are the works by \citet{Larson69}, \citet{Penston69} and \citet{Shu77}. These are commonly refered to as `outside-in' and `inside-out' models, indicating the succession of the spherical collapse of isothermal clouds, which describe the gas flows (immediately) prior to and after development of a protostar (singularity), respectively. The density profiles of \citet{Larson69} and \citet{Penston69} exhibits a $r^{-2}$ relation while the density profile of the inner free-falling and outer static envelopes of \citet{Shu77} model follow $r^{-1.5}$ and $r^{-2}$, separately. 
On the other hand, when turbulent pressure is taken into account to explain the observed linewidth-size scaling relation, the logatropic (nonisothermal) gas follows, a flatter profile proportional to $r^{-1}$ (e.g. \citealt{MP97}) in the outer region.
Recently, the process of spherically symmetric cloud collapse has been revisited extensively: \citet{Coughlin17} present solutions for arbitrary initial density profiles, extending to non-self-similar regime; work by \citet{Murray15}, \citet{Murray17}
considers turbulent energy generated from gravitational collapse and a highly dynamic system (as compared to hydrostatic equilibrium assumed by \citet{Shu77}).  Furthermore, due to significant heating sources and high opacities associated with massive star-forming clouds, the assumption of isothermality might break down and a polytropic equation of state (EOS) needs to be introduced which quantifies the balance of gas cooling and heating and can incorporate turbulent pressure (\citealt{Curry00}). The polytropic index $\gamma$ (with $T\,\propto\,$$\rho^{\gamma-1}$) has been shown to have a decisive effect on the dynamical evolution of molecular clouds (\citealt{Passot98}, \citealt{Spaans00}) and eventually on the IMF (e.g. \citealt{Klessen07}, \citealt{Jappsen}).   
Moreover, recent works have demonstrated that a simple EOS assumption for the gas evolution might fail to explain the invariability of the peak of the IMF, while (proto)stellar radiative heating, a process that is not fully captured by the EOS, seems to play a crucial role in setting the characteristic mass scale (e.g. \citealt{Bate09}, \citealt{KKM11}, \citealt{Guz16, Guz17}). 
Given these theoretical developments, it is timely to measure with observations the detailed gas temperature and density distribution inside massive clumps. 

Understanding how the mass of massive clumps is distributed over different density regimes is fundamental to understanding the evolution of the star formation rate (SFR) and star formation efficiency (SFE) on larger physical scales (e.g. \citealt{Lee15}, \citealt{Parmentier19}). On cloud scales ($\gtrsim$10 pc), the gas column density distribution follows a log-normal probability function (N-PDF) in a turbulent medium while it develops a power-law tail(s) 
as significant gravitational collapse commences in high density regimes (e.g. \citealt{Klessen00}, \citealt{Kritsuk11}). The relevant scales are readily resolved in nearby star-forming clouds and OB cluster forming regions (e.g. \citealt{Kainulainen09}, \citealt{Lin16, Lin17}). The power-law shape is suggested to originate from power-law density profiles (\citealt{FCK12}, \citealt{Myers15}). Hence, measurements of clump density profiles can provide insights on how the dense gas of molecular clouds lead to the power-law excess of N-PDFs. 

Most previous works on the density structure of massive clumps are based on single dish observations, of both continuum and molecular lines. Works discussing samples of sources include but are not limited to, e.g. \citet{vdt00}, \citet{Mueller02}, \citet{Beuther02a}, \citealt{Hatchell03}, \citet{Rolffs11}, \citet{Williams05}, \citet{Palau14}. With the advent of wide-band receivers, especially those equipping interferometers, spatially resolved multi-line observations have become efficient (e.g. \citealt{Beuther07}, \citealt{Li19}, \citealt{Gieser21}), which are indispensable to measure the broad density and temperature ranges associated with massive star formation. 

\begin{figure*}
    \centering
     \includegraphics[scale=0.25]{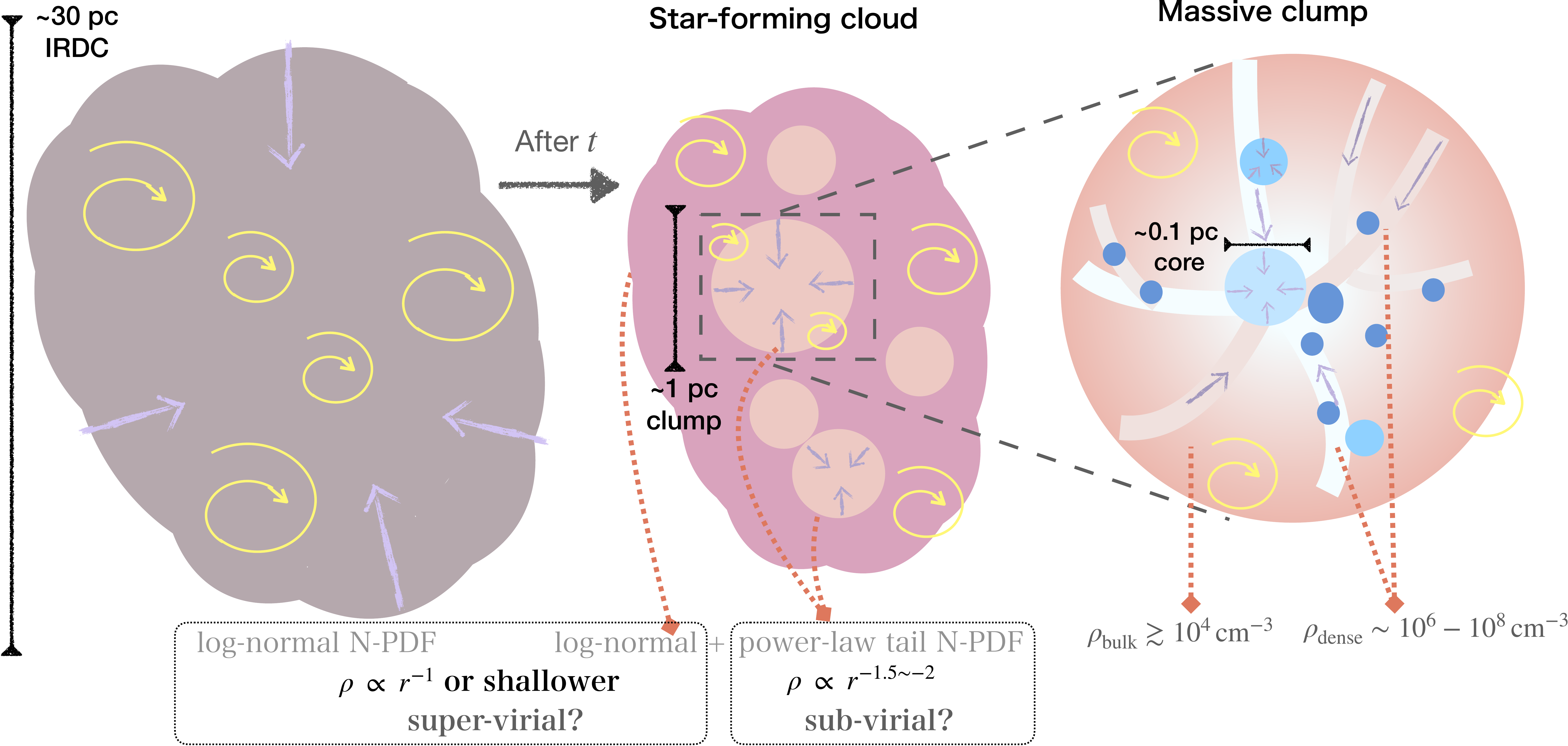}
    \caption{Schematic picture of molecular cloud structure over spatial scales of $>$10 pc to $\sim$0.1 pc: from cloud to core scale. A molecular cloud starts contraction from an initial stage that appears to be an infrared dark cloud (IRDC) and evolves into a star-forming one, embedding a number of molecular clumps. The massive clumps, of $\sim$1 pc in size, are generally composed of filamentary structures and cores at different evolutionary stages. In all figures, yellow curved arrows represent turbulent motions and purple arrows indicate gravitational contraction or gas inflows (along filaments). In the rightmost figure, thick lines show filaments and blue ovals indicate cores of different masses; the color gradient of the clump indicate a density gradient of the bulk gas. The characteristics of different structures are linked to texts by dotted arrows.}
    \label{fig:schematic_struct}
\end{figure*}

We have conducted a pilot survey of eight massive clumps with the Submillimeter Array (SMA) and the APEX telescope. For the target clump selection, we followed the criterion elaborated in Sect.~\ref{sec:sselect}.
Main molecular lines of interest are listed in Table \ref{tab:lines}, which include multiple efficient thermometers and densitometers for massive clumps, as suggested by single dish observations 
towards a statistically large sample (\citealt{Giannetti17}, \citealt{Leurini04, Leurini07}). We use various modeling methods to quantify the clump density and temperature structure using these lines.
Throughout the paper, We follow the existing nomenclature in the literature (e.g., \citealt{Williams00}, \citealt{Zhang09}; \citealt{Liu2012ApJ}, \citealt{Motte18}). In this way, massive molecular clumps refer to structures with sizes of $\sim$0.5-1 pc, massive molecular cores refer to the $<$0.1 pc size structures embedded within a clump, and condensations refer to the distinct molecular substructures within a core. In Figure \ref{fig:schematic_struct} we provide a schematic picture of different scales of a molecular cloud. The physical characteristics across the scales, as elaborated in the above, are marked for individual structures. We are interested in understanding the physical structure of massive clumps, which have a vast range of gas densities, and are the building blocks of the star-forming clouds; particularly they compose the high-density tail of the cloud N-PDF.

The paper is organised as follows: in Sect.~\ref{sec:observation} we describe the observations and data reduction. 
In Sect. ~\ref{sec:ana}, we describe the radiative transfer modeling procedure and elaborate on the analysis of both line and continuum observations, to derive the temperature, density, linewidth and virial parameter profiles, and abundances of multiple species. Sect.~\ref{sec:outline} gives a general outline of the radiative modeling methods and procedures we adopted.
Sect. ~\ref{sec:mom0} provides an overview of the distribution of the molecular lines used as thermometers and densitometers in this paper. Sect.~\ref{sec:xclass}, Sect.~\ref{sec:radex_nh2}, in addition to Appendix ~\ref{app:radmc}, ~\ref{app:lime} focus on the radiative transfer modeling procedures and results of continuum and molecular lines. 
In Section ~\ref{sec:sma_cont} the properties of the sources extracted from SMA 1.2 mm continuum are presented. 
Analysis of some complementary lines is presented in Appendix ~\ref{app:otherlines}. In Section ~\ref{sec:discussion}
we discuss the outcome of the modeling results, with a comparison between target sources and the physical implications. Finally in Section ~\ref{sec:conclusion} we concluded on our findings.  


\begin{figure}
\hspace{-0.25cm}\includegraphics[scale=0.46]{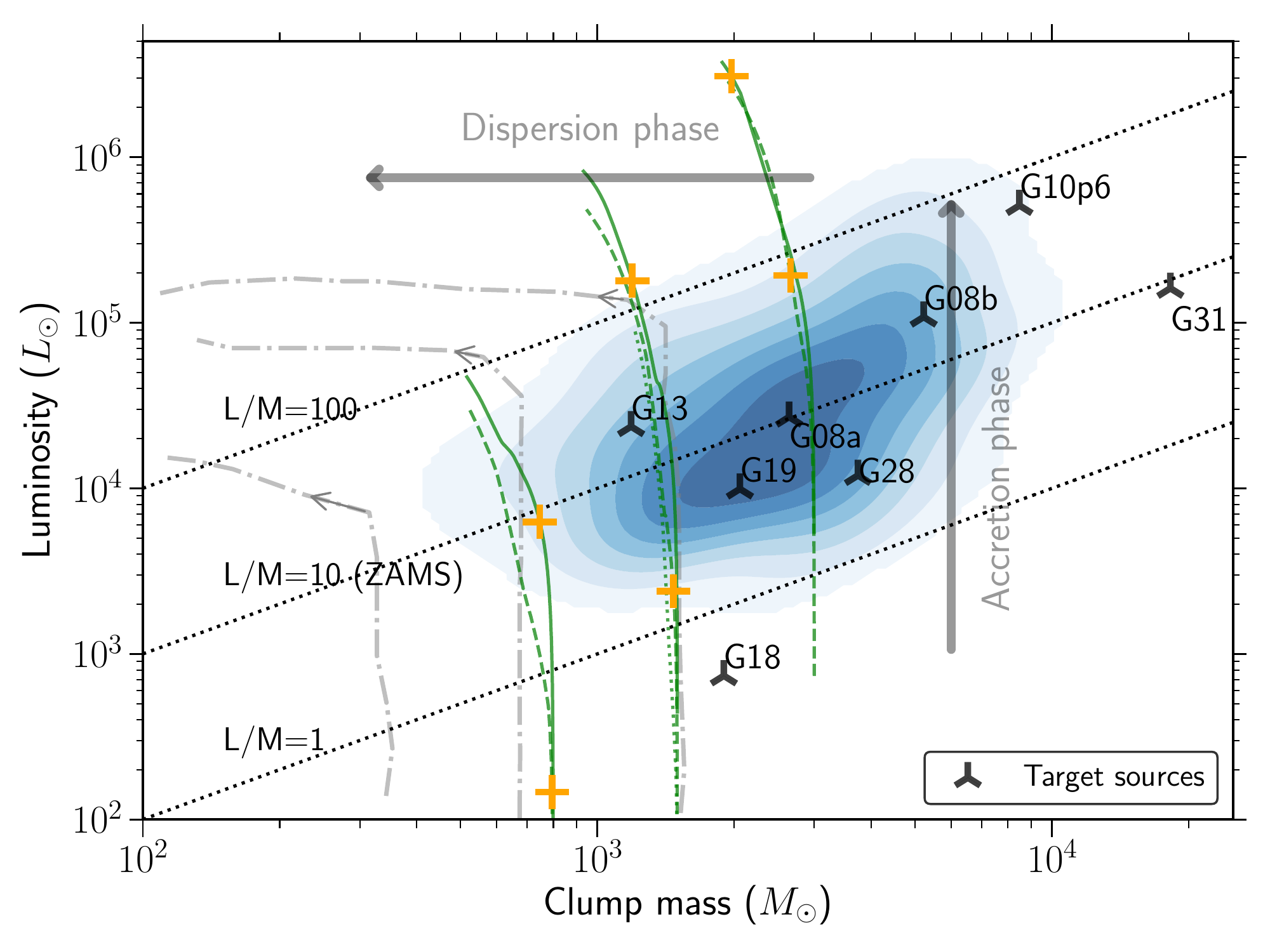}
\caption{The luminosity-mass diagram of target sources (three-branched triangle markers). The evolutionary tracks of massive clumps of different envelope masses and will from a cluster of stars with different accretion rates are shown in green lines (dotted and solid); gray lines (dash-dotted) with arrows are showing the evolutionary track of clumps of different envelope masses, but are assumed to form a single massive star (\citealt{Molinari08}, for more details see Sect. \ref{sec:sselect}.)}
\label{fig:lm_sma}
\end{figure}

\section{Observations and data reduction}\label{sec:observation}

\subsection{Source selection}\label{sec:sselect}
The target sources are selected from APEX telescope Large Survey of the Galaxy (ATLASGAL) survey (\citealt{Schuller09}) and are located at a distance of 4-6 kpc (\citealt{Urquhart18}). They cover different evolutionary stages, suggested by different luminosity-to-mass ratios (Figure \ref{fig:lm_sma}) and different signposts of star-formation activity (see further detail in Appendix \ref{app:target_sources}). 
For comparison, in Figure \ref{fig:lm_sma} we present the distribution of luminosity and mass for ATLASGAL sources that (1) are located at a distance within 4-8 kpc with a radius of $<$2 pc (\citealt{Urquhart18}), and (2) have masses and peak fluxes higher than 300 $M_{\odot}$ and peak flux $\sim$2 Jy/beam, respectively. The background contours illustrate the distribution of ATLASGAL sources in distance range of 4-8 kpc, with masses over 300 $M_{\odot}$ and peak flux $\gtrsim$2 Jy/beam with a radius of less than 2 pc.  

In Fig. \ref{fig:lm_sma}, we also include several evolutionary tracks: Gray dashed-dotted lines are the empirical evolutionary tracks for massive clumps with envelope masses of 80, 140, 350, 700, and 2000 $M_{\odot}$ which will form a single protostar with varying accretion rates based on turbulent core model (\citealt{MckeeTan03}), as derived in \citet{Molinari08}.
green lines indicate tracks of massive clumps having constant accretion rates of 10$^{-5}$ M$_{\odot} yr^{-1}$ (dotted), 10$^{-4}$ M$_{\odot} yr^{-1}$ (dashed) and 10$^{-3}$ M$_{\odot} yr^{-1}$ (solid), for the most massive star in the cluster. The other stellar members follow an equal accretion stopping probability, with an accretion rate $\propto$ $M^{1.5}$. The orange pluses mark the time epoch of 2$\times$10$^{4}$ yr for each accretion track. The SFE is assumed to be 30$\%$ and the underlying stellar population follows canonical IMF (\citealt{Kroupa93}). Accretion luminosities are estimated by interpolating massive protostar models in \citet{Hosokawa09}. 

The lower mass limit of 300 $M_{\odot}$ corresponds to the mass of a massive clump in which at least one $>$ 8 $M_{\odot}$ star will form according to the normal IMF with an assumed star formation efficiency (SFE) of 30$\%$ (\citealt{Kroupa93}, \citealt{Sanhueza17}). The peak flux density of 2 Jy/beam with beam full-width at half-maximum (FWHM) $\sim$20$''$ at 870 $\mu$m from ATLASGAL survey, considering a distance of 6 kpc, implies a mass of $>$100 $M_{\odot}$ concentrated in the clump central $\sim$0.5 pc region, assuming a dust temperature of 50 K and dust opacity of 1.8 cm$^{2}$g$^{-1}$ with a gas-to-dust ratio of 100. Thus, this selection criterion therefore yields a sample of eight massive clumps (Table \ref{tab:sources}) with moderate to high central concentration. We note that our target sample is representative of relatively more evolved sources with respect to those that fulfill the aforementioned criteria. These sources can be easily detected with 1 mm lines given their favorable excitation conditions, and we further complement the sample with an infrared dark source G18.

\subsection{SMA observations} \label{subsec:sma}
We performed SMA observations in the  $\sim$1 mm band towards seven 
clumps in the subcompact array configuration on 2017 June 11, and in the compact array configuration on 2017 August 28 (Project 2017A-S030, PI: Yuxin Lin), which covered baseline lengths of 9.5-45 meters and 16-77 meters, respectively.
The selected target sources are summarized in Table \ref{tab:sources}. 
Detailed information about each target source from previous studies is summarized in Appendix \ref{app:target_sources}.
We used the dual receivers mode supported with the SMA Wideband Astronomical ROACH2 Machine (SWARM) backend:
The RxA receivers covered the frequency ranges of 188.4-196.7 GHz and 204.4-212.7 GHz in the lower and upper sidebands, respectively; the RxB receivers covered the frequency ranges of 238.5-246.8 GHz and 254.5-262.8 GHz, respectively.
The intrinsic spectral channel width was 140 kHz.
The molecular line transitions we covered are summarized in Table \ref{tab:lines}. 

In addition, we retrieved archival SMA data towards the luminous ultra compact (UC) H\textsc{ii} region G10.624-0.38, which remains deeply embedded in a $M_{\mbox{\scriptsize gas}}=$10$^{3}$-10$^{4}$ $M_{\odot}$ molecular clump and harbors a cluster of OB stars.
These observations covered the CH$_{3}$OH $J$=5-4 and $J$=7-6 and the CH$_{3}$CN $J$=19-18 line multiplets.
We refer to \citet{Liu2010ApJ}, \citet{Liu2011ApJ}, and \citet{Liu2012ApJ} for details of these observations.

We followed the standard SMA data calibration strategy.
The application of system temperature (T$_{\mbox{\tiny sys}}$) information and the absolute flux, passband, and gain calibrations were carried out using the MIR IDL software package \citep{Qi2003}. 
The absolute flux scalings were derived by comparing the visibility amplitudes of the gain calibrators with those of the absolute flux standard sources of the SMA, which were Callisto and Neptune for the subcompact and compact array observations, respectively.
After calibration, we performed zeroth-order fitting of continuum levels from line-free channels and the joint weighted imaging adopting robust weighting of all continuum data were performed using the \textsc{Miriad} software package \citep{Sault1995}. 
The resultant synthesised beam is typically 4$\as$5 at 241 GHz. The sensitivity (3$\sigma$) of continuum observation is $\sim$0.04 Jy beam$^{-1}$ and of lines $\sim$0.5 K. For clump G18, we do not obtain robust detection of the thermometer lines of CH$_{3}$CCH, H$_{2}$CS and CH$_{3}$CN (Table 3) with this achieved line sensitivity. We used the previously published result from IRAM 30m telescope observations of 3 mm CH$_{3}$CCH and CH$_{3}$CN lines (\citealt{Giannetti17}) instead.

\begin{figure*}
\begin{tabular}{p{0.245\linewidth}p{0.245\linewidth}p{0.245\linewidth}p{0.245\linewidth}}
\hspace{-.4cm}\includegraphics[scale=0.2]{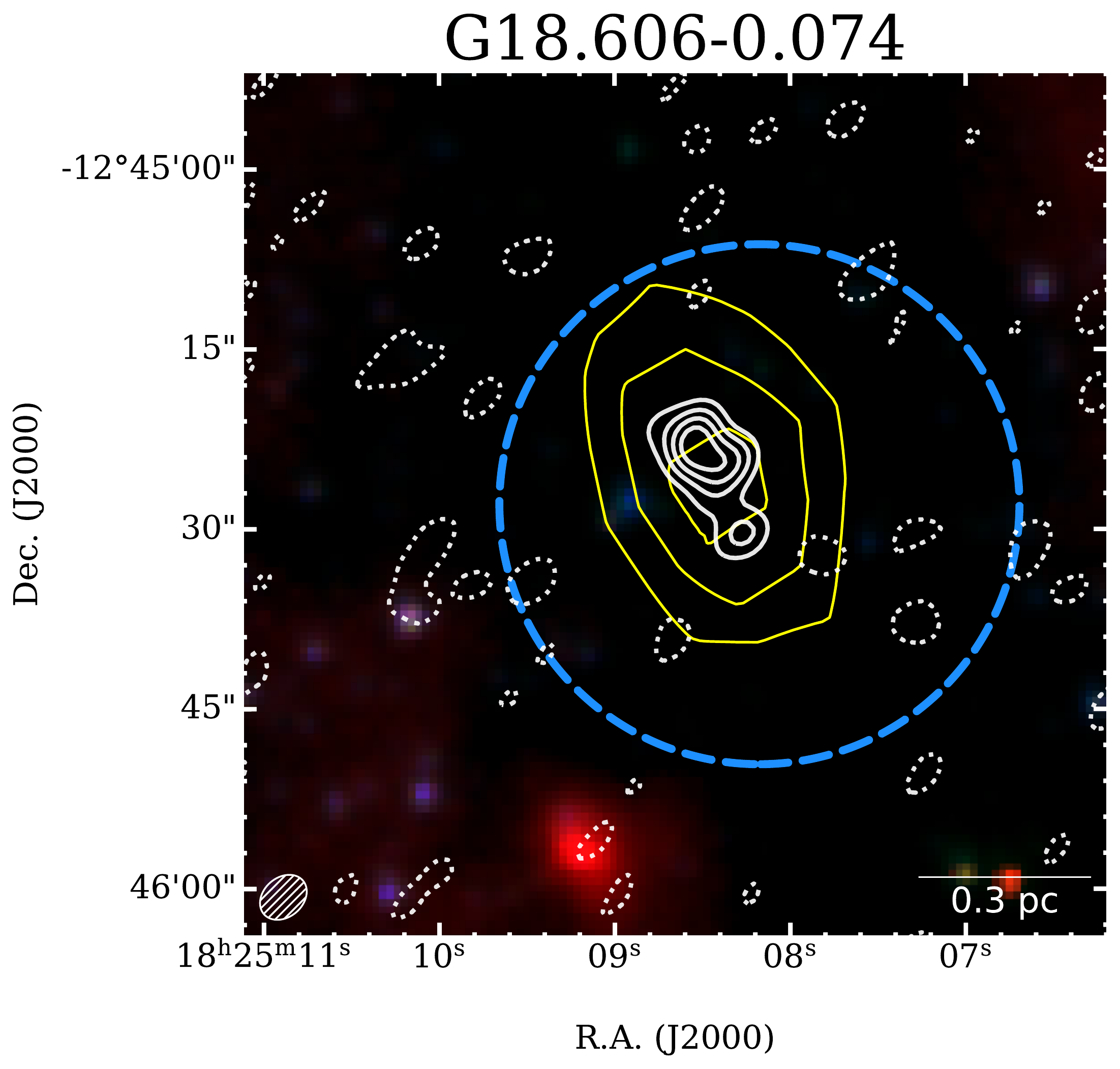}&\hspace{-0.35cm}\includegraphics[scale=0.2]{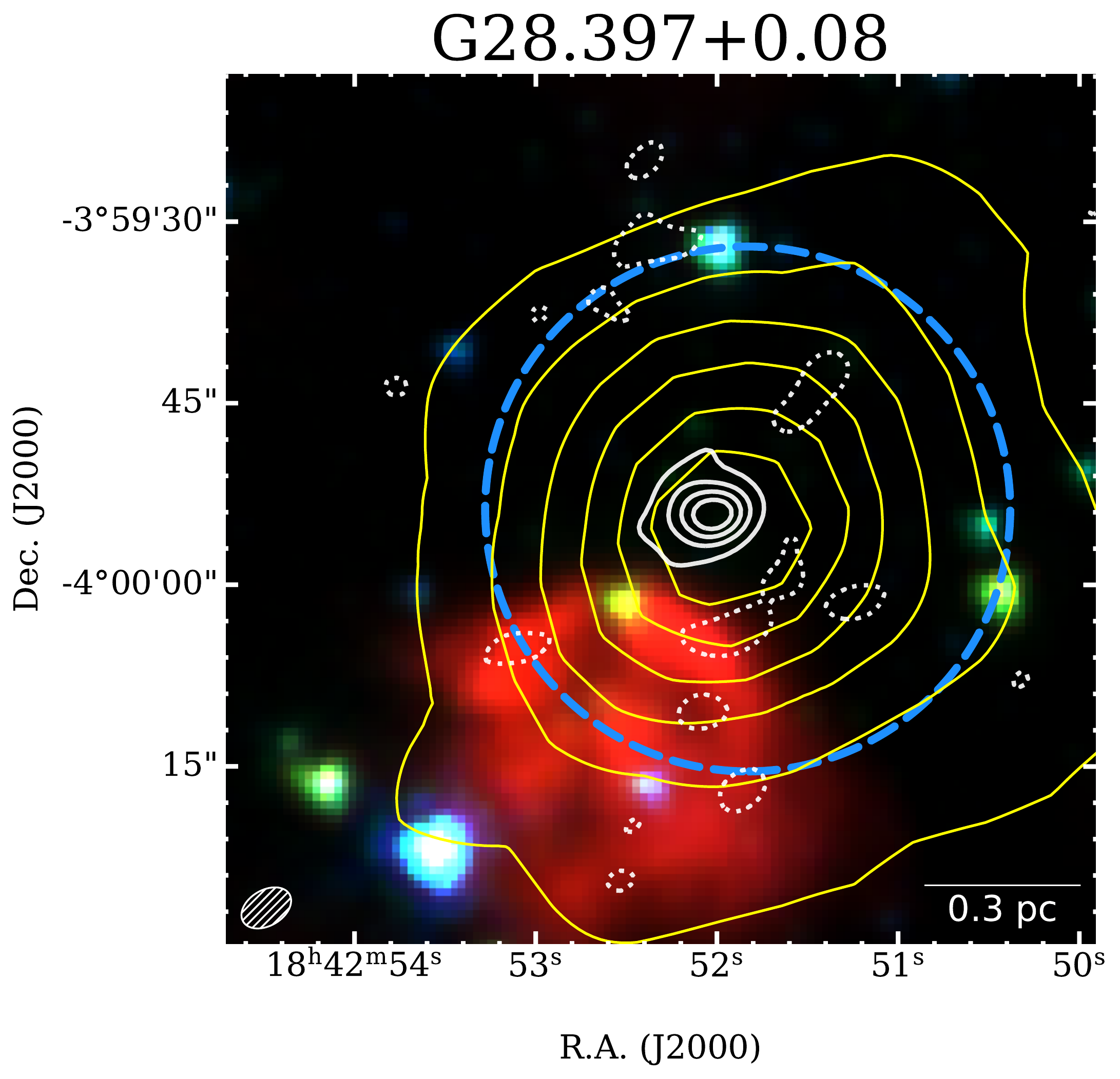}&\hspace{-0.3cm}\includegraphics[scale=0.2]{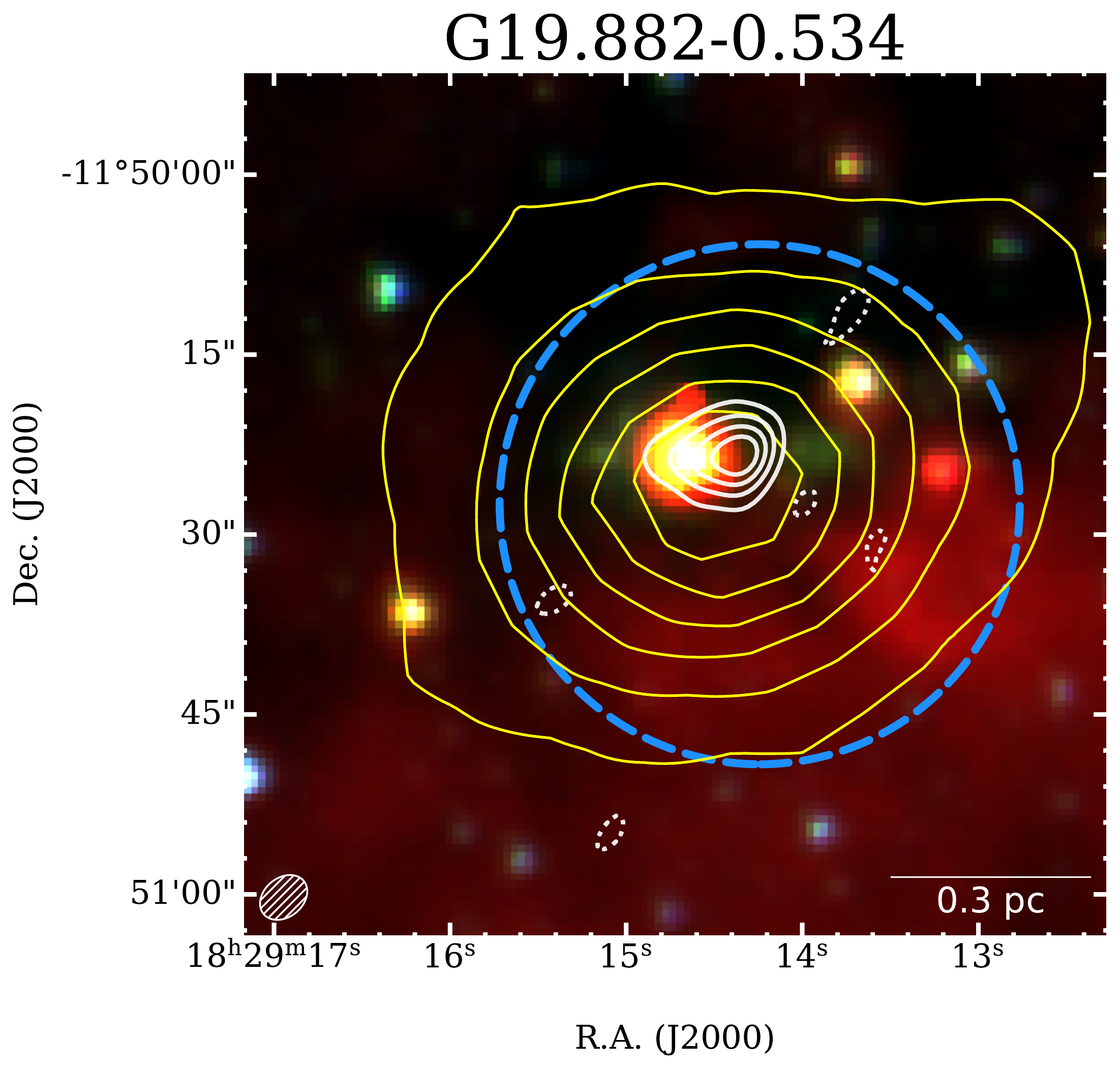}&\hspace{-0.3cm}\includegraphics[scale=0.2]{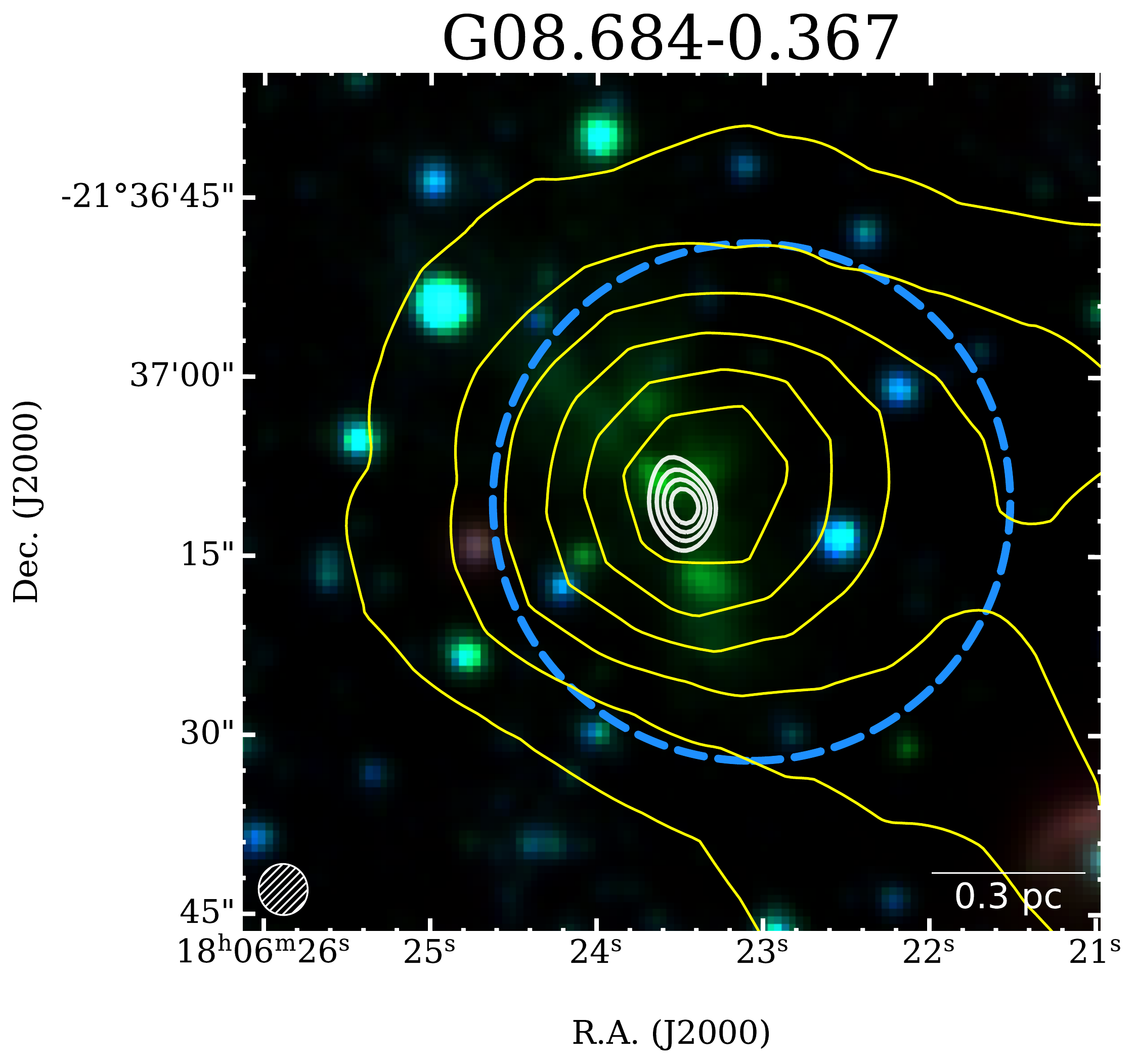}\\
\hspace{-.4cm}\includegraphics[scale=0.2]{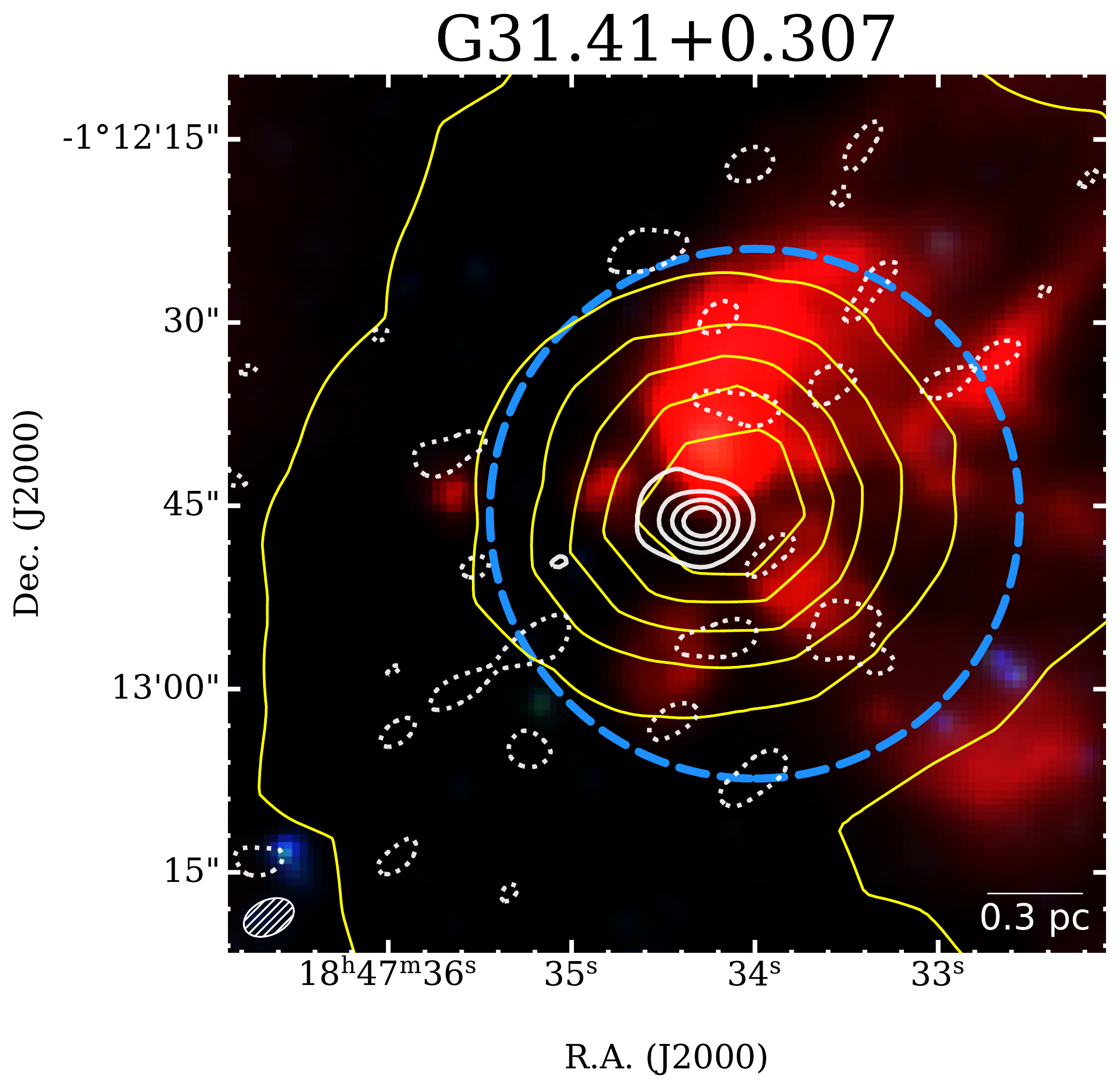}&
\hspace{-0.35cm}\includegraphics[scale=0.2]{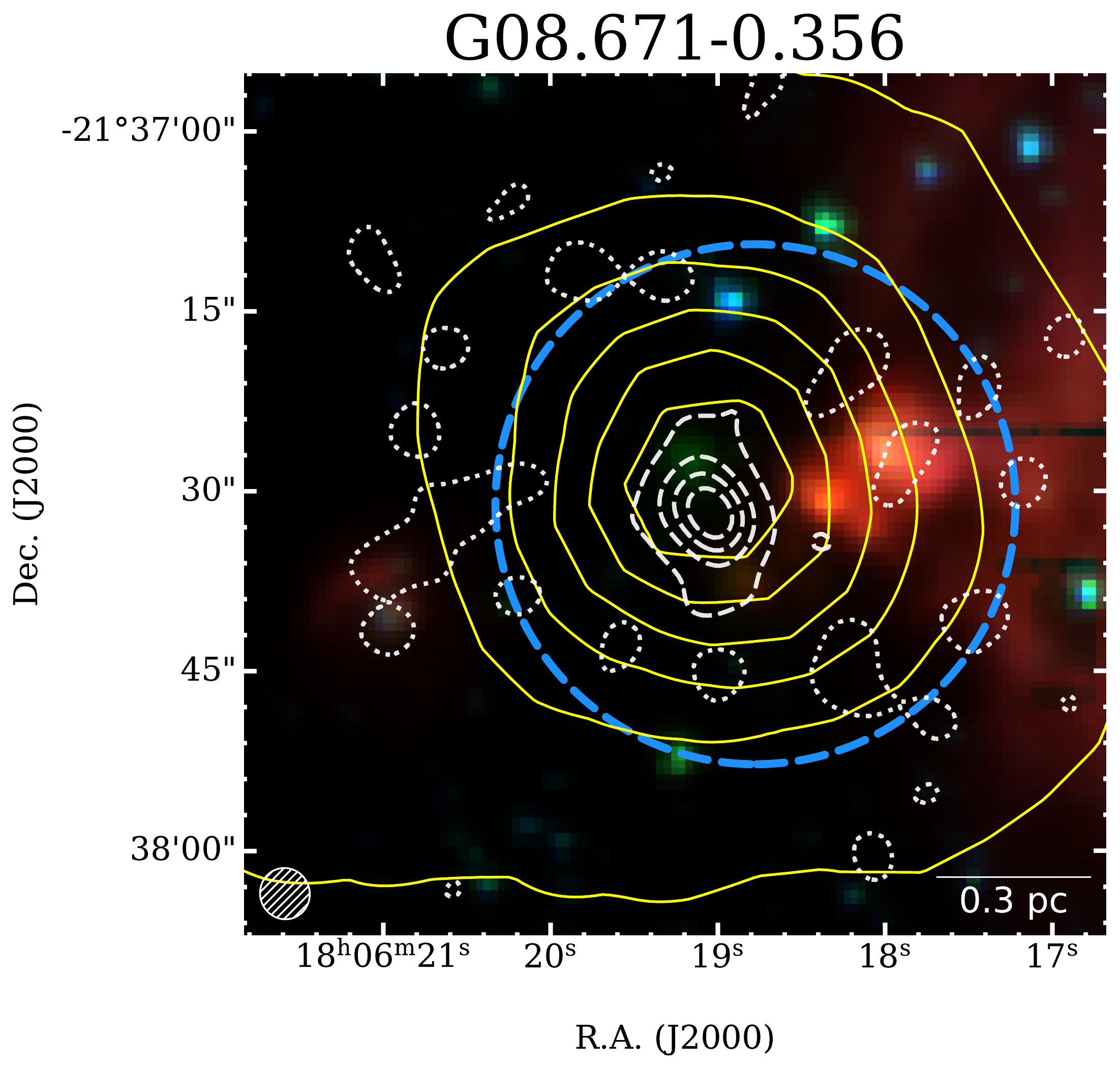}&\hspace{-0.3cm}\includegraphics[scale=0.2]{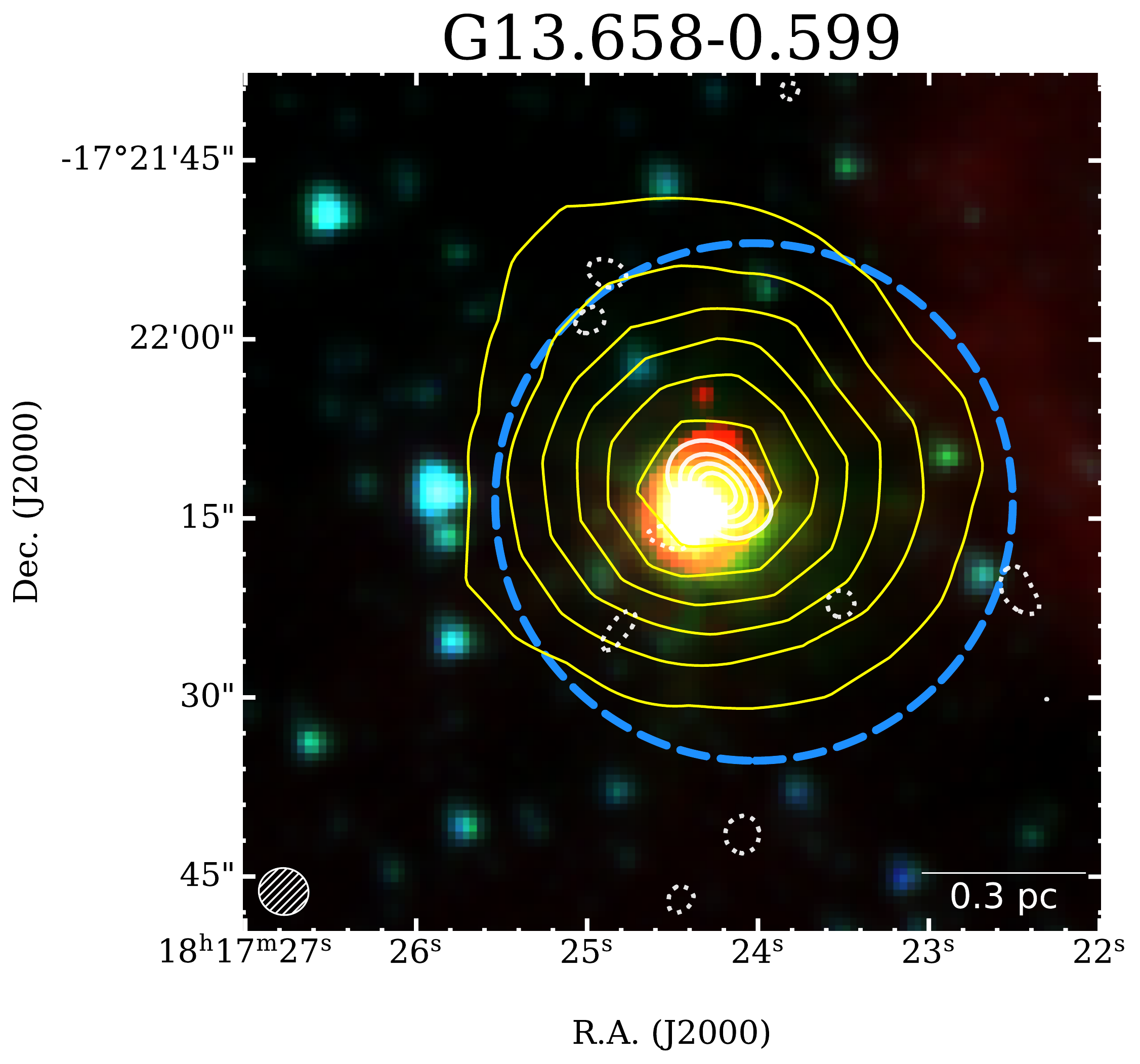}&\hspace{-0.3cm}\includegraphics[scale=0.2]{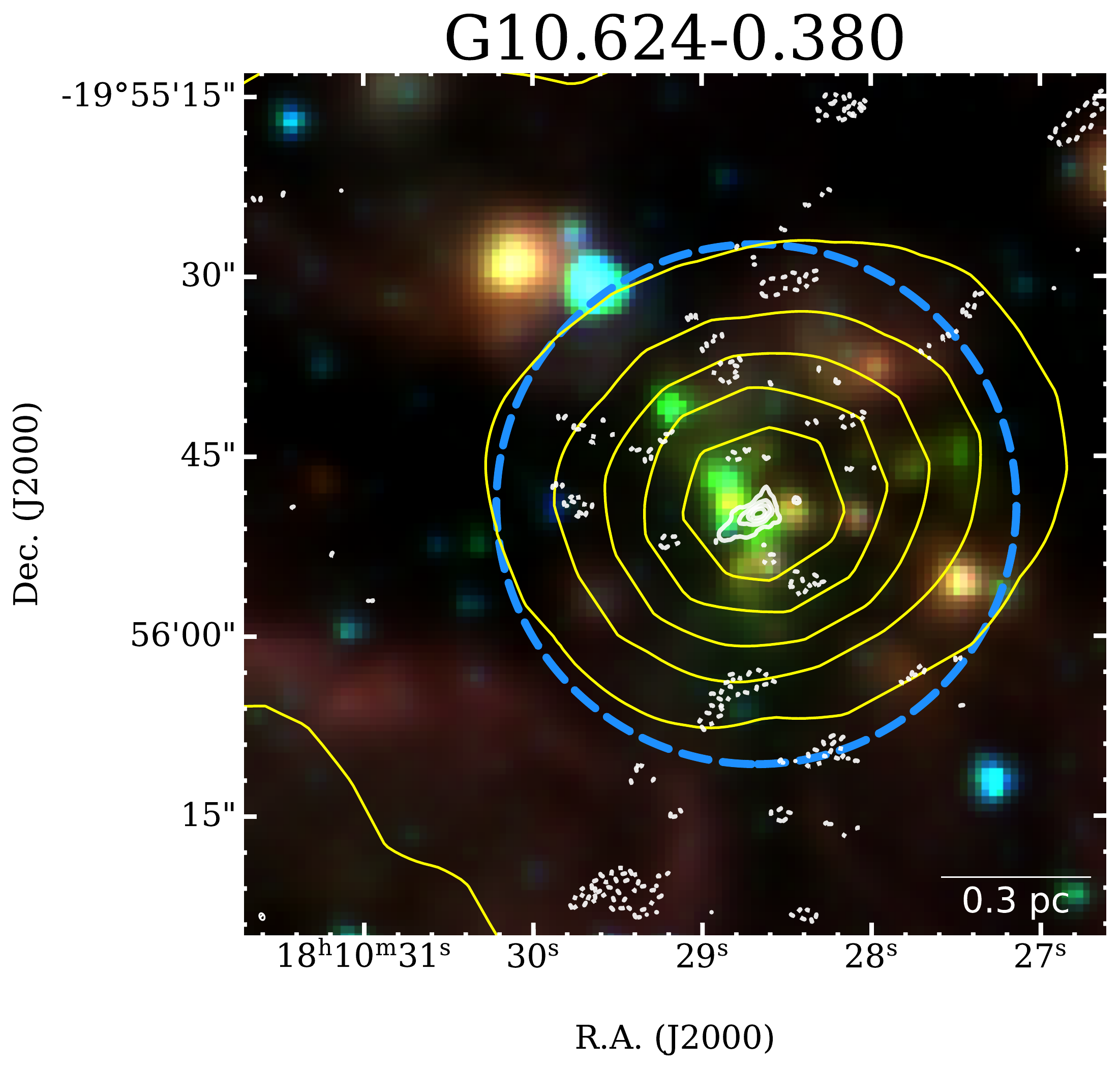}\\
\end{tabular}
\centering\caption{{\it{Spitzer}} IRAC RGBs (R: 8.0 $\mu$m; G: 4.5 $\mu$m B: 3.6 $\mu$m) maps of the target sources. Yellow contours show the ATLASGAL 870 $\mu$m emission from 1 Jy/beam to the peak flux for each source, using 7 levels with uniform spacing. White contours show SMA 1.2 mm emission from 3$\sigma$ to the peak flux 
using 5 levels with uniform spacings. 
Negative flux levels of the 1.2 mm continuum are shown in contours of dotted lines, from -$1\sigma$ to the minimum negative flux with 2 levels.
The beam of the SMA continuum is shown in the lower left corner of each plot. The beam size of the archival data for source G10 is much smaller than other sources (Section ~\ref{subsec:sma}). The primary beam size is shown in each plot as blue dashed circle.}
\label{fig:rgb}
\end{figure*}

\begin{table*}
\centering
\begin{threeparttable}
\caption{Target sources.}
\label{tab:sources}
\begin{tabular}{cccccccc}
\toprule
   Source$^{a}$  &  R.A. &  Decl.  &  Distance$^{b}$ & 
   Gas mass$^{c}$ & Luminosity$^{c}$ &  $L/M$ & Category \\
   & (J2000) & (J2000) & (kpc) & (10$^{2}$ $M_{\odot}$) & (10$^{3}$ $L_{\odot}$)&($L_{\odot}$/$M_{\odot}$)& \\
   \midrule
            G18.606--00.074 (G18)  & 18\rah25\ras08\ras.27  & -12\decd45\decm22\decs7  & 3.7  & 19 & 0.75 & 0.4 &  IR weak      \\
            G28.397+00.080 (G28) & 18\rah42\ras52\ras.08  & -03\decd59\decm53\decs7  & 4.8  & 37  & 12  & 3.2   &  IR bright    \\
            G19.882--00.534 (G19) & 18\rah29\ras14\ras.19  & -11\decd50\decm28\decs4  & 3.7  & 20 & 99   & 4.8   &  IR bright    \\
            G08.684--00.367 (G08a)  & 18\rah06\ras23\ras.35  & -21\decd37\decm05\decs2  & 4.8  & 27  & 27   & 10.2   &  IR weak      \\
            G31.412+00.307 (G31) & 18\rah47\ras34\ras.32  & -01\decd12\decm45\decs5  & 7.9  & 182  & 161   & 9.0   &  H\textsc{ii} \\
            G08.671--00.356  (G08b) & 18\rah06\ras19\ras.23  & -21\decd37\decm26\decs8  & 4.8  & 52  & 11   & 20.7   &  H\textsc{ii} \\
            G13.658--00.599 (G13) & 18\rah17\ras23\ras.46  & -17\decd22\decm09\decs2  & 4.5  & 12 & 24   & 19.9   &  IR bright    \\
            G10.624--00.380 (G10) & 18\rah10\ras28\ras.638  & -19\decd55\decm49\decs1 &4.95      &84   &  511    &60   & H\textsc{ii}     \\
            \bottomrule
            \end{tabular}
            \begin{tablenotes}
\small
\item a: The abbreviated source names are given in brackets.  
\item b: The references for the source distances are: \citet{GreenMcClure11} for G08a, G08b and G19; \citet{Sanna14} for G10; \citet{Wienen15} for G13; \citet{Churchwell90} for G31; \citet{Carey00} for G28; \citet{Sridharan05} for G18.
\item c: The gas mass and luminosity of the clump is taken from \citet{Urquhart18}, but scaled to the adopted distance of the clump.
\end{tablenotes}
  \end{threeparttable}
\end{table*}

\begin{table*}
\centering
\begin{threeparttable}
\caption{Source properties from 1.2 mm SMA continuum.}
\label{tab:sma_cont_direct}
\begin{tabular}{lcccccc}
\toprule
  Source  &  $S_{\mathrm{1.2mm}}$$^{a}$ &  $R_{\mathrm{eff}}$ $^{b}$ & $\bar{T_{\mathrm{d}}}$$^{c}$ & $M_{\mathrm{core}}$&$M^{\mathrm{Abel}}_{\mathrm{core}}$&$\bar{\rho}_{\mathrm{core}}$$^{d}$\\
  & (Jy) & (pc) & (K) & ($M_{\odot}$) & ($M_{\odot}$) &(10$^{4}$ cm$^{-3}$)\\
  \midrule
            G18$^{e}$  &0.13 &0.10 &36&24.0&22.1&6.0\\
                 &0.02&0.04&19&5.5&-&-\\
            G28  &0.69&0.12 &49&93.5&103.7&19.5\\
	        G19& 1.07 &0.10  &37&124.1&65.3&23.6\\
            G08a&0.35 &0.10 &50&49.1&18.6&8.0\\
            G31  & 5.12&0.17&94&943.8&914.0&61.0\\
            G08b  &1.91 &0.16&56&223.2&188.1&15.0\\
            G13 & 0.51 &0.11 &48&59.0&58.5&15.2\\
            G10&1.85 &0.04&132&102.0&67.8&210.0\\
\bottomrule
\end{tabular}
\begin{tablenotes}
\item a: Total flux above 5$\sigma$. For source G08b, G31 and G10, subtraction of free-free emission is considered (details in Appendix \ref{app:cont_cont}). 
\item b: Effective radius is defined as $\pi R_{\mathrm{eff}}^{2}=$ Area, where Area stands for the emission region above 5$\sigma$ for each 1.2 mm map, i.e., non-deconvolved averaged size. 
\item c: Dust temperature is assumed to be equal to gas temperature $T(r)$ obtained and refined in Sec \ref{sec:xclass}, and the average temperature $\bar{T_{\mathrm{d}}}$ is calculated by averaging over pixels that have continuum emission above 5$\sigma$. 
\item d: Average density calculated from $M^{\mathrm{Abel}}_{\mathrm{core}}$ and $R_{\mathrm{eff}}$.
\item e: for clump G18, the calculation for two cores are given. In the Abel inversion calculation which assumes spherical symmetry, only the central, more massive core is considered. So $M^{\mathrm{Abel}}_{\mathrm{core}}$ and $\bar{\rho}_{\mathrm{core}}$ are omitted for the secondary core.
 \end{tablenotes}
  \end{threeparttable}
\end{table*}

\begin{table*}
\centering
\begin{threeparttable}
\caption{Molecular lines of interest covered by the SMA observations. Information is taken from CDMS database (\citealt{Mueller01}).}
\label{tab:lines}
\begin{tabular}{lcclcc}
\toprule
  Transitions  &  Rest frequency &  $E_{\mbox{\scriptsize{up}}}$&Transitions  & Rest frequency &  $E_{\mbox{\scriptsize{up}}}$ \\
  & (GHz) & (K)&& (GHz) & (K) \\
  
  \midrule
    \multicolumn{3}{c}{(RxA lower sideband)} &                    \multicolumn{3}{c}{(RxB lower sideband)} \\ 
    C$^{34}$S 5-4   &   241.016 &    27.8 &                  CH$_{3}$OH $\varv_{t}$=1 5$_{-1,5}$-4$_{-1,4}$$E$&241.203&326.2 \\     
    CS        5-4      & 244.935  &  35.3    &               CH$_{3}$OH $\varv_{t}$=1 5$_{2,4}$-4$_{2,3}$$A$&241.192&333.4 \\
    \multicolumn{3}{c}{(RxA upper sideband)} &               CH$_{3}$OH $\varv_{t}$=1 5$_{2,3}$-4$_{2,2}$$A$&241.196&333.4 \\
    CH$_{3}$CCH 12$_{0}$-11$_{0}$ &   205.081 &   64.0 &           CH$_{3}$OH $\varv_{t}$=1 5$_{0,5}$-4$_{0,4}$$E$&241.206&333.5 \\
    CH$_{3}$CCH 12$_{1}$-11$_{1}$  & 205.076 &    71.1&           CH$_{3}$OH $\varv_{t}$=1 5$_{3,3}$-4$_{3,2}$$E$&241.180&357.4 \\
    CH$_{3}$CCH  12$_{2}$-11$_{2}$&   205.065 &    92.5&           CH$_{3}$OH $\varv_{t}$=1 5$_{-3,2}$-4$_{-3,1}$$E$&241.180&357.4 \\
    CH$_{3}$CCH 12$_{3}$-11$_{3}$ &   205.045 &    128.2&          CH$_{3}$CN 13$_{0}$-12$_{0}$ &   239.138    &80.3 \\
    CH$_{3}$CCH 12$_{4}$-11$_{4}$ &   205.018 &    178.2&          CH$_{3}$CN 13$_{1}$-12$_{1}$&   239.133    &87.5 \\
    H$_{2}$CS 6$_{0,6}$-5$_{0,5}$&205.99&35.6&               CH$_{3}$CN 13$_{2}$-12$_{2}$ &   239.119    &108.9 \\
    H$_{2}$CS 6$_{2,5}$-5$_{2,4}$&206.05&87.3&               CH$_{3}$CN 13$_{3}$-12$_{3}$ &   239.096    &144.6 \\
    H$_{2}$CS 6$_{2,4}$-5$_{2,3}$&206.16&87.3&               CH$_{3}$CN 13$_{4}$-12$_{4}$ &   239.064    &194.6 \\
    H$_{2}$CS 6$_{3,4}$-5$_{3,3}$&206.05&153.0&              CH$_{3}$CN 13$_{5}$-12$_{5}$ &   239.023    &258.9\\
    H$_{2}$CS 6$_{3,3}$-5$_{3,2}$&206.05&153.0&              CH$_{3}$CCH 14$_{0}$-13$_{0}$ &   239.252 &   82.1 \\
    H$_{2}$CS 6$_{4,3}$-5$_{4,2}$&206.00&244.9&              CH$_{3}$CCH 14$_{1}$-13$_{1}$ &   239.248 &    93.3\\
    H$_{2}$CS 6$_{4,2}$-5$_{4,1}$&206.00&244.8&              CH$_{3}$CCH 14$_{2}$-13$_{2}$ &   239.234 &    114.7\\
    SO 4$_{5}$-3$_{4}$&206.176 &38.6&                        CH$_{3}$CCH 14$_{3}$-13$_{3}$ &   239.211 &    150.3\\
   SO$_{2}$ 3$_{2,2}$-2$_{1,1}$&208.700&15.3&                CH$_{3}$CCH 14$_{4}$-13$_{4}$ &   239.179 &    200.3\\
   \multicolumn{3}{c}{(RxB lower sideband)} &                H$_{2}$CS 7$_{0,7}$-6$_{0,6}$  &240.267&46.1 \\
    CH$_{3}$OH 5$_{0,5}$-4$_{0,4}$$A$&   241.791& 34.8  &    H$_{2}$CS 7$_{2,6}$-6$_{2,5}$  &240.382&98.8 \\
    CH$_{3}$OH 5$_{1,5}$-4$_{1,4}$$E$&   241.767& 40.4  &    H$_{2}$CS 7$_{3,5}$-6$_{3,4}$  &240.392&164.6 \\
    CH$_{3}$OH 5$_{0,5}$-4$_{0,4}$$E$&   241.700& 47.9  &    H$_{2}$CS 7$_{3,4}$-6$_{3,3}$  &240.393&164.6 \\
    CH$_{3}$OH 5$_{-1,4}$-4$_{-1,3}$$E$&   241.879& 55.9 &     H$_{2}$CS 7$_{4,4}$-6$_{4,3}$  &240.332&256.5 \\                  
    CH$_{3}$OH 5$_{-2,4}$-4$_{-2,3}$$E$&   241.905&57.1  &    H$_{2}$CS 7$_{4,3}$-6$_{4,2}$  &240.332&256.5 \\                  
    CH$_{3}$OH 5$_{2,3}$-4$_{2,3}$$E$&   241.904& 60.7  &   SO$_{2}$ 14$_{0,14}$-13$_{1,13}$&244.254&93.9\\
    CH$_{3}$OH 5$_{2,4}$-4$_{2,3}$$A$&   241.842& 72.5  &   \multicolumn{3}{c}{(RxB upper sideband)} \\
    CH$_{3}$OH 5$_{2,3}$-4$_{2,2}$$A$&   241.887& 72.5  &    C$_{2}$H  $N$=3-2, J=7/2-5/2, F=3-2 &  262.01     &  25.1  \\
    CH$_{3}$OH 5$_{-3,3}$-4$_{-3,2}$$E$&   241.844& 82.5 &     C$_{2}$H $N$=3-2, J=7/2-5/2, F=4-3 &     262.00  &   25.1 \\      
    CH$_{3}$OH 5$_{3,3}$-4$_{3,2}$$A$&   241.833& 84.7  &    C$_{2}$H  $N$=3-2, J=5/2-3/2, F=3-2 &   262.06    &    25.1\\
    CH$_{3}$OH 5$_{3,2}$-4$_{3,1}$$A$&   241.833& 84.7  &    C$_{2}$H  $N$=3-2, J=5/2-3/2, F=2-2 &   262.08    &    25.2\\
    CH$_{3}$OH 5$_{3,3}$-4$_{3,2}$$E$&   241.852& 97.5 &     C$_{2}$H $N$=3-2, J=5/2-3/2, F=2-1 &    262.07   &   25.2 \\
    CH$_{3}$OH 5$_{3,3}$-4$_{3,2}$$A$&   241.807& 115.2  &   CH$_{3}$CCH 15$_{0}$-14$_{0}$ &   256.337 &   98.4 \\                   
    CH$_{3}$OH 5$_{4,1}$-4$_{4,0}$$A$&   241.807& 115.2 &    CH$_{3}$CCH 15$_{1}$-14$_{1}$ &   256.331 &    105.6\\
    CH$_{3}$OH 5$_{4,2}$-4$_{4,1}$$E$&   241.813& 122.7  &    CH$_{3}$CCH 15$_{2}$-14$_{2}$ &   256.317 &    127.0\\                  
    CH$_{3}$OH 5$_{-4,1}$-4$_{-4,0}$$E$&   241.830& 130.8  &   CH$_{3}$CCH 15$_{3}$-14$_{3}$ &   256.292 &    162.7\\                   
    &&&CH$_{3}$CCH 15$_{4}$-14$_{4}$ &   256.258 &    212.6\\                  
    &&&H$^{13}$CO$^+$ 3-2 &260.255&25.0\\                                
    &&&SO 6$_{6}$-5$_{5}$&258.256&56.5\\                                 
    &&&SO 7$_{6}$-6$_{5}$&261.843&47.6\\                                
                                          
\bottomrule
\end{tabular}
 \end{threeparttable}
\end{table*}


\subsection{APEX observations} \label{subsec:apex}
Single-dish observations at 1 mm towards our target sources were performed with the MPIfR principal investigator (PI) instrument PI230 on the the Atacama Pathfinder Experiment 12-meter submillimeter telescope (APEX, \citealt{Guesten06}), between March to September 2017 and July 2018 (Project M-099.F-9513A-2017, PI: Yuxin Lin). The PI230 receiver is a dual polarisation sideband separating heterodyne system with a total of 32 GHz bandwidth working at 230 GHz, and can cover the spectral range of 200-270 GHz. We conducted observations with two spectral setups, covering frequency ranges of 202.2-210.0 GHz, 218.0-225.8 GHz and 239.2-247.0 GHz, 255.0-262.8 GHz, respectively. For each target source a region of 3$'$$\times$3$'$ centered at the source position was mapped in the On-The-Fly (OTF) mode with both setups.

During the observations, the typical precipitable water vapor (PWV) was $\sim$1.5 to 2.5 mm. The pointing was determined by continuum observations on Saturn when available, or CO $J$=2-1 observations on bright nearby evolved stars (e.g. RAFGL2135, R-Dor). The pointing accuracy was found to be within 3$''$.  Focus was checked on Saturn every 2-4 hours. 
The main beam efficiency ($\eta_{\mbox{\scriptsize{mb}}}$) for the PI230 instrument varies over the observing period, with a range of $\sim$63$\%$$-$72$\%$\footnote{\url{http://www.apex-telescope.org/telescope/efficiency/}}. The calibration uncertainty is typically within 20$\%$, estimated based on the flux measurement of the pointing sources.

Basic data reductions were done with the GILDAS software package \footnote{\url{http://www.iram.fr/IRAMFR/GILDAS }}, including flagging of bad spectra, baseline subtraction, unit conversion ($T^{\ast}_{\mbox{\scriptsize{A}}}$ to $T_{\mbox{\scriptsize{ mb}}}$), and building spectral cubes. Final spectral cubes are re-sampled to 0.5 km/s spectral resolution for all lines.

\subsection{SMA-APEX combination and imaging} \label{subsec:combine}
For our primary target lines covered by both SMA and APEX, which have extended emission, namely CH$_{3}$CCH $J$=12-11, CH$_{3}$OH $J$=5-4, C$_{2}$H $J$=3-2, H$^{13}$CO$^+$ $J$=3-2, CS J=5-4, C$^{34}$S $J$=5-4 and SO $J$ = 4$_{5}$-3$_{4}$, we combined the two dataset in the Fourier domain ({\it{uv}}-domain) with \textsc{Miriad}. This combination is essentially imaging together the pseudo-visibilities generated from single-dish observations and interferometer measurements, so that the short-spacing information including zero baseline which is obtained with single dish can be complemented to interferometry data; the method is commonly referred to as joint deconvolution or joint reconstruction method (\citealt{Kurono09}, \citealt{Koda11}). 

In the combination procedure, we first deconvolved the APEX data from the single-dish Gaussian beam (FWHM$\sim$30$''$) and then multiplied the resultant image with the primary beam of the SMA observations. The obtained image is then used to generate a single-dish {\it{uv}} model, i.e. the pseudo-visibilities, by randomly sampling a visibility distribution to match that of the single-dish beam pattern. The zero spacing visibility is additionally sampled and added to the produced pseudo-visibilities. Finally, the pseudo- and interferometric visibilities are imaged together to produce the combined image. In the final imaging step, we apply a Gaussian tapering function of FWHM$\sim$2$''$ to increase the detectability of extended emission. In the end we adopt a final step to linearly combine the product of this standard joint deconvolution method with the single-dish image in the Fourier domain, using {\tt{immerge}} in the {\textsc{Miriad}} package. This step is necessary and found to be preserving the single-dish overall fluxes better than using solely the joint deconvolution method, due to the fact that the deconvolution method is not flux conserving; a similar procedure has been adopted in e.g. \citet{Monsch18}.
The combined images have comparable total fluxes to the APEX data, within a difference of 15$\%$.

For details on the proper weighting scheme in the joint deconvolution method and the impact of sensitivities of single-dish and interferometry data, we refer to \citet{Kurono09}. 

\subsection{Ancillary data: mid-/far-infrared and submm-continuum data from multiple single-dish telescopes}\label{sec:submm_data}
We used the single-dish mid- and far-infrared, and submm continuum data to constrain the bulk gas structures and construct the SEDs (Figure \ref{fig:flowchart}). 

Besides the 870$\,\mu$m data from ATLASGAL survey (\citealt{Schuller09}, \citealt{Csengeri16}) obtained by APEX-LABOCA (\citealt{Siringo09}),  we also adopted 350$\,\mu$m data obtained by CSO-SHARC2 or APEX-SABOCA instrument. The information of the observations and data reduction procedure are detailed in \citet{Lin17} and \citet{Lin19}. 

For sources without available 350$\,\mu$m from ground-based telescope (of 10$''$ angular resolution), we used the available observations from James Clerk Maxwell Telescope (JCMT)\footnote{The James Clerk Maxwell Telescope is operated by the East Asian Observatory on behalf of The National Astronomical Observatory of Japan, Academia Sinica Institute of Astronomy and Astrophysics, the Korea Astronomy and Space Science Institute, the National Astronomical Observatories of China and the Chinese Academy of Sciences (Grant No. XDB09000000), with additional funding support from the Science and Technology Facilities Council of the United Kingdom and participating universities in the United Kingdom and Canada. The James Clerk Maxwell Telescope has historically been operated by the Joint Astronomy Centre on behalf of the Science and Technology Facilities Council of the United Kingdom, the National Research Council of Canada and the Netherlands Organization for Scientific Research. Additional funds for the construction of SCUBA-2 were provided by the Canada Foundation for Innovation.} Submillimetre Common-User Bolometer Array 2 (SCUBA2) (\citealt{Dempsey13},\citealt{Holland13}) at 450 $\mu$m from the online data archive.

We also retrieved the archival Herschel\footnote{Herschel is an ESA space observatory with science instruments provided by European-led Principal Investigator consortia and with important participation from NASA.} images at 70/160$\,\mu$m and 250/350/500$\,\mu$m from the Herschel Infrared Galactic Plane (Hi-GAL) survey (\citealt{Molinari10}) taken by the PACS (\citealt{Poglitsch10}) and SPIRE instrument (\citealt{Griffin10}). For the mid-infrared data, we used the 24$\,\mu$m images from the MIPS Galactic Plane Survey (MIPSGAL, \citealt{Carey09}) taken by {\emph{Spitzer}} telescope.

\section{Results and analysis}\label{sec:ana}

\subsection{Outline of the modeling and analysis procedure}\label{sec:outline}

\begin{table}
\centering
\begin{threeparttable}
\caption{Critical density for transitions of interest.}
\label{tab:lines_more}
\begin{tabular}{lc}
\toprule
 Transitions  &  Critical density $^{a}$\\
  & (cm$^{-3}$) \\
  
  \midrule
    CS        5-4        &  1.1$\times$10$^{6}$  \\
    CH$_{3}$CCH 12$_{0}$-11$_{0}$ & 2.7$\times$10$^{4}$ $^{b}$  \\
    H$_{2}$CS 6$_{0,6}$-5$_{0,5}$ & 2.6$\times$10$^{5}$\\
    CH$_{3}$OH 5$_{1,5}$-4$_{1,4}$    & 6.8$\times$10$^{5}$\\
    CH$_{3}$OH  5$_{4,5}$-4$_{4,4}$   & 6.2$\times$10$^{7}$\\
    CH$_{3}$CN 13$_{0}$-12$_{0}$ &  3.5$\times$10$^{6}$ \\
    CH$_{3}$CCH 14$_{0}$-13$_{0}$ &4.3$\times$10$^{4}$ $^{b}$ \\
    H$_{2}$CS 7$_{0,7}$-6$_{0,6}$ & 4.3$\times$10$^{5}$ \\
    C$_{2}$H N = 3-2, J=7/2-5/2, F=4-3  &  4.1$\times$10$^{5}$ \\
    SO 4$_{5}$-3$_{4}$& 2.4$\times$10$^{5}$\\
    SO 6$_{6}$-5$_{5}$& 6.4$\times$10$^{5}$\\
    SO 7$_{6}$-6$_{5}$& 1.0$\times$10$^{6}$\\ 
    H$^{13}$CO$^+$ 3-2 &1.1$\times$10$^{6}$\\
    SO$_{2}$ 14$_{0,14}$-13$_{1,13}$ & 3.9$\times$10$^{5}$\\
    SO$_{2}$ 3$_{2,2}$-2$_{1,1}$ & 1.7$\times$10$^{5}$\\
    CH$_{3}$CCH 15$_{0}$-14$_{0}$& 5.3$\times$10$^{4}$ $^{b}$ \\
\bottomrule
\end{tabular}
    \begin{tablenotes}
      \small
      \item a: Calculated following definition in \citet{Shirley15} in the optically thin limit ($\tau$ $\sim$ 0.1) at 50 K, considering a multi-level energy system whenever necessary.
     \item b: Calculated using collisional coefficients of CH$_{3}$CN.
      \end{tablenotes}
  \end{threeparttable}
\end{table}

\begin{figure*}[htb]
\usetikzlibrary{
  arrows.meta,
  calc,
  fit,
  positioning,
  quotes
}
\tikzset{
  meta box/.style={
    draw,
    black,
    very thick,
    text centered
  },
  punkt/.style={
    meta box,
    rectangle,
    rounded corners,
    inner sep=5pt,
    minimum height=2em,
    minimum width=6em,
    align=center,
    text width=6em
  },
  round box/.style={
    meta box,
    circle
  },
  every fit/.style={
    draw,
    thick,
    dashed,
    gray,
    inner sep=6pt
  }
}
\tikzstyle{format}=[rectangle,draw,thin,fill=white]  
\tikzstyle{test}=[diamond,aspect=2,draw,thin]  
\tikzstyle{point}=[coordinate,on grid,]  
\tikzstyle{startstop} = [rectangle, rounded corners, minimum width = 2cm, minimum height=1cm,text centered, draw = black]
\tikzstyle{io} = [trapezium, trapezium left angle=70, trapezium right angle=110, minimum width=1.5cm, minimum height=1cm, text centered, draw=black]
\tikzstyle{process} = [rectangle, minimum width=3cm, minimum height=1cm, text centered, draw=black]
\tikzstyle{decision} = [diamond, aspect = 3, text centered, draw=black]
\tikzstyle{arrow} = [->,>=stealth]
\begin{tikzpicture}[node distance=10pt]
\node[startstop,align=center](start){All input: SD continuum, \\
CH$_{3}$CCH, CH$_{3}$OH, CH$_{3}$CN, H$_{2}$CS line cubes};
\node[draw, below=of start, align=center,xshift=-1.5cm]   (step 0){One-component \\dust SED modeling\\(pixel-by-pixel)};
\node[draw, below=of start, align=center,xshift=3.cm]                         (step 1)  {One-component LTE modeling of \\CH$_{3}$CCH, H$_{2}$CS and CH$_{3}$CN \\(pixel-by-pixel) (Section ~\ref{sec:xclass})};
  \node[io, aspect=0.5, below=of start, yshift=-2 cm,align=center]              (step 2) {T$\rm_{dust}$ map, \\multiple T$\rm_{rot}$ maps (Fig. \ref{fig:xclass_maps})};
  \node[draw, below=of step 2, align=center]                      (step 3) {Derive radial temperature profile $T(r)$ \\(Section ~\ref{sec:T_rho_profiles})};
  \node[draw, below=of step 3,align=center]                        (step 4)  {One-component non-LTE modeling of CH$_{3}$OH \\ with T$\rm_{kin}$ fixed to $T(r)$\\(pixel-by-pixel) (Sec.\ref{sec:radex_nh2}, Appendix \ref{app:radex_mcmc_interp})};
  \node[draw, right of=step 4, xshift = 6cm,align=center]        (frd step 1)  {RADMC modeling of SD continuum \\ with $T(r)$ fixed (Appendix \ref{app:radmc})};
    \coordinate (point2) at (3cm, -4cm);
  \node[io,aspect=1,below=of frd step 1,align=center]           (frd step 2)  {$\rho$(r) for bulk gas,\\ $\rho\rm_{bulk}$(r) (Table \ref{tab:radmc_para})};
  \node[draw, below=of frd step 2, align=center]   (frd step 3)  {LIME modeling of CH$_{3}$OH and CH$_{3}$CCH \\ 
  with $T(r)$ and $\rho$(r) fixed, $\rho$(r) = $\rho\rm_{bulk}$(r) (Appendix \ref{app:lime})};
  \node[io,aspect=1,below=of frd step 3, align=center]  (frd output 1) {Best-fit model cube \\and abundance profiles (Table \ref{tab:lime_radmcpara})};
 \node[io, aspect=0.2, below=of step 4]              (step 5) { $n(\mathrm{H_{\mathrm 2}})$ map (Fig. \ref{fig:nmaps})};
 \node[draw, below=of step 5,align=center]                   (step 6) {Derive n(r) for dense gas,\\ $\rho\rm_{dense}$(r)$\rm_{RADEX}$ \\ (Section ~\ref{sec:T_rho_profiles})};
 \node[draw, below=of step 6, align=center]                  (step 7) {LIME modeling of CH$_{3}$OH and CH$_{3}$CCH\\ with $T(r)$ fixed, n(r) manually adjusted \\according to $\rho\rm_{dense}$(r)$\rm_{RADEX}$ (Appendix \ref{app:lime})};
  \coordinate (point1) at (-4.5cm, -5.5cm);
\node[io,aspect=1,below=of step 7,align=center]  (frd output 2) {Best-fit model cube, \\$\rho\rm_{dense}$(r)$\rm_{LIME}$ and abundance profiles (Table \ref{tab:lime_radmcpara})};
 \node[draw,aspect=1,below=of frd output 1, yshift=-0.1cm, align=center] (frd compare) {Compare two set of\\ LIME best-fit models};
 \node[startstop,aspect=1,right=of frd compare,xshift=.6cm,align=center]        (output) {Output:\\ $\rho\rm_{dense}$(r),$\rho\rm_{bulk}$(r),$T(r)$};
  \draw[->] (start) -- (step 1);
    \draw[->] (start) -- (step 0);
 \draw[->] (step 1) -- (step 2);
  \draw[->] (step 0) -- (step 2);

    \draw[->] (step 1) -- (step 2);
    \draw[->] (step 2) -- (step 3);
    \draw[->] (step 3) -| (frd step 1);

    \draw[->] (frd step 1) -- (frd step 2);
    \draw[->] (frd step 2) -- (frd step 3);
        \draw[->] (frd step 2) -| (output);
\draw[->] (frd step 3) -- (frd output 1);
 \draw[->] (step 3) -- (step 4);
\draw[->] (step 4) -- (step 5);
\draw[->] (step 5) -- (step 6);
\draw[->] (step 6) -- (step 7);
\draw[->] (step 7) -- (frd output 2);
\draw[->] (frd output 2) -| (frd compare);
\draw[->] (frd output 1) -- (frd compare);
\draw[->] (frd compare) -- (output);
\draw [dashed] (step 7) -| node [gray,below] {Benchmark $T(r)$} (point1);
\draw [dashed,->] (point1) |- (step 3);
  \node[
    fit=(frd step 1) (frd step 2),
    label={[gray,anchor=south]north east:Benchmark and refinement of $T(r)$}] {};
\end{tikzpicture}
  \caption{Overall work flow showing the radiative transfer modeling procedure followed in this work.}
  \label{fig:flowchart}
\end{figure*}
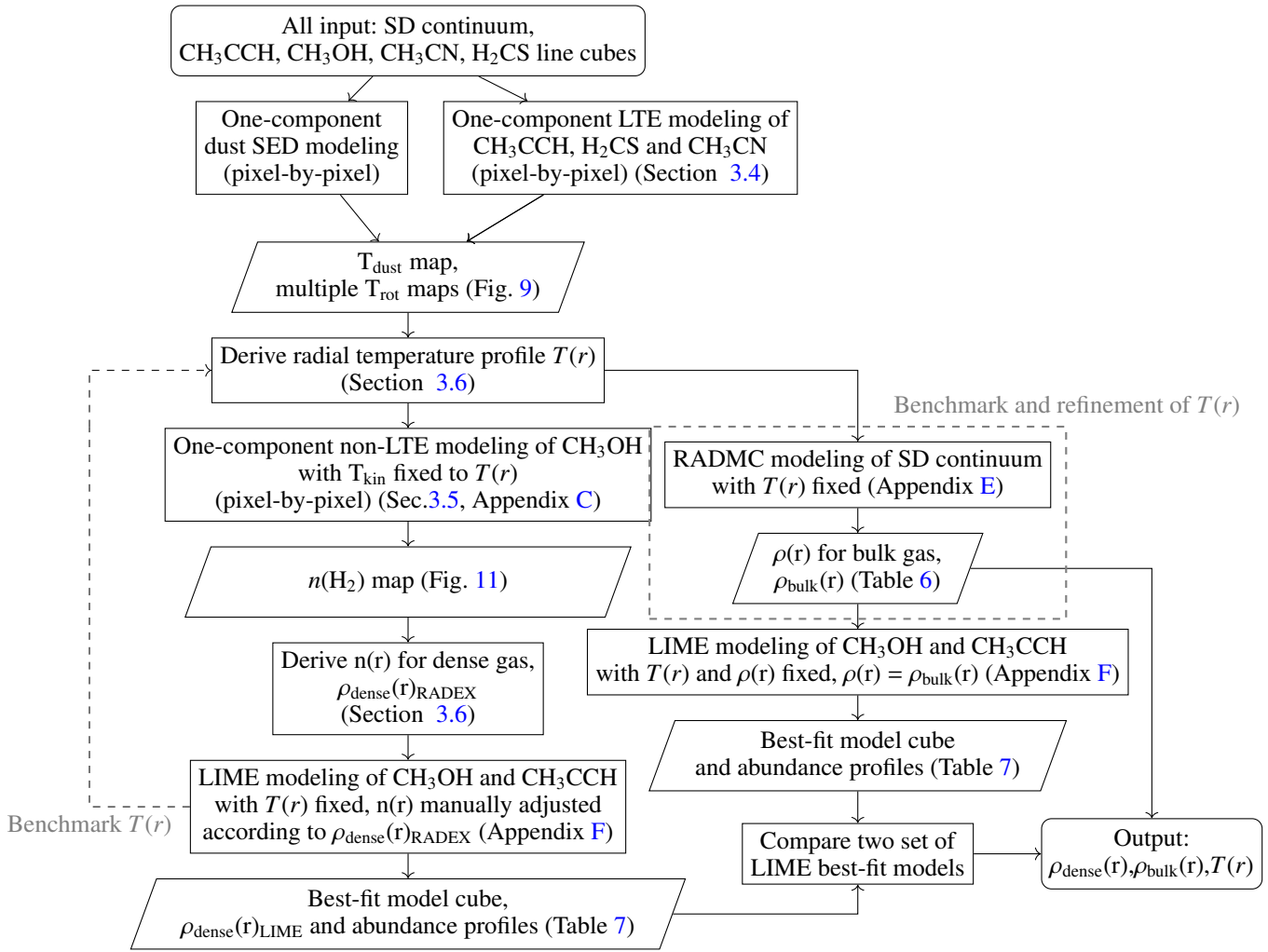

In this work, we aim to provide a description of the gas density and temperature of massive clumps by performing radiative transfer calculations of molecular lines and multi-wavelength dust continuum. In this section we introduce the workflow of the modeling, starting by basic definitions of molecular line excitation. The modeling steps (shown in Figure \ref{fig:flowchart}) are explained in more detail in Section ~\ref{sec:xclass}-~\ref{sec:T_rho_profiles} as well as in Appendix \ref{app:radmc} to \ref{app:lime}. The results of the radiative transfer models are discussed in Section ~\ref{sec:discussion}. 

Massive clumps have average molecular hydrogen densities of typically $\sim$10$^{4}$ cm$^{-3}$ (\citealt{Csengeri14}, \citealt{Urquhart18}); the collisional partner participating in the de- and excitation of molecular lines considered in this paper is primarily hydrogen gas. 
The critical density ($n_{\mathrm{crit}}$) (Table \ref{tab:lines_more}) defines the way in which a molecule in an excited state decays to ground state. When the main collisional partner is hydrogen, it stands for the critical hydrogen density at which timescales of radiative decay and collisional de-excitation are comparable, i.e. the net radiative decay rate from level $J$ to a certain lower level equals the rate of collisional de-population out of the upper level $J$, for a multilevel system (e.g.,\citealt{Wilson}).

With gas densites close to and well above $n_{\mathrm{crit}}$, the thermalisation of energy levels is achieved, such that the excitation temperature ($T_{\mathrm{ex}}$) can approximate the gas kinetic temperature ($T_{\mathrm{kin}}$), with the population of energy levels reaching Boltzmann prediction (local thermodynamic equilibrium, LTE).  
On the other hand, if gas densities are below $n_{\mathrm{crit}}$ (sub-thermal excitation), then the population of the upper energy level is sensitive to varying gas densities. Observations of multiple transitions with different $n_{\mathrm{crit}}$ can probe a range of gas densities, by showing rather different ratios of line intensities. In particular, if the energy levels associated with these transitions are of similar energy, then the dependence of line ratios on temperature is minimised, and so it does the degeneracy of the mutual effect of gas temperature and density in determining level populations. We take advantage of these radiative properties to use selected molecular transitions as `densitometers' of our target sources. 

In a simple view, massive star-forming clumps may be considered as multi-layered gas structures showing centrally peaked gas density profiles. This is a natural outcome under self-gravity. From the outermost layer to the innermost region, transitions of higher and higher $n_{\mathrm{crit}}$ are thermalised progressively. Using a combination of thermometers of different $n_{\mathrm{crit}}$, based on LTE assumption, can constrain gas temperatures over continuous spatial scales (with respect to the clump center). Analogous to a `densitometer', a `thermometer' is defined here as a set of molecular lines of a certain species whose level population is only (or dominantly) sensitive to gas temperature, which arise from energy levels spanning a wide energy range and are connected ideally only through collisions, provided e.g. by K-ladder lines of symmetric top molecules. 

Considering the gas density regime of massive star-forming clumps, and based on previous single-dish experiments (\citealt{Giannetti17}) we have identified CH$_{3}$CCH, H$_{2}$CS and CH$_{3}$CN lines at 1 mm band (listed in Table \ref{tab:lines}) as ideal tracers for measuring temperature profiles of massive clumps. These tracers have $n_{\mathrm{crit}}$ of several 10$^{4}$, 10$^{5}$ and 10$^{6}$ cm$^{-3}$, respectively (Table \ref{tab:lines_more}). On the other hand, the combination of distinct $n_{\mathrm{crit}}$ triggers a filtering effect, such that with each thermometer, the region it probes is limited to the gas density regimes ranging around and above its $n_{\mathrm {crit}}$. Contamination by fore- and background gas layers of lower density to the observed emission is therefore negligible. This means that the line-of-sight (averaging) effect is reduced to gas component of a limited density range.  

In addition, optical depths ($\tau$) are low when typical abundances and excitation conditions are considered for these molecules, which implies that line ratios probe the gas kinetic temperature at the inner location of the gas layer. With these properties in mind, we derive the rotational temperature ($T_{\mathrm{rot}}$) maps under the LTE assumption using multiple thermometers in Section ~\ref{sec:xclass}. Temperature measurement of the outer regions are obtained using the extended CH$_{3}$CCH and H$_{2}$CS emission, and combined with temperatures derived from CH$_{3}$CN which is confined to the central region of the clumps. This combination allows us to establish the full radial temperature profile of the clumps. 
We also use multi-wavelength single-dish dust emission (SD continuum, as in Figure \ref{fig:flowchart}) to derive dust temperature maps by building spectral energy distributions (SEDs) assuming single-component modified black-body emission (\citealt{Lin16, Lin19}). The dust temperatures at the outer layer of clumps, are used to complete the temperature profile at larger radii for the clumps.  
With the simple one-component LTE modeling and one-component dust SED construction, we derive the projected radial profile of the obtained multiple temperature maps as the radial profiles, denoted as $T(r)$. With this approximation, a natural difference caused by line-of-sight (LOS) projection effects may appear as a function of radius due to density-weighted emission. 
However, as previously mentioned, due to the density-filtering effect by combination of multiple thermometers, the difference between the two profiles is largely minimised. 
Moreover, the projected radial temperature profile used as radial temperature profile is further benchmarked and refined by SED construction from full radiative transfer modeling of dust based on a density profile adopted for the clump (Sect.~\ref{sec:T_rho_profiles}, Appendix \ref{app:radmc}), and further shown to be able to produce the observed CH$_{3}$CCH lines and their spatial variation by full line radiative transfer models (Appendix \ref{app:lime}).
  
To probe the gas density, we rely on CH$_{3}$OH line series in the 1 mm band as a densitometer. We adopt one-component non-LTE models to derive the hydrogen volume density ($n(\mathrm{H_{\mathrm 2}})$) maps and benchmark the results using full non-LTE radiative transfer modeling (Sect.~\ref{sec:T_rho_profiles}, Appendix \ref{app:lime}). The highest and lowest $n_{\mathrm{crit}}$ of the adopted line series are listed in Table \ref{tab:lines_more}. Moreover, with measured radial temperature profiles, the degeneracy of temperature and density can be further reduced by introducing $T_{\mathrm{kin}}$ ($T(r)$) in the non-LTE modeling, to constrain solely $n(\mathrm{H_{\mathrm 2}})$. We adopt this strategy in Section ~\ref{sec:radex_nh2} (see also Appendix \ref{app:radex_mcmc_interp}). 
In Section ~\ref{sec:radex_nh2} we use the one-component non-LTE model of CH$_{3}$OH lines to constrain $n(\mathrm{H_{\mathrm 2}})$. As mentioned before, the excitation of molecular lines has a selective effect on gas densities. Since we are also interested in the bulk gas density structure of massive clumps, we complement the density distribution measure with single-dish multi-wavelength dust emission.
We conduct full radiative transfer modeling of dust to fit with these data (Appendix \ref{app:radmc}), again incorporating the $T(r)$ initially measured from thermometers. To benchmark the one-component non-LTE models, as well as to understand the difference of gas density results between modeling of dust emission and simple non-LTE modeling of CH$_{3}$OH, we utilise non-LTE full radiative transfer calculation of lines in Appendix \ref{app:lime}, for CH$_{3}$OH and CH$_{3}$CCH lines. 
This also helps to examine the possibility of spatial abundance variations of these lines as an additional factor in affecting the distribution of line emission. In this effort, particularly, the full non-LTE modeling of CH$_{3}$CCH provides a sanity check on the measured radial temperature profile $T(r)$ from rotational temperature maps.  The workflow of the whole procedure is graphically summarized in Figure \ref{fig:flowchart}. 

 \begin{figure*}
 \begin{tabular}{p{0.95\linewidth}}
 \hspace{.5cm}\includegraphics[scale=.385]{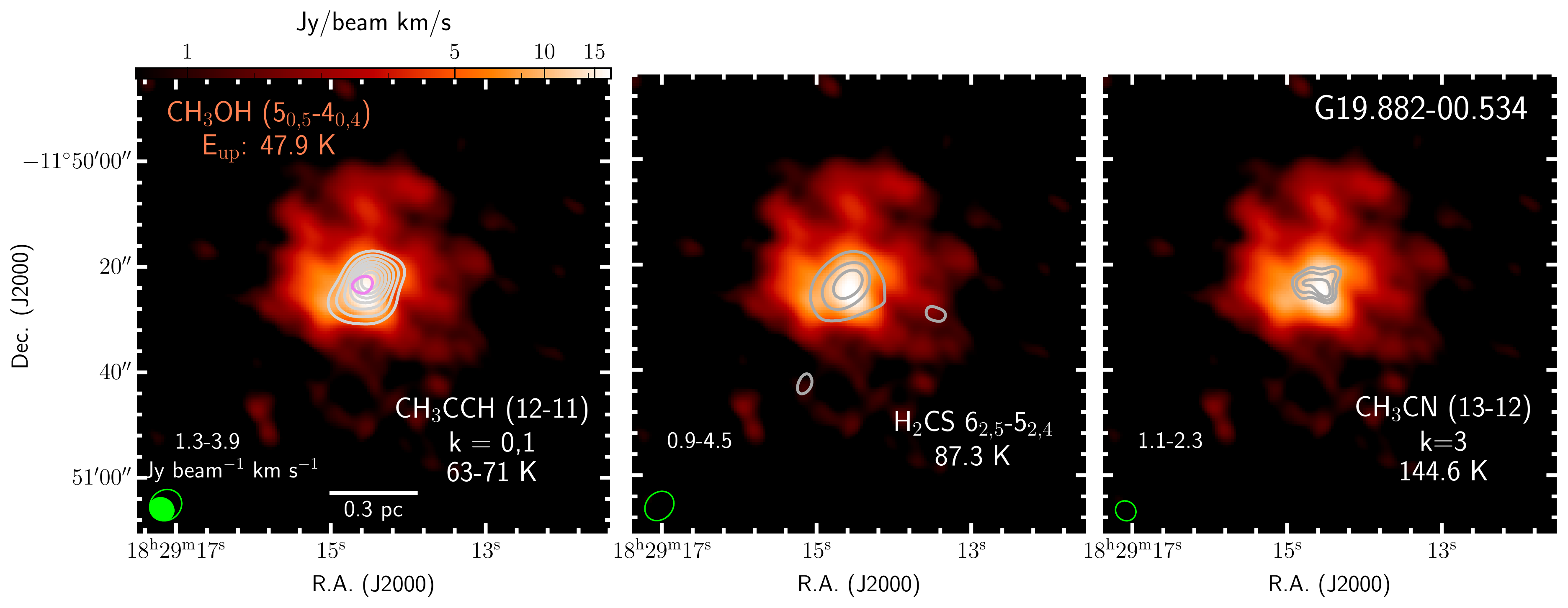}\\
 \hspace{.5cm}\includegraphics[scale=0.385]{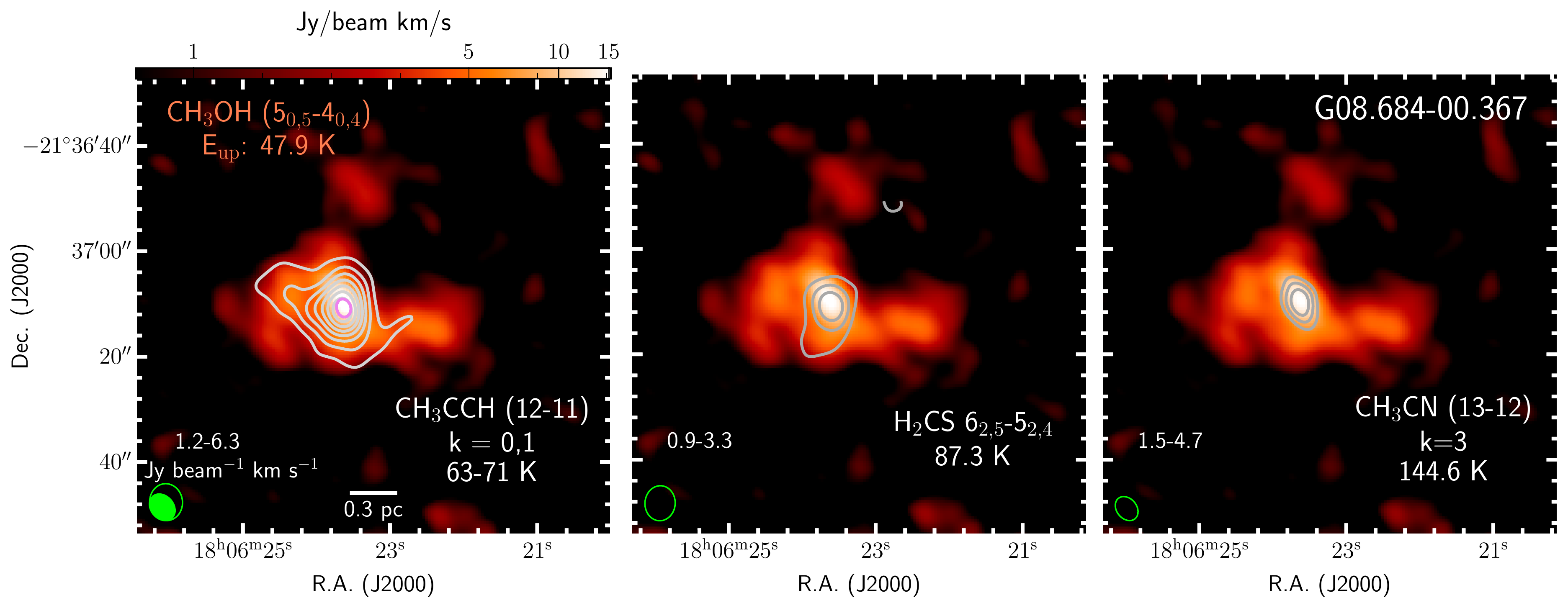}\\
\hspace{.5cm}\includegraphics[scale=0.385]{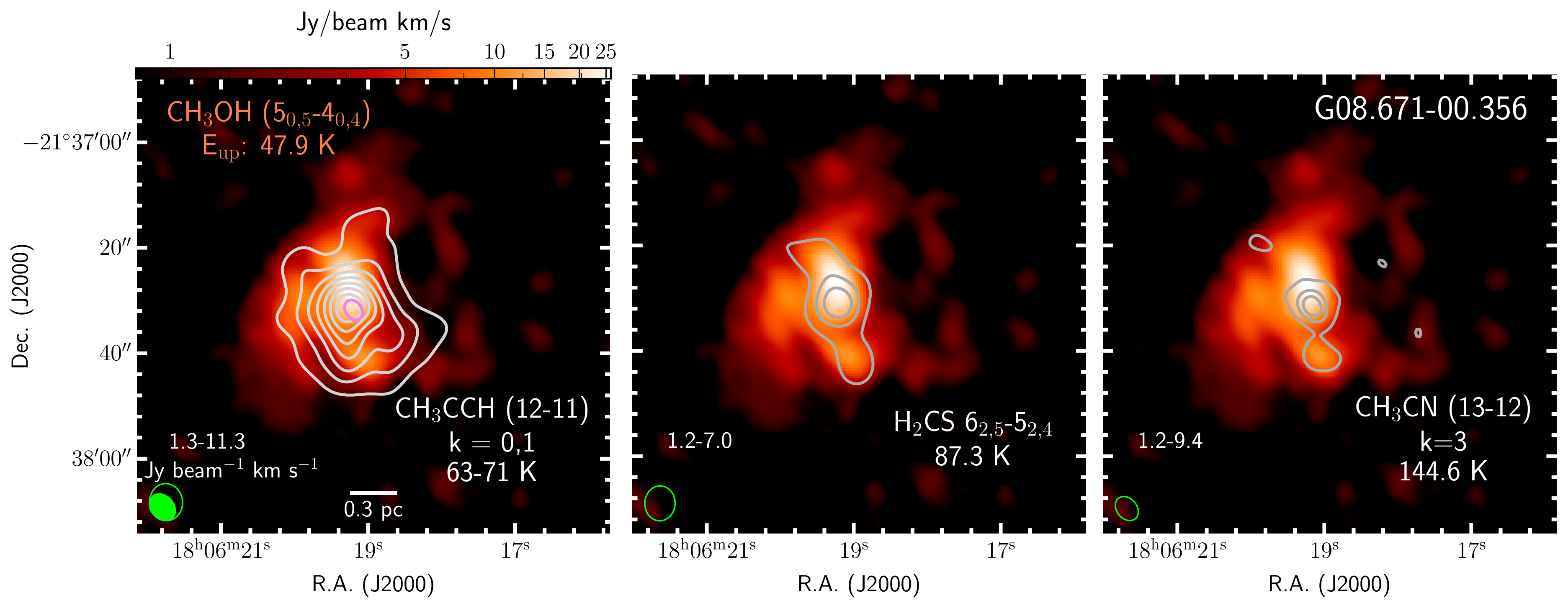}\\
\end{tabular}
\caption{Integrated intensity maps (gray contours) of CH$_{3}$CCH, H$_{2}$CS and CH$_{3}$CN toward sources G19, G08a and G08b. Integrated intensity of CH$_{3}$OH 5$_{0,5}$-4$_{0,4}$ ([$v_{\rm lsr}-3$, $v_{\rm lsr}+3$] km/s) is shown in color scale. Gray contours show the intensity levels with uniform intervals from 5$\sigma$ up to the peak flux, with the emission range (Jy beam$^{-1}$ km s$^{-1}$) indicated in the lower left corner of each panel. Magenta contour in the {\emph{left panel}} shows the location of 0.8$\times$ peak emission of the 1.2 mm SMA continuum image. The green ellipses indicate beams of corresponding molecular lines (void) and CH$_{3}$OH 5$_{0,5}$-4$_{0,4}$ line (filled).}
\label{fig:m0}
\end{figure*}

 \begin{figure*}
 \begin{tabular}{p{0.95\linewidth}}
 \hspace{.5cm}\includegraphics[scale=.385]{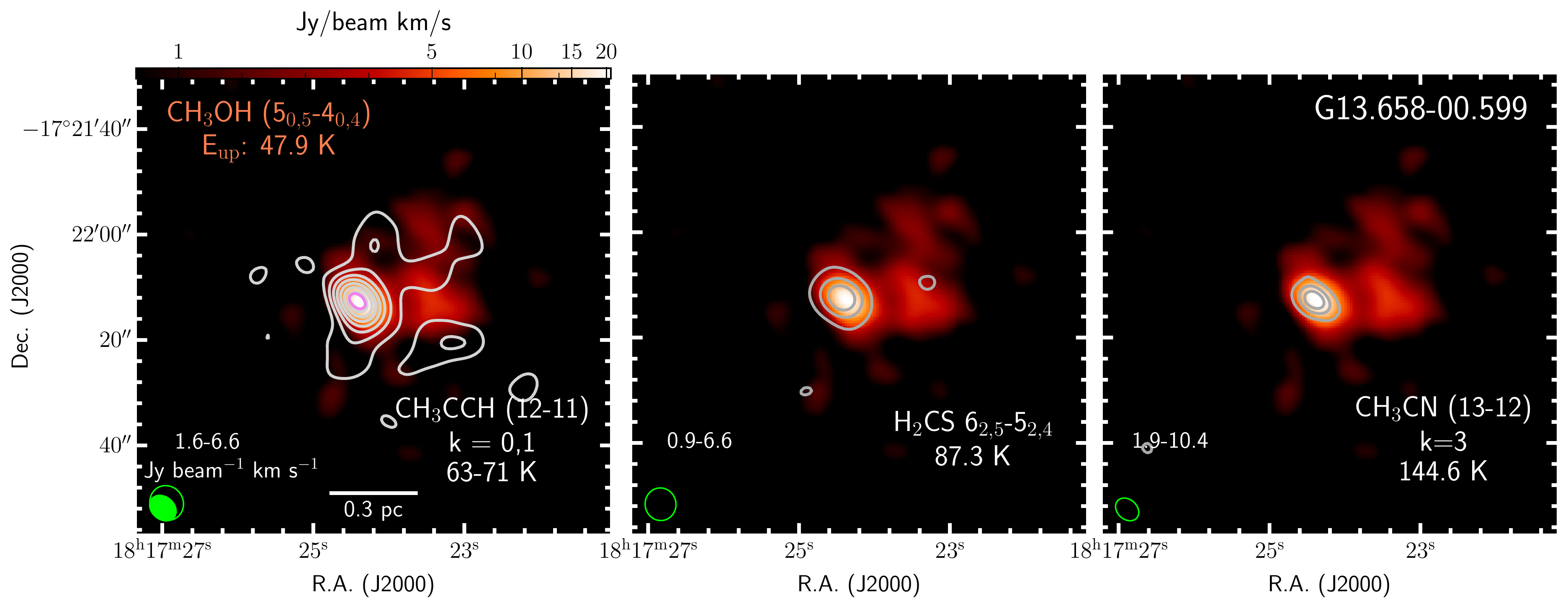}\\
 \hspace{.5cm}\includegraphics[scale=0.385]{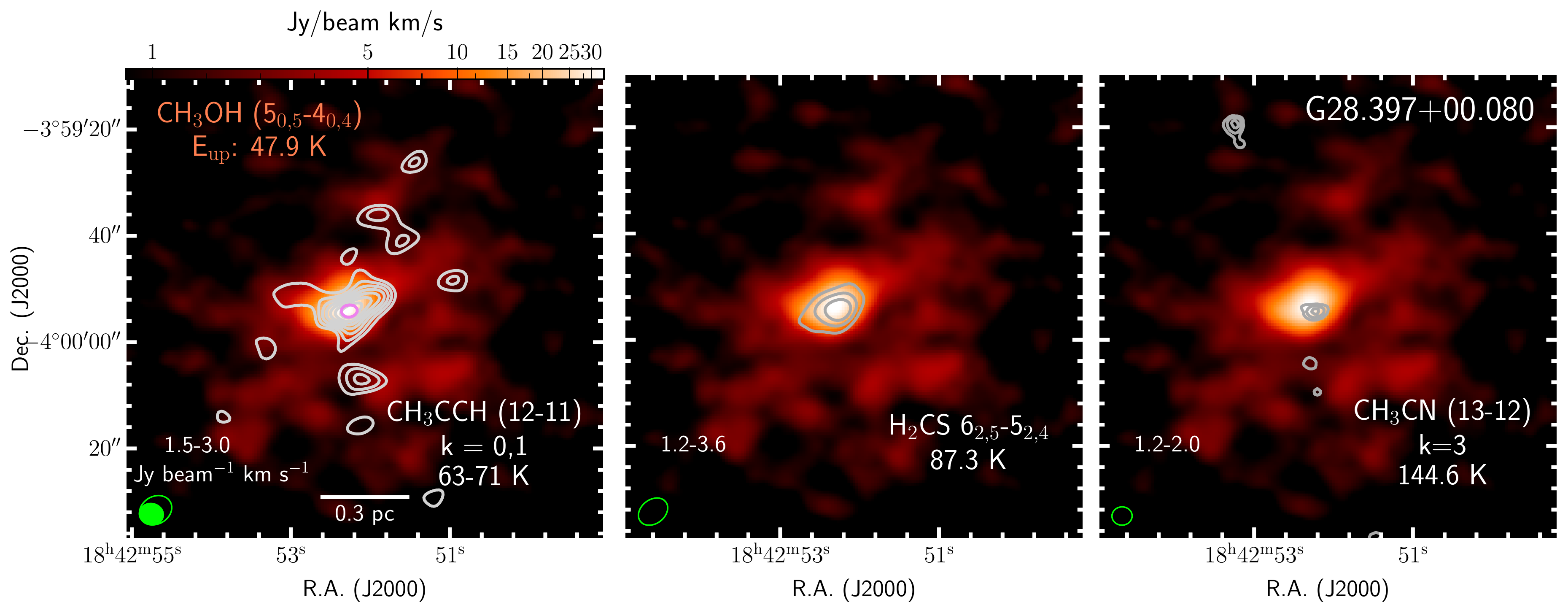}\\
\hspace{.5cm}\includegraphics[scale=0.385]{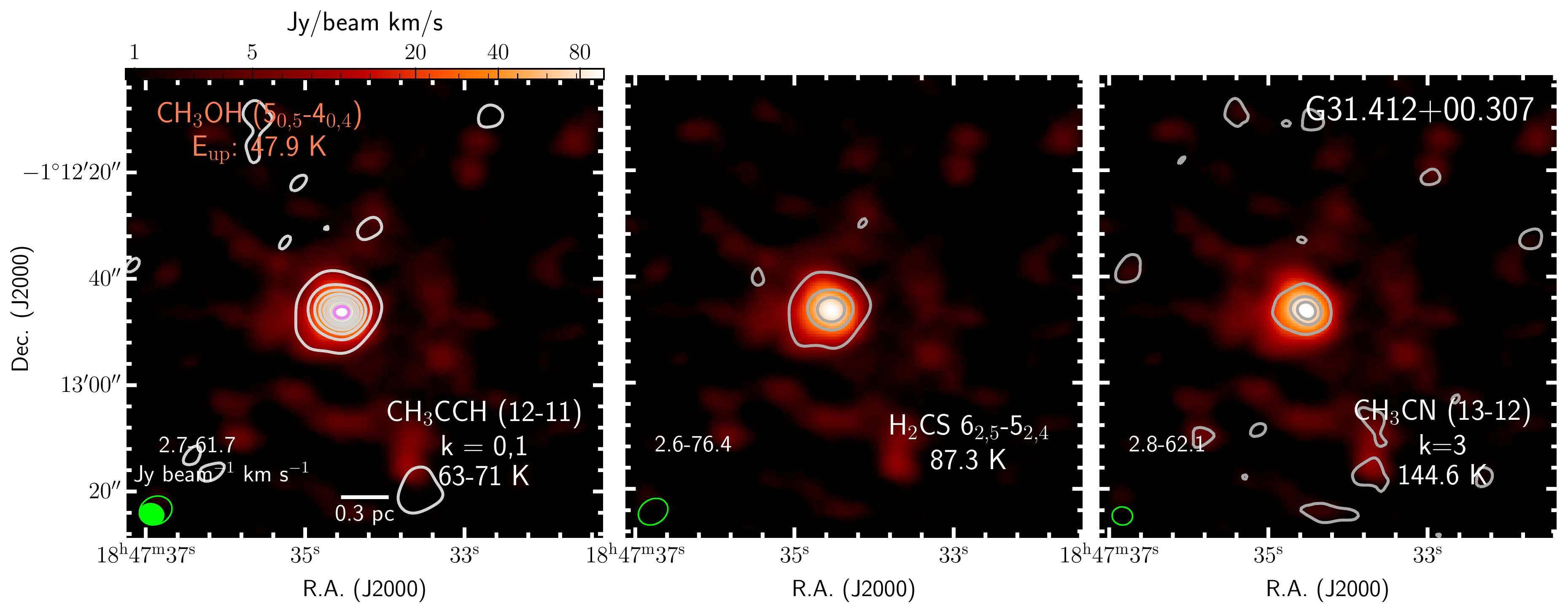}\\
\end{tabular}
\caption{Same as Fig. \ref{fig:m0}, but for target clump G13, G28 and G31.}
\label{fig:m0_more}
\end{figure*}

\subsection{SMA 1.2 mm continuum}\label{sec:sma_cont}
The SMA 1.2 mm dust continuum images resolved two compact sources (separated by $\sim$7$\as$2, $\sim$0.15 pc) in G18, and resolved isolated compact sources in the rest of the samples (Figure \ref{fig:rgb}). Hereafer we refer to these compact sources as core structures. 
Before any further analyses, we utilised the archival centimeter band data to subtract free-free contamination in G08b, G31, and G10, assuming optically thin emission (i.e., $S_{\nu}^{\mathrm{free-free}}$ $\propto$ $\nu^{-0.1}$; for details see Appendix \ref{app:cont_cont}). Then the core radius is defined as the area above 5$\sigma$ emission contours of the 1.2 mm images. The core effective radius, peak flux and integrated flux are listed in Table \ref{tab:sma_cont_direct}.

We assumed that dust emission in all cores is optically thin at 1.2 mm.
Based on the OH5 opacity model (i.e., $\kappa_{\mathrm{1.2 mm}}$ = 0.81 cm$^{2}$g$^{-1}$; \citealt{OH94}), we converted the continuum intensity detected at $>$5$\sigma$ to dust mass surface density, which was subsequently converted to gas mass surface density by assuming that the gas-to-dust mass ratio is 100.
In these mass estimates, we assumed that dust temperature is identical to the gas temperature $T(r)$ which we derived (and refined, Equation \ref{eq:tformula}) (c.f. Section \ref{sec:xclass}, \ref{sec:T_rho_profiles} and Appendix \ref{app:radmc}).

There is a subtlety in the way we applied $T(r)$, which is related to the assumption of the thermal and density structures of the cores.
We compared two ways of applying $T(r)$.
In the first, we defined a mean core gas temperature $\bar{T_{\mathrm{core}}}$ by making averages of $T(r)$ within the core size.
For each pixel, we then adopted a dust temperature which is equal to $min\{$$T(r=\ell)$, $\bar{T_{\mathrm{core}}}$$\}$ when deriving $n(\mathrm{H_{\mathrm 2}})$ where $\ell$ is the projected distance from the pixel to the 1.2 mm continuum peak (i.e., centers of the sources).
This means that we use the smaller value of the two temperatures of the average core temperature and the radial temperature at the each pixel position, to estimate the gas mass probed by dust emission. This is a reasonable assumption since dust emission is sampling all the gas component along LOS and the average mass temperature is likely dominated by the outer, colder gas component.
Given that the projected radius $\ell$ is always smaller than the radius $r$, this approach still tends to overestimate the dust temperatures at small projected radii although it is alleviated.
This in turn results in an underestimate of $n(\mathrm{H_{\mathrm 2}})$.
%

In a second approach, we assumed that the cores are spherically symmetric and optically thin.
We used Abel transformation to convert the observed azimuthally averaged intensity profile of 1.2 mm emission to gas density $\rho(r)$ (for more details see \citealt{Roy13}), as
\begin{equation}
    \rho(r) = -\frac{1}{\pi\kappa_{\nu}B_{\nu}[T(r)]}\int_{r}^{r_{\mathrm{eff}}} \frac{\mathrm{d} I_{\nu}}{\mathrm{d}b}\frac{\mathrm{d}b}{\sqrt{b^{2}-r^{2}}},
\end{equation}
where $r_{\mathrm{eff}}$ is the core effective radius.
We then integrated $\rho(r)$ over the line-of-sight to obtain another version of $n(\mathrm{H_{\mathrm 2}})$ map.
The two versions of $n(\mathrm{H_{\mathrm 2}})$ maps agree within a factor of 1.5-2. 
The average dust/gas temperatures within the core, two sets of mass estimates $M_{\mathrm{core}}$ and $M^{\mathrm{Abel}}_{\mathrm{core}}$, and average core density are summarized in Table \ref{tab:sma_cont_direct}.

\subsection{The distribution of the emission from CH$_{3}$CCH, H$_{2}$CS, CH$_{3}$CN, CH$_{3}$OH lines and 1.2 mm continuum}\label{sec:mom0}
Figure \ref{fig:rgb} shows the 1.2 mm dust continuum images taken with the SMA.  
We resolved two compact sources (separated by $\sim$7$\as$2, $\sim$0.15 pc) in G18, and resolved isolated compact sources in the rest of the samples.
Figure \ref{fig:m0} shows the integrated intensity maps of CH$_{3}$CCH, H$_{2}$CS, and CH$_{3}$CN which are overlaid on the integrated intensity maps of CH$_{3}$OH.
In general, the CH$_{3}$OH lines and the lower $K$ ladders of CH$_{3}$CCH were resolved on 0.3-0.4 pc scales while the CH$_{3}$CN lines and higher $K$ ladders of H$_{2}$CS were resolved on 0.1-0.2 pc scales.
The results of our quantitative analyses are presented in the following subsections.

\subsection{Deriving pixel-based gas rotational temperature maps with LTE modelling for multiple thermometers}\label{sec:xclass}

\subsubsection{Thermometers}
CH$_{3}$CCH and CH$_{3}$CN are symmetric top molecules.
Their $K$ ladder populations at a certain $J$ level are determined primarily through collisions. 
Therefore, they have been regarded as thermometers for molecular clouds (\citealt{Kuiper84}, \citealt{Bergin94}).
Given their similar geometry and molecular weight, CH$_{3}$CCH and CH$_{3}$CN are often assumed to have the same collisional coefficients,
while CH$_{3}$CN has higher dipole moments than CH$_{3}$CCH.
Due to this, a molecular clump can exhibit brighter CH$_{3}$CN line emission than CH$_{3}$CCH even in the case that the excitation of the CH$_{3}$CN molecules is limited to small pockets of dense gas, e.g. confined to the hot core region.

The CH$_{3}$CN lines have been very commonly observed (\citealt{Cummins83}, \citealt{Sutton86}, \citealt{Fayolle15}).
They have been regarded as good tracers of hot molecular cores owing to the fact that they were mainly detected around significantly heated regions.
On the other hand, CH$_{3}$CCH has been detected in spatially more extended, lower temperature regions (e.\ g.\ \citealt{Bergin94}, \citealt{Oberg14}) and is therefore particularly advantageous for probing sources in relatively early evolutionary stages (\citealt{Molinari16}), prior to hot core formation.  
\citet{Giannetti17} showed that among various thermometers, the kinetic temperature constrained by CH$_{3}$CCH is representative of bulk gas temperature of massive clumps.

The H$_{2}$CS molecule is a near-prolate rotor.
Its transitions between levels at various $K$ ladders are also excellent indicators of the gas kinetic temperature (\citealt{Blake94}).
As a sulfur-bearing species, the gas phase H$_{2}$CS abundance can be enhanced either by direct evaporation or by outflow/shock sputtering (e.g. \citealt{Bachiller97}, \citealt{Minh11}).
Previous observations have also revealed that the H$_{2}$CS emission originates from extended warm regions surrounding compact hot cores (e.g. \citealt{HelmichVDT97}). 

\begin{figure*}
\begin{tabular}{p{0.88\linewidth}}
\includegraphics[scale=0.33]{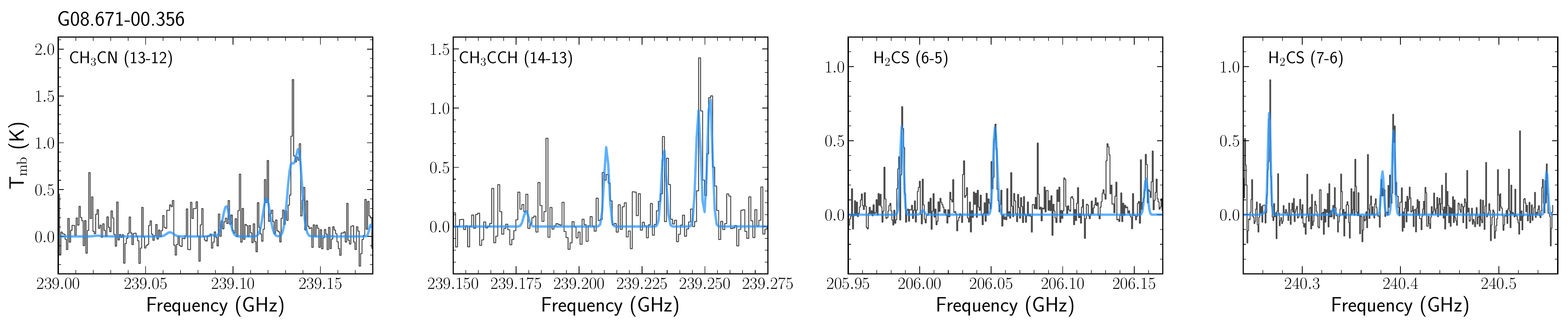}\\
\includegraphics[scale=0.33]{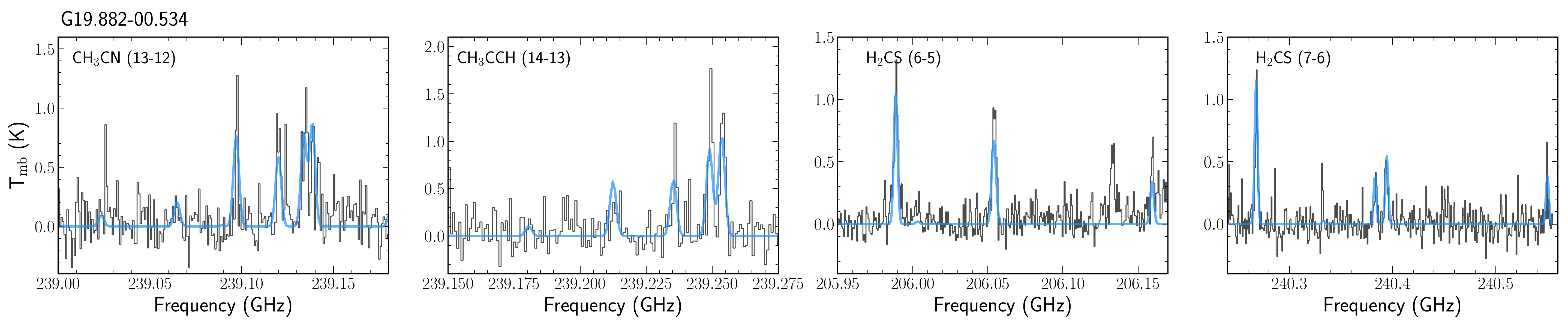}\\
\includegraphics[scale=0.33]{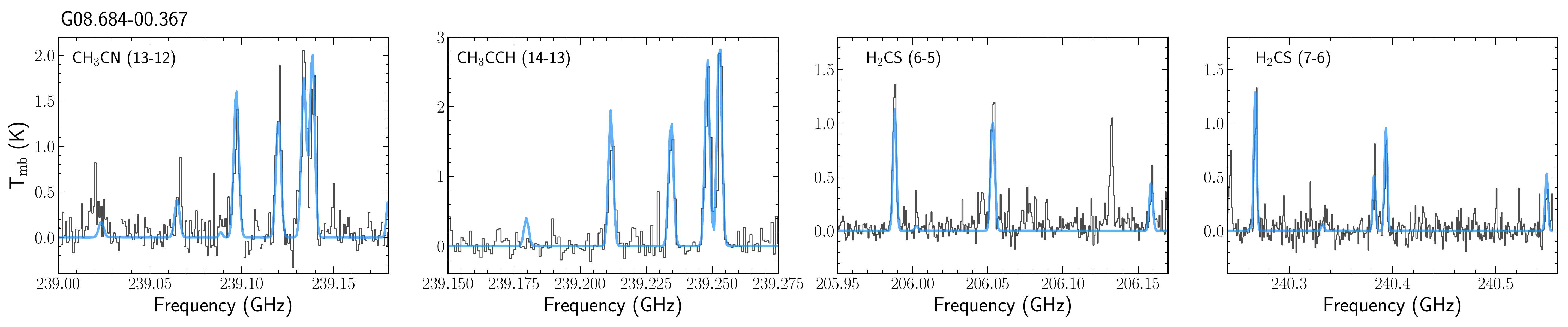}\\
\includegraphics[scale=0.33]{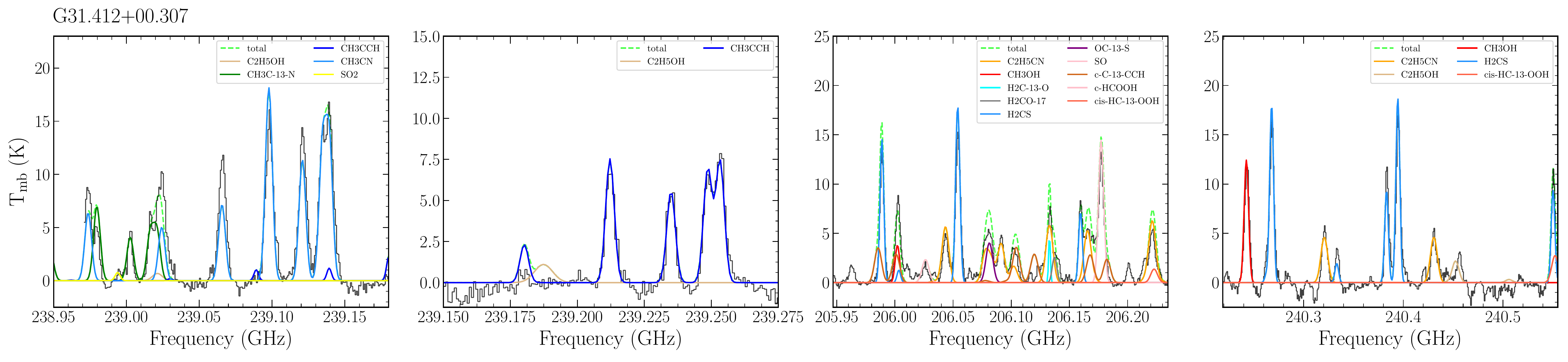}\\

\end{tabular}
\caption{Example spectra of thermometer lines CH$_{3}$CN $J=$13-12, CH$_{3}$CCH $J$=14-13, H$_{2}$CS $J$=6-5, H$_{2}$CS $J$=7-6 at the continuum peak of the target source; blue profiles show the XCLASS LTE fitting results. For source G31.412$+$0.307 which presents significant line blending from other species, the fittings also included those species/transitions that can potentially make prominent contributions to the spectrum.}
\label{fig:xclass_spectra}
\end{figure*}

\begin{figure*}[htb]
\begin{tabular}{p{0.88\linewidth}}
\hspace{-0.4cm}\includegraphics[scale=0.33]{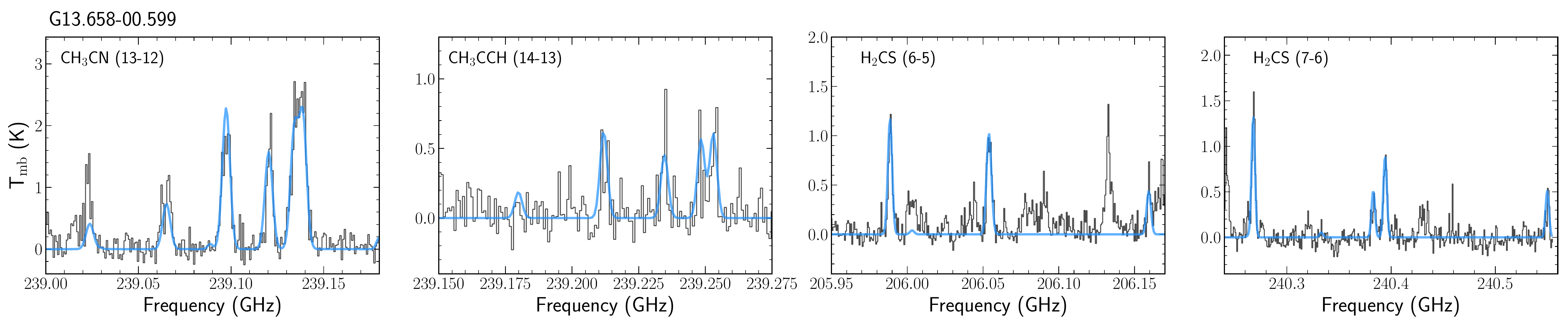}\\
\hspace{-0.9cm}\includegraphics[scale=0.34]{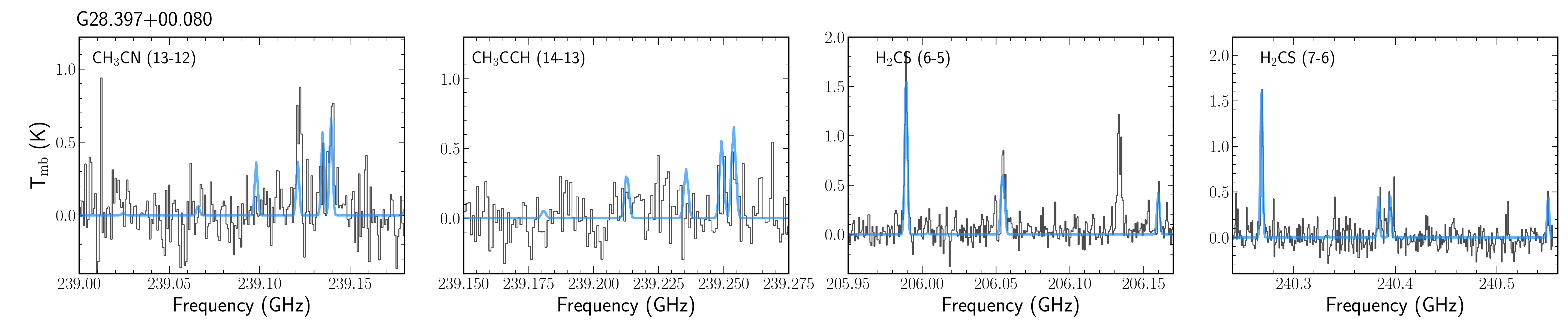}\\
\end{tabular}
\caption{Same as Figure \ref{fig:xclass_spectra}, continued.}
\label{fig:xclass_spectra1}
\end{figure*}

\begin{figure*}[htb]
\begin{tabular}{p{0.25\linewidth}p{0.25\linewidth}p{0.25\linewidth}p{0.25\linewidth}}
\hspace{-0.5cm}\includegraphics[scale=0.35]{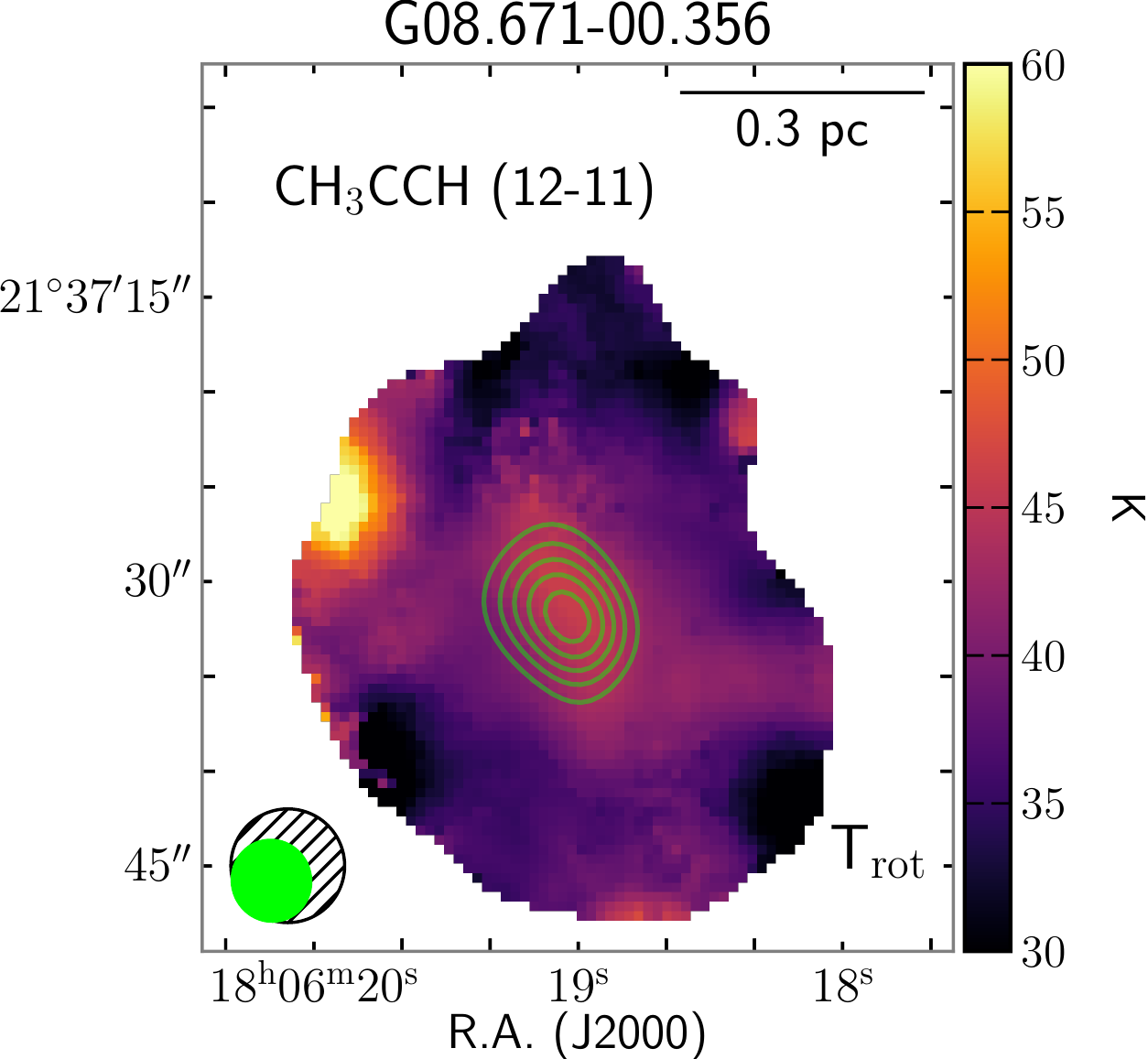}&\hspace{.2cm}\includegraphics[scale=0.35]{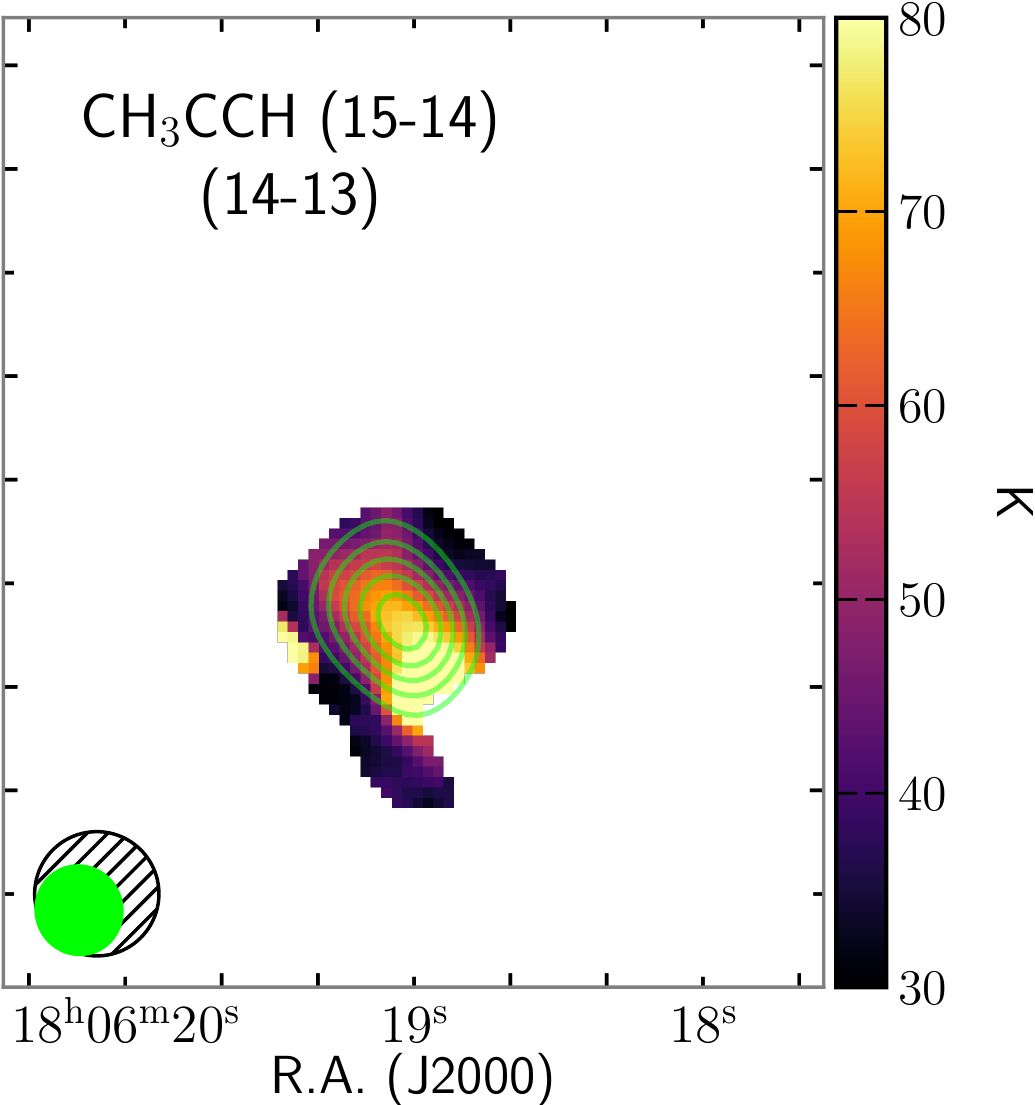}&\hspace{-.2cm}\includegraphics[scale=0.35]{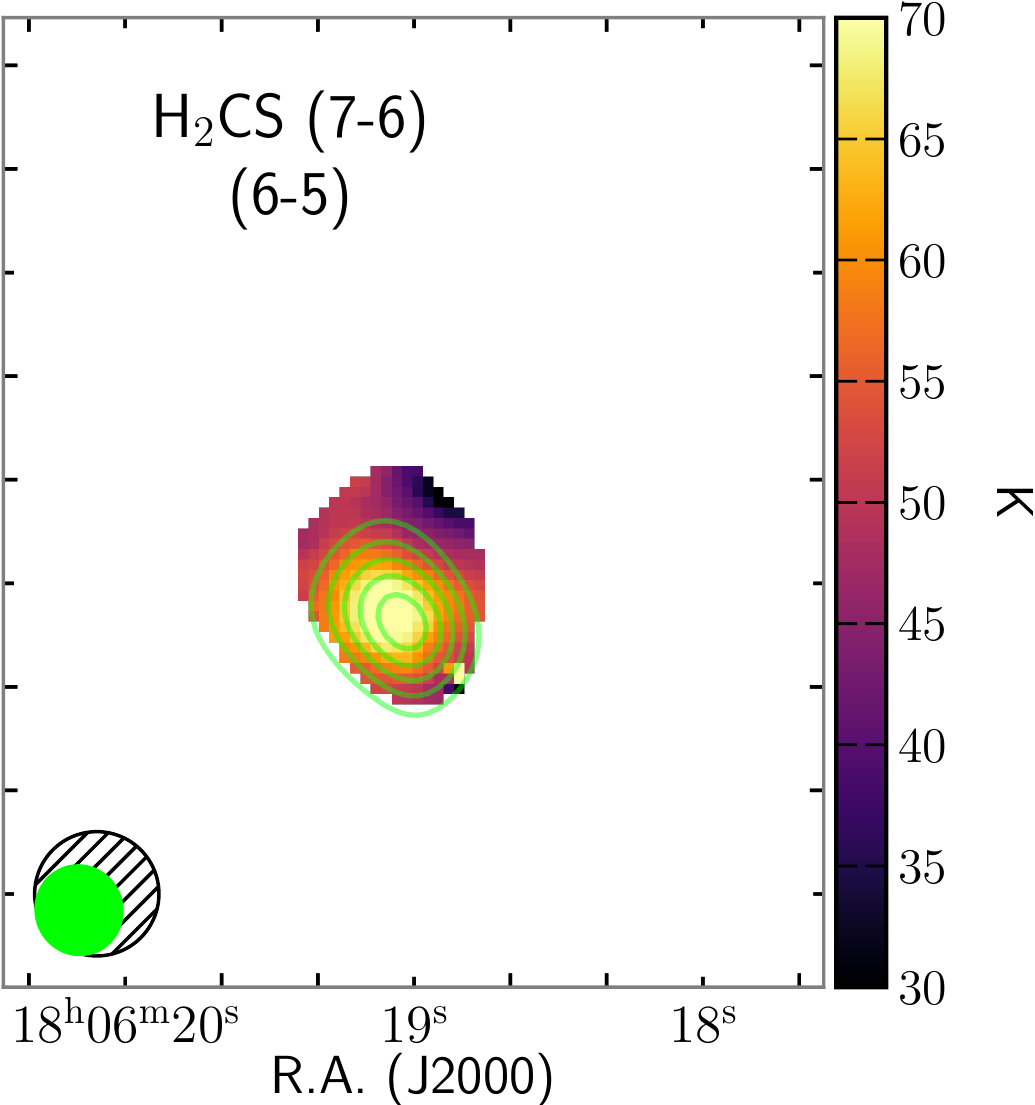}&\hspace{-.5cm}\includegraphics[scale=0.35]{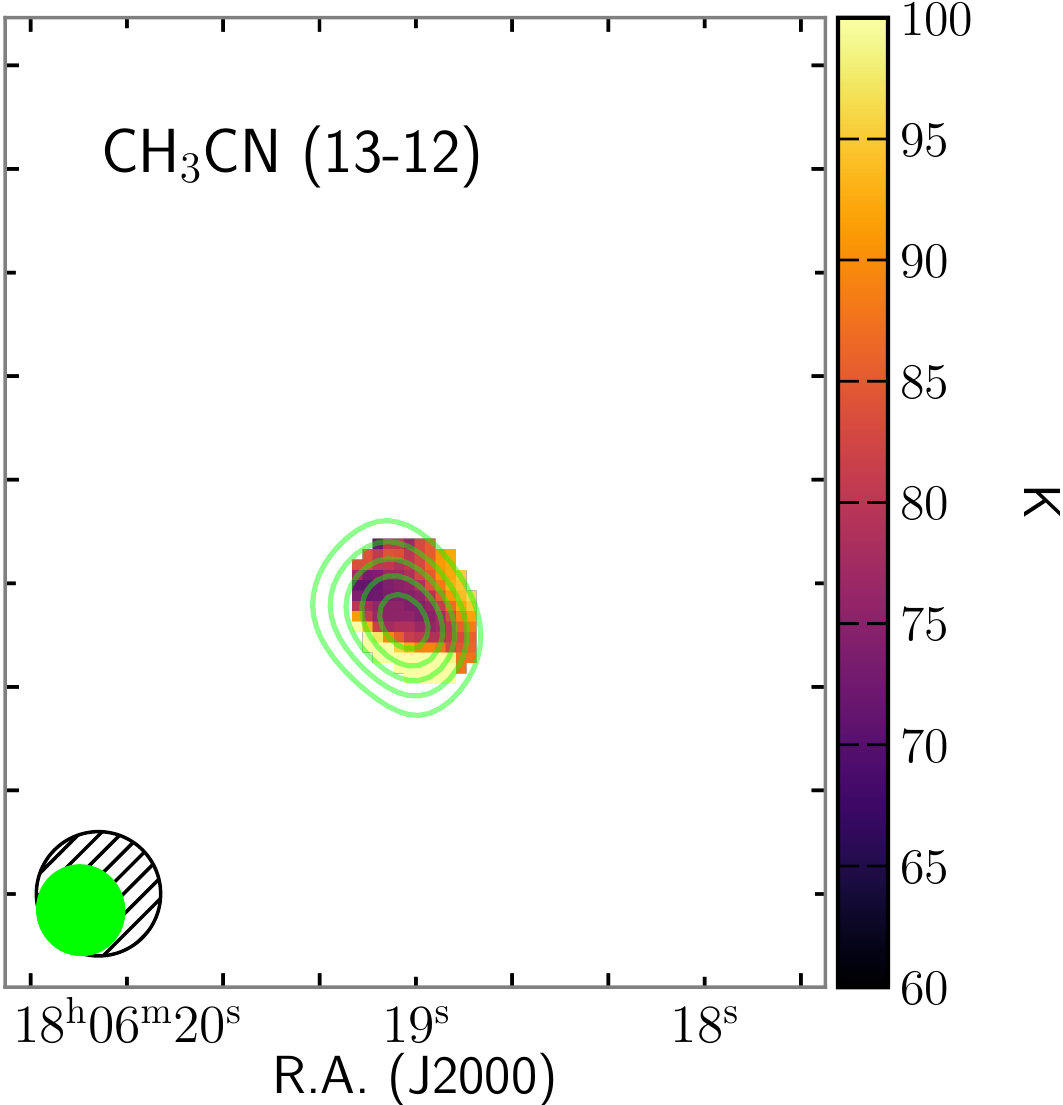}\\
\end{tabular}
\centering\caption{Rotational temperature maps of source G08b derived from multiple thermometer lines using the XCLASS package (Section ~\ref{sec:xclass}). Green contours indicate SMA 1.2 mm continuum levels from 0.3 to 0.9$\times$ peak flux (Table \ref{tab:sma_cont_direct}) by 5 levels of uniform interval. The beams of continuum and respective lines are shown in the lower left corner, as green and hatched ellipses.}
\label{fig:xclass_maps}
\end{figure*}

\begin{figure*}
\begin{tabular}{p{0.33\linewidth}p{0.33\linewidth}p{0.33\linewidth}p{0.33\linewidth}}
\hspace{-0.6cm}\includegraphics[scale=0.42]{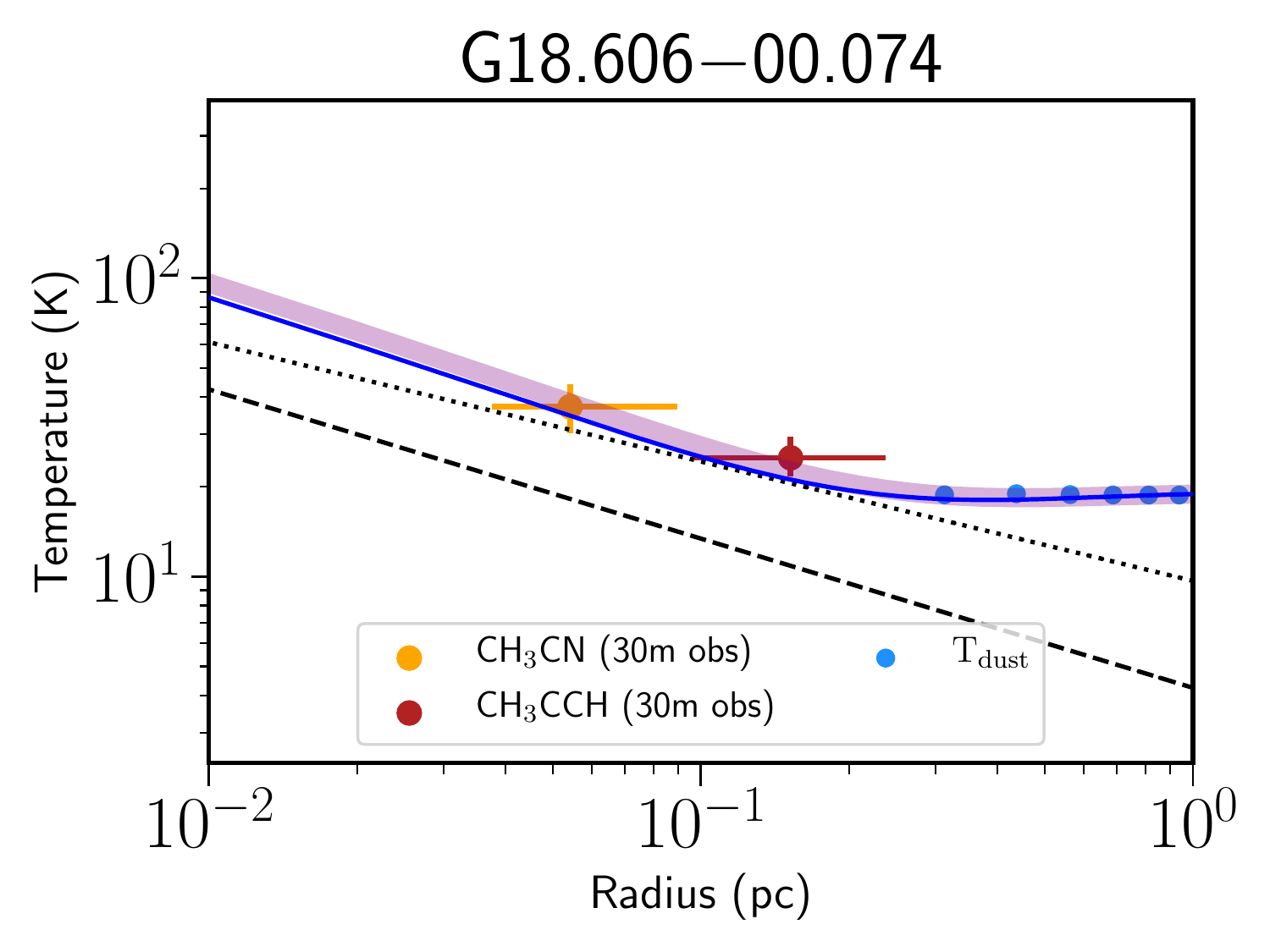}&\hspace{-0.5cm}\includegraphics[scale=0.42]{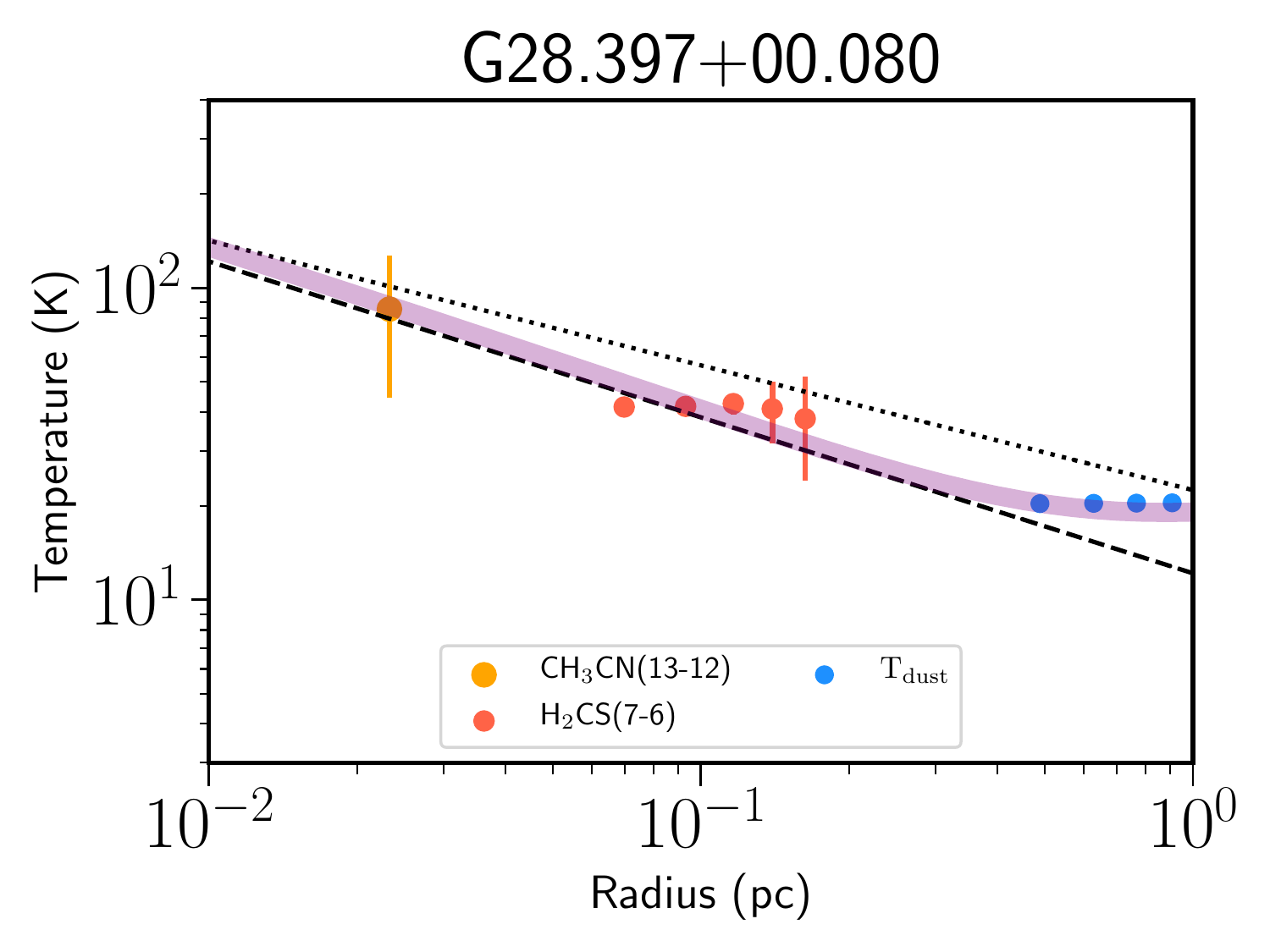}&\hspace{-0.3cm}\includegraphics[scale=0.42]{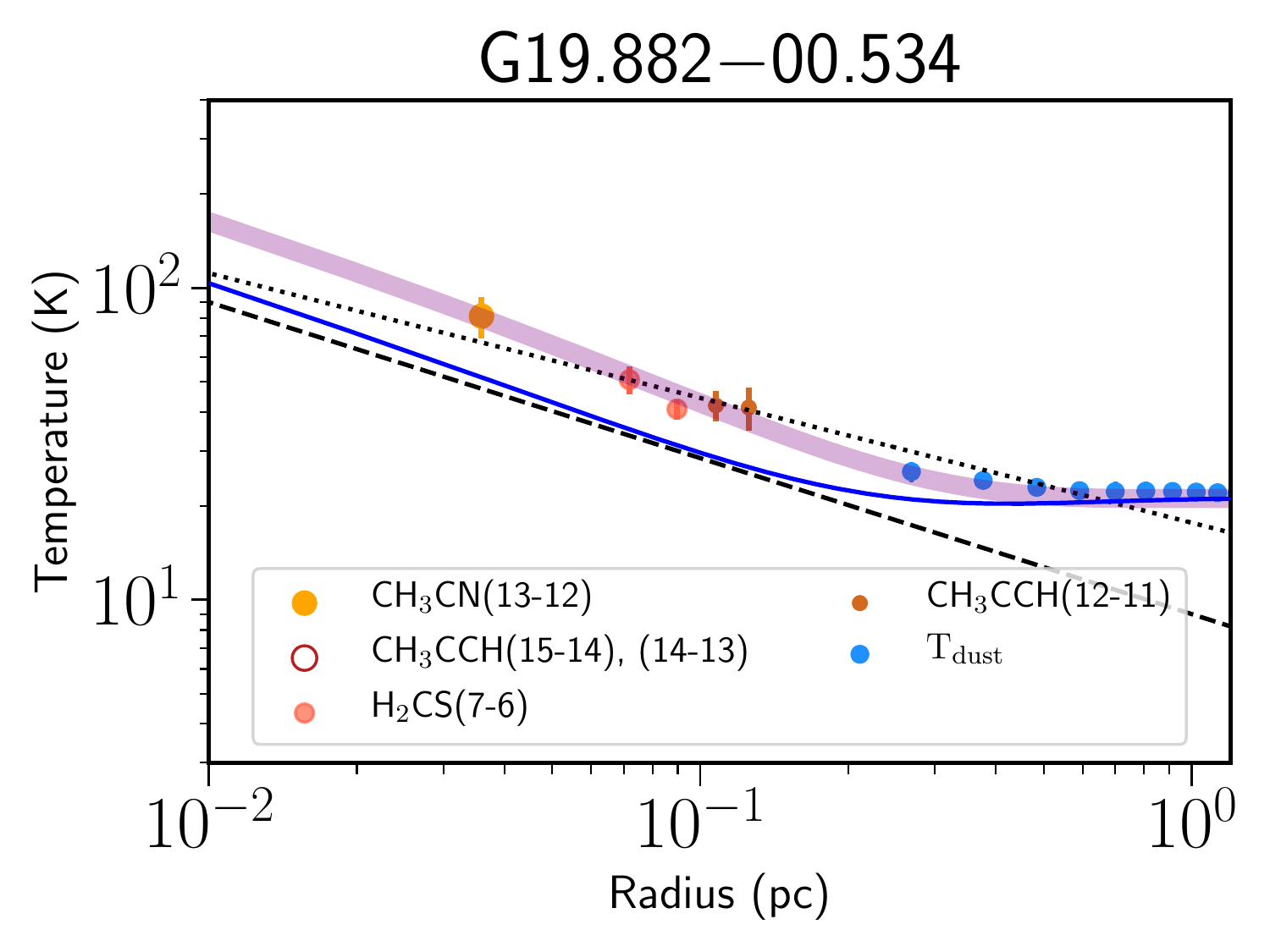}\\
\hspace{-0.6cm}\includegraphics[scale=0.42]{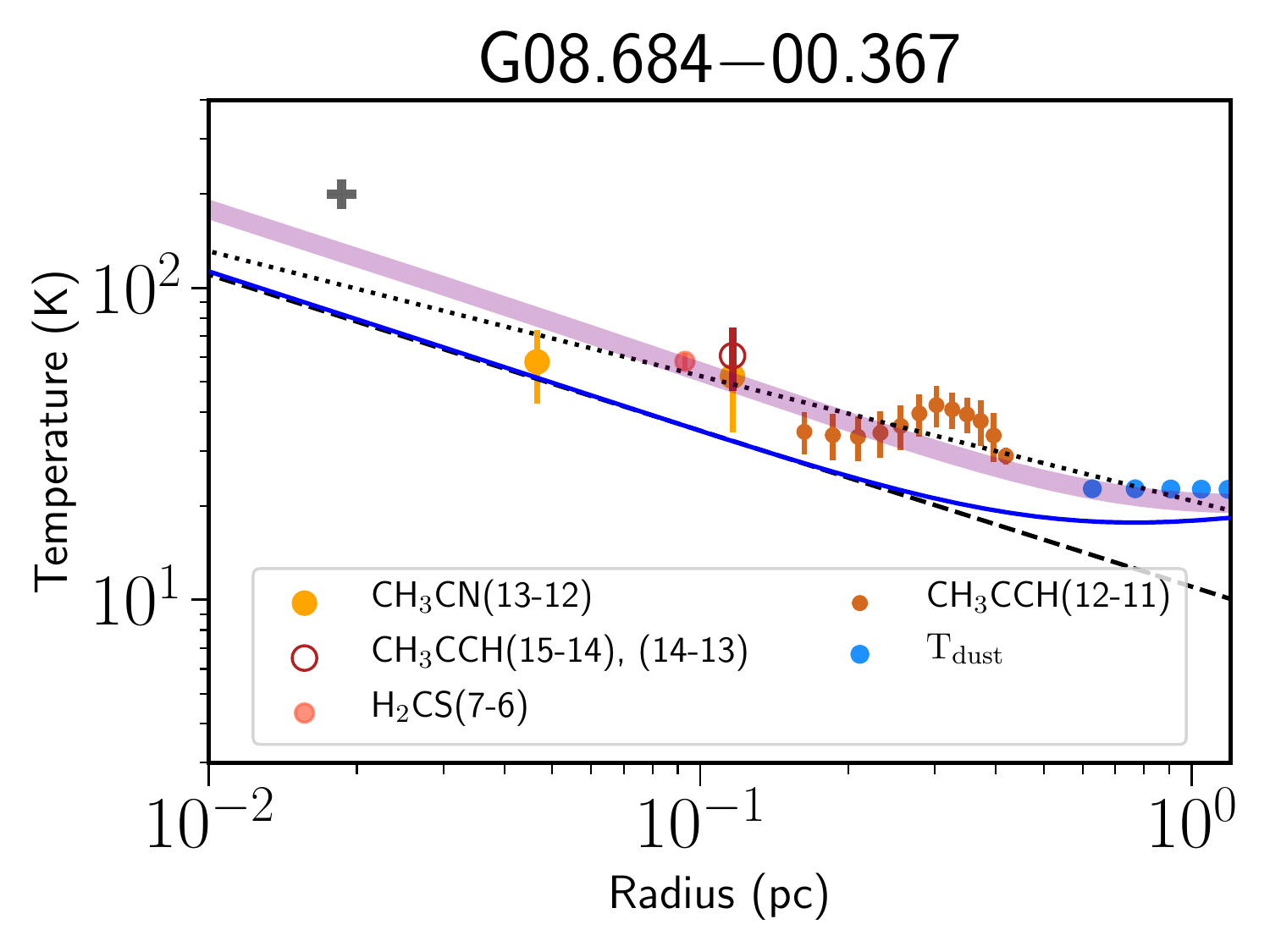}&\hspace{-0.5cm}\includegraphics[scale=0.42]{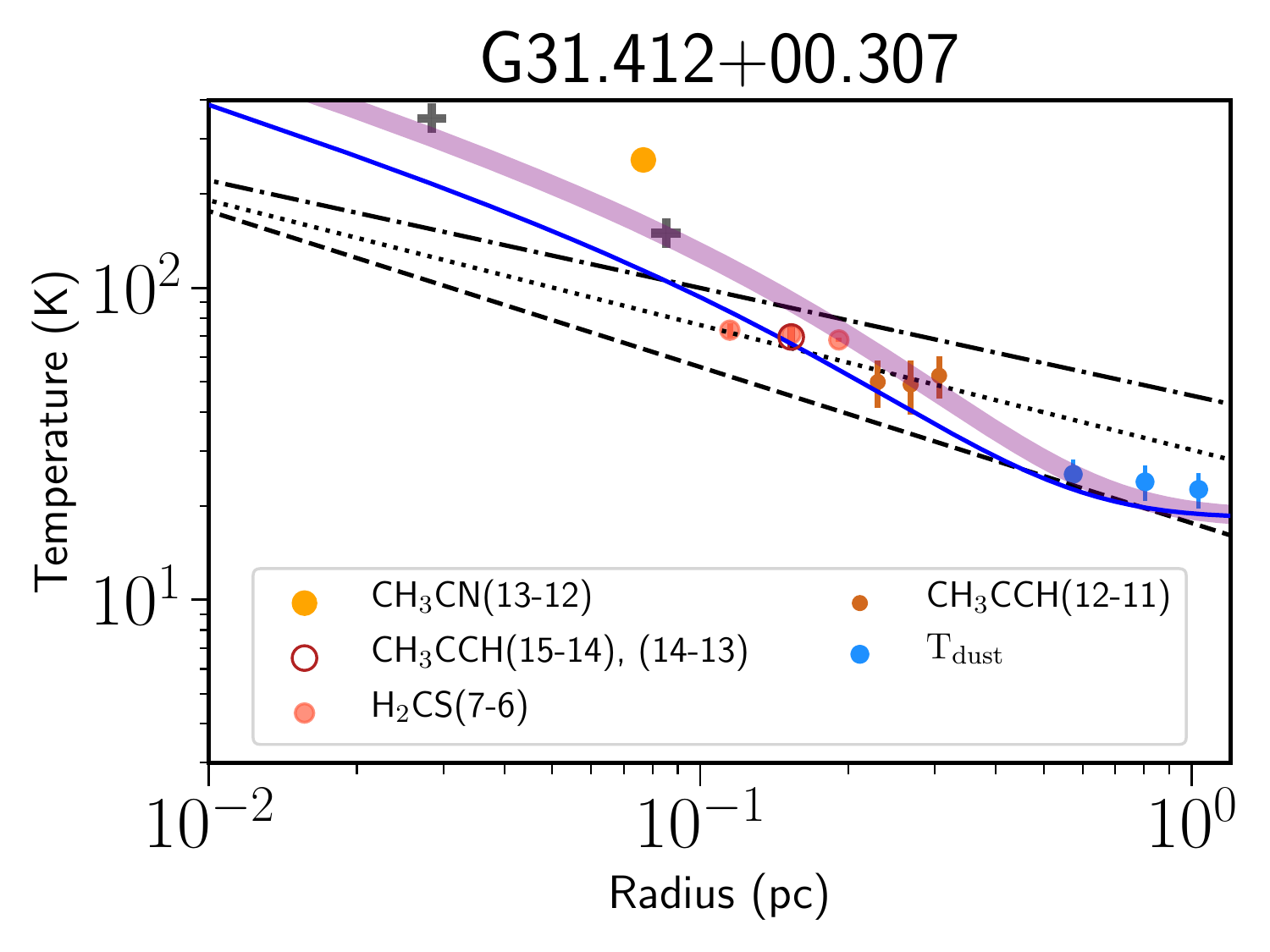}&\hspace{-0.3cm}\includegraphics[scale=0.42]{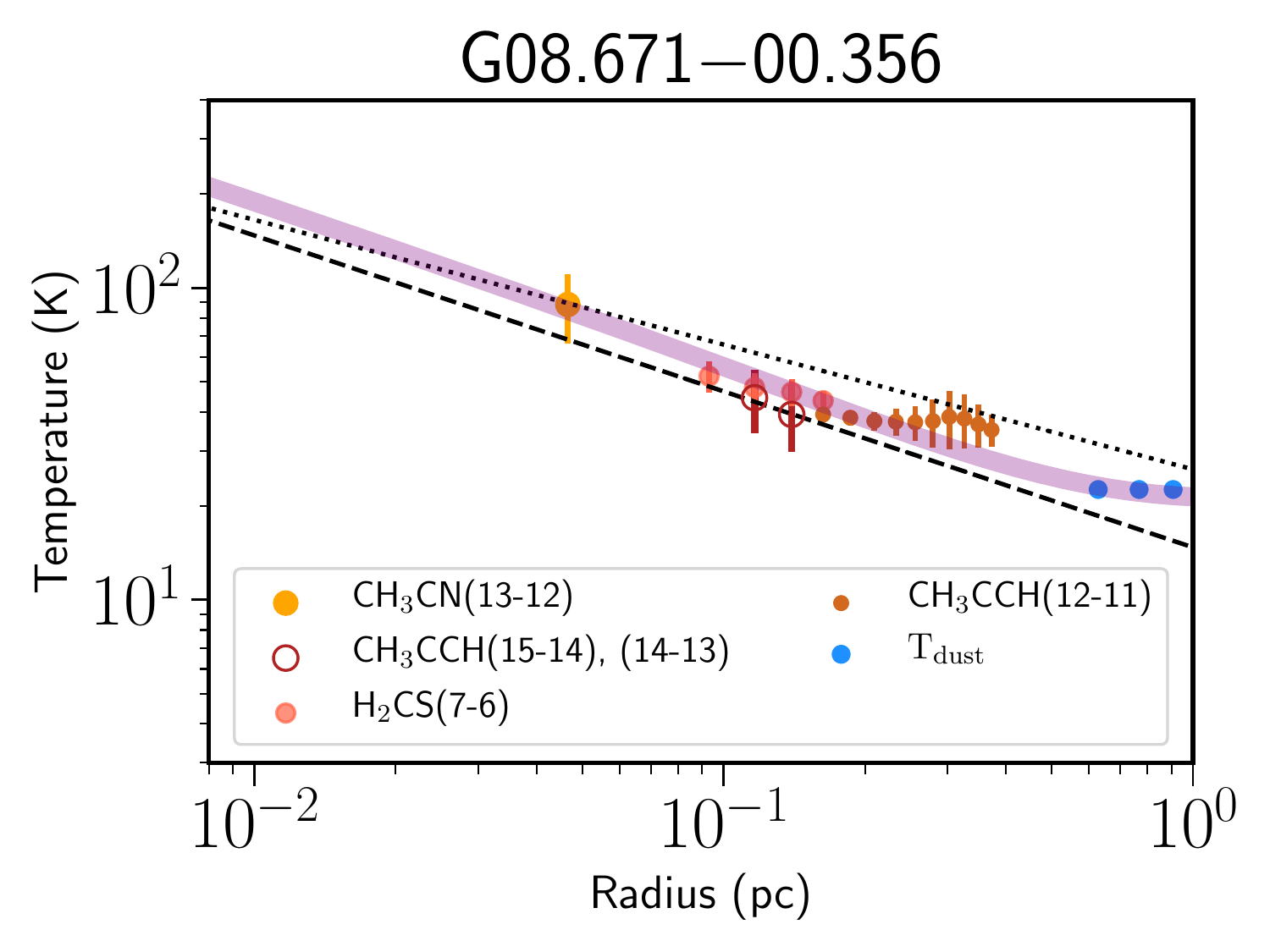}\\
\hspace{-0.6cm}\includegraphics[scale=0.42]{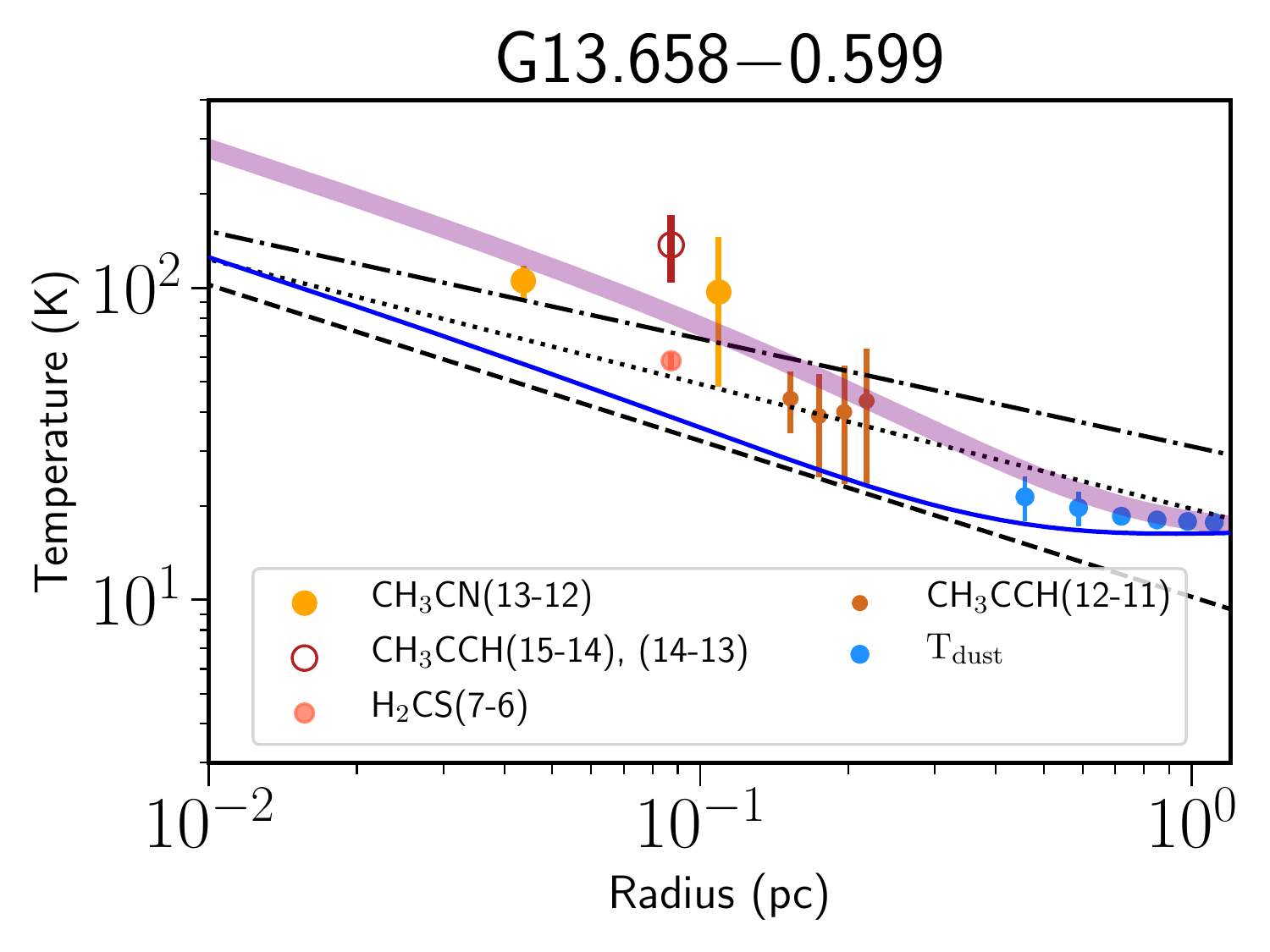}&\hspace{-0.5cm}\includegraphics[scale=0.42]{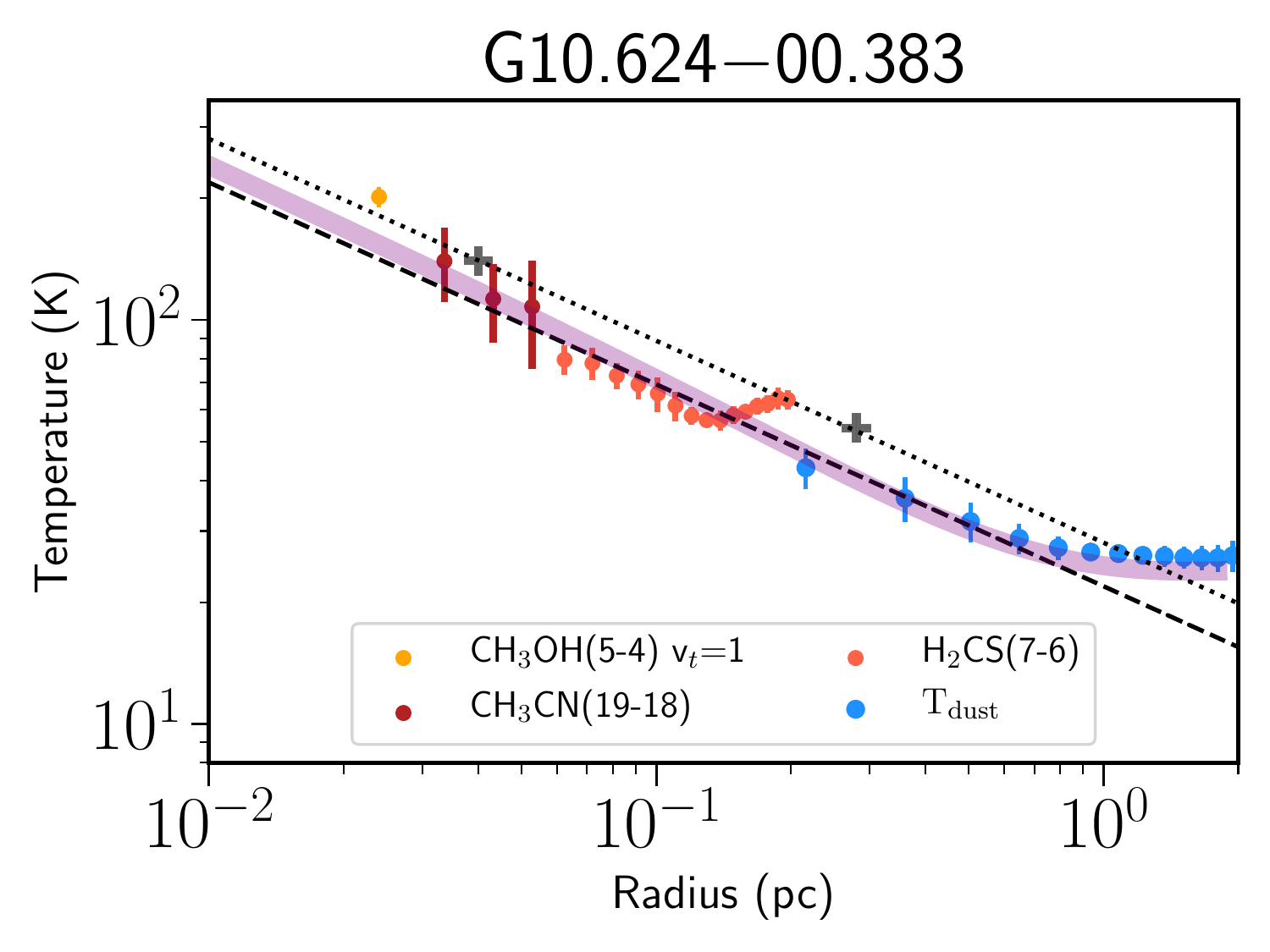}&\hspace{-0.2cm}\includegraphics[scale=0.42]{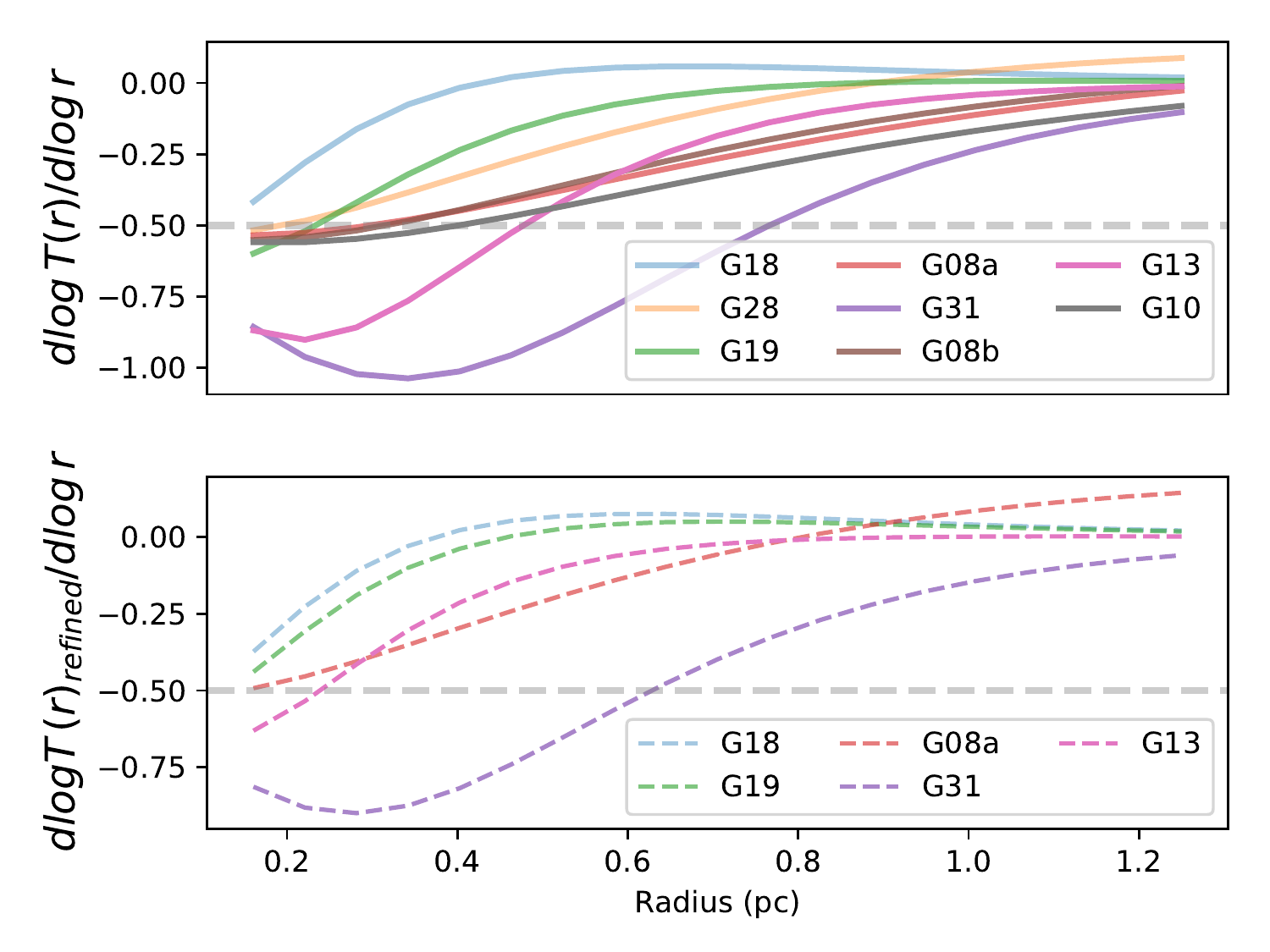}\\
\end{tabular}
\centering\caption{Derived radial averaged temperature profiles of the target sources from multiple thermometers. Error bars are showing the standard deviations for each annular average. Dashed and dotted show a radial temperature profile that follows $\propto$$L^{0.25}$$r^{-0.5}$ ($\beta$ = 0),  $L^{0.2}$$r^{-0.4}$ ($\beta$ = 1) respectively (more details see Sect.~\ref{sec:T_rho_profiles}). For G13 and G31, an additional radial temperature profile of $\propto$$L^{0.17}$$r^{-0.34}$ ($\beta$ = 1.8, \citealt{Adams91}) is shown (dashed dotted line). Whenever available, temperature measurement from higher angular resolution observations from previous work is included in the plots as gray crosses. Thick purple line indicates the fitted temperature profile $T(r)$ described in Section ~\ref{sec:T_rho_profiles}. Blue thin line indicates the refined temperature profile by varying $r_{\mathrm{in}}$ in $T(r)$ (Equation \ref{eq:tformula}) to fit with dust SED (refined $T(r)$, Appendix \ref{app:radmc}). Plot in the bottom right panel shows the first derivative, $d$ log$T$/$d$log$R$, for all the sources, calculated from the fitted profile $T(r)$ and refined $T(r)$ (for 5 sources), in the upper and lower panel, respectively.}
\label{fig:tprofiles}
\end{figure*}



\begin{table*}
\centering
\begin{threeparttable}
\caption{Parameters of CH$_{3}$OH derived radial density $\rho_{\mathrm{dense}}(r)$ and multi-thermometer derived temperature profiles $T(r)$.}
 \label{tab:nprofile_fits}
\begin{tabular}{l c c c c c c c c}
\toprule
   Source  &  
   $\rho_{\mathrm{0\, 0.1 pc}}$ &
   Power-law slope $q_{\mathrm{dense}}$ &
 $R_{\mathrm{eff}}^{a}$&
$R_{\mathrm{max}}^{b}$ &
 $T_{\mathrm{in}}$ &
  $r_{\mathrm{in}}$& 
  $T_{\mathrm{out}}$&
  $r_{\mathrm{out}}$\\
  &$\times$10$^{6}$ (cm$^{-3}$) &
  &(pc) &
  (pc)&
  (K)&
  (pc)&
  (K)&
  (pc)  \\
  \midrule
            G18  & 2.9(0.2)&-0.26(0.07) &0.08&0.10&64.7(1.1)&0.016$^{e}$&19.2(0.1)&0.25(0.05)\\
            G28  &6.3(0.4) &-0.83(0.07) &0.20&0.30&96.1(1.2)&0.02&21.5(0.1)&0.80(0.10) \\
	    G19 &  5.3(0.3)& -0.61(0.05) &0.20&0.25&119.5(2.5)&0.008$^{e}$&21.2(0.1)&0.25(0.03)\\
            G08a  &  10.3(1.2)& -1.33(0.12)&0.30&0.50&128.6(1.8)&0.008$^{e}$&22.4(0.1)&1.1(0.14)  \\
            G31$^{c}$ &  8660 (3000)& -3.22(0.37)&0.34&0.42&400.0$^{d}$&0.01$^{e}$&18.3(1.4)&0.27(0.03)\\
            G08b  & 22.9(0.7)&-1.35(0.02)&0.34&0.50&135.0(1.4)&0.02&22.3(0.1)&0.80(0.08)\\
            G13 &  9.1(1.8)&-1.67(0.20) &0.20&0.40&229.6(23.1)&0.004$^{e}$&17.0(1.5)&0.20(0.07)\\
            G10 & 51.6(2.6)&-1.07(0.03) &0.28&0.42&172.4(4.8)&0.02&24.7(0.6)&0.90(0.31)\\
\bottomrule
\end{tabular}
    \begin{tablenotes}
      \small

\item             $^{a}$: Effective radius $R_{\mathrm{eff}}$ is defined as the $\pi$ $R_{\mathrm{eff}}$$^{2}$ = A, in which the A is the CH$_{3}$OH emission area where $n(\mathrm{H_{\mathrm 2}})$ can be reliably derived.
\item            $^{b}$: The largest radius (distance to the center) of $n(\mathrm{H_{\mathrm 2}})$ map derived by CH$_{3}$OH; this is due to the irregular shape of the emission area.
\item           $^{c}$: The density profile of source G31 is better described by a two-component power-law form consisting of a shallow slope in the inner region followed by a steep slope of $-$5.22(0.045) in the outer region. 
\item           $^{d}$: Upper limit is set to 400 K in the fit.
\item            $^{e}$: For sources G18, G19, G08a, G31 and G13, $r_{\mathrm{in}}$ is a re-adjusted parameter based on SED calculation and comparison presented in Figure \ref{fig:sed_radmc} (temperature profile shown as blue lines in Figure \ref{fig:tprofiles}). 

\end{tablenotes}
  \end{threeparttable}
\end{table*}

\subsubsection{Modeling procedure}
Using the XCLASS package (\citealt{Moeller17}), we have established a pixel-by-pixel Local Thermodynamic Equilibrium (LTE) model-fitting procedure for the observed CH$_{3}$CCH, H$_{2}$CS, CH$_{3}$CN, and CH$_{3}$OH $\nu_{t}=$1 (for G10) lines which returns the best fit of source size, rotational temperature ($T_{\mathrm{rot}}$), molecular column density ($N_{\mathrm{mol}}$), line width ($\Delta\,V$) and the source velocity ($V_{\mathrm{source}}$).
In this specific implementation, we fixed the source size to the synthesized beam size (i.e., assuming beam filling factor of 1) and optimized the rest of free parameters.
The optimization procedure employed an initial global parameter search using the {\tt bees} algorithm (\citealt{Pham06}) which was followed by Levenberg-Marquardt iterations.
Such a procedure helps to avoid trapping in local minima.
We fit the $J$ = 15-14 and $J$ = 14-13 ladders of CH$_{3}$CCH together, and the $J$ = 12-11 ladders separately, given that the former lines show less extended emission and appear to trace hotter gas. 
Examples of the fitted spectra are presented in Figure \ref{fig:xclass_spectra}-\ref{fig:xclass_spectra1}. Examples of the obtained rotational temperature maps are shown in Figs. \ref{fig:xclass_maps}. When deriving these rotational temperature maps, we use pixels where the third lowest energy transition in consideration has intensity larger than our 3$\sigma$ detection limit.
For the hot molecular core G31, besides the aforementioned few molecular species that we targeted, the fittings also considered several other species that can potentially make a prominent contribution in our spectra. They are shown in Figure \ref{fig:xclass_spectra} for G31 in different colors.
For clump G18, we do not have robust detection of these thermometer lines from our SMA observations and we rely on previous IRAM 30m telescope observations (\citealt{Giannetti17}) to describe the temperature profile, with same thermometers but their lower transitions. These pointed observations from IRAM 30m telescope did not give information on relevant physical scales, rather, based on a fixed temperature profile ($T(r)\,\propto\,R^{0.4}L^{-0.25}$, \citealt{Giannetti17}) the radius of a certain measured rotational temperature was deduced. Following the same workflow (Figure \ref{fig:flowchart}), these measurements are combined with one-component dust temperatures (\citealt{Lin19}) in the outer region of the clump, to compose an initial temperature profile $T(r)$ to be refined later by SED comparison using {\tt{RADMC-3D}}. 

\subsection{Deriving the pixel-based hydrogen volume density maps with non-LTE RADEX model}\label{sec:radex_nh2}
\subsubsection{Methanol lines}
Methanol (CH$_{3}$OH) is a slightly asymmetric top molecule.
It has three types of symmetry, which are denoted as A, E1 and E2, respectively. 
The E1 and E2 states can be considered as doubly degenerate states of the E symmetry where the quantum number of the angular momentum along the symmetry axis of the CH$_{3}$ group ($k$) can take either positive and negative values.  
The torsional ground state {\it{E}}$-$CH$_{3}$OH $5_{k, 5}$$\,-\,$$4$$_{k, 4}$, $K\,=$ 0, $\pm$1,  $\pm$2, $\pm$3, $\pm$4 ($\varv_{t}=0$) transitions were found to be a good densitometer for gas denser than 10$^{4}$\, cm$^{-3}$ \citep{Leurini04, Leurini07}. 

The excitation of these $K$ ladders is usually observed to be sub-thermal.
These $K$ = 0 and $K$ = $\pm$1 ladders occupy a rather narrow range of upper level energies ($E_{\mathrm{up}}\,\sim\,$40-55 K).
At the same time, they cover a wide range of critical densities ($\sim$10$^{5}$ to $\gtrsim$10$^{7}$ cm$^{-3}$) (Table \ref{tab:lines_more}), which implies that the line ratios of two $K$ components can be good density probes.
The higher $K$ components ($K$$\,\geq\,$3, E$_{up}$$\,>\,$80 K) are generally excited in hot regions where the gas volume densities are close to or higher than the critical densities. 
Hence, ratios of the $K\,\geq\,$3 components provide additionally constraints on kinetic temperature.
Apart from the high abundance of CH$_{3}$OH, it is this property of the methanol energy system and the relatively low upper level energies of the $K\,<\,$3 transitions that makes this line series sensitive to gas density for a broad range of physical conditions in molecular clouds (\citealt{Leurini04}). 

As illustrated in Figure \ref{fig:m0}, the CH$_{3}$OH emission appears clumpy and exhibits elongated structures, extending for up to 0.5 pc with respect to the continuum peak.    
The $K\,<\,$2 transitions of {\it{E}}$-$CH$_{3}$OH (5-4) ($\nu_{t}$ = 0) are excited over an extended region, while the emission of the $K\,>\,$2 lines are confined to the central region of the clumps.

\subsubsection{Modeling procedure}\label{sec:radex}
\begin{figure*}
\begin{tabular}{p{0.31\linewidth}p{0.31\linewidth}p{0.31\linewidth}}
\hspace{-1.0cm}\includegraphics[scale=0.33]{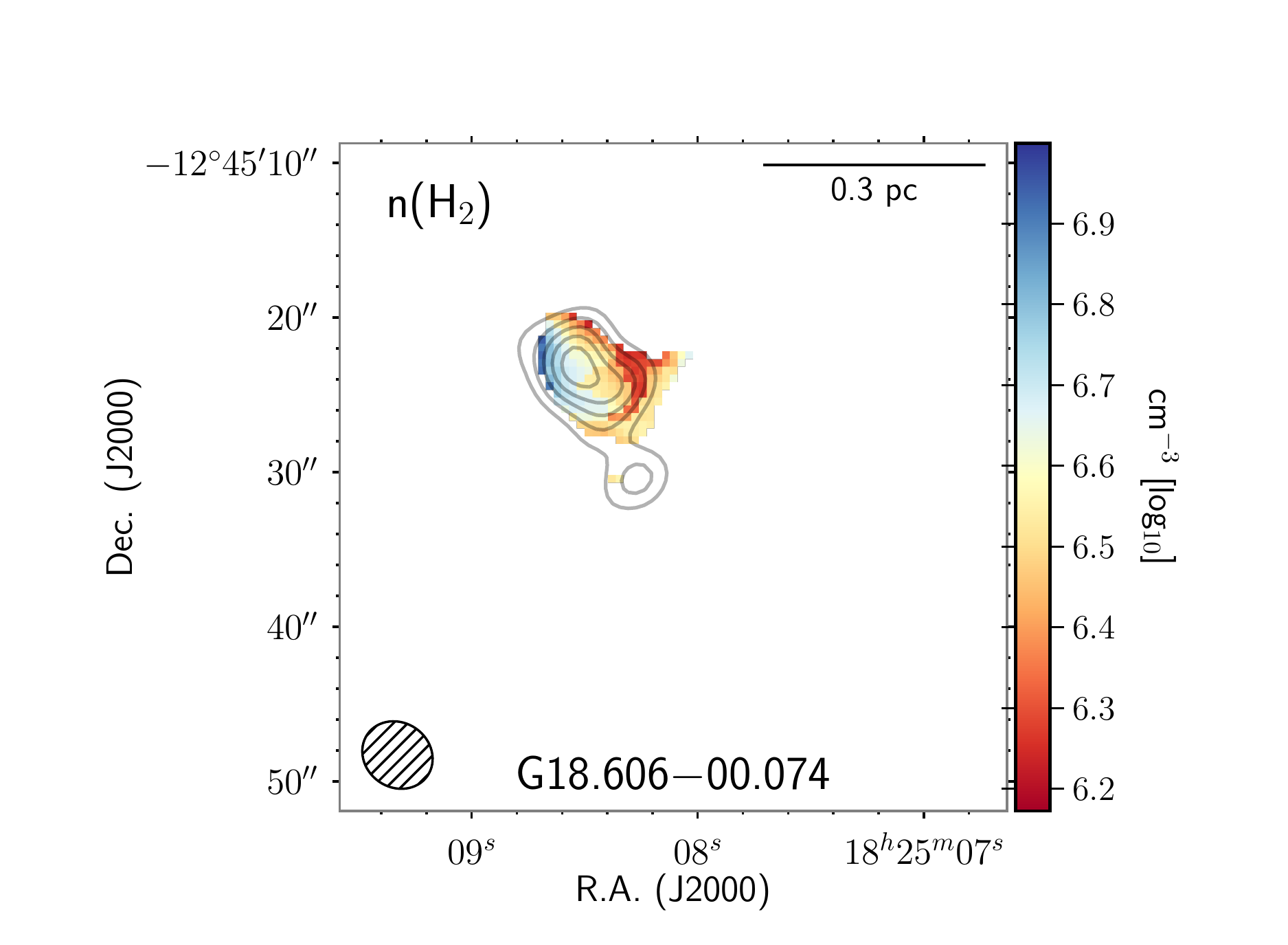}&\hspace{-0.1cm}\includegraphics[scale=0.33]{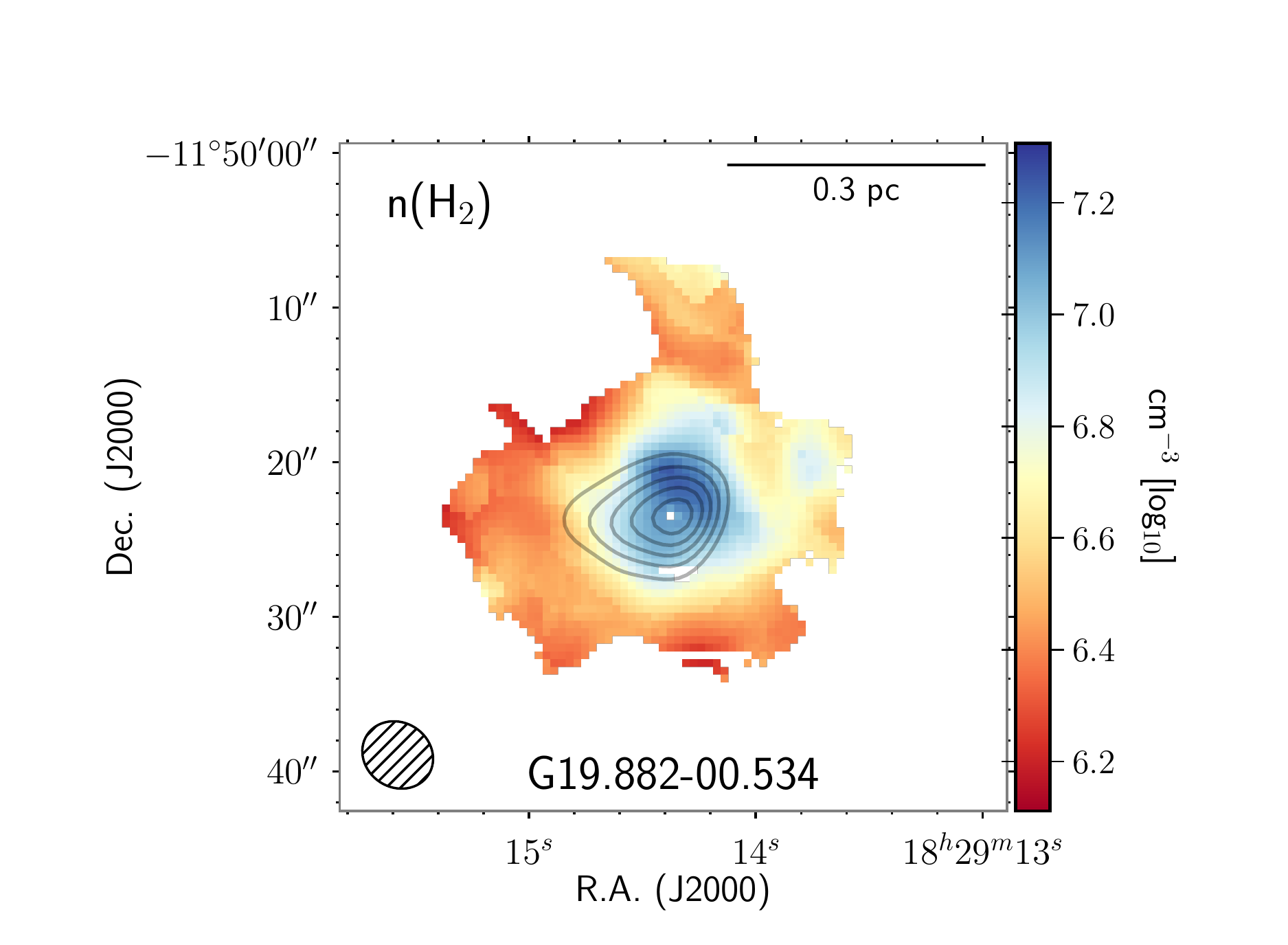}&\hspace{0.2cm}\includegraphics[scale=0.33]{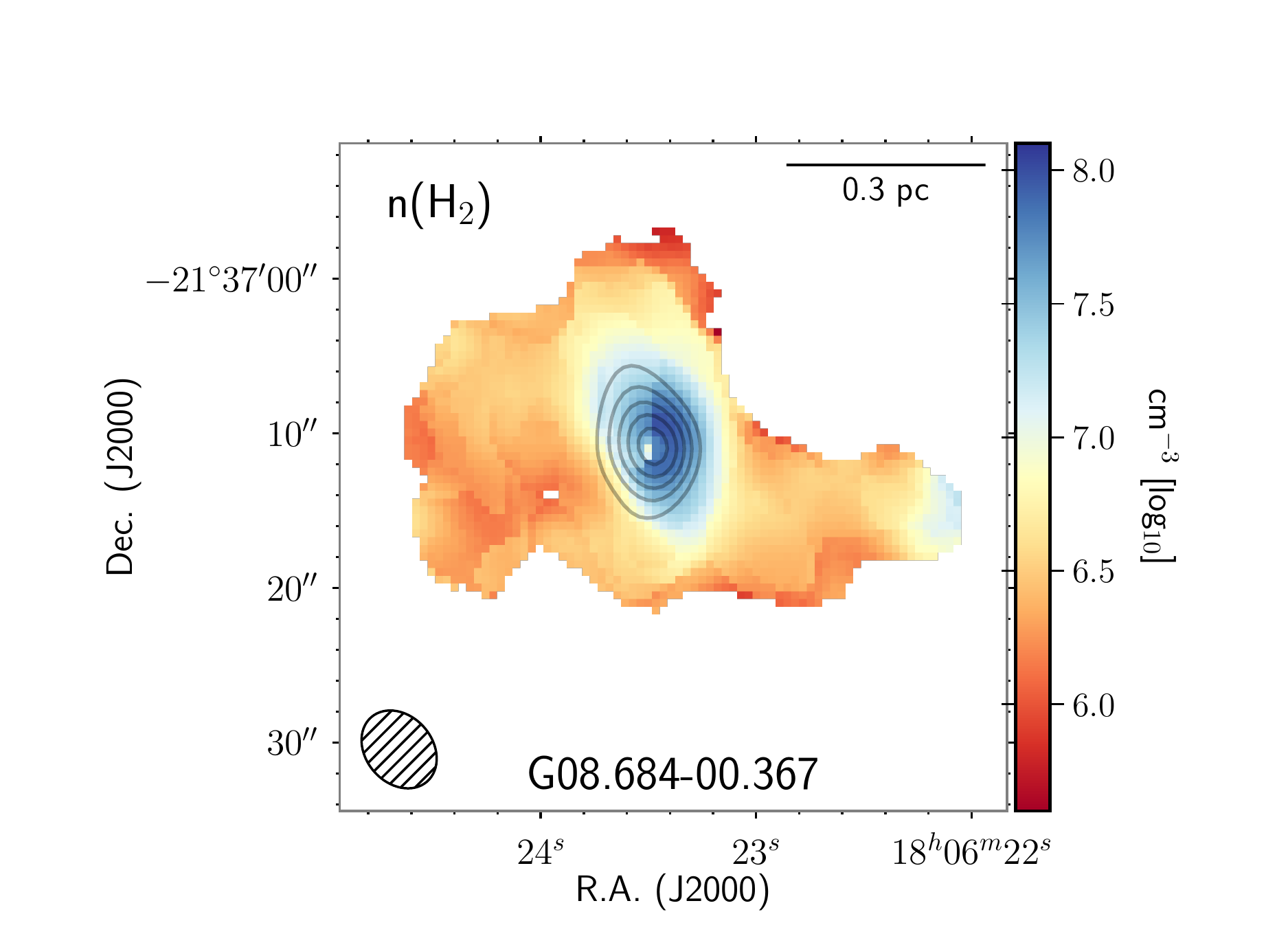}\\
\hspace{-1.0cm}\includegraphics[scale=0.33]{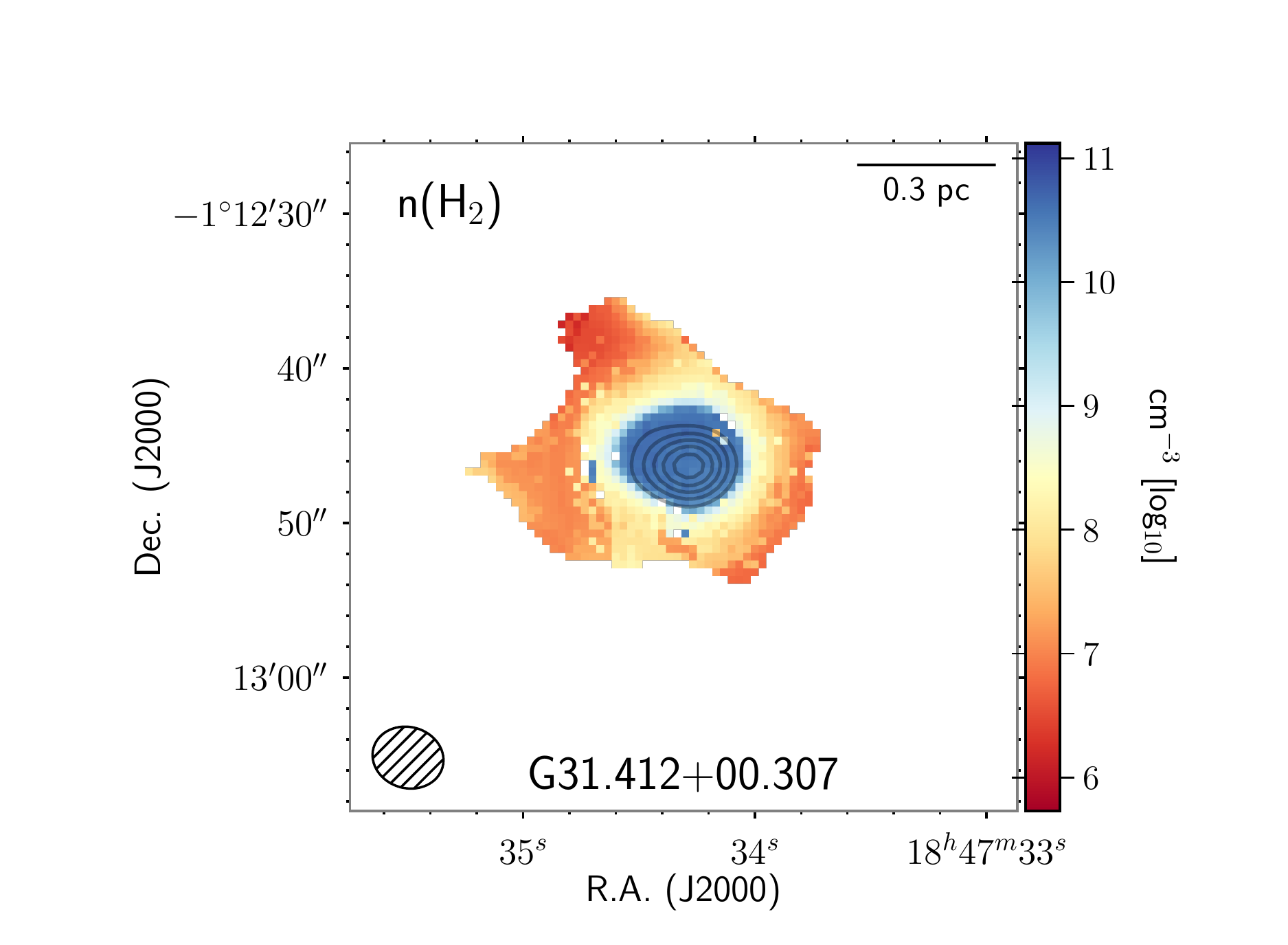}&\hspace{-0.1cm}\includegraphics[scale=0.33]{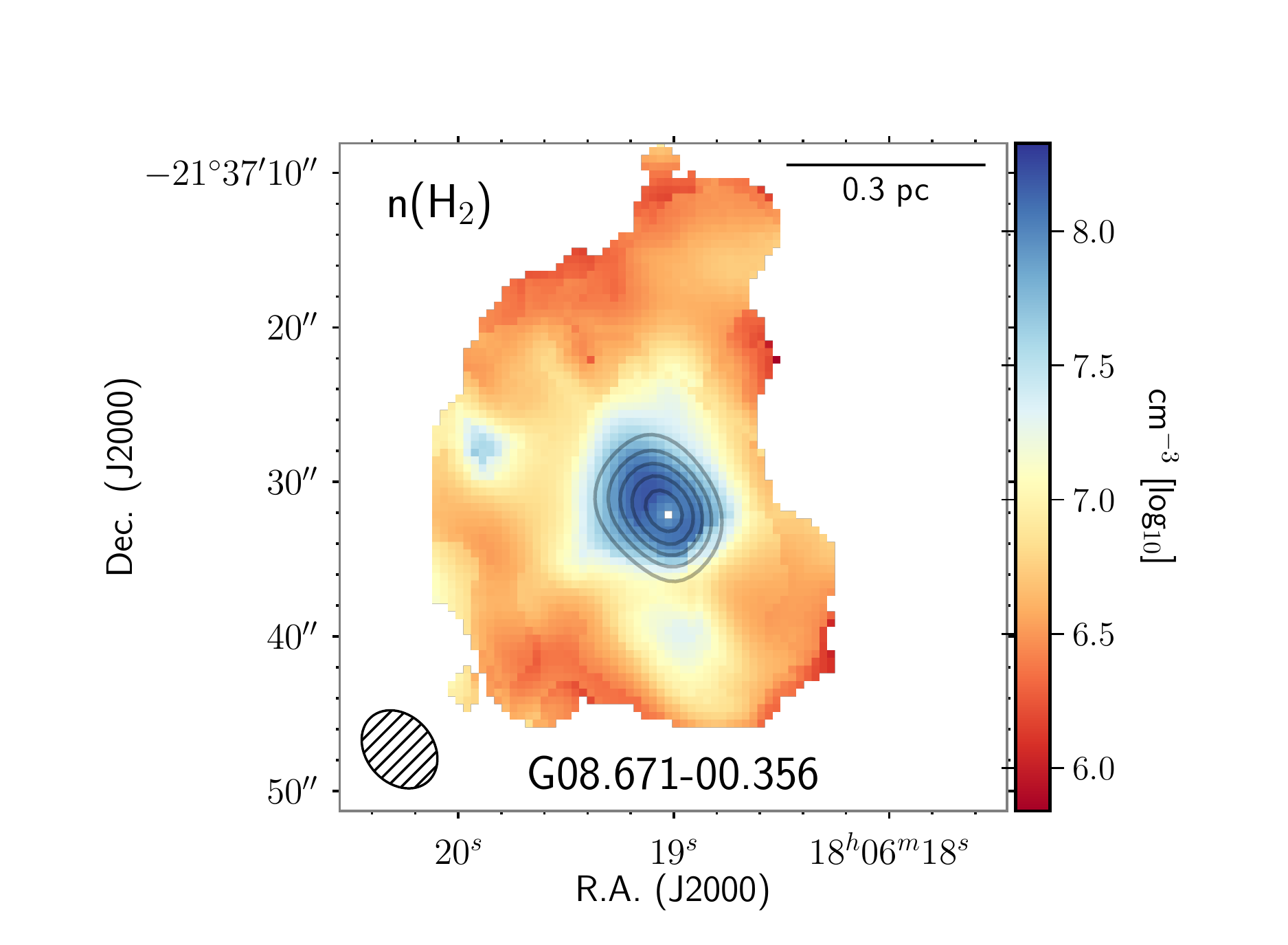}&\hspace{0.2cm}\includegraphics[scale=0.33]{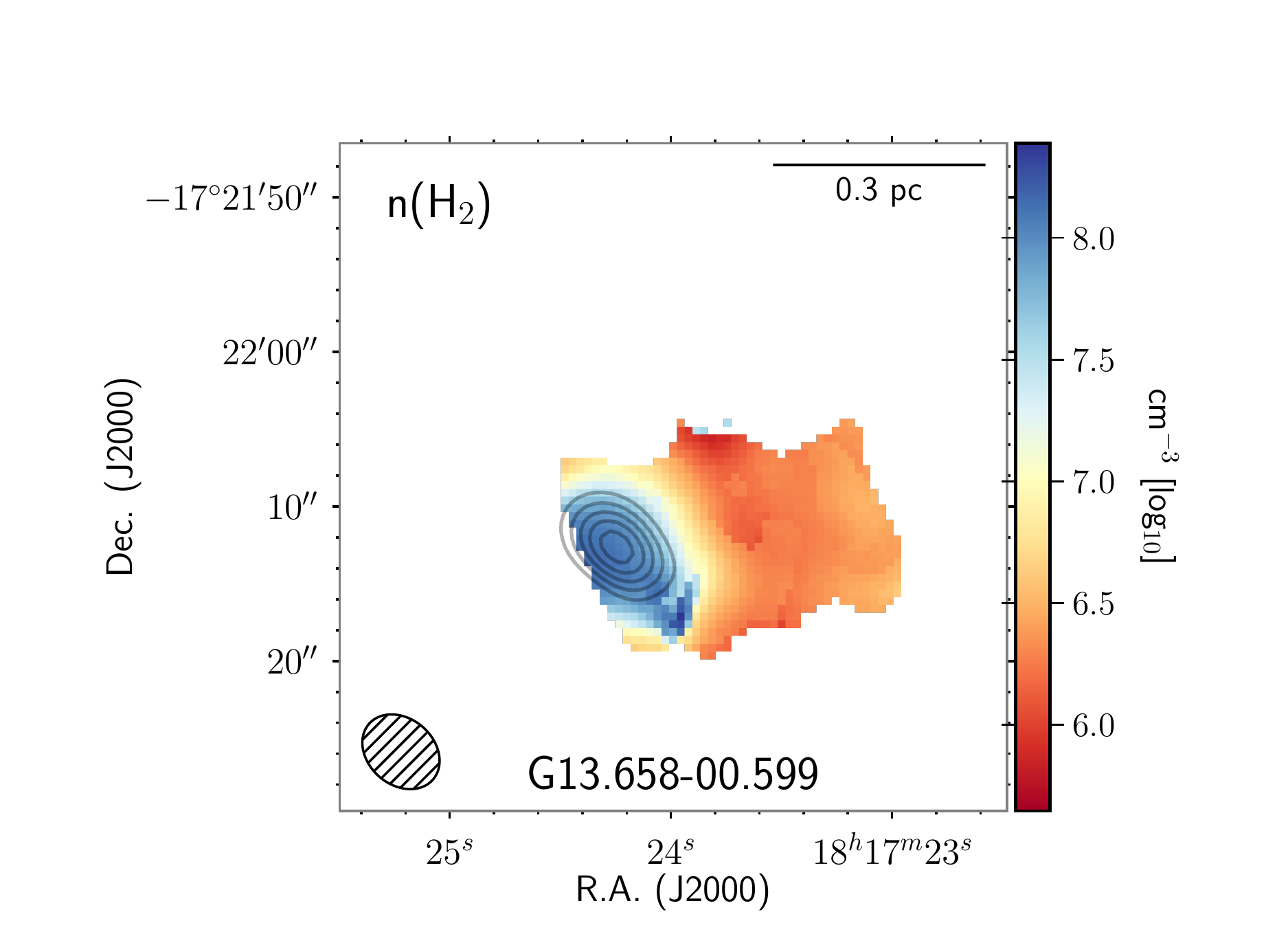}\\
\hspace{-1.0cm}\includegraphics[scale=0.33]{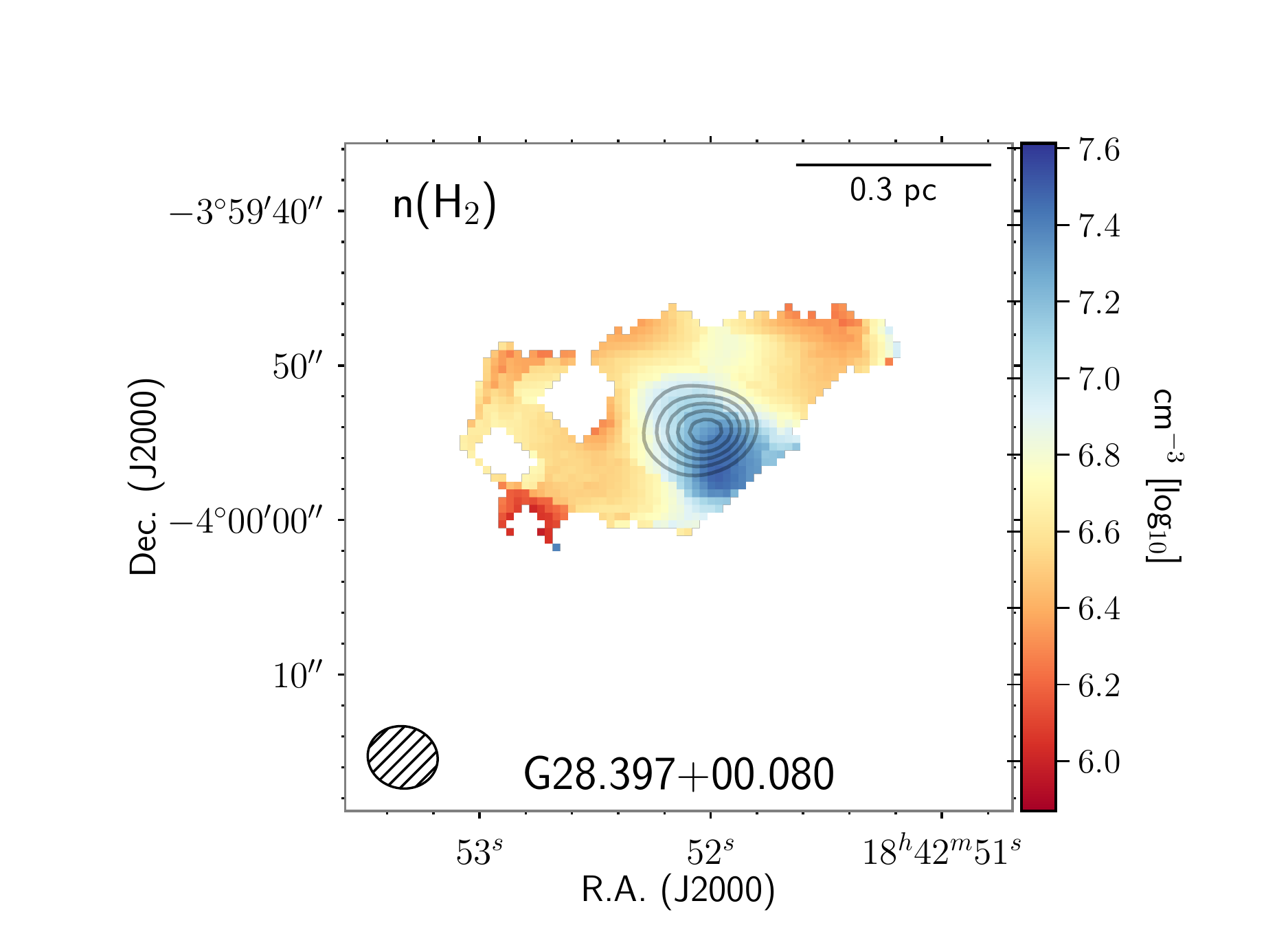}&\hspace{-0.1cm}\includegraphics[scale=0.33]{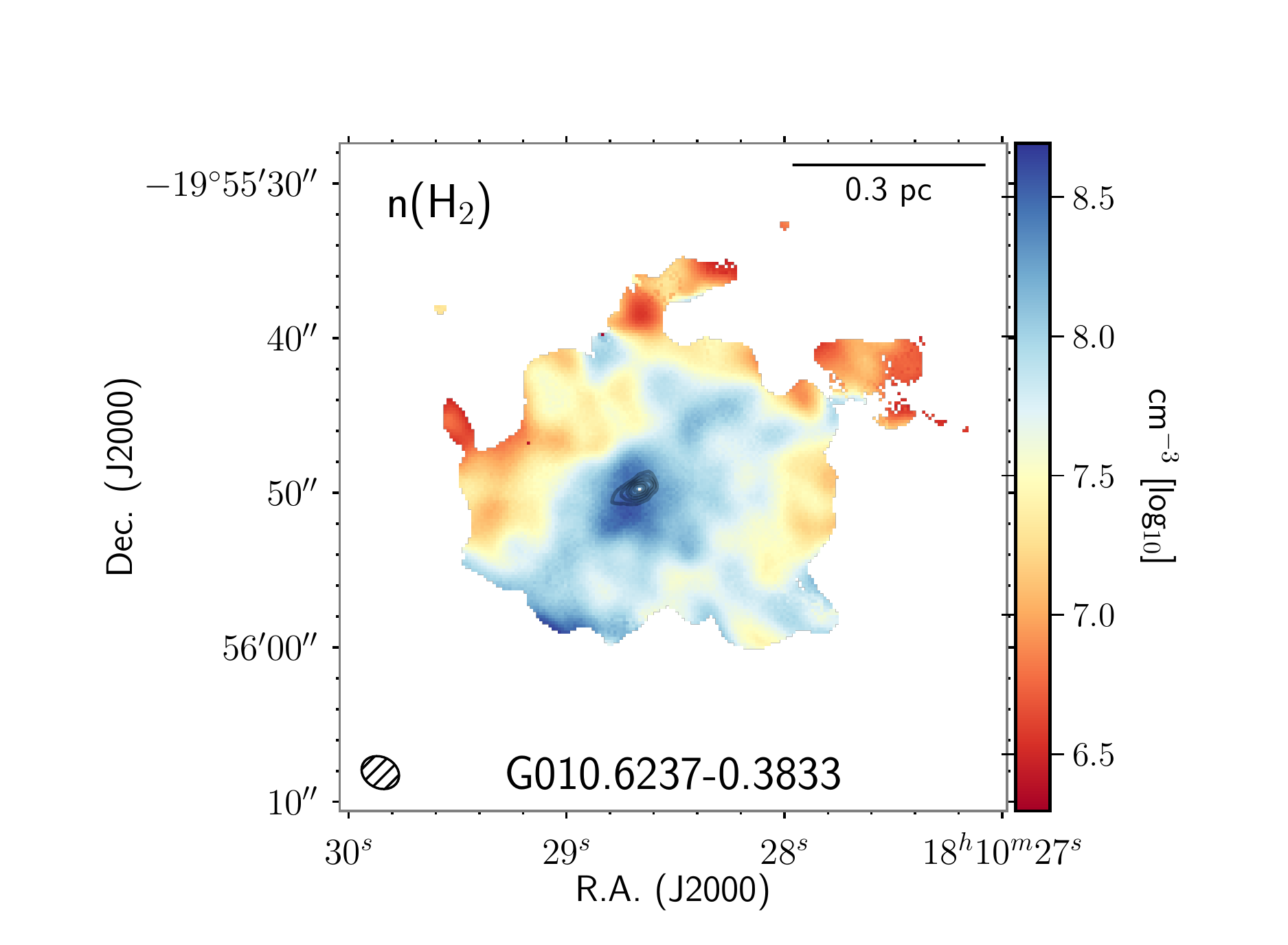}&\\
\end{tabular}
\centering\caption{CH$_{3}$OH derived $n(\mathrm{H_{\mathrm 2}})$ maps from RADEX modeling, of all target sources. The beam of CH$_{3}$OH 5$_{-1}$-4$_{-1}$ $E$ line is indicated in the bottom left corner. Gray contours indicates the SMA 1.2 mm continuum level from 0.1 to 0.9$\times$peak flux represented by 5 levels of uniform interval.}
\label{fig:nmaps}
\end{figure*}

We produced a series of large velocity gradient (LVG) RADEX models \citep{vdt07} to search for the best fits of $n(\mathrm{H_{\mathrm 2}})$, CH$_{3}$OH column density, N(CH$_{3}$OH-$E$)/N(CH$_{3}$OH-$A$), and kinetic temperature ($T_{\mathrm{kin}}$) to the observed CH$_{3}$OH ($J$ = 5-4, $\nu_{\mathrm{t}}$ = 0) lines.  
We took the collisional rates from \citet{Rabli10} which were evaluated for temperatures from 10 to 200 K. We adopt a Markov Chains Monte Carlo (MCMCs) method to derive the parameters and estimate the associated uncertainties, taking into consideration the upper limits for weakly detected line components. The details of the modeling procedure are elaborated in Appendix \ref{app:radex_mcmc_interp} where the formulas used for the likelihood function are given. 
In the fitting, for each pixel we enforce the posterior distribution of $T_{\mathrm{kin}}$ to be a narrow Gaussian distribution centralised at $T(r)$ (more in Sect.~\ref{sec:T_rho_profiles}) as measured in Section ~\ref{sec:xclass} from the multiple rotational temperature maps. Although the ratios between the lower $K$ ladders of CH$_{3}$OH lines depend only weakly on the kinetic temperature, having a fixed term helps to avoid randomly converged parameters, which is useful to ensure that the resultant parameter maps are continuous. The obtained $n(\mathrm{H_{\mathrm 2}})$ maps are shown in Figure \ref{fig:nmaps}. The CH$_{3}$OH column density maps are shown in Figure \ref{fig:cdmaps}. 

\begin{table*}
\centering
\begin{threeparttable}
\caption{Best-fit parameters from {\tt{RADMC-3D}} modelling of the dust continuum in 350/450 $\mu$m and 870 $\mu$m. }
\label{tab:radmc_para}
\begin{tabular}{ l c c c c c c c}
\toprule
  
  Source  & 
  $\bar{\rho}^{a}$ &  
  $q^{b}$ & 
  $R_{\mathrm{clump}}$ & $\rho_{\mathrm{0\,0.1\,pc}}^{c}$ &  $M_{\mathrm{tot}}$  &  
  $L_{\mathrm{bol}}^{d}$ &
  $M_{\mathrm{tot}}$ ($<$ 0.5 pc)$^{e}$ \\
  &  (10$^{3}$ cm$^{-3}$) & & (pc) & (10$^{4}$ cm$^{-3}$) & (10$^{3}$ $M_{\odot}$) & (10$^{4}$ $L_{\odot}$) & (10$^{3}$ $M_{\odot}$)\\
  \midrule
            G18  & 14.8 & -0.57 &0.8&3.81&1.9&0.8&0.7\\
            G28  &12.3&-0.66 &1.25&5.1&6.9&6.1&0.8\\
	        G19&7.8&-1.37  &1.04&10.4&2.5&3.1&0.7\\
	        G08a&13.2 &-1.19 &1.00&12.3&3.8&3.5&1.1\\
            G31  &11.7 &-1.22&2.18&29.9&35.3&57.0&2.6\\
            G08b  &3.6&-1.66&2.00&22.9&8.2&24.3&1.3\\
            G13 & 9.5&-1.41 &0.88&10.8&1.9&3.9&0.8\\
            G10 & 7.8 & -1.52 & 1.58 & 23.4 & 8.9 & 65.3 & 1.6\\
\bottomrule
\end{tabular}
    \begin{tablenotes}
      \small
\item $^{a}$: Average density within clump radius of $R_{\mathrm{clump}}$. 
\item $^{b}$: Density power-law slope. 
\item $^{c}$: Density at 0.1 pc.
\item $^{d}$: Bolometric luminosity calculated by using the SED profile shown in Figure \ref{fig:sed_radmc} (blue lines for source G18, G19, G08a, G31 and G13). 
\item $^{e}$: Total mass within 0.5 pc from clump center.
 \end{tablenotes}
  \end{threeparttable}
\end{table*}

\begin{table*}
\centering
\begin{threeparttable}
\caption{Best-fit CH$_{3}$OH and CH$_{3}$CCH abundance results of {\tt{LIME}} modeling based on density model from A: {\tt{RADMC}} continuum modeling as listed in Table \ref{tab:radmc_para}; B: manually-adjusted RADEX radial density profile as listed in Table \ref{tab:nprofile_fits}.}
\label{tab:lime_radmcpara}
\begin{tabular}{l c c c c c c c c c c c c c c }
\toprule
   \multicolumn{6}{c}{A}&
   \multicolumn{8}{c}{B}\\
   &
   \multicolumn{3}{c}{CH$_{3}$OH} &  \multicolumn{3}{c}{CH$_{3}$CCH}&\multicolumn{1}{c}{}&\multicolumn{3}{c}{CH$_{3}$OH}&\multicolumn{3}{c}{CH$_{3}$CCH}&\multicolumn{1}{c}{}\\
  Source  &  $X_{\mathrm{out}}$ &  $T_{\mathrm{jump}}$ & $f_{\mathrm{inc}}$$^{a}$ & $X_{\mathrm{out}}$ &  $T_{\mathrm{jump}}$ & $f_{\mathrm{inc}}$& 
$f_{\mathrm{r}}$$^{b}$ & $X_{\mathrm{out}}$ &  $T_{\mathrm{jump}}$ & $f_{\mathrm{inc}}$&  $X_{\mathrm{out}}$ &  $T_{\mathrm{jump}}$ & $f_{\mathrm{inc}}$ & $ff_{\mathrm{dens}}$$^{c}$\\
  & ($\times$10$^{-10}$) & (K) & &($\times$10$^{-10}$) & (K)& &&($\times$10$^{-10}$) & (K) & &($\times$10$^{-10}$)& (K)  &&
  \\
  \midrule
  G28  &10&30&10&10&30&10 &3&1.5&80&150&0.018&30&200&0.13\\
  G19 &  15 &80&10&75&$-$&1&3&1.5&80&100&5&80&30&0.19\\
  G08a & 10&80&20 &100&80&20&3&0.84&80&35&6&$-$&1&0.06\\
  G31 &2.5&80&10&20&80&10 &50&4.0&120&8&12&120&2&0.09\\
  G08b &10 &80&5&100&$-$&1&5&0.6&80&40&5&80&70&0.05\\
  G13 &10&80&5&20&80&10&3&0.67&80&4&1.9&80&3&0.28\\
  G10 &10&80&5&$-$&$-$&$-$&7.5&0.54&40, 100&4, 75&$-$&$-$&$-$&0.05 \\
\bottomrule
\end{tabular}
    \begin{tablenotes}
      \small
\item Marker ``$-$'' denotes parameter invalid or not available.
\item a: $f_{\mathrm{inc}}$ represents increase factor of the abundance jump model (Equation \ref{eq:xmol}).
\item b: Reduction factor applied to RADEX density results in the modeling (Equation \ref{eq:limeB}).
\item c: Dense gas volume filling factor defined in Appendix \ref{app:lime}.
\end{tablenotes}
  \end{threeparttable}
\end{table*}

\subsection{Radial density and temperature profiles used in full radiative transfer models}\label{sec:T_rho_profiles}
We use the {\tt{RADMC-3D}} code (\citealt{radmc3d}) in our full radiative transfer analyses (Figure \ref{fig:flowchart}; Sect. \ref{sec:outline}) for multi-wavelength dust continuum (Appendix E). We assumed that the gas density profile for the bulk gas ($\rho_{\mathrm{bulk}}(r)$) is described by the following functional form:
\begin{equation}
   \rho_{\mathrm{bulk}}(r) = \left\{\begin{array}{rcl} \bar{\rho}\cdot(\frac{r}{r_{\mathrm{c}} })^{q} \,\,\,\,\, (r \le R_{\mathrm{clump}}),  \\
   0  \,\,\,\,\, (r > R_{\mathrm{clump}} ),\label{eq:rhobulk}
\end{array}\right.
\end{equation}
where $\bar{\rho}$ is the mean hydrogen gas number density, and $r_{\mathrm{c}}$ is the radius where $\rho(r)=\bar{\rho}$, $R_{\mathrm{clump}}$ is the assumed outer radii of the clumps which were fixed to the FHWM measured from the ATLASGAL 870 $\mu m$ maps (c.f., \citealt{Contreras13}).
When converting gas density to mass density, we assume that the mass per hydrogen molecule is 2.8$\,m_{\mathrm{H}}$, where $m_{\mathrm{H}}$ is the hydrogen atom mass.
We assumed that the gas-to-dust mass ratio is 100.

We parameterised the measured temperature profiles $T(r)$ by
\begin{equation}
 T(r) = \omega T_{\mathrm{in}}(\frac{r}{r_{\mathrm{in}}})^{-0.5}+(1-\omega)T_{\mathrm{out}}, \label{eq:tformula}
\end{equation}
where $\omega\,=\,e^{-\frac{r}{r_{\mathrm{out}}}}$ is an exponential tapering function characterized by outer radius $r_{\mathrm{out}}$; $T_{\mathrm{in}}$ and $T_{\mathrm{out}}$ are the characteristic temperatures at the radius $r_{\mathrm{in}}$ and at asymptotically large radii, respectively.
In this equation, the first term describes radiative heating by the centrally embedded stars while the second term can be attributed to the ambient radiation fields of the massive clumps (\citealt{Liu18}).
The multiplicative factors $\omega$ and $(1-\omega)$ prescribe the transition from one heating regime to the other.

\begin{figure*}
    \begin{tabular}{p{0.48\linewidth}p{0.48\linewidth}}
    \includegraphics[scale=0.28]{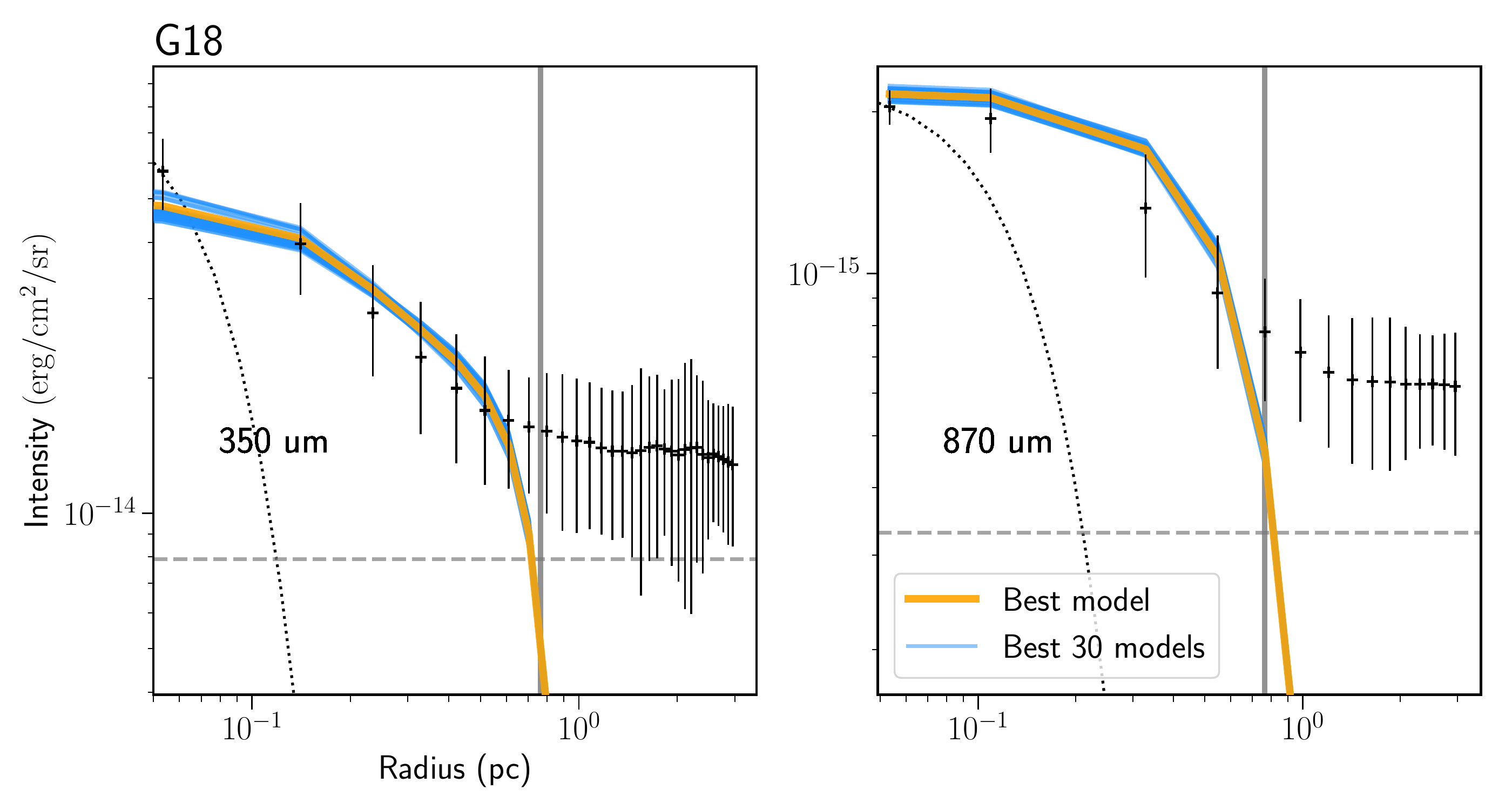}&
    \includegraphics[scale=0.28]{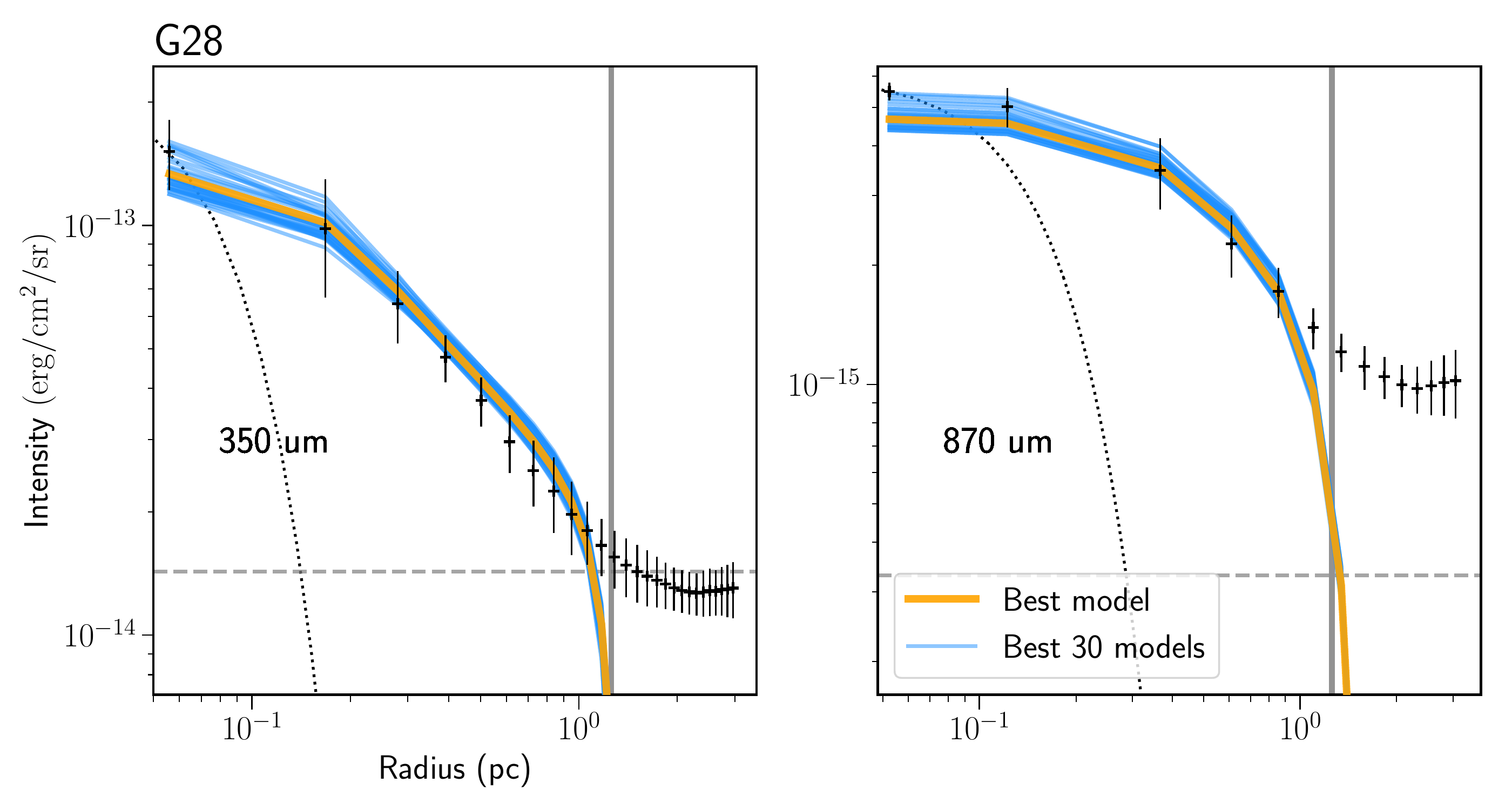}\\
    \includegraphics[scale=0.28]{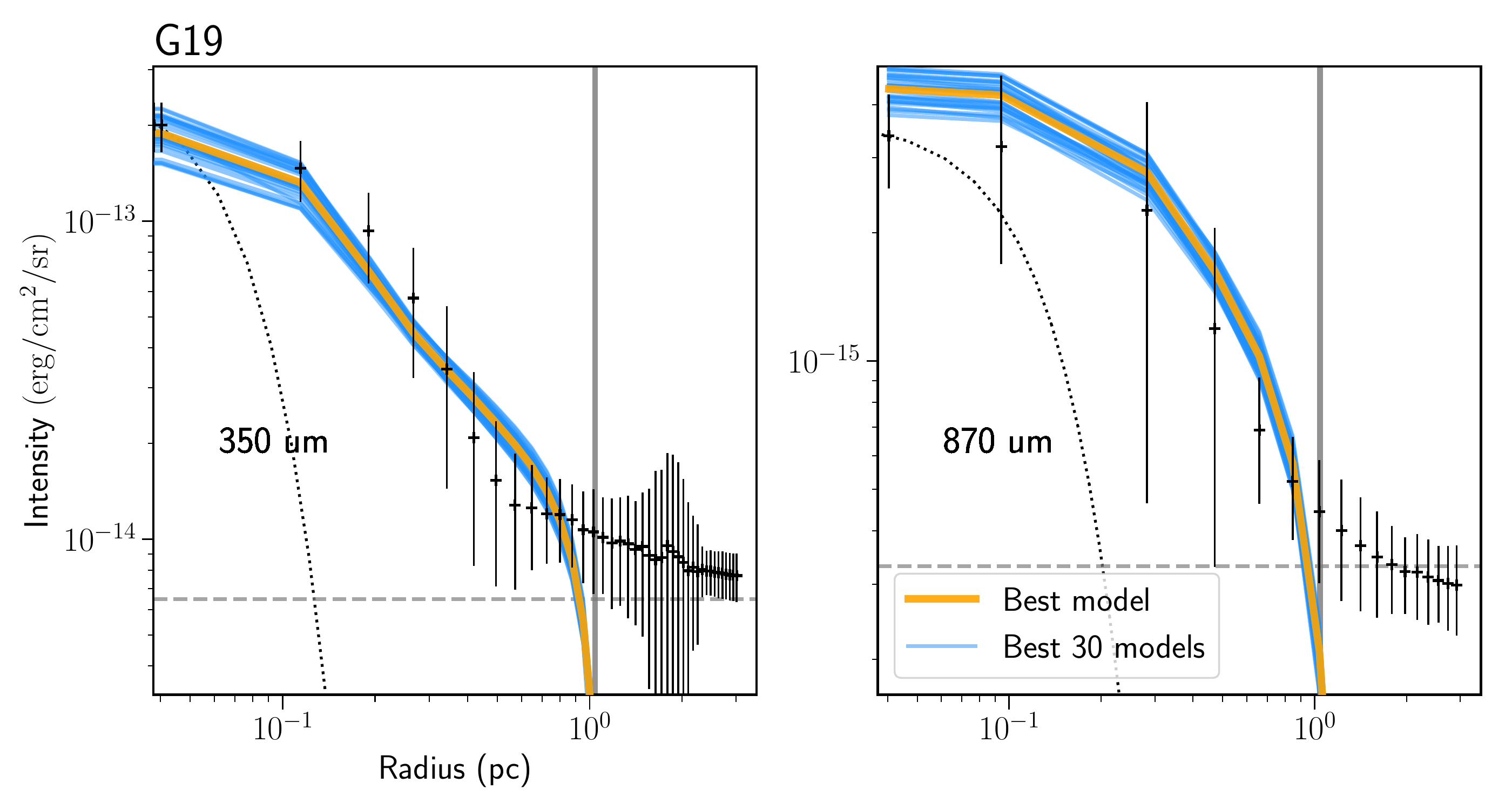}&
    \includegraphics[scale=0.28]{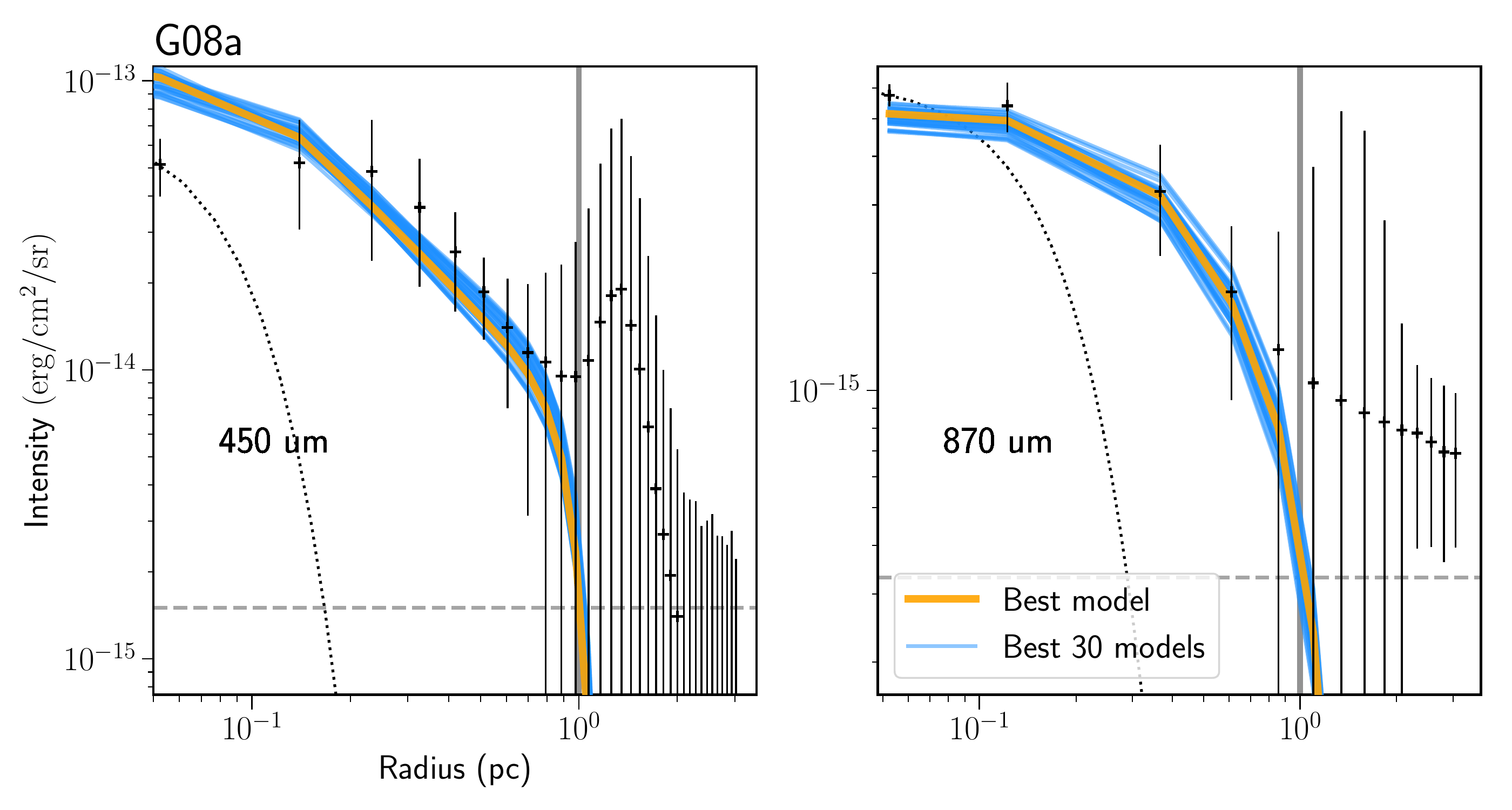}\\
    \includegraphics[scale=0.28]{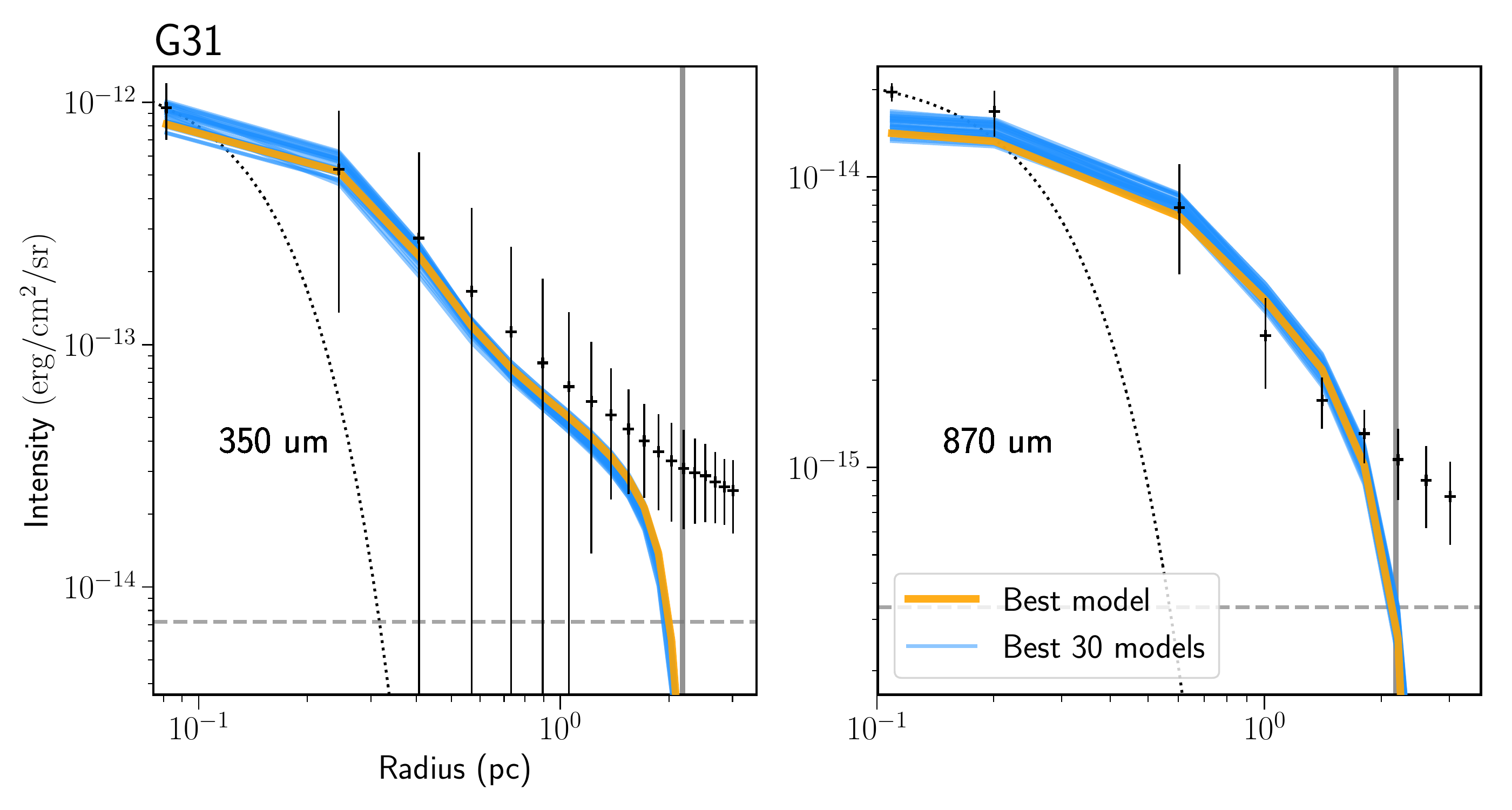}&
    \includegraphics[scale=0.28]{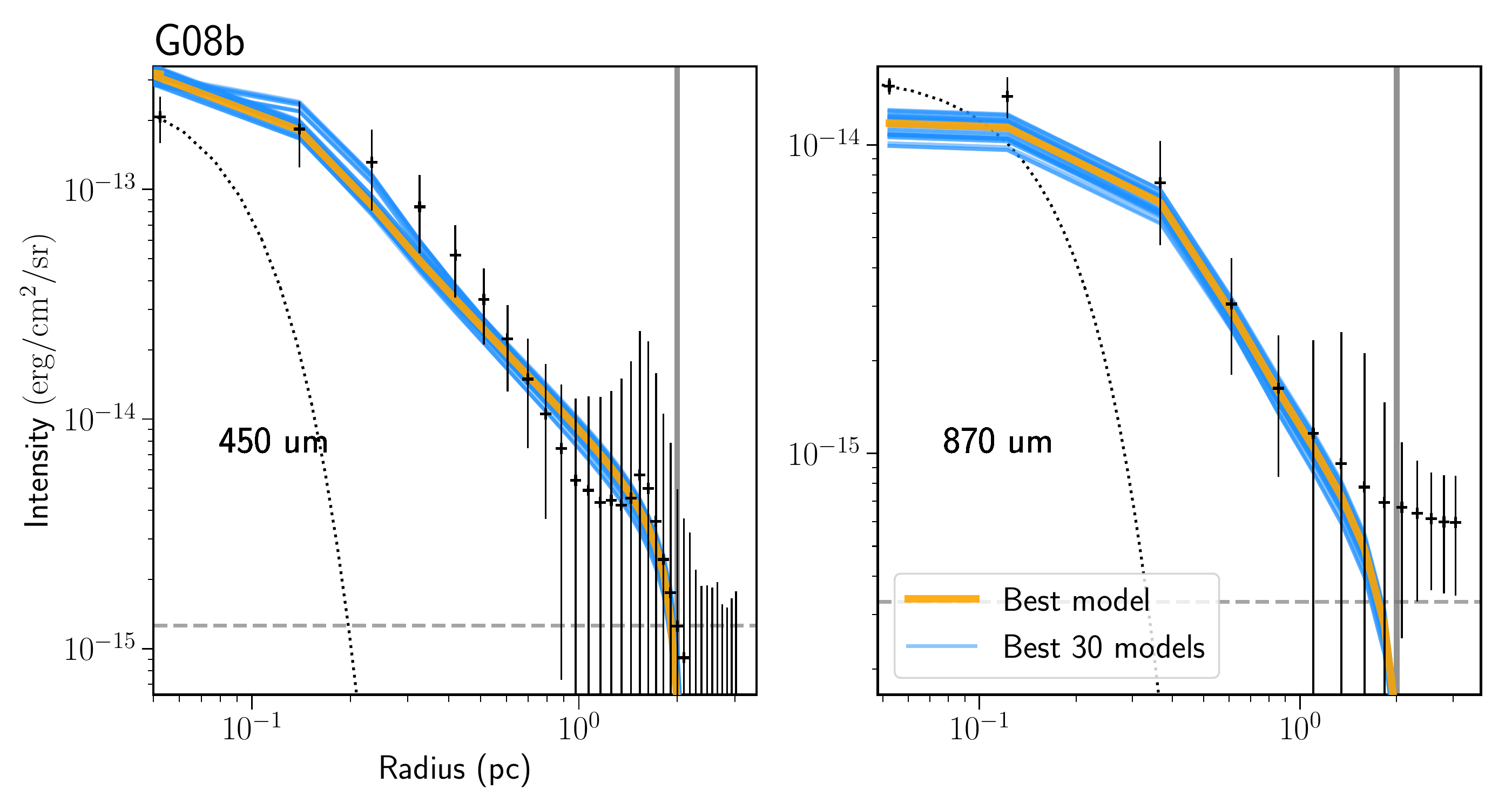}\\
   \includegraphics[scale=0.28]{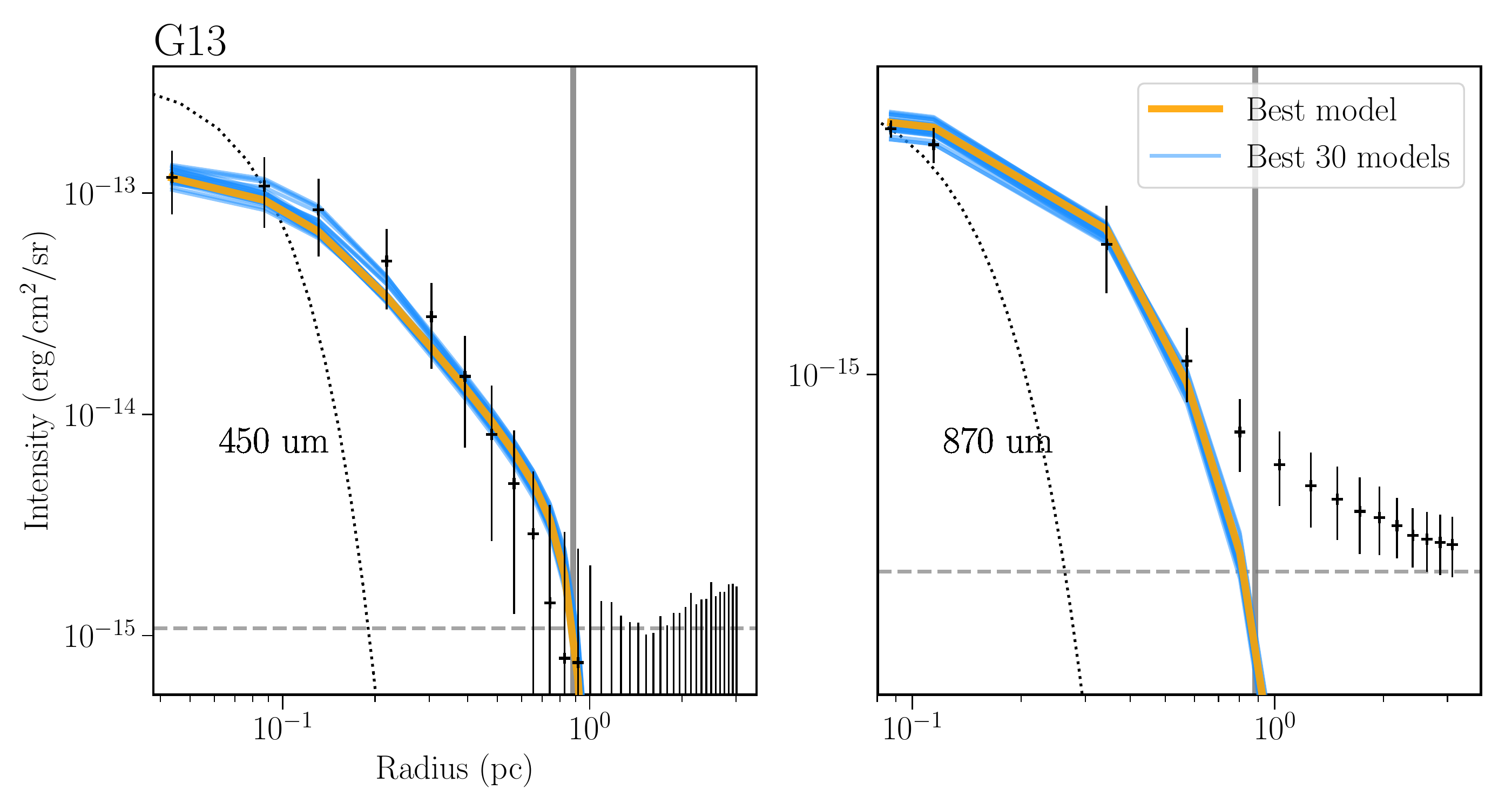}&
   \includegraphics[scale=0.28]{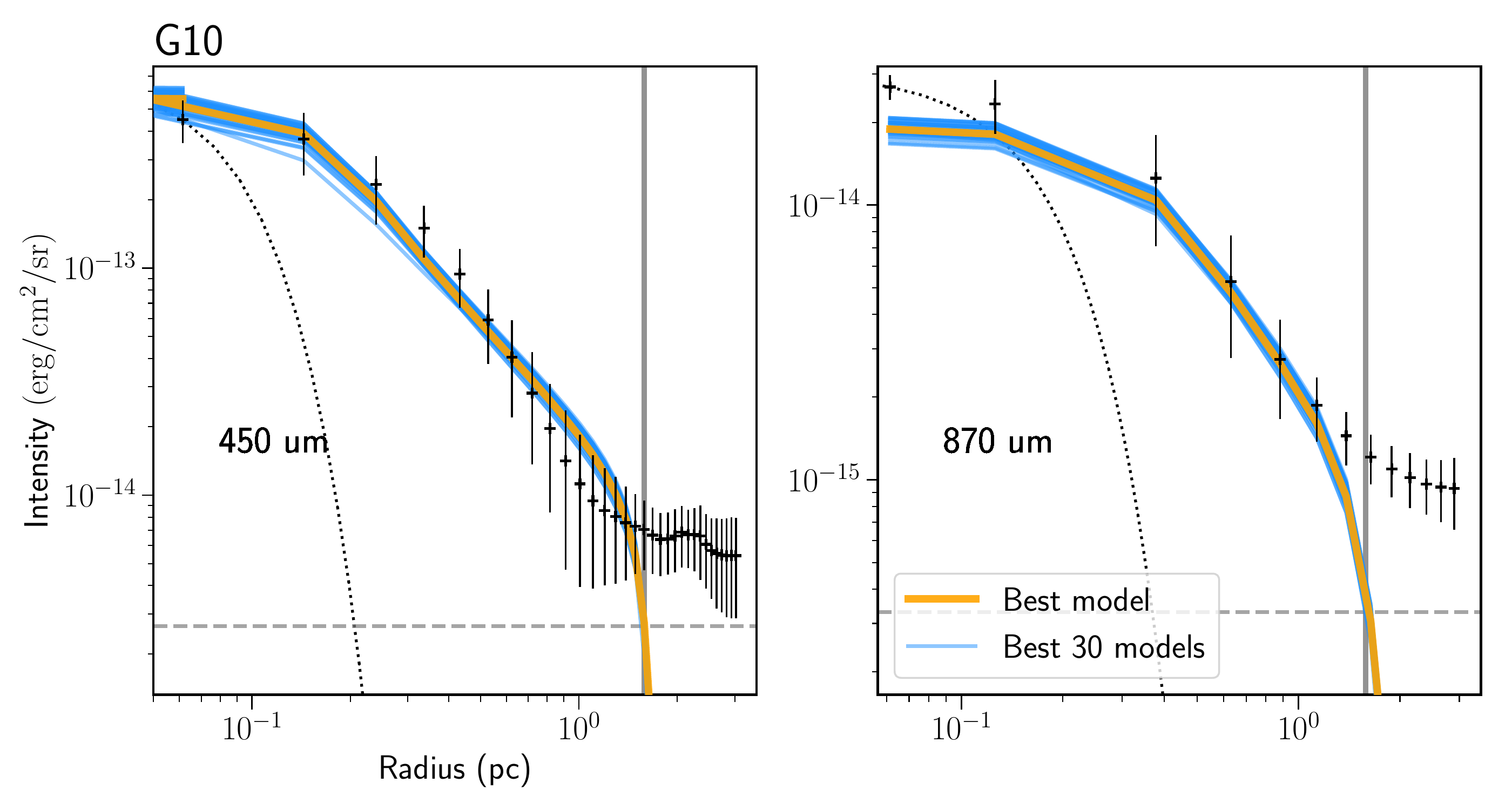}\\
    \end{tabular}
     \centering\caption{Radial intensity profile comparisons between observations and best-fit {\tt{RADMC-3D}} models. Gray horizontal dashed lines indicate the noise level (3$\sigma$). Gray vertical lines indicate the clump radius used in the modeling. Dotted line indicates beam shape in each plot. For source G18, G19, G08a, G13 and G31, model fit after re-adjusting $T(r)$ is shown. The gray vertical line indicates the clump radius $R_{\mathrm{clump}}$.}
    \label{fig:radmc1}
\end{figure*}

\begin{figure*}
    \centering
    \hspace{-1cm}\includegraphics[scale=0.43]{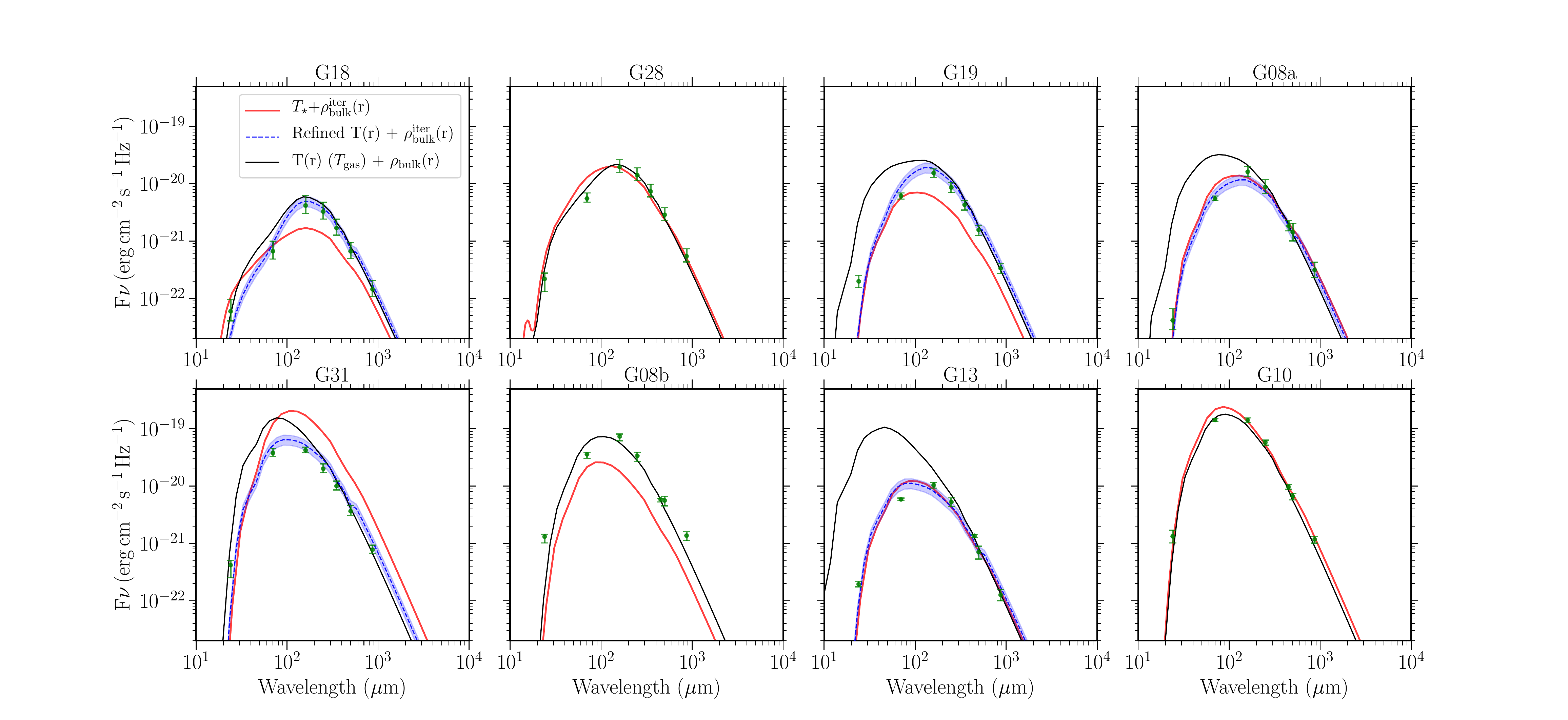}
    \caption{Comparison of SEDs of the best-fit {\tt{RADMC-3D}} models with measured multi-wavelength fluxes (in green dots with error bars indicating 0.8 and 1.2 times the flux level) for each source. Black line indicates the SED generated from assumed $T(r)$ and the corresponding best density profile fits. Blue dashed line indicates the SED generated from refined $T(r)$ and the re-iterated best density profile fits. Blue shaded regions indicates 20$\%$ difference around the blue dashed SED profile. Red line shows the SED generated by self-consistently calculating the dust temperature adopting a central heating ZAMS star plus the re-iterated best density profile (for more details see Appendix \ref{app:radmc}).}
    \label{fig:sed_radmc}
\end{figure*}

Based on the multiple rotational temperature maps, we derived the (projected) radially averaged temperature profile and obtained best-fit parameters $T_{\mathrm{in}}$, $T_{\mathrm{out}}$ and $r_{\mathrm{out}}$, while $r_{\mathrm{in}}$ is initially kept as a fiducial value of 0.02 pc. 
Figures \ref{fig:radmc1} and \ref{fig:sed_radmc} show the comparison between these multi-wavelength radial intensity profiles and the SEDs evaluated from best-fit {\tt{RADMC-3D}} (\citealt{radmc3d}) models. 
Based on this comparison, we re-adjust the $r_{\mathrm{in}}$ of Equation \ref{eq:tformula} in the {\tt{RADMC-3D}} modeling to obtain an SED profile consistent with the observed data points.  $T(r)$ is then updated by the refined temperature profile. The parameters $T_{\mathrm{in}}$, $T_{\mathrm{out}}$, $r_{\mathrm{in}}$ and $r_{\mathrm{out}}$ that define $T(r)$ are listed in Table \ref{tab:nprofile_fits}. With $T(r)$ defined, we fix the dust temperature profile in the multi-wavelength continuum modeling for the bulk gas and obtain $\rho_{\mathrm{bulk}}(r)$; the best-fit parameters $\bar\rho$ and $q$ as in Equation \ref{eq:rhobulk} are listed in Table \ref{tab:radmc_para}.

\begin{figure}[htb]
\begin{tabular}{p{0.5\linewidth}p{0.5\linewidth}}
\hspace{-0.45cm}\includegraphics[scale=0.31]{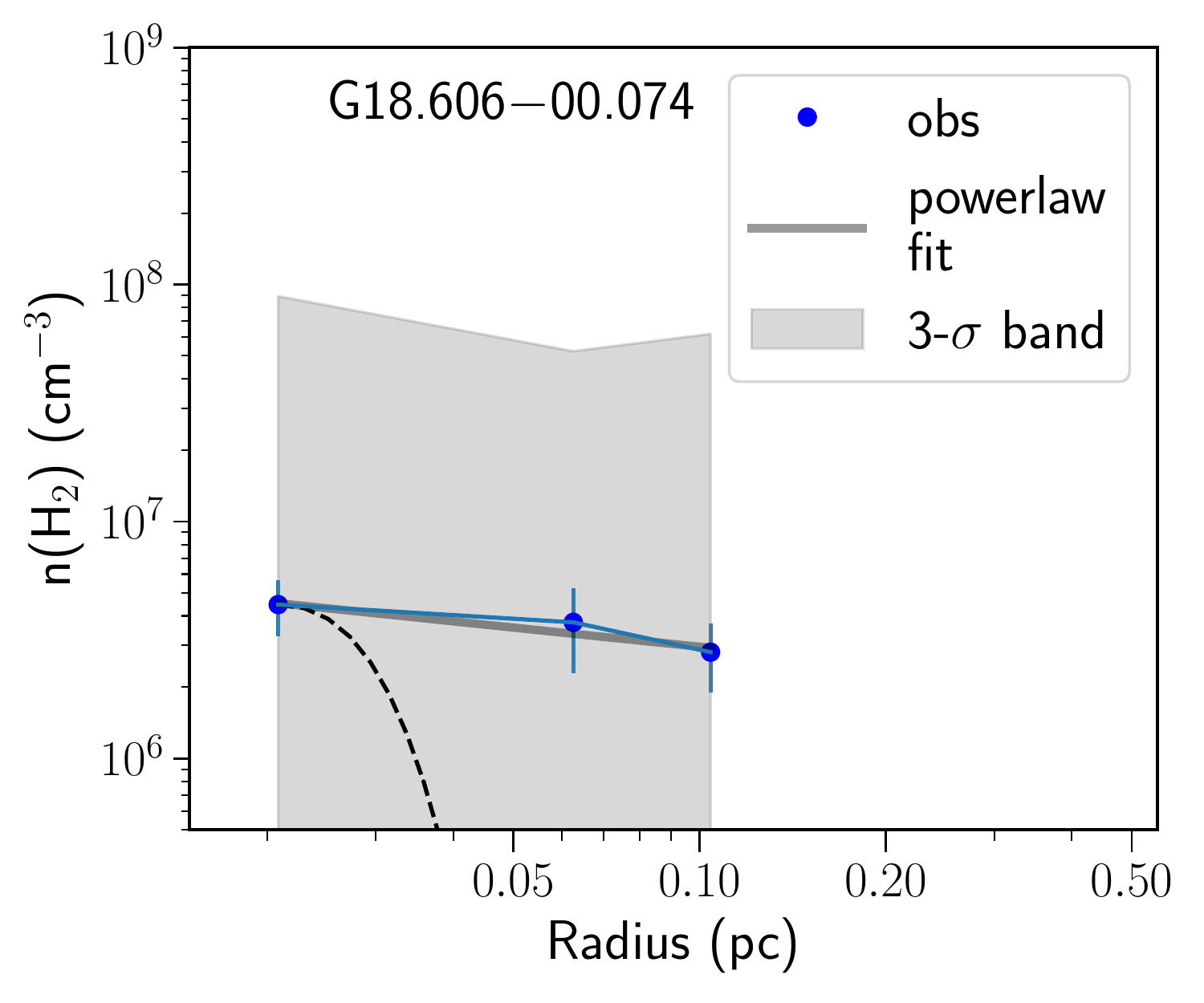}&\hspace{-0.55cm}\includegraphics[scale=0.31]{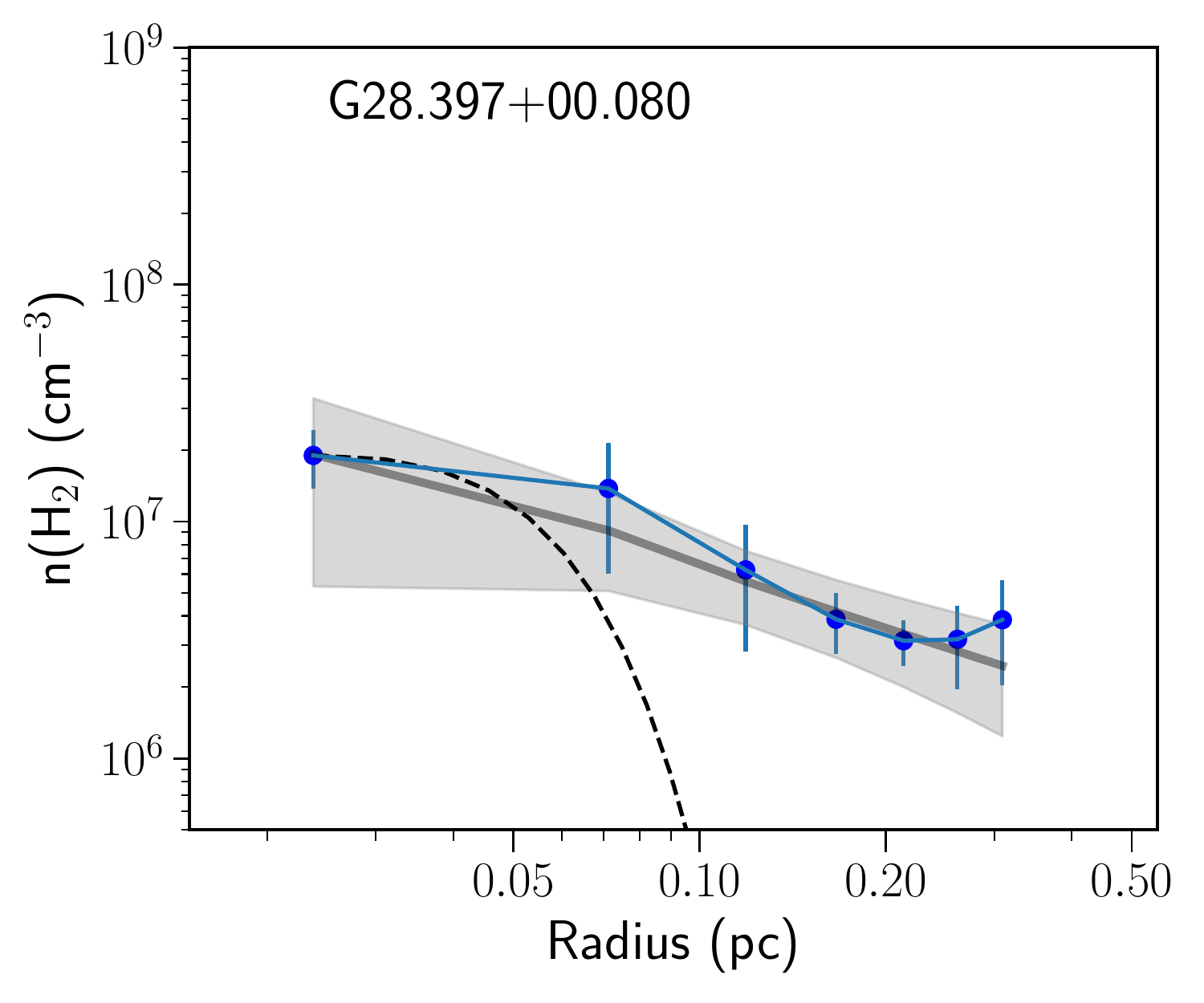}\\
\hspace{-0.45cm}\includegraphics[scale=0.31]{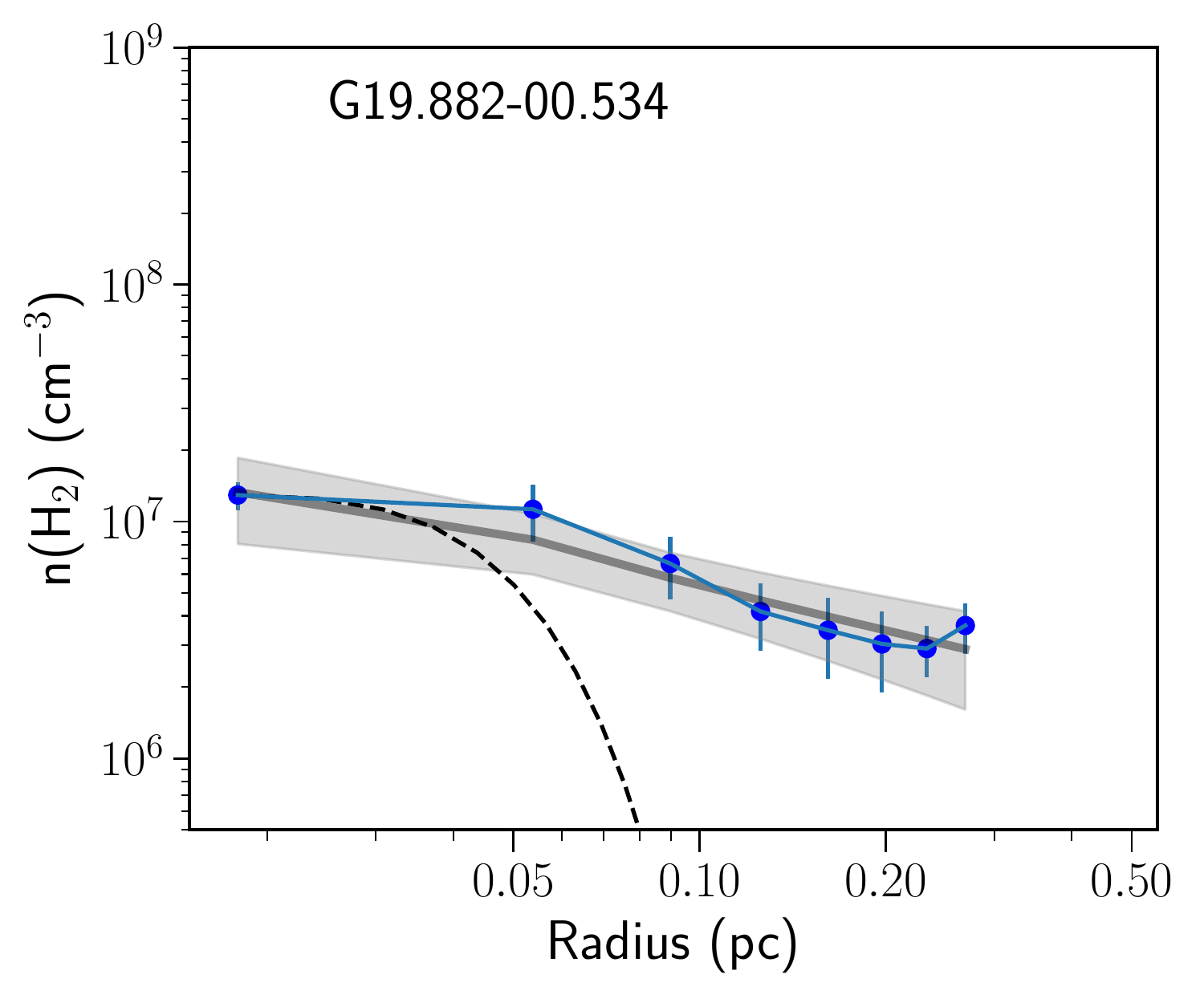}&\hspace{-0.5cm}\includegraphics[scale=0.31]{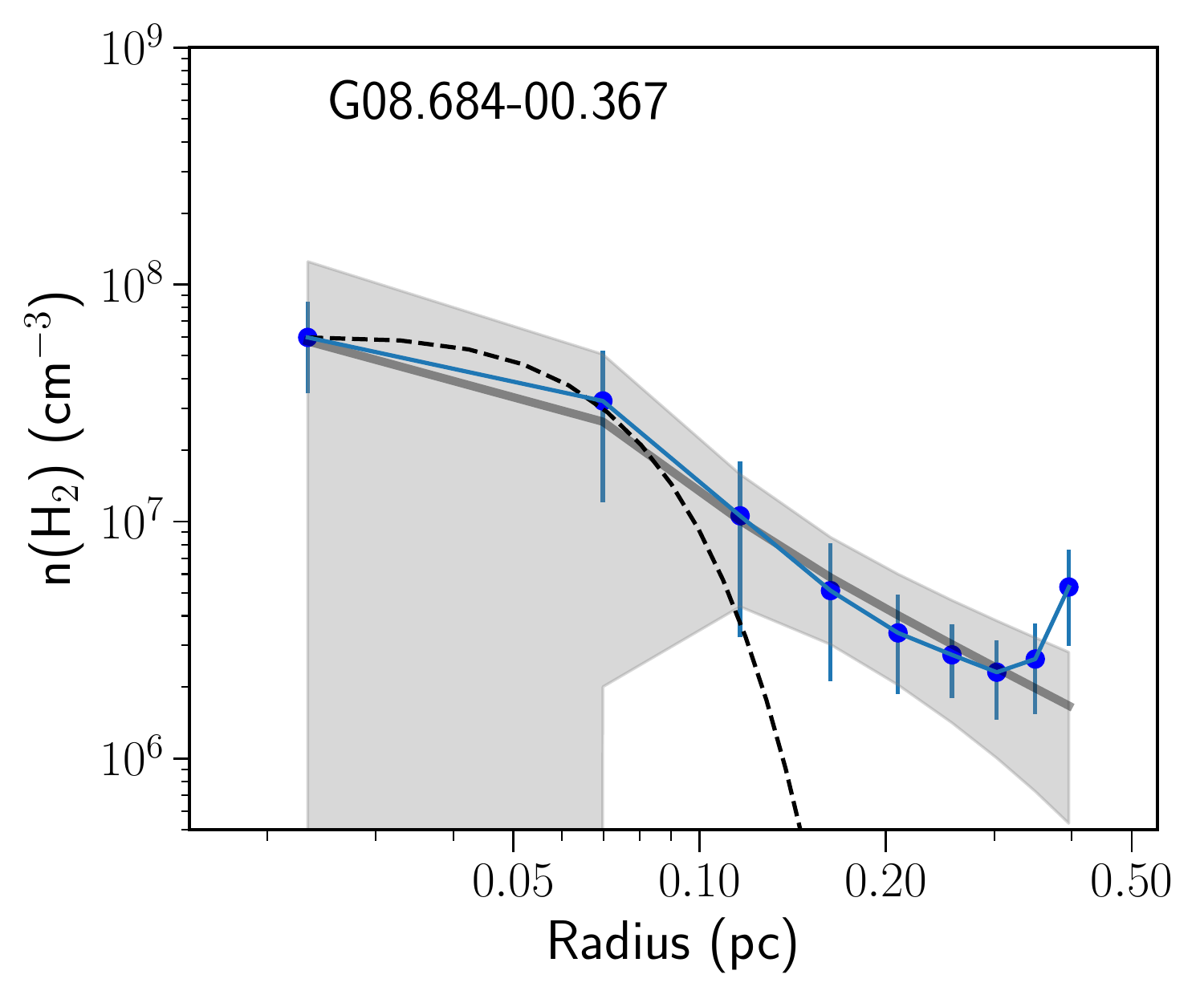}\\
\hspace{-0.45cm}\includegraphics[scale=0.31]{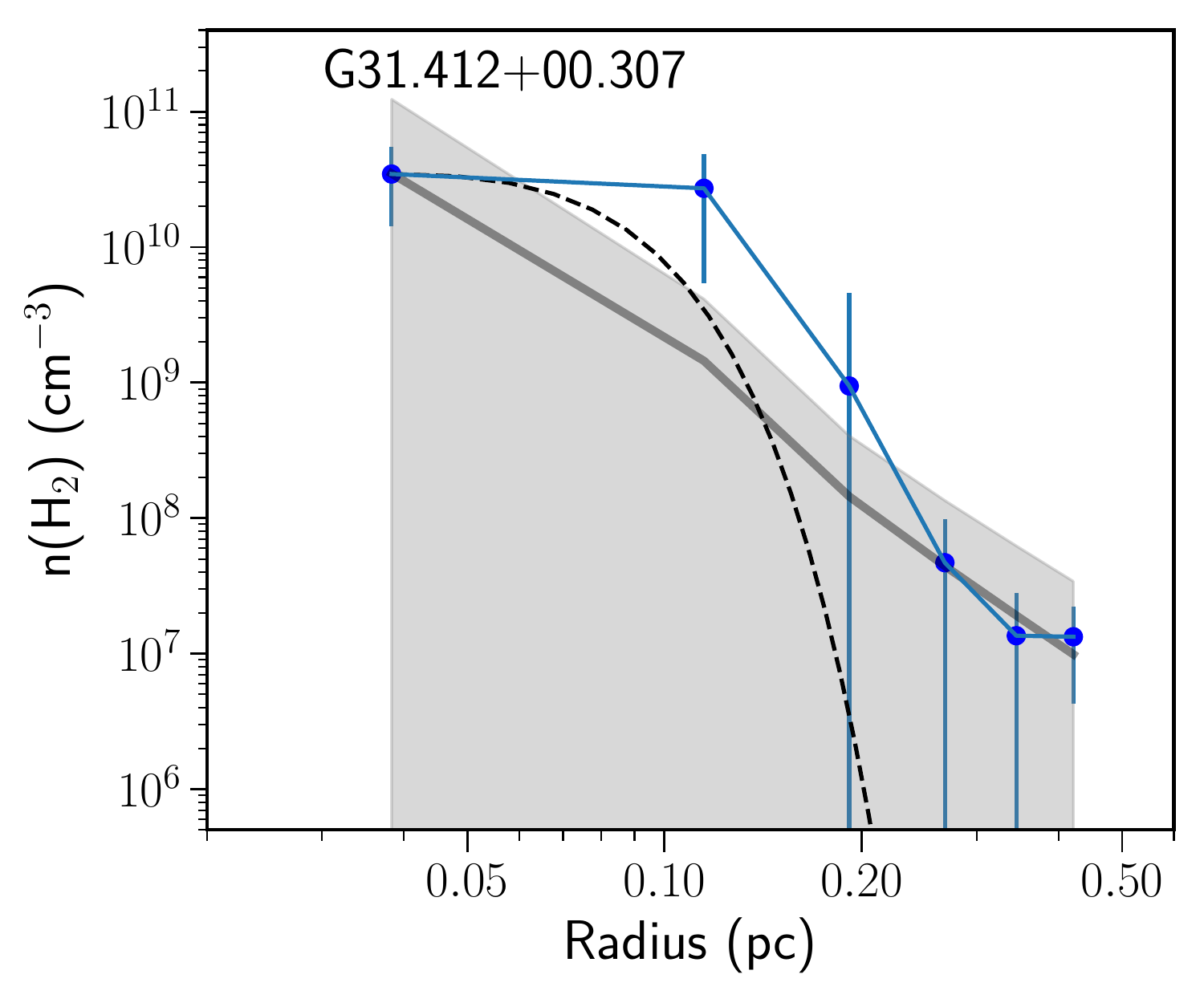}&\hspace{-0.5cm}\includegraphics[scale=0.31]{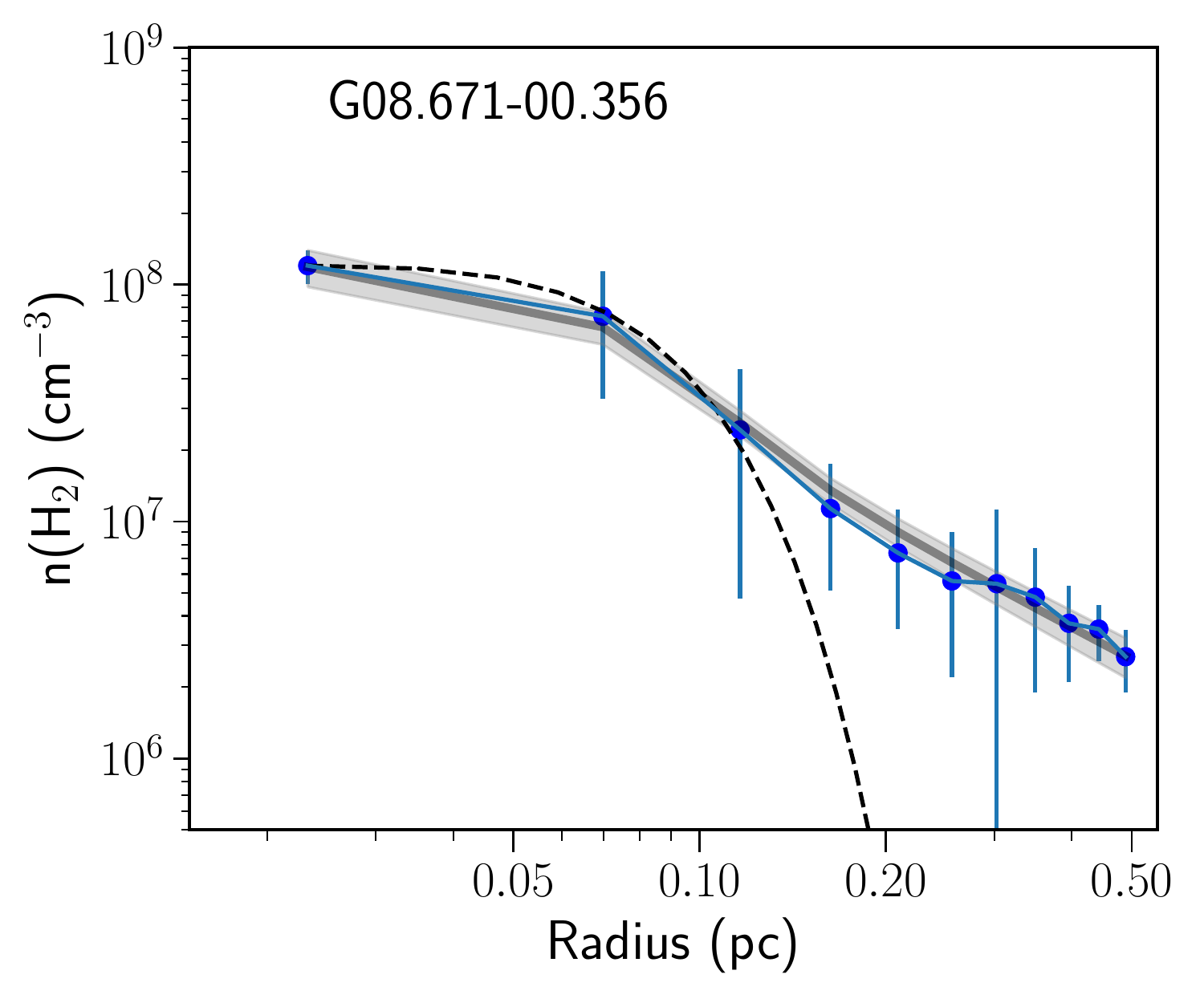}\\
\hspace{-0.45cm}\includegraphics[scale=0.31]{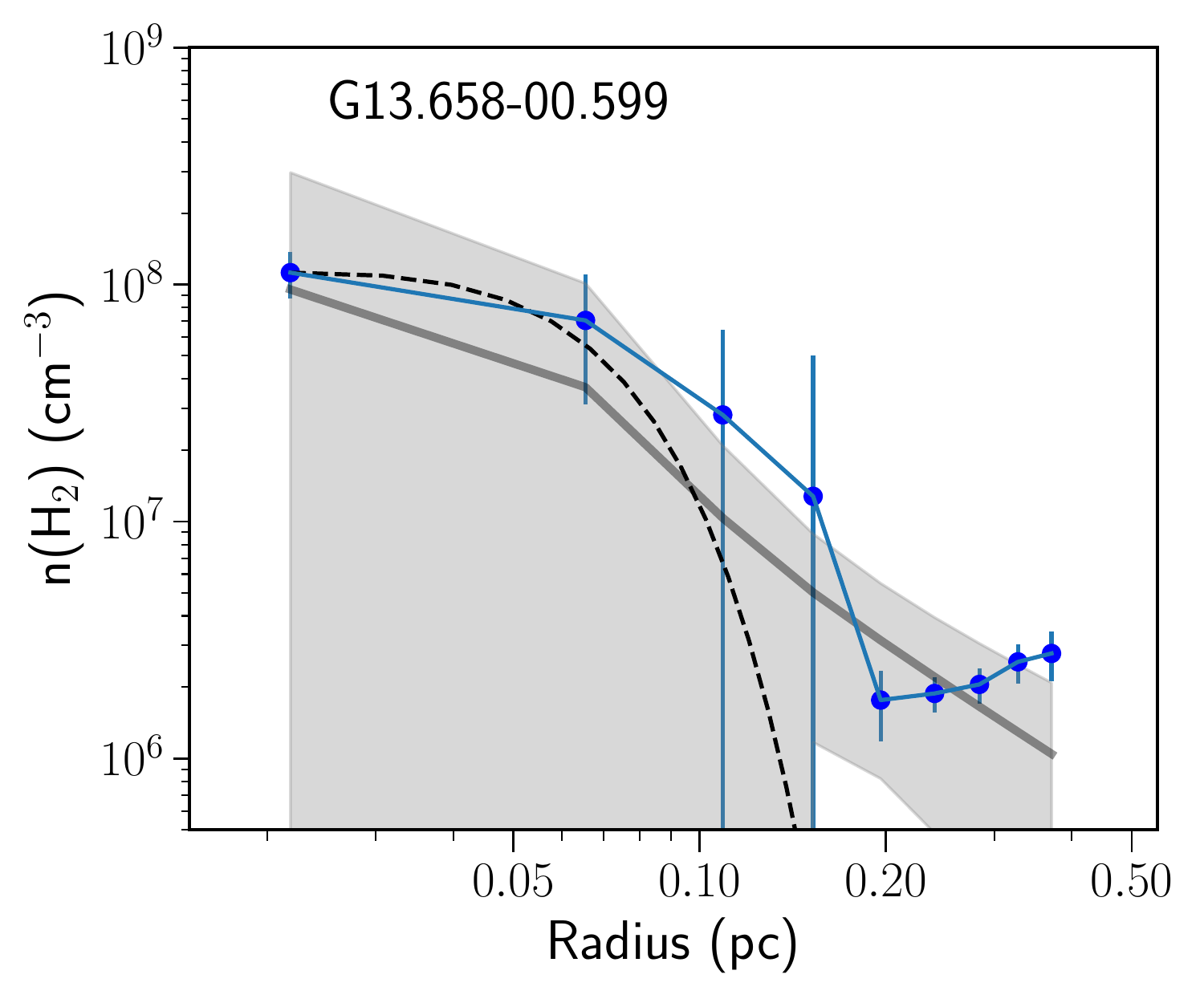}&\hspace{-0.6cm}\includegraphics[scale=0.31]{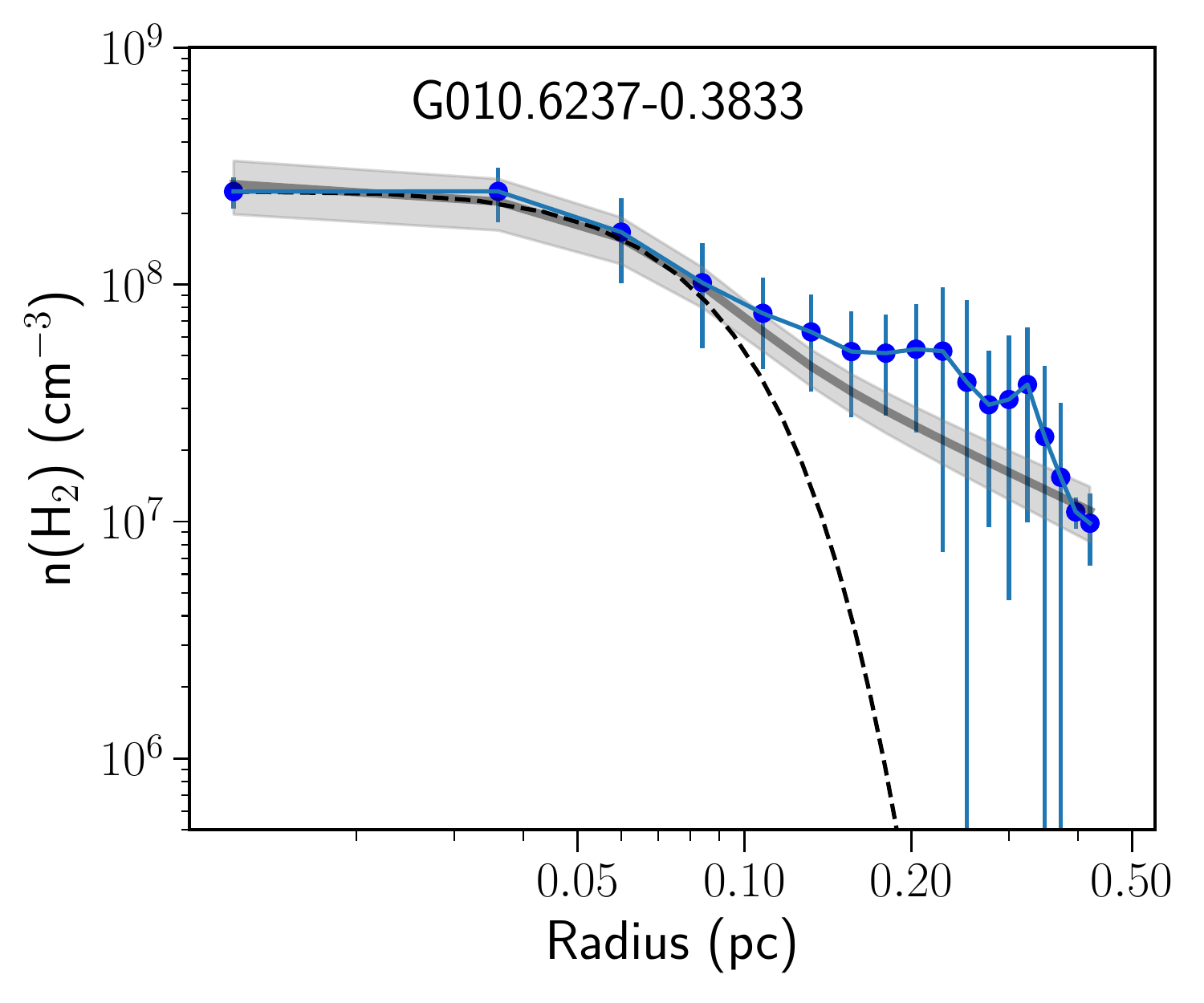}\\
\end{tabular}
\centering\caption{Projected radial averaged $n(\mathrm{H_{\mathrm 2}})$ radial profiles derived from $n(\mathrm{H_{\mathrm 2}})$  maps shown in Figure \ref{fig:nmaps}. Thick gray line indicates the best-fit single power-law model (beam convolution considered). Gray shadowed band indicates the 3$\sigma$ confidence band of the best-fit model. The model parameters and 1$\sigma$ errors are listed in Table \ref{tab:nprofile_fits}.}
\label{fig:nprofiles}
\end{figure}

From RADEX modeling of CH$_{3}$OH lines we constrain the radial density profiles for the dense gas, $\rho_{\mathrm{dense}}(r)$, from the obtained $n(\mathrm{H_{\mathrm 2}})$ maps. Similarly, when deriving the $n(\mathrm{H_{\mathrm 2}})$ maps (Figure \ref{fig:nmaps}) we fix the gas kinetic temperature in the modeling to $T(r)$ for each pixel. We adopt a single power-law form as Equation \ref{eq:rhobulk} to characterize the dense gas density profiles, as
\begin{equation}
    \rho_{\mathrm{dense}}(r) = \rho_{\mathrm{0\,0.1 pc}}(\frac{r}{0.1 pc})^{q_{\mathrm{radex}}},\label{eq:rhoradex}
\end{equation}
where $\rho_{\mathrm{0\,0.1 pc}}$ is the reference gas density at 0.1 pc. The description is valid up to a maximum scale of $R_{\mathrm{max}}$, which is determined from the largest radius where $n(\mathrm{H_{\mathrm 2}})$ can be robustly estimated. These parameters are also listed in Table \ref{tab:nprofile_fits}. Figure \ref{fig:nprofiles} shows the comparison between the model fits and the observed radial profiles.

\begin{figure*}[htb]
\begin{tabular}{p{0.95\linewidth}}
\hspace{-2.75cm}\includegraphics[scale=0.32]{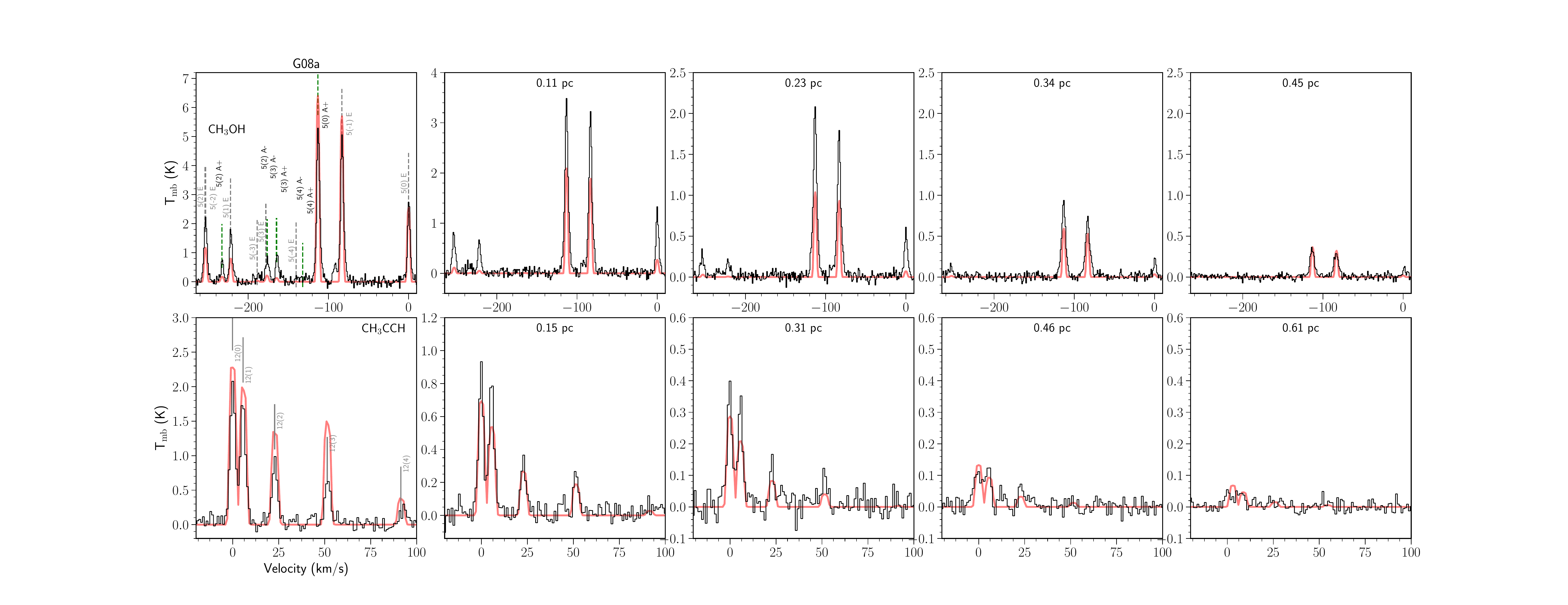}\\
\hspace{-2.75cm}\includegraphics[scale=0.32]{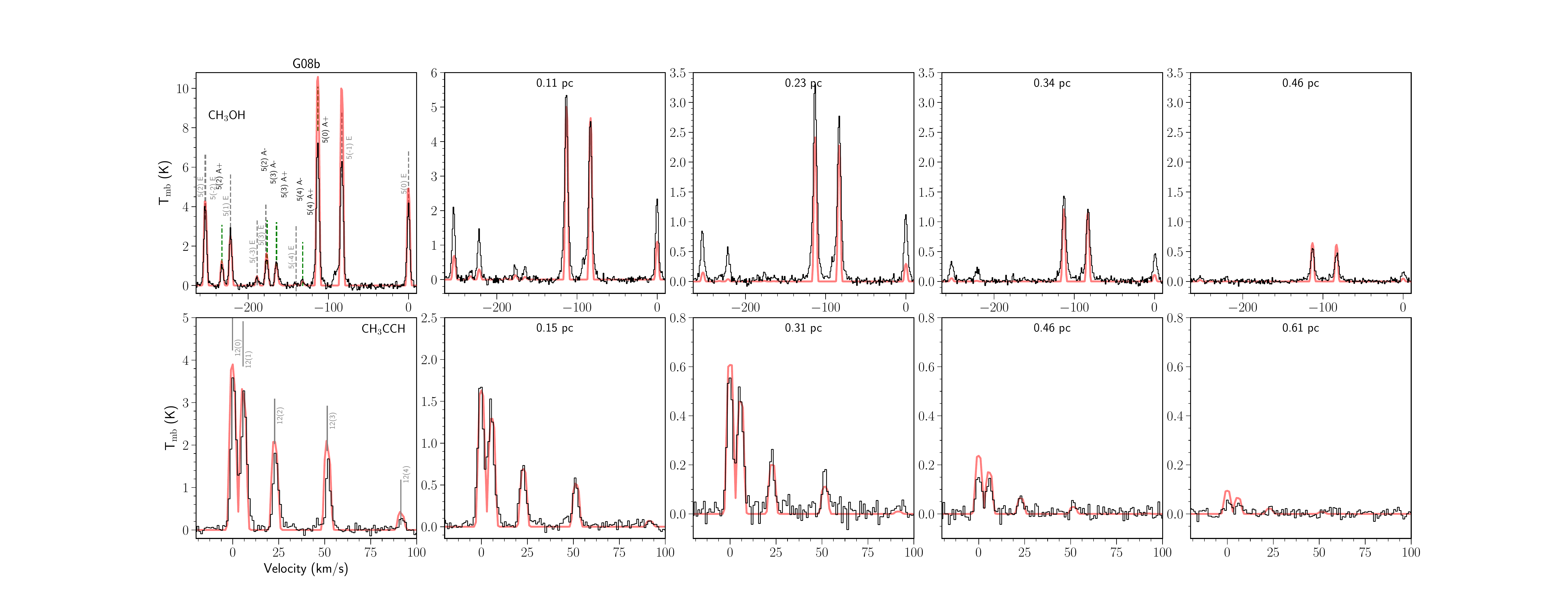}
\end{tabular}
\centering\caption{{\tt{LIME}} modeling result (best-fit parameters listed in Table \ref{tab:lime_radmcpara}, column A) based on best-fit density model from {\tt{RADMC-3D}} continuum modeling. From {\it{left}} to {\it{right}}: annular beam-averaged spectra from the continuum center to the outer envelope. Considering the typical beam FWHM of our observations: the distance from the center of each annular region to the center of the source is marked on top of each spectra. The line components of $A$ and $E$-type CH$_{3}$OH are indicated with short dashed vertical lines in green and gray, respectively.}
\label{fig:lime_radmc_dens}
\end{figure*}

\begin{figure*}[htb]
\begin{tabular}{p{0.95\linewidth}}
\hspace{-2.75cm}\includegraphics[scale=0.3]{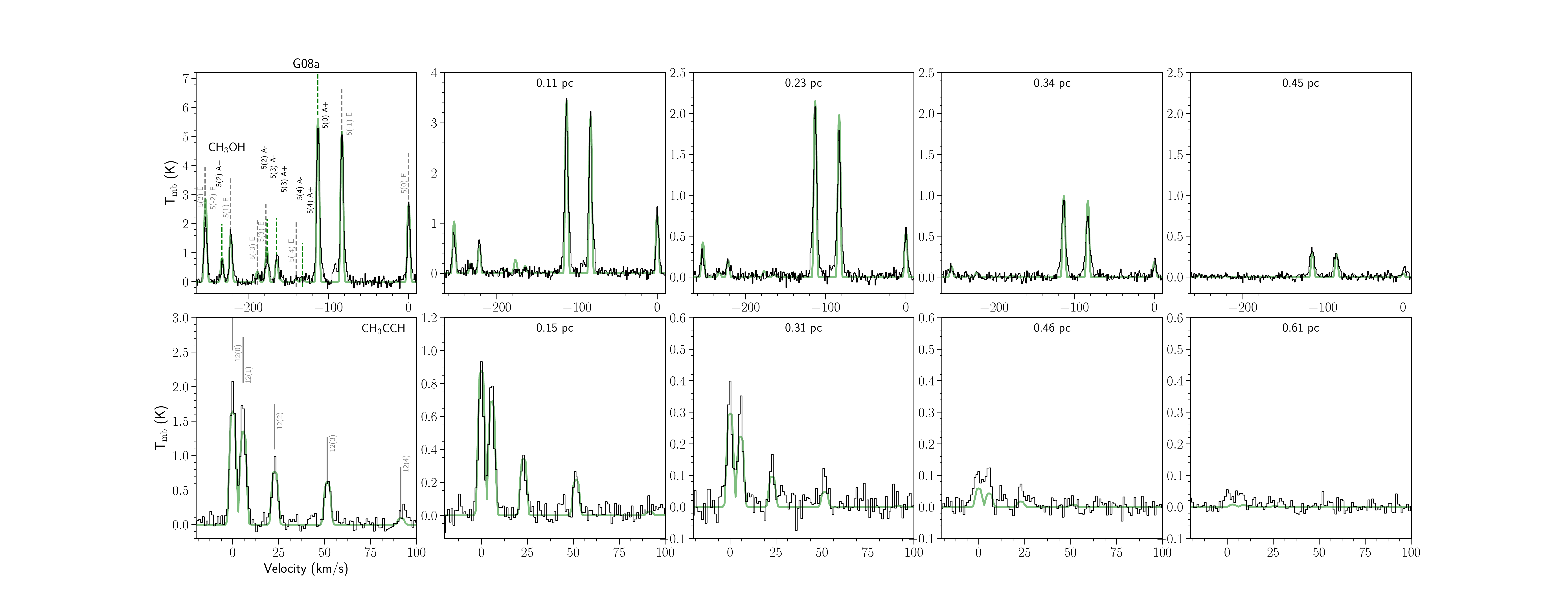}\\
\hspace{-2.75cm}\includegraphics[scale=0.3]{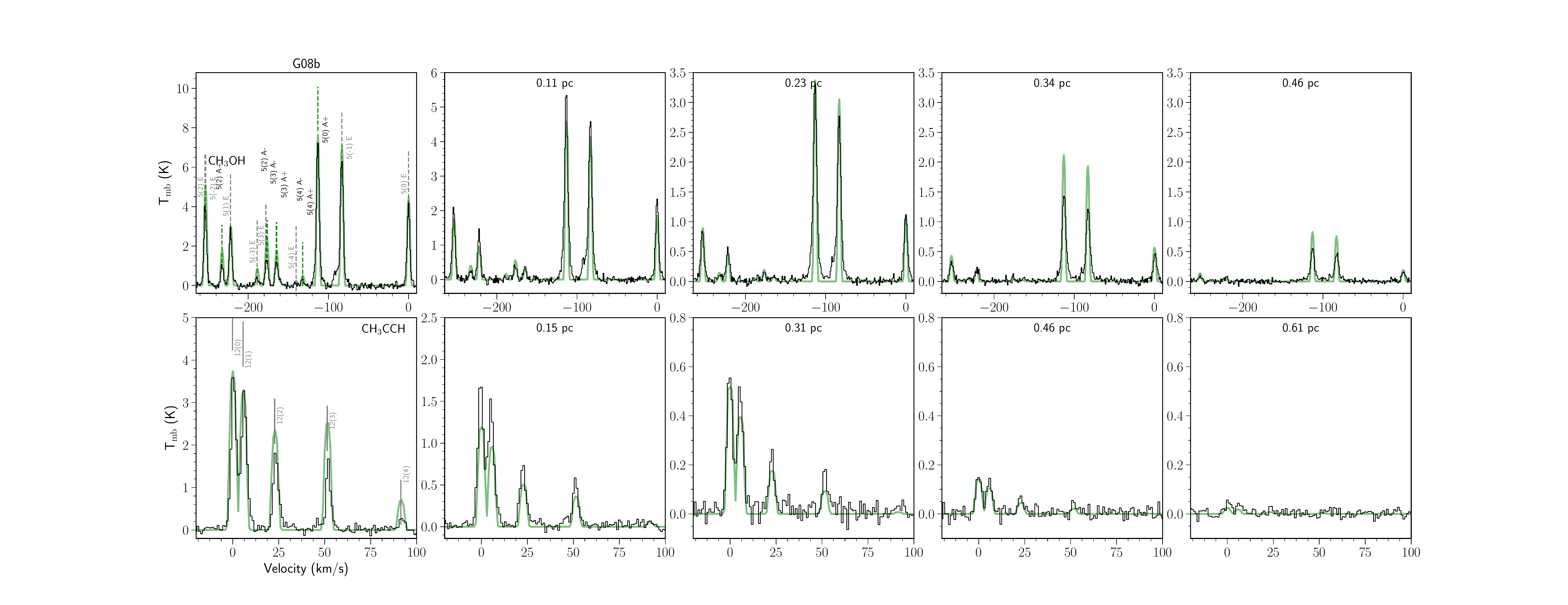}\\
\end{tabular}
\centering\caption{{\tt{LIME}} modeling result (best-fit parameters listed in Table \ref{tab:lime_radmcpara}, column B) after manually adjusting the density profile obtained from RADEX modeling (Eq. \ref{eq:rhoradex}), which is prescribed as a piecewise power-law (Eq. \ref{eq:limeB}). From {\it{left}} to {\it{right}}: annular beam-averaged spectra from the continuum center to the outer envelope. Considering the typical beam FWHM of our observations: the distance from the center of each annular region to the center of the source is marked on top of each spectra.}
\label{fig:lime_radmc_dens_radexadj}
\end{figure*}

We then conduct full radiative transfer modeling with {\tt{LIME}} (\citealt{Brinch10}) to benchmark and refine these results. In the {\tt{LIME}} modeling of CH$_{3}$OH and CH$_{3}$CCH lines, we first adopted the bulk gas density profile $\rho_{\mathrm{bulk}}(r)$ constrained from single-dish dust continuum modeling, and $T(r)$ with assumed abundance profiles to find the best-fit models. 
We parameterized the molecular abundance profiles ($X_{\mathrm{mol}}(r)$) as:
\begin{equation}\label{eq:xmol}
    X_{\mathrm{\mathrm{mol}}}(r) = \left\{
\begin{array}{rcl}
X_{\mathrm{out}} && (T(r) < T_{\mathrm{jump}})\\
X_{\mathrm{in}} \equiv f_{\mathrm{inc}}X_{\mathrm{out}} && (T(r) > T_{\mathrm{jump}}),\\
\end{array} \right.
\end{equation}
where $T_{\mathrm{jump}}$ is a threshold temperature chosen to be either 30 or 80 K, $X_{\mathrm{out}}$ is the abundance at outer radii, and $f_{\mathrm{inc}}$ is an increment factor to characterize the abundance enhancement in inner regions of higher temperature. This form is driven by previous chemical models of CH$_{3}$OH and CH$_{3}$CCH, in which prominent abundance enhancement is seen around the two $T_{\mathrm{jump}}$ temperatures (see also Appendix \ref{app:lime}).
The best-fit parameters $T_{\mathrm{jump}}$, $X_{\mathrm{out}}$ and $f_{\mathrm {inc}}$ for this model (hereafter model A) are listed in Table \ref{tab:lime_radmcpara} (column A). 
For all sources, we find that with the assumed density profile of $\rho_{\mathrm{bulk}} (r)$ the models cannot reproduce the observed high ratios between the higher and lower $K$ components of CH$_{3}$OH lines, as shown in Figure \ref{fig:lime_radmc_dens} (presenting the comparisons between modelled results and observations towards clump G08a and G08b), which points to, as also indicated from the RADEX modeling results, a much higher gas density regime from which these CH$_{3}$OH higher $K$ components originate.
Therefore, we complemented the {\tt{LIME}} models with density profiles of the dense gas component $\rho_{\mathrm{dense}}(r)$ as in Equation \ref{eq:rhoradex} (the RADEX results of gas density radial profiles), following:
\begin{equation}\label{eq:limeB}
    \rho_{\mathrm{dense}} (r) = \left\{
\begin{array}{rcl}
\frac{1}{f_{\mathrm r}}\rho_{0} && (r < r_{0}) \\ 
\frac{1}{f_{\mathrm r}}\rho_{0}(r/r_{0})^{q_{\mathrm{radex}}} && (r_{0}< r < R_{\mathrm{max}}),\\
\end{array} \right. 
\end{equation}
where $r_{\mathrm{0}}$ denotes reference radius of 0.1 pc, or 0.05 pc (for G13 and G31); $\rho_{\mathrm{0}}$ is the reference density at $r_{\mathrm{0}}$. These values, together with $R_{\mathrm{max}}$ were derived by RADEX modeling. Here $f_{\mathrm{r}}$ is a reduction factor applied to $\rho_{0}$. This parameter is empirically added to the density profile so that the {\tt{LIME}} models better match with the observed data. In essence, it means that RADEX results of one-component non-LTE modeling tend to overestimate the (projected) radial density in a 3D-structure clump. We manually adjusted $f_{\mathrm{r}}$ and $X_{\mathrm{mol}}(r)$ to seek for better fits to the observational data. In what remains, we refer to $q_{\mathrm{radex}}$ as $q_{\mathrm{dense}}$ as this slope is fitted based on $n(\mathrm{H_{\mathrm 2}})$ maps of CH$_{3}$OH RADEX modeling and retained as the slope for the dense gas profile in full radiative transfer {\tt{LIME}} models. 
The best-fit model parameters $T_{\mathrm{jump}}$, $X_{\mathrm{out}}$, $f_{\mathrm{inc}}$ and $f_{\mathrm r}$ for this model (hereafter model B) are summarized in Table \ref{tab:lime_radmcpara} (column B).
Figure \ref{fig:lime_radmc_dens_radexadj} shows a comparison between the CH$_{3}$OH line profiles reproduced from model B and the observations towards clump G08a and G08b. The comparisons between model A, model B and observations for other target sources are shown in Appendix \ref{app:other_sps}.

\subsection{A comparison of the samples: density and temperature structures}\label{sec:comp_rough}
\begin{figure*}[htb]
\begin{tabular}{p{0.5\linewidth}p{0.5\linewidth}}
\hspace{-0.5cm}\includegraphics[scale=0.52]{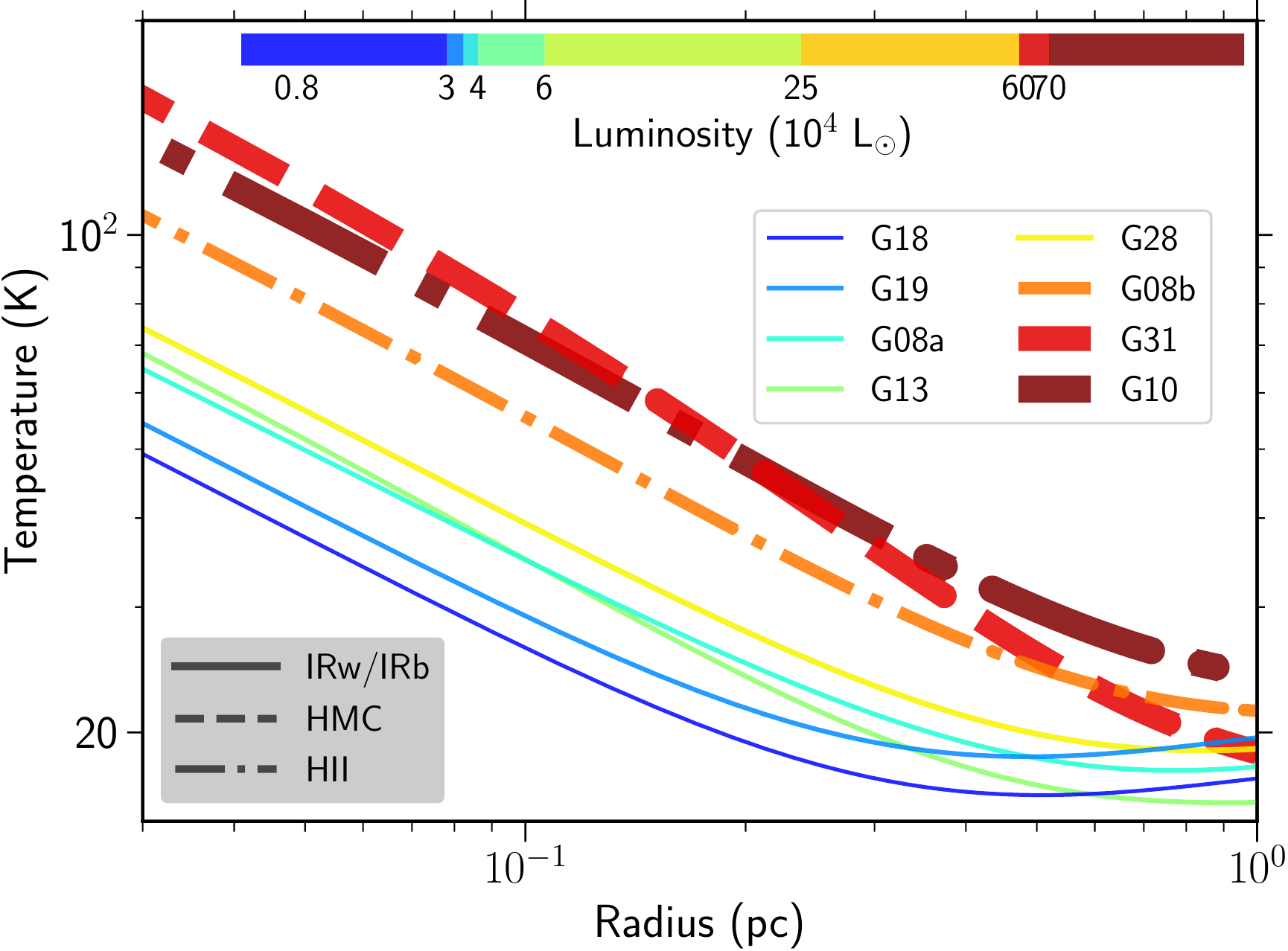}&
\hspace{-0.5cm}\includegraphics[scale=0.475]{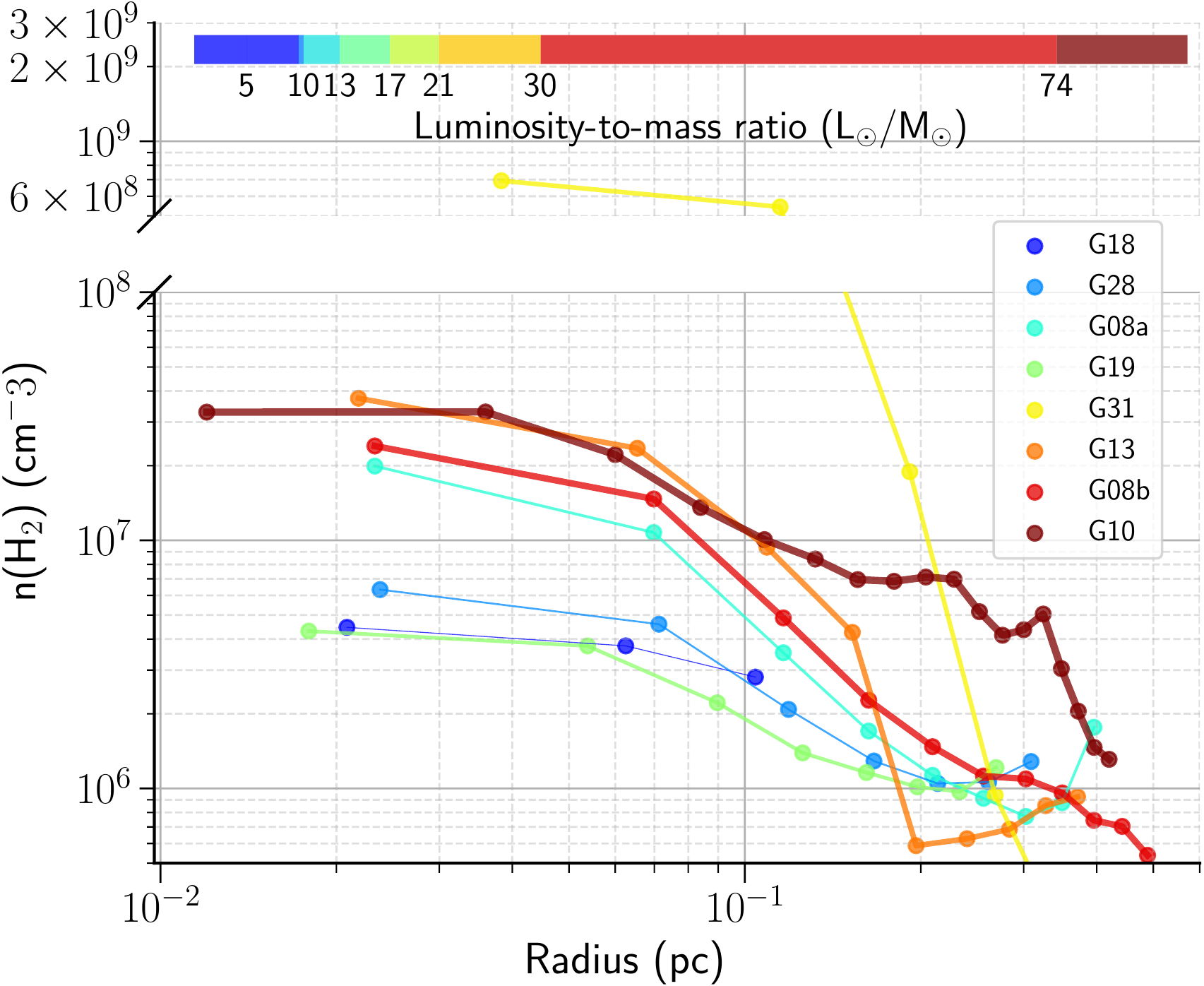}\\
\end{tabular}
\centering\caption{{\it{Left panel:}} Radial temperature profiles $T(r)$ for all the target sources (Equation \ref{eq:tformula}, refined $T(r)$ is used for the relevant source). The thickness of the lines increases with increasing luminosity. {\it{Right panel}:} The radial (projected radial averaged) density profile of the dense gas ($\rho_{\mathrm{dense}}$) of all the target sources, from $n(\mathrm{H_{\mathrm 2}})$ maps derived by RADEX modeling of CH$_{3}$OH lines. The range of the y axis is trimmed to increase contrast. For both panels, the luminosity and clump mass are calculated from {\tt{RADMC-3D}} best-fit model (Table \ref{tab:radmc_para}), which takes into account all the gas component present in the clumps.}
\label{fig:evo_trend}
\end{figure*}

We make a comparison of the fitted and refined $T(r)$ profiles (Equation \ref{eq:tformula}) of all sources in Figure \ref{fig:evo_trend} (left panel).
We can see that, for all the clumps, at 0.1 pc the resolved gas temperatures range from 30-80 K, and at 1 pc at around 20-30 K. 
Moreover, the gas temperature at a certain clump radius is not a monotonic function of the bolometric luminosity of the clump. Particularly, the hot massive core G31 and the source G13 display higher temperatures in the inner regions than their immediate more luminous sources in the sample. We discuss these temperature profiles in more detail in Section \ref{sec:temp_prof}. 

In Figure \ref{fig:evo_trend} (right panel), we also compare the derived radial gas density profile of the dense gas from CH$_{3}$OH modeling with RADEX/{\tt{LIME}}. It can be seen that $n(\mathrm{H_{\mathrm 2}})$ is several $\sim$10$^{5}$ to 10$^{7}$ cm$^{-3}$ at $\sim$0.2-0.3 pc (projected) radii.
In the inner $\sim$0.1 pc where the SMA identified continuum cores (Sect. ~\ref{sec:sma_cont}),  $n(\mathrm{H_{\mathrm 2}})$ ranges between several $\sim$10$^{6}$ to 10$^{8}$ cm$^{-3}$ . 
There is exceptionally high $n(\mathrm{H_{\mathrm 2}})$ values at the center of G31. Although we have verified the high level of dense gas of this source compared with the rest of the sample by full radiative transfer modeling of CH$_{3}$OH lines (Appendix \ref{app:lime}), we caution that in this density regime CH$_{3}$OH (5-4) lines are becoming heavily optically thick and the critical density for the considered line transitions is reached (e.g., n$>$n$_{\mbox{\scriptsize crit}}$, Table \ref{tab:lines_more}), such that the relative differences between the level populations do not serve as ideal densitometers anymore. Nonetheless, we can safely argue that the hot massive core G31 has much higher gas densities in its inner region than other sources, which is also reflected by the very monolithic nature of its central core from higher angular resolution observations ($\sim$2000 au, c.f., \citealt{Beltran18}). 
For source G10, there is prominently higher gas densities at extended radii of 0.1-0.4 pc than other sources, which is related to the presence of a large disk-like flattened structure. We discuss further on the density profiles among the sample in Section \ref{sec:dens}.  

\begin{figure*}[htb]
\begin{tabular}{p{0.5\linewidth}p{0.5\linewidth}}
\hspace{-.5cm}\includegraphics[scale=0.5]{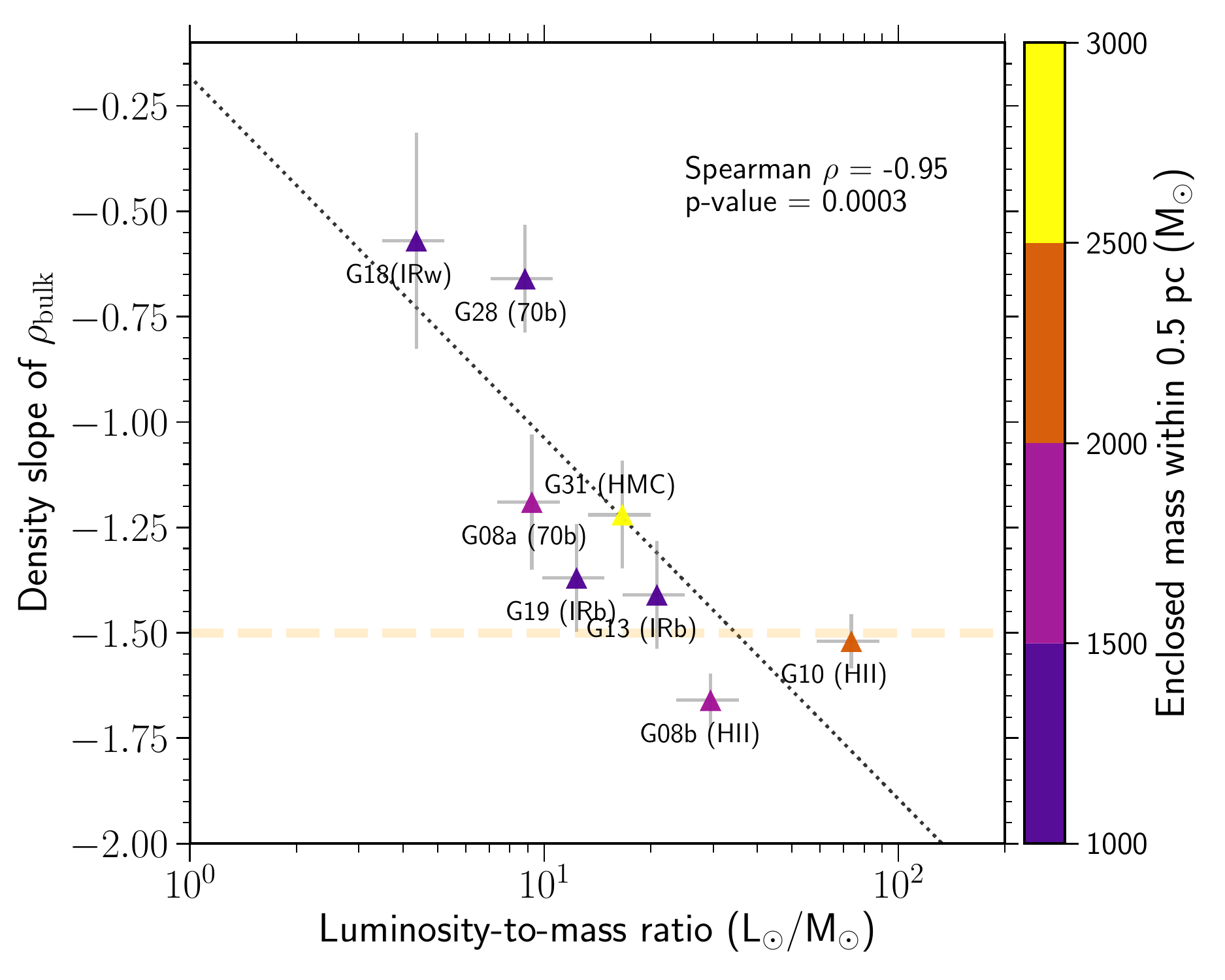}&
\hspace{-.4cm}\includegraphics[scale=0.5]{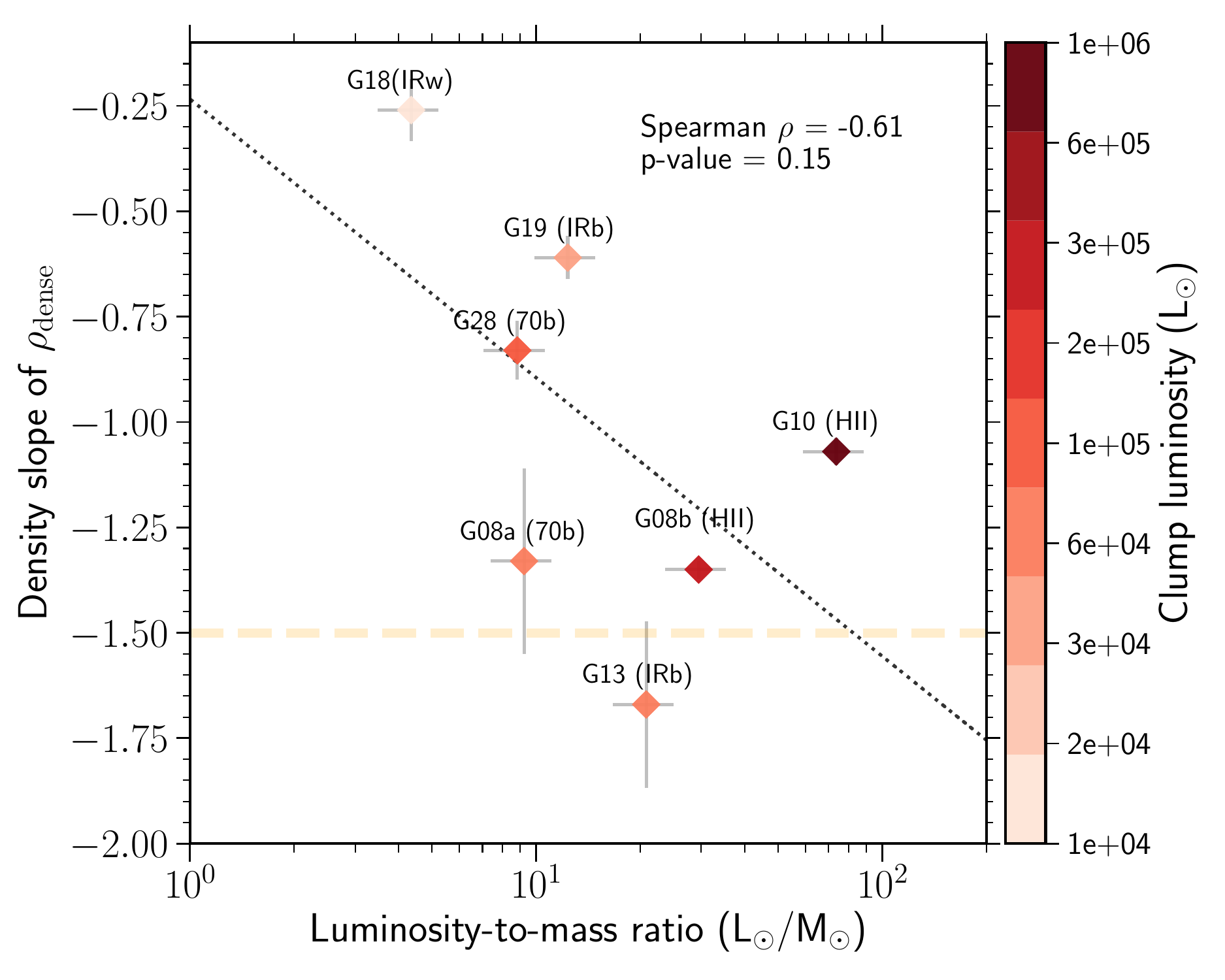}\\
\end{tabular}
\centering\caption{{\it{Left panel:}} Density power-law slope derived from continuum ($q$) based on {\tt{RADMC-3D}} modeling detailed in Appendix \ref{app:radmc}. The luminosity, clump mass and enclosed mass within 0.5 pc are calculated from {\tt{RADMC-3D}} best-fit model, as listed in Table \ref{tab:radmc_para}.  {\it{Right panel:}} Density power-law slope derived from $n(\mathrm{H_{\mathrm 2}})$ maps ($q_{\mathrm{radex}}$) from CH$_{3}$OH RADEX modeling detailed in Sect.~\ref{sec:radex_nh2}. Yellow horizontal line in both plots shows a slope of -1.5, indicating the free-falling density profile of an singular isothermal sphere (with an initial density slope of -2) as in \citealt{Shu77}, and the attractor solution of the gravo-turbulent collapsing in \citet{Murray17}.}
\label{fig:slopes_lm}
\end{figure*}

We compare the steepness of the radial gas density profiles ($q_{\mathrm{bulk}}$ and $q_{\mathrm{dense}}$, for $\rho_{\mathrm{bulk}}$ and $\rho_{\mathrm{dense}}$) as a function of source evolutionary stages, which is indicated by the clump bolometric luminosity to mass ratio $L/M$ (Figure \ref{fig:slopes_lm}). Using Spearman correlation measure, we find that there are positive correlations (correlation coefficient $\rho$ = -0.95 and -0.65) between the density power-law slopes with $L/M$, for both the dense gas component and the bulk gas structures, although the significance of the correlation of the former is low (p-value = 0.15). The slopes range from -0.6 to -1.7 for the bulk gas, and -0.25 to -1.7 for the dense gas, for $L/M$ spanning from 10 to $\sim$100 ($L_{\odot}/M_{\odot}$) of all sources. The correlation between $L/M$ and the density slope representing the bulk gas distribution is clearly stronger. 
A similar evolutionary trend was reported by \citet{Beuther02a} based on analyses of 1.2 mm dust continuum emission of a sample of massive clumps, in which the bulk gas density structure was probed. 
Comparably, other works on the density structures of massive clumps typically derived power-law slopes ranging in $-$2.25-$-$0.75 and peaking at $-$1.8-$-$1.6 (\citealt{Mueller02}, \citealt{Beuther02a}, \citealt{vdt00}).
In the early-stage sources (L/M < 20), the slopes we derived are relatively shallow ($>$-1.0) for both the bulk gas and dense gas density structures. We note that the slope derived for the dense gas structure of the early-stage source G18 is valid for a confined region of $\sim$0.1 pc (just above the beam size), which merely reflects a pocket of dense gas that is rather compact and remains unresolved. Yet, for other early-stage sources, i.e. G28 and G19, the statistics for determining the density slope are rather adequate, and these two sources do exhibit shallow slopes of $\sim$-0.6.  
We note that different analysis methods of density structure could result in systematic biases in the derived density slopes. 
In addition, the analyses of dust continuum that were based on the optically thin assumption, instead of relying on full radiative transfer models, suffer from the degeneracy of density and temperature profiles in determining the radial intensity profiles.
Moreover, close to the source center, the optically thin assumption for dust emission may also break down. Although qualitative comparison can be made, a careful gauge between different analysis conducted are necessary for a stringent comparison between different works. In Sect.~\ref{sec:evo} we discuss further on the relation between density profiles of massive clumps and statistics on cloud structure. We also elaborate on physical implications by comparing $\rho_{\mathrm{bulk}}$ and $\rho_{\mathrm{dense}}$, among the sample.

\subsection{Molecular linewidths and virial parameter}\label{sec:velo_width}
To understand the dynamic states of the target clumps, we examined how the linewidths and virial parameters vary with clump radii.
Part of these analyses were based on the thermometer lines, CH$_{3}$CN, H$_{2}$CS and CH$_{3}$CCH.
They primarily trace the dense gas close to the centers (0.1-0.4 pc) of the clumps.
In addition, we examined the CS and C$^{34}$S (5-4) and H$^{13}$CO$^+$ (3-2) lines which can trace spatially more extended clump structures due to their lower excitation conditions.
We performed single component Gaussian fits to the CS, C$^{34}$S (5-4) and H$^{13}$CO$^+$ (3-2) line cubes in a pixel-by-pixel manner to obtain the linewidth maps.
For the analysis, we trimmed the pixels that have fitting errors of linewidth larger than 2 times the velocity channel widths ($\Delta\,V$$<$2 km/s).

The virial parameter $\alpha_{\mathrm{vir}}$ characterizes an important aspect of the physical states of the molecular clumps.
The ordinary definition of $\alpha_{\mathrm{vir}}$ (i.e., ignoring magnetic field; c.f., \citealt{Bertoldi92}) is
\begin{equation}
    \alpha_{\mathrm{vir}} = a_{1}\frac{2T}{|W|},\label{eq:ve}
\end{equation}
where $T$ $=$ $\frac{3}{2}$$M_{\mathrm{enc}}$$\sigma_{\mathrm{rms}}^{2}$ is the kinetic energy, $W$ $=$ -$\frac{3}{5}$$a_{1}$$\frac{GM_{\mathrm{enc}}^{2}}{R}$ is the gravitational potential energy, $M_{\mathrm{enc}}$ is the enclosed mass, and $a_{1}$ is a geometric factor which accounts for the inhomogeneity of the density distribution (e.g. \citealt{Bertoldi92}, \citealt{McKee99}).
For a spherical clump that has a $\propto r^{-q}$ radial gas density profile, $a_{1}$ = $\frac{1+q/3}{1+2q/5}$. 
With this definition, a source in energy equipartition (T$\sim$$|W|$) has a critical virial parameter of $\alpha_{\mathrm{cr}}\,=\,2a_{1}$.
In a virialized source (2T$\sim$$|W|$), it stands that $\alpha_{\mathrm{vir}}$ = $a_{1}$ with $\alpha_{\mathrm{vir}}/\alpha_{\mathrm{cr}}\,=\,0.5$.  In the following we refer to the states of $\alpha_{\mathrm{vir}}/\alpha_{\mathrm{cr}}\,<\,0.5$, $\sim$0.5-1 and $>$1 as sub-virial, virial and super-virial state, respectively. 

When deriving $\alpha_{\mathrm{vir}}$, it is critical that the tracers observed for the measurement of $M_{\mathrm{enc}}$ and $\sigma_{\mathrm{rms}}$ are predominantly emanated from the same gas entity (\citealt{Traficante18}).
The mass tracer we adopted, which is the dust continuum emission, traces a broad range of gas volume density distributed in a wide range of radius.
Our selected tracers to indicate linewidths, as the way the temperature profile is measured, show emission of progressively larger radii, which are complemented with two more extended tracers. 
We can now examine spatial variation of linewidths and $\alpha_{\mathrm{vir}}$ based on multiple tracers that cover distinct critical densities (Table \ref{tab:lines_more}), and hence different spatial scales.
We evaluated how $\alpha_{\mathrm{vir}}/\alpha_{\mathrm{cr}}$ varies with radius using the best-fit density models from the {\tt{RADMC-3D}} continuum modeling (Appendix \ref{app:radmc}) to obtain $M_{\mathrm{enc}}$, $a_{1}$ and the linewidth maps from aforementioned tracers.
Note that as compared with the SMA observations, the {\tt{RADMC-3D}} models constrained by the coarser resolution single-dish continuum data systematically under-predicted the 1.2 mm fluxes in the inner radii.
To avoid this bias, we adopt the $M_{\mathrm{enc}}$ as $M^{\mathrm{Abel}}_{\mathrm{core}}$ (Sect.~\ref{sec:sma_cont}) for the inner regions. We recall that $M^{\mathrm{Abel}}_{\mathrm{core}}$ is calculated by applying the derived $T(r)$ to SMA 1.2 mm continuum. We discuss the obtained radial profiles of linewidth and virial parameter in Sect.~\ref{sec:velo_width_dis}.

\subsection{Molecular abundance and abundance ratios}\label{sec:ab_cal}
\begin{figure*}
    \centering
    \begin{tabular}{p{0.5\linewidth}p{0.5\linewidth}}
    \includegraphics[scale=0.42]{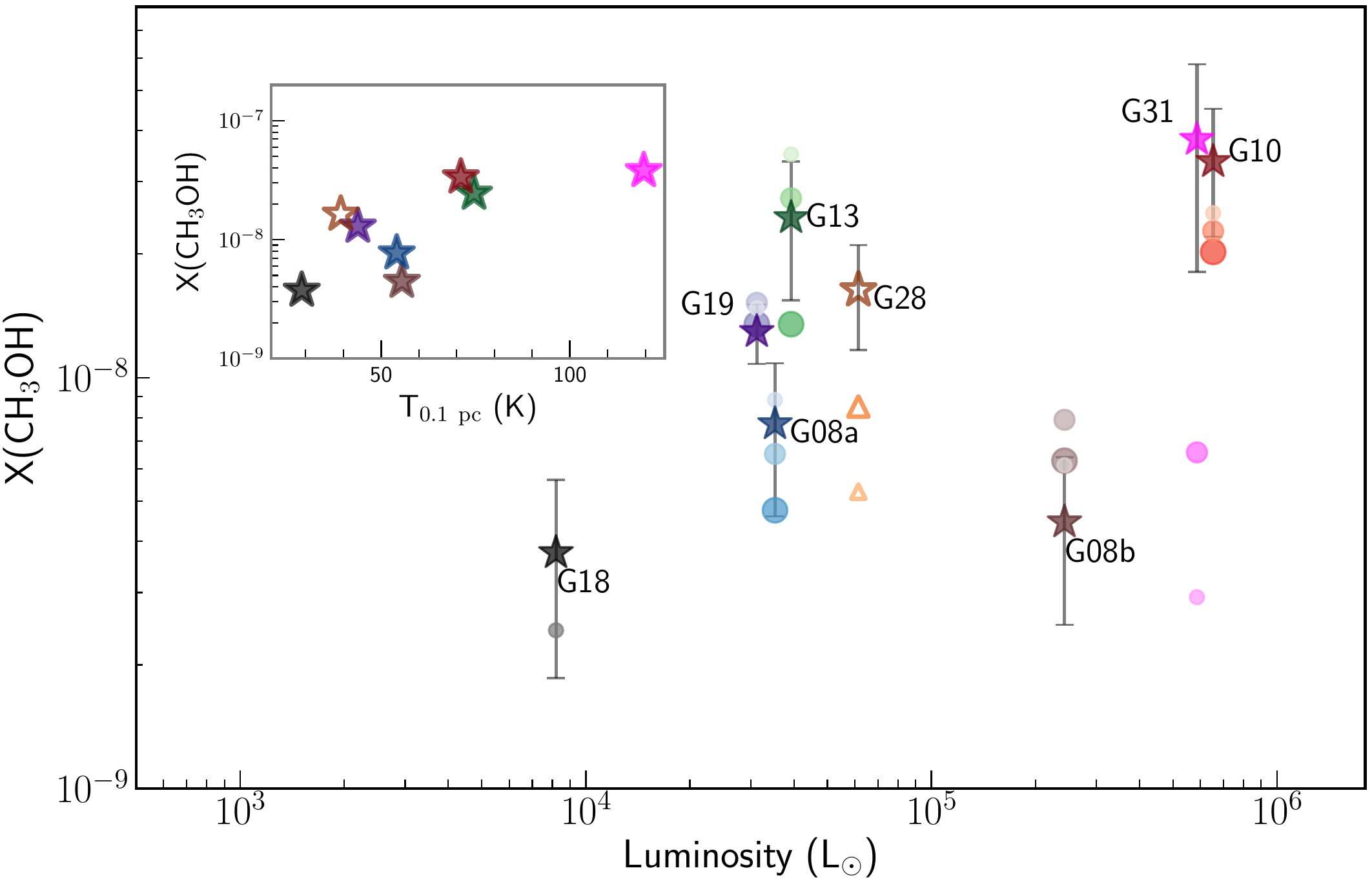}&\includegraphics[scale=0.42]{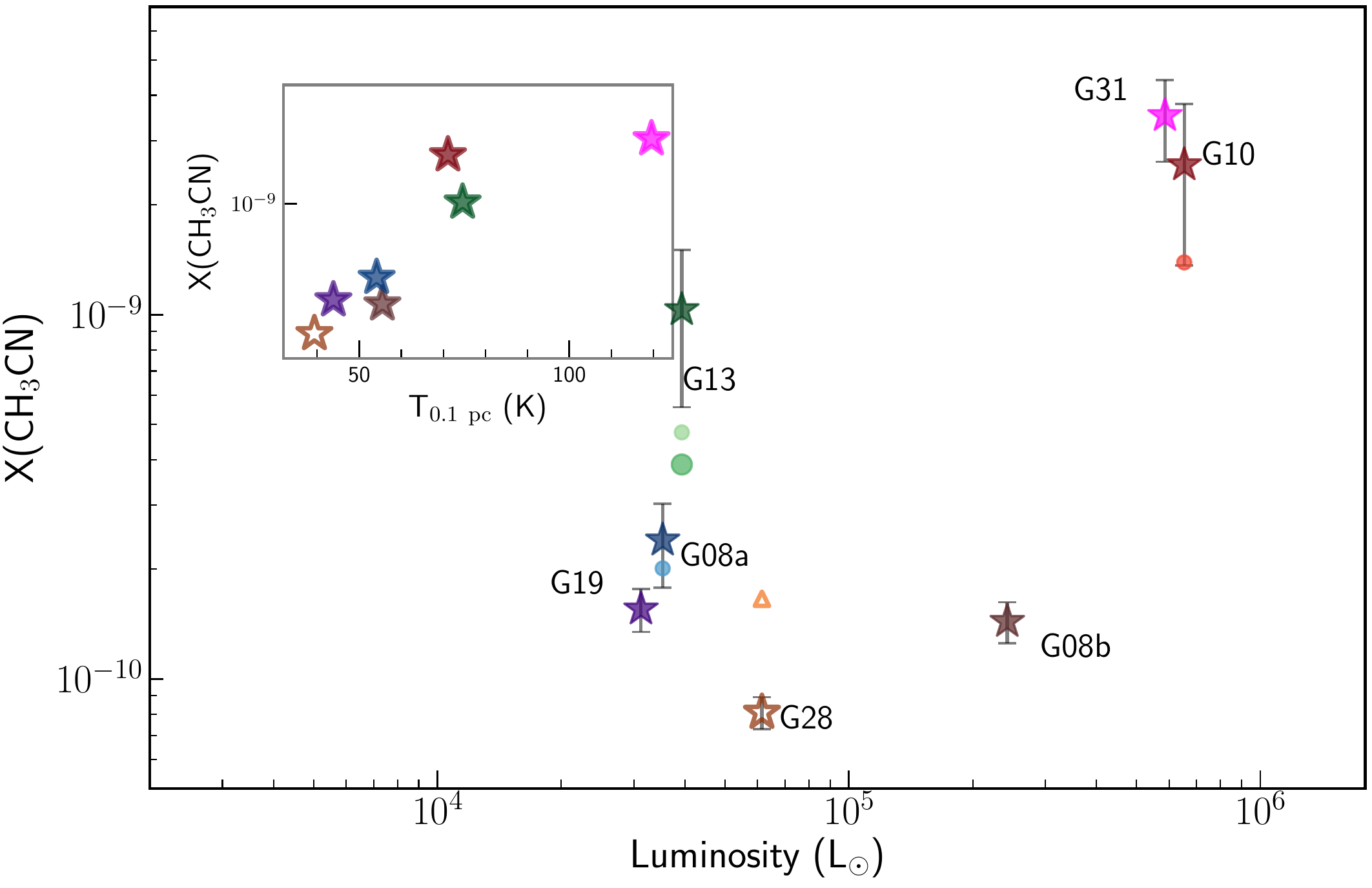}\\
    \includegraphics[scale=0.42]{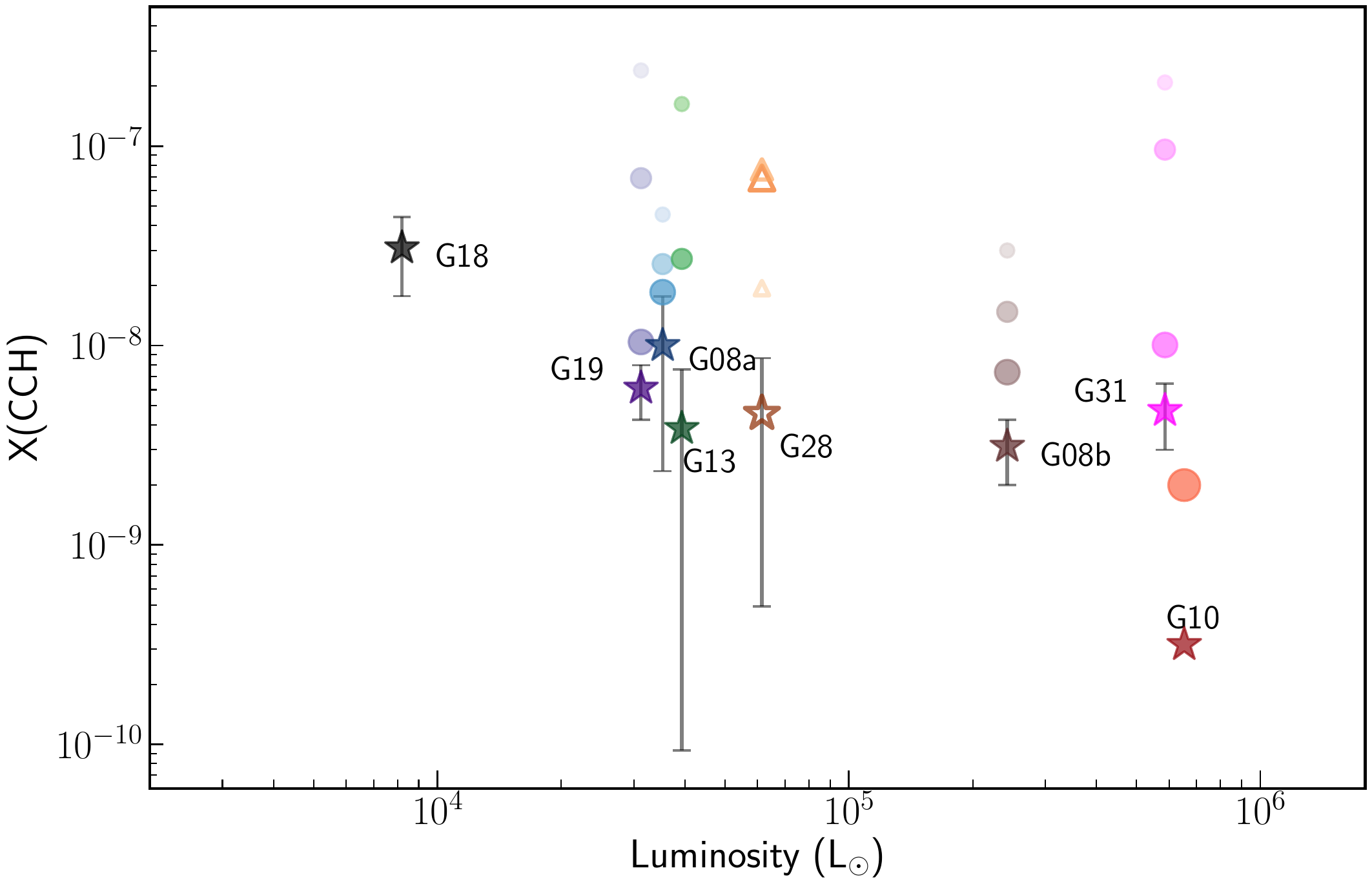}&\includegraphics[scale=0.42]{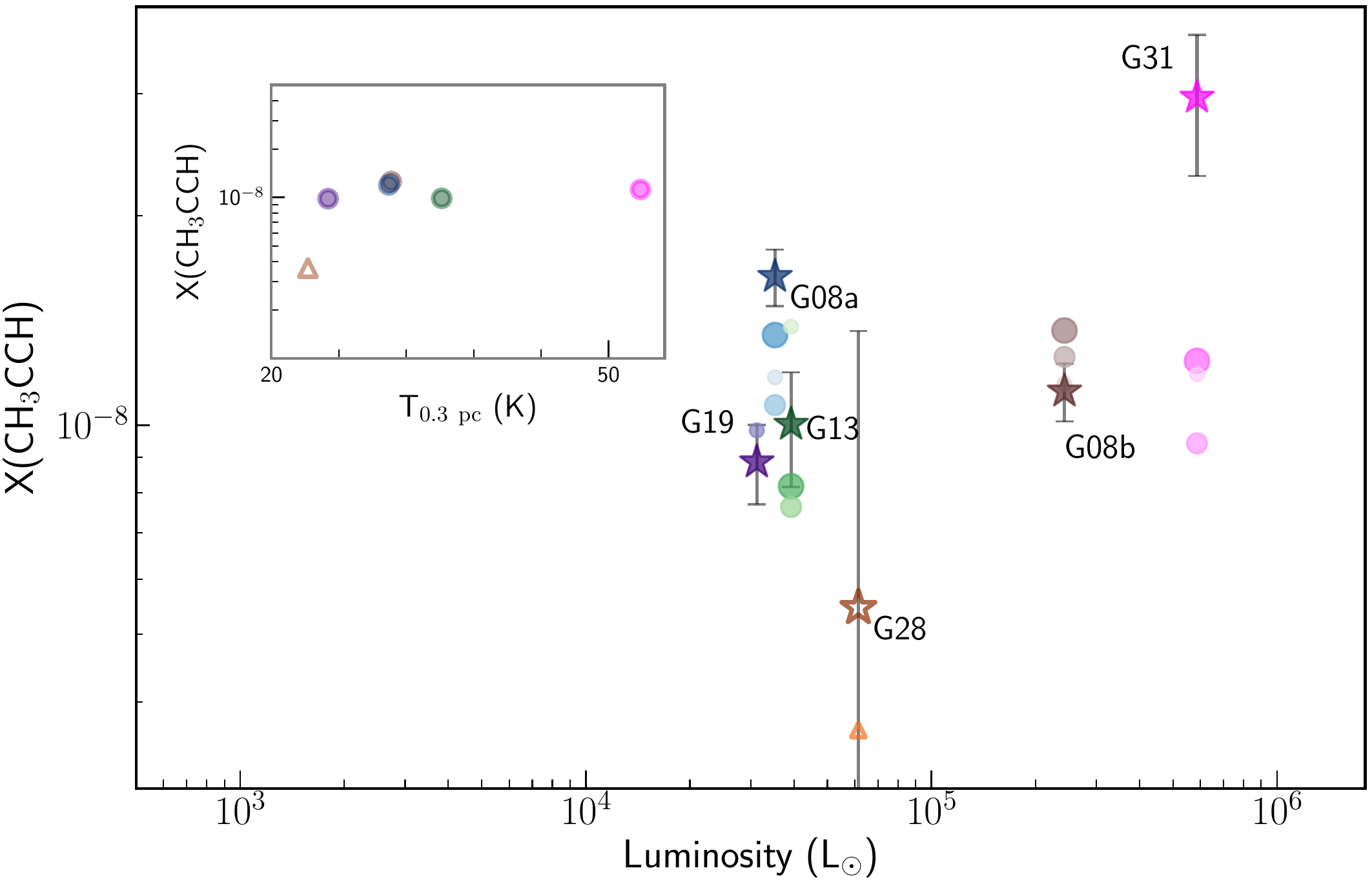}\\
    \end{tabular}
    \caption{Abundance profiles of molecules CH$_{3}$OH, CH$_{3}$CN, CCH and CH$_{3}$CCH towards target sources. Markers are color-coded based on relative distance to the clump center (the 1.2 mm continuum peak). Stars represent abundances at the clump centre, whereas hollow circles or triangles indicate abundances obtained from different radii: the larger and darker the latter markers, the closer distance to the continuum peak they represent. 
    For clump G28, the data points at outer radii are marked as triangles, whereas those for other clumps are marked as circles, to further avoid confusion.
    X(CCH) for source G10 is taken from \citet{Jiang15}. The inset plot for CH$_{3}$CN shows the central abundance vs. gas temperature at 0.1 pc; for CH$_{3}$CCH shows the envelope abundance vs. gas temperature at 0.3 pc.}
    \label{fig:ab_profiles}
\end{figure*}

\begin{figure*}
    \centering
    \begin{tabular}{p{0.5\linewidth}p{0.5\linewidth}}
    \includegraphics[scale=0.42]{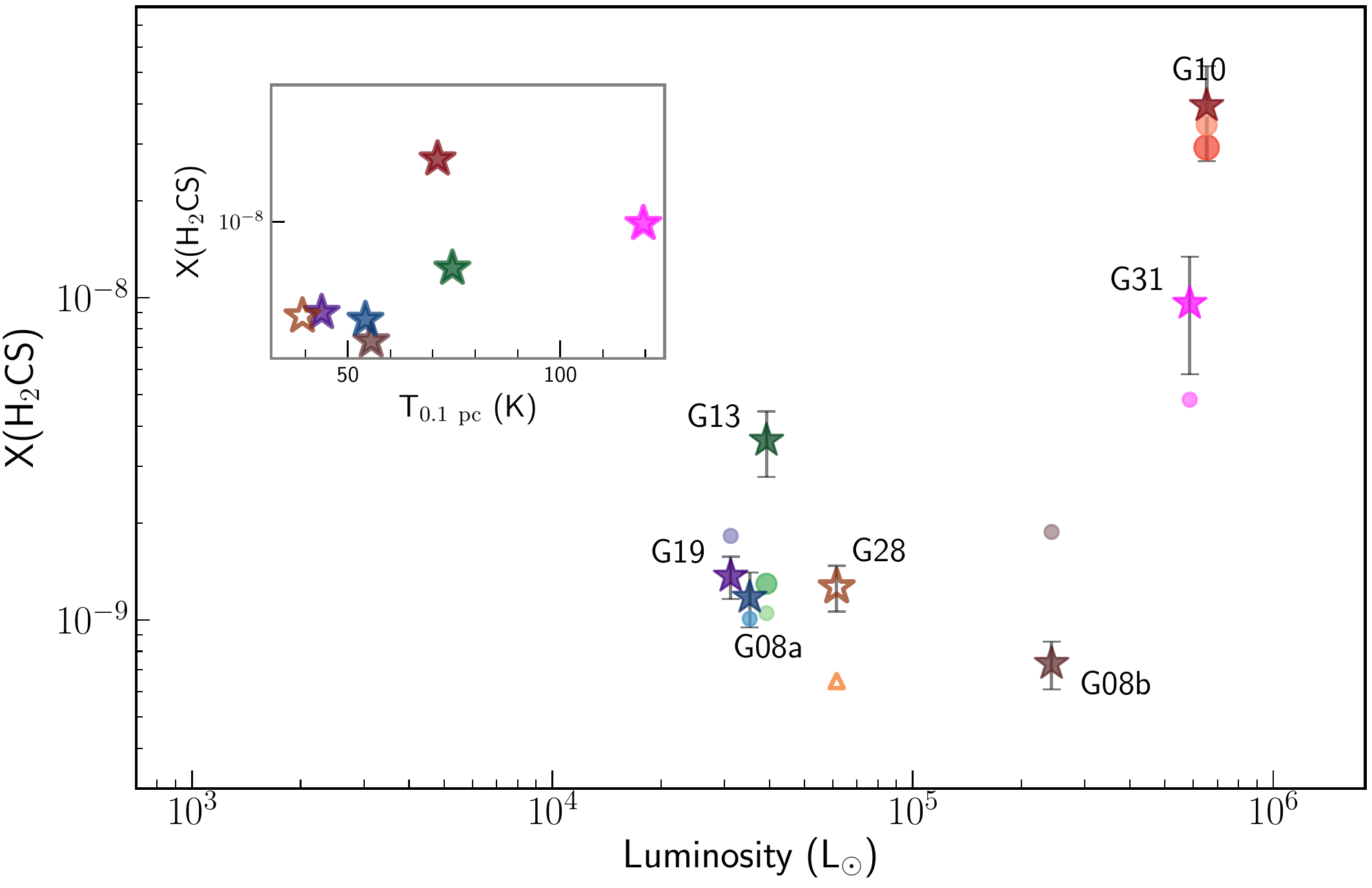}&\includegraphics[scale=0.42]{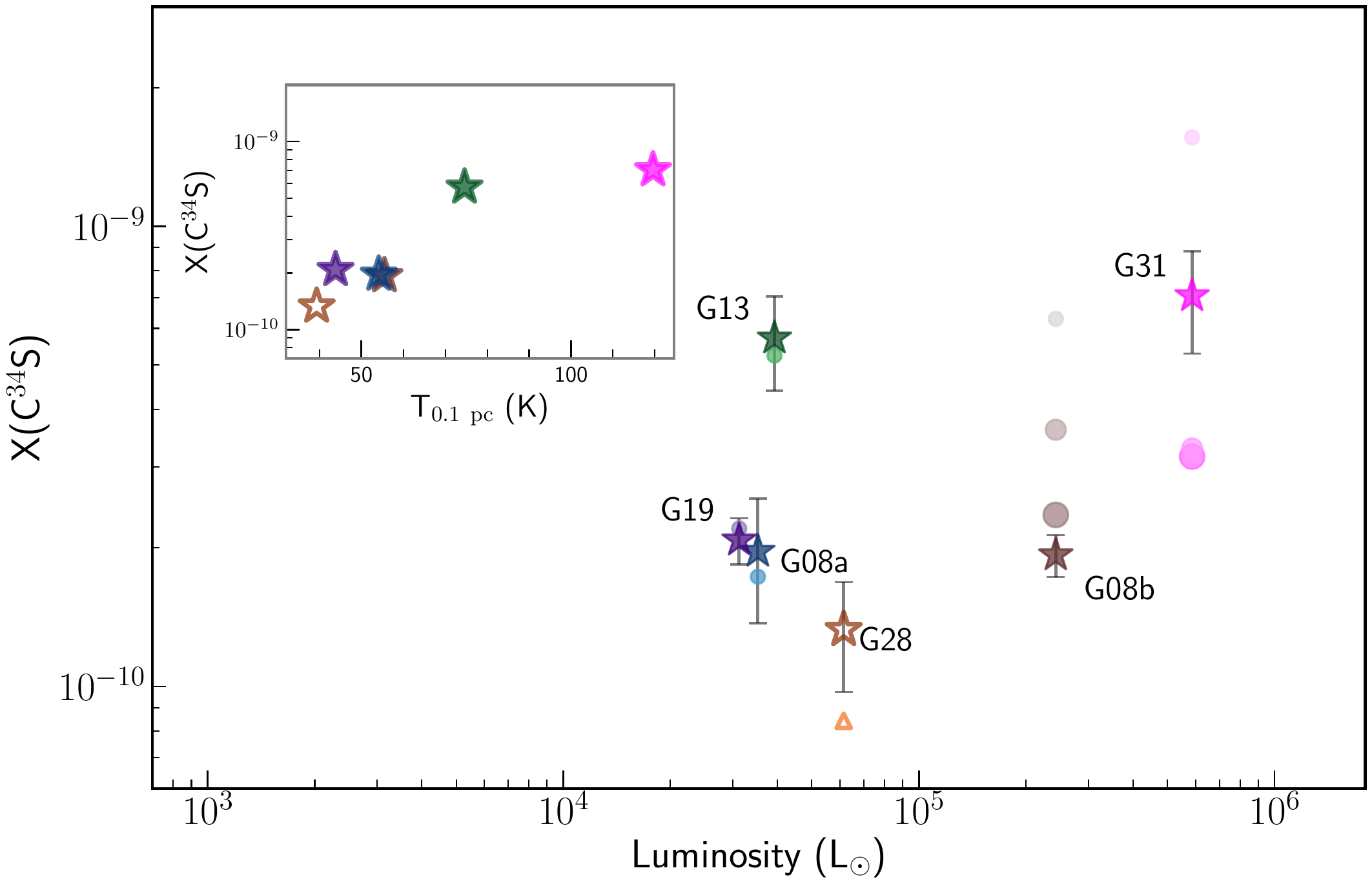}\\
    \includegraphics[scale=0.42]{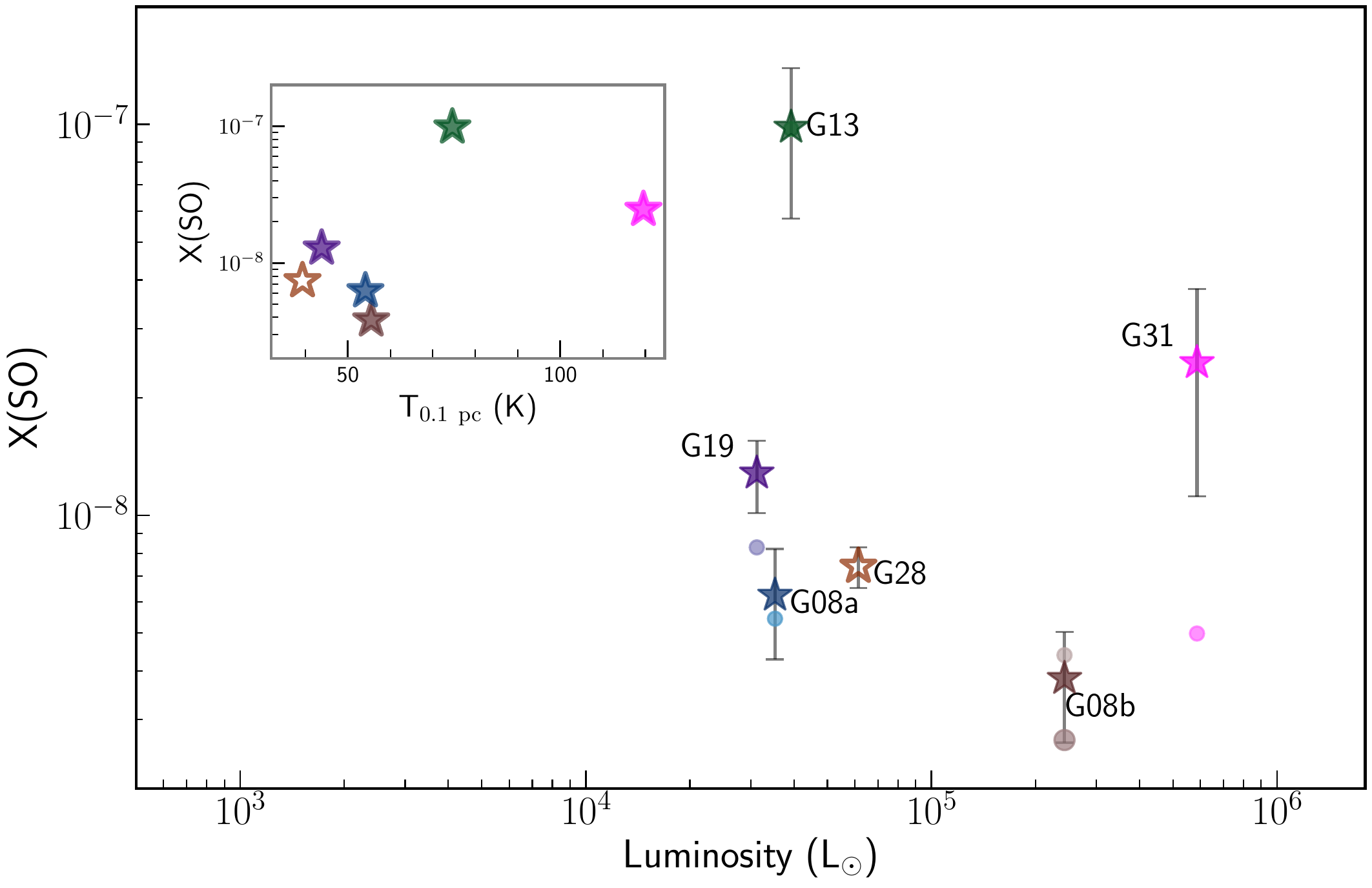}&\includegraphics[scale=0.42]{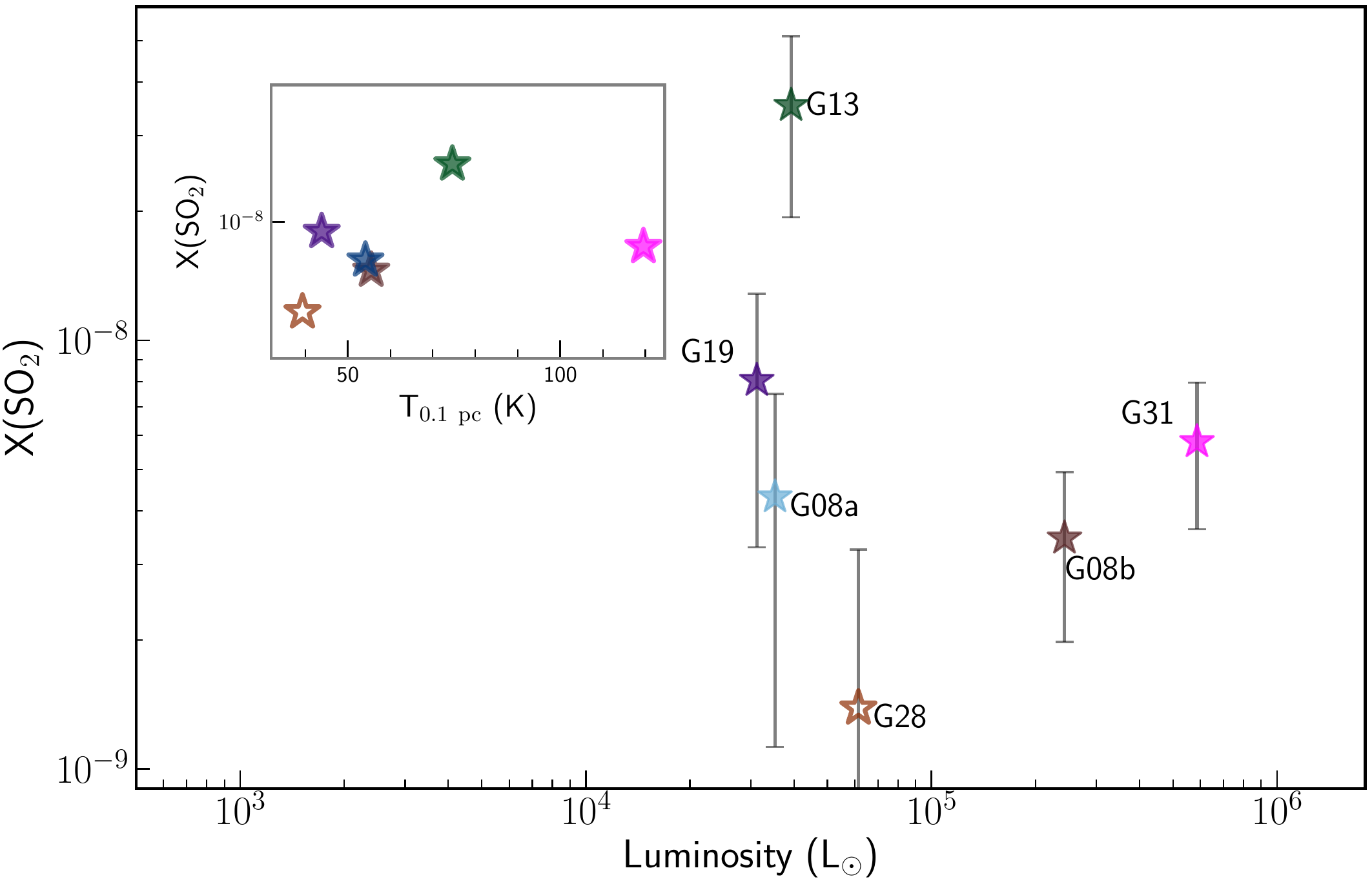}\\ 
    \end{tabular}
    \caption{Same as Figure \ref{fig:ab_profiles}, but for H$_{2}$CS, C$^{34}$S, SO and SO$_{2}$.}
    \label{fig:ab_profiles1}
\end{figure*}

To facilitate the analysis on clump evolutionary stages, we derived the LOS integrated abundance maps ($N_{\mathrm{mol}}$/$N(\mathrm{H_{2}}$)) for some relevant molecular species for all sources.
The bulk gas density profiles ($\rho_{\mathrm{bulk}}$$(r)$, Sect.~\ref{sec:T_rho_profiles}, and Appendix \ref{app:radmc}) were adopted and smoothed to the angular resolution of the specific line transition when deriving $N_{\mathrm{mol}}$/$N(\mathrm{H_{2}}$).
The calculation of $N_{\mathrm{mol}}$ for CH$_{3}$CCH, CH$_{3}$CN, H$_{2}$CS and CH$_{3}$OH is introduced in Sect.~\ref{sec:xclass} and Section ~\ref{sec:radex_nh2}. The calculations of $N_{\mathrm{mol}}$ maps of CS/C$^{34}$S, SO, SO$_{2}$ and CCH lines were based on LTE assumption and are detailed in Appendix \ref{app:otherlines}.
We then derived the projected radial averaged abundance profiles for each molecule.
Naturally, the projected radial averaging suppresses the contrast in the spatial variations of the abundance for the molecules that are enriched in the clump center or other localised regions (which reduces $N(\mathrm{H_{2}})$ to localised values rather than integration along the LOS extension).
Nevertheless, this does not qualitatively change the overall radial trends as long as the abundance is increasing or decreasing with radius monotonically, with a steeper profile than that of the column density, while the latter is rather shallow, following $\Sigma$ $\propto$ $\rho r$ $\propto$ $\rho^{\mathrm{1+q}}$ $<$ $\rho^{-0.7}$.  
In Figures \ref{fig:ab_profiles}-\ref{fig:ab_profiles1}, we show the relations between the clump bolometric luminosity and the radial abundance variations of the carbon-chain and sulfur-bearing molecules in consideration, respectively.

From Figure \ref{fig:ab_profiles} we note that the abundances of CH$_{3}$OH, CH$_{3}$CCH and CH$_{3}$CN show similar behaviors. 
They appear largest in the hot massive cores G31; in the rest of the sources, the abundances of  CH$_{3}$OH and CH$_{3}$CCH are in the range of several 10$^{-9}$-10$^{-8}$, while the abundances of CH$_{3}$CN are in the range of 10$^{-10}$-10$^{-9}$.
These abundances are slightly positively correlated with the clump luminosity.
The abundances of CH$_{3}$CN and CH$_{3}$OH appear more tightly correlated with the source temperatures in the inner $\sim$0.1 pc regions (see the insets in Figure \ref{fig:ab_profiles}).
These trends are consistent with theoretical predictions that the de-sorption of these molecules from grain surface is more efficient with higher temperature.
We note that these results cannot be obtained if the gas temperature distributions were derived based merely on the assumptions of bolometric luminosity scaling instead of being derived based on multi-transition rotational temperature maps, since we have previously seen that the gas temperatures at certain radii do not necessarily increase monotonically with clump luminosity (Figure \ref{fig:evo_trend}, left panel).
This result demonstrates the importance of measuring detailed temperature profiles when studying chemical evolution.

We observed a weak correlation of abundance of CH$_{3}$CCH and bulk gas temperature, which is reminiscent of the small variation of the CH$_{3}$CCH abundance towards massive clumps of various evolutionary stages reported by \citet{Giannetti17} (see also \citealt{Oberg14}). 
Other higher angular resolution observations of this species indicated a mixed behaviour of its spatial distribution, depending on whether the emission coincides with the localised hot cores or appears offset and/or showing more extended structures (\citealt{Bogelund19}, \citealt{Oberg14}, \citealt{Fayolle15}). 
Comparing rotational temperature maps of CH$_{3}$CCH (12-11), the temperatures of G13 and G31 in the core region is 20-30 K higher than those of G19, G08a and G08b. 
In addition, the higher J transitions CH$_{3}$CCH (14-13), (15-14) trace systematically higher temperatures in the core region towards all sources.
This evidence indicates that the emission of CH$_{3}$CCH does have contribution from gas components residing inside or in the vicinity of hot cores.  
As for the result of CCH, except for the earliest-stage source G18, its abundance appears enhanced at outer radii.
We can also see that the abundance of CCH measured in the inner $\sim$0.1-0.2 pc region is anti-correlated with the clump luminosity.
These results are consistent with previous observations that show shell-like CCH emission towards late-stage massive star-forming regions (\citealt{Beuther08}, \citealt{Jiang15}).

\begin{figure*}
\includegraphics[scale=0.5]{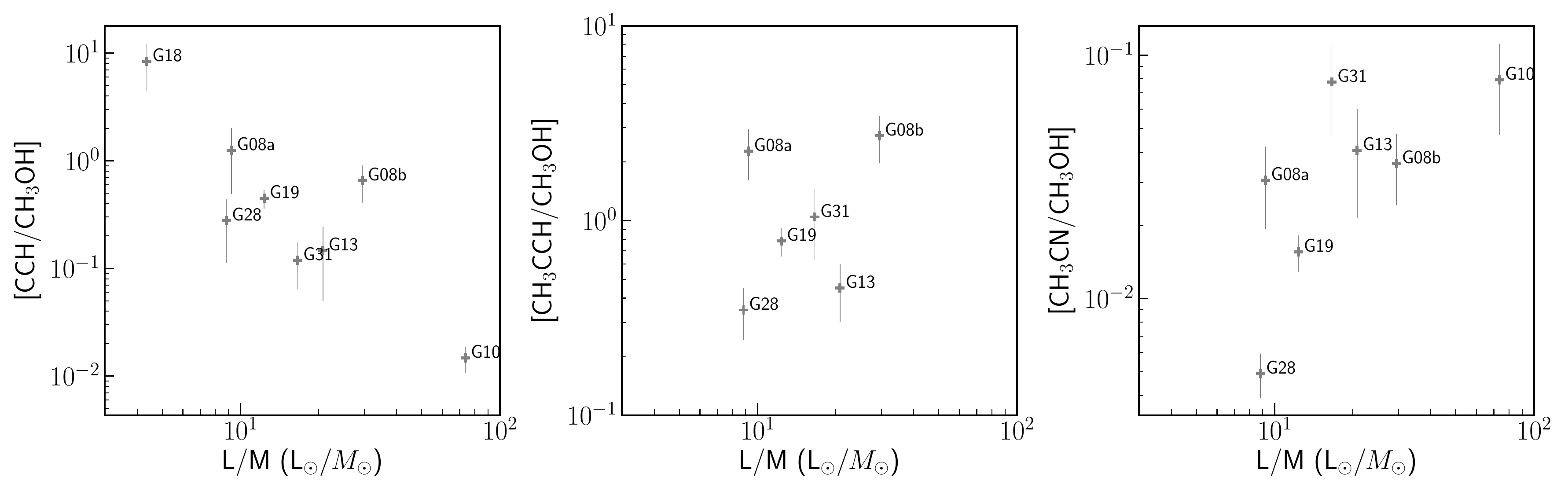}
\includegraphics[scale=0.5]{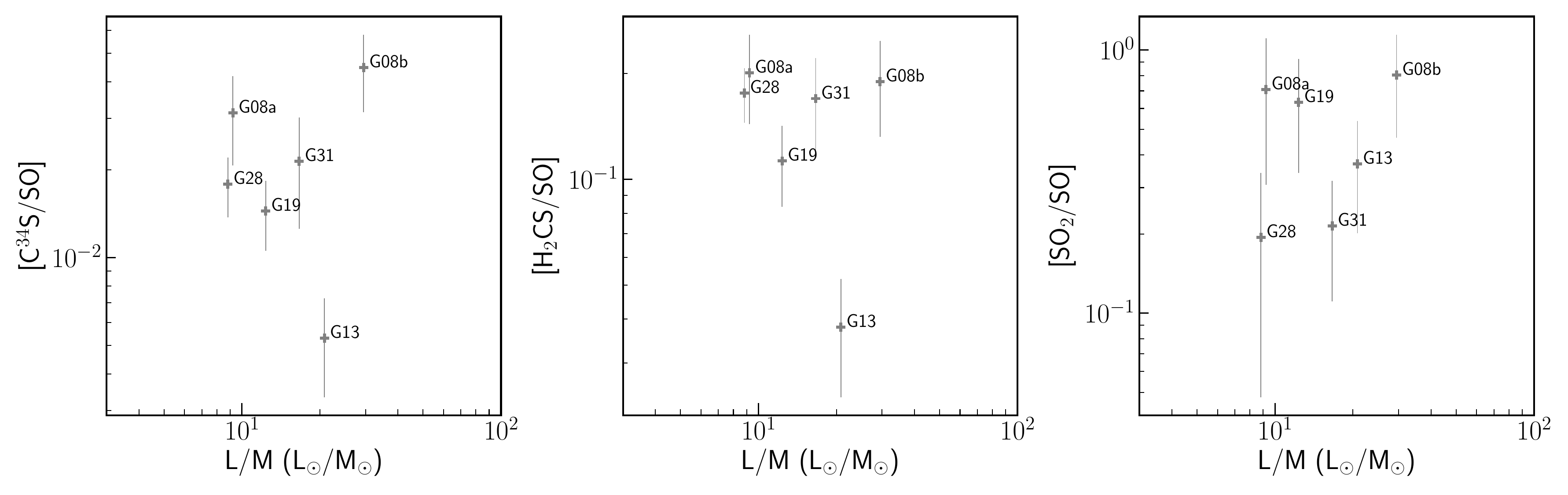}

\caption{{\it{Top: }}Abundance ratios of carbon-chain molecules; {\it{Bottom:}}Abundance ratios of sulfur-bearing molecules
at clump center (0.1-0.15 pc, beam averaged), shown against source $L/M$ ratios.}
\label{fig:com_ratios} 
\end{figure*}

The abundance ratios of CCH, CH$_{3}$CCH and CH$_{3}$CN with CH$_{3}$OH in the clump center (0.1-0.15 pc) are shown in Figure \ref{fig:com_ratios}. 
Here, normalisation with CH$_{3}$OH allows exploring the chemical evolution or initial condition by eliminating the effect of different desorption levels among the sample. There are substantial variations of [CCH]/[CH$_{3}$OH] and [CH$_{3}$CN/CH$_{3}$OH] across $L/M$, while variations of [CH$_{3}$CCH/CH$_{3}$OH] appear moderate. 

The abundances of sulfur-bearing species in the central regions of the clumps also show correlation with temperatures (Figure \ref{fig:ab_profiles1}). 
Comparing the relative abundance of C$^{34}$S and H$_{2}$CS with respect to SO as a function of $L/M$ (Figure \ref{fig:com_ratios}, bottom panel), it seems there is a mixed behaviour: except for source G13, the other sources show an increasing trend. On the other hand, for X(SO$_{2}$)/X(SO) there is a consistently increasing trend with source $L/M$. We discuss the abundance variations further in Sect.~\ref{sec:ab_dis} in a broader context, with comparisons with published results from chemical modeling.

\section{Discussion}\label{sec:discussion}
\subsection{Temperature structure and heating mechanisms of massive star-forming clumps}\label{sec:temp_prof}
The temperature measurement from multiple $T_{\mathrm{rot}}$ maps and the fitted and refined temperature profile $T(r)$ (Equation \ref{eq:tformula}) are shown in Figure \ref{fig:tprofiles}.
The $d\,$log$T$/$d\,$log$R$ profiles are also summarized in the bottom right panel.
Except for sources G13 and G31, the $d\,$log$T$/$d\,$log$R$ profiles asymptotes from -0.5 to zero from inner to outer radii.
The levelling-off of the temperature at the outer radius of the clump is expected since all these massive star-forming clumps are immersed in intense interstellar radiation fields. 
At gas densities $>$10$^{4.5}$-10$^{5}$ cm$^{-3}$, thermal coupling between gas and dust can be quickly achieved due to collisions (\citealt{Goldsmith01, Glover12}). 
However, the 10$^{4.5}$ cm$^{-3}$ density threshold is not met in the outer layers of the sources which have lower masses.
As we will explain below, the thermal decoupling between gas and dust is seen in some of these sources.
The outer envelope dust temperature for all these sources flattens around 18-25 K. 
These values are consistent with an elevated infrared radiation field associated with these regions. Based on these temperatures, the scaling factors of the local ISRF (\citealt{Mathis83}) to characterise the radiation field surrounding these clumps are roughly $\gtrsim$10$^{2.5}$-10$^{4}$ (estimated in the optically thick limit, \citealt{Krumholz14}), which are typical values found in the vicinity of Galactic massive star-forming complexes (\citealt{Binder18}). 

The $T(r)$ and $d\,$log$T$/$d\,$log$R$ in the inner regions may be approximated by the analytic temperature profile of thermally balanced dust grains distributed around a central heating point source (e.g. \citealt{Adams91}).
With optically thin condition and a submillimeter dust opacity spectral index ($\beta$), the model of \citet{Adams91} predicted that the radial temperature profile for dust grains in thermal balance around a central heating point source is proportional to $r^{-2/(4+\beta)}$.
If $\beta$ has no spatial variations, then $d\,$log$T$/$d\,$log$R$ should be a constant of radius.
In the diffuse interstellar medium $\beta$ is around 1.8 (for a review see \citealt{Hildebrand83}), which yields a temperature slope of -$\sim$0.35.
In high-density regions, $\beta$ may become lower due to dust growth (e.g. \citealt{OH94}), resulting in a steeper temperature profile.
Values of $\beta$ lower than 1.8 have been have been reported by some previous observations towards Class 0-II young stellar objects, and towards protostellar and prestellar cores (e.g. \citealt{Beckwith1991}, \citealt{Jorgensen07}, \citealt{Bracco2017}, and references therein).
It should be caveated that the previous (sub)millimeter observations of dust growth might have systematically underestimated $\beta$ values owing to underestimating optical depths (e.g., \citealt{Li17}), neglecting the effect of dust scattering opacity (e.g., \citealt{Liu19}), as well as the effects of temperature mixing when performing SED fittings (e.g. \citealt{Juvela17}).
From a modern point of view, on the spatial scales of molecular clumps, there might not yet be a solid evidence for the presence of $\beta<$1, i.e. exhibiting a flattened SED curve at longer wavelengths. We note that observations revealing a prevalent excess of 3 mm emission compared to the generic modified blackbody model have been reported (e.g. \citealt{Lowe21}), however, the origin of such a feature remains uncertain.

In Figure \ref{fig:tprofiles}, we also compare the observed temperature profiles with the centrally heated models evaluated for $\beta=$1 (i.e., $T=$ 2.70$(\frac{L}{L_{\odot}})^{0.2}$~$(\frac{R}{\mathrm{1\,pc}})^{-0.4}$; this is based on the assumption of a dust sublimation temperature of 1.1$\times$10$^{3}$ K in the derivation, \citealt{Adams91}, for more calculation details see their appendix), and with the simple expectation based on the Stefan-Boltzmann law, i.e. T $=$ 0.86$(\frac{L}{L_{\odot}})^{0.25}$~$(\frac{R}{\mathrm{1\,pc}})^{-0.5}$. 
We found that these profiles qualitatively agree with the measurements of $T(r)$ except for G13 and G31.

From Figure \ref{fig:tprofiles}, it is seen that the temperature profiles of G13 and G31 deviate from the form of a single power-law.
Specifically, they show abrupt changes as well as elevated temperatures in the ranges of 0.1-0.3 pc and 0.1-0.5 pc radii, respectively.
After we adjusted $T(r)$ (Appendix \ref{app:radmc}) according to dust SED profiles, G13 shows a less prominent temperature enhancement in the center while G31 still stands out.
\citet{Beltran18} also noted the steep temperature profile of G31 within the central 0.1 pc.
Their measurement consistently falls onto the functional form we fitted (i.e. the thick purple line in Figure \ref{fig:tprofiles}) which was based on our independent measurements of $T_{\mathrm{rot}}$ on the larger spatial scales. 

The rapid decrement of radial temperatures observed in G31 and G13 may be explained by their density profiles of the embedded dense gas structures (i.e., $\rho_{\mathrm{dense}}(r)$).
In Figure \ref{fig:nprofiles}, it can be seen that these two sources have the most steeply decreasing $\rho_{\mathrm{dense}}(r)$. 
In addition, their central $\sim$0.1 pc regions show prominent high-density plateaus.
The optically thin assumption of \citet{Adams91} may break down in the central region of the G31 and G13. 
Such a higher concentration of dense gas may steepen the temperature distribution in the inner envelope according to $\propto$ r$^{-\frac{1-q}{4-\beta}}$ (e.g. \citealt{Adams85}, \citealt{Rolffs11}), where $q$ is the power-law index of radial density profile. This is sclosed to the diffusion approximation, which effectively means that a higher optical depth gas would mimic the lower value of $\beta$ in determining the temperature structure.
The presence of flattened (protostellar) disks could also induce a steeper gradient ($\sim$-0.75) of temperature variations due to the gas heating by infall and accretion shocks (e.g. \citealt{LyndenBell74}, \citealt{Walch09}). Interestingly, hydrodynamic calculations of protostellar collapse demonstrate that a centrally flattened density profile results in a transitory phase of energetic accretion (\citealt{FC93}, see also \citealt{Henriksen97}), which may also be tentatively related to the elevated temperatures of G13 and G31.

Observations towards dense massive cores and envelopes of YSOs and discs generally find temperature slopes in the range of $\sim$[-0.35,-0.7] (e.g. \citealt{Palau14}, \citealt{Beuther07}, \citealt{Persson16}, \citealt{Jacobsen18J}, \citealt{Vanthoff20}, \citealt{Gieser21}). However, the exceptionally steep temperature profile of slope steeper than -0.9 is seen at 1 000-2 000 au around massive YSO object W3IRS4 (\citealt{Mottram20}), reminiscent of the result of G31 based on observations of similar spatial scales (\citealt{Beltran18}). 

Strong radiative feedback has been invoked as a possible mechanism resisting over-fragmentation, which favors the formation of massive stars and can be influential in ultra dense environment (e.g. \citealt{Krumholz12}).
When observed with $\sim$2000 au resolution, source G31 consists of two cores with one major core dominating the emission ($\sim$60 times flux difference at 1 mm continuum, \citealt{Beltran18}). 
The elevated temperature in the inner 0.1 pc might have suppressed the fragmentation of the envelope structure of the main core, leaving with only one satellite core surrounding it. The highly concentrated dense gas structure of G13 traced by methanol lines (Figure \ref{fig:nmaps}) that does not extend further beyond its continuum emission (except to the west direction) may also indicate a featureless fragmentation; high resolution (1$''$) MIR imaging by \citet{Varricatt18} reveal a binary system, although the mass contrast between the two YSO objects embedded is not as drastic as that of G31.

 \subsection{Density structure evolution: comparison with theoretical predictions}\label{sec:dens}


Figure \ref{fig:nmaps} shows the $n(\mathrm{H_{\mathrm 2}})$ maps which were derived based on the RADEX modeling for the CH$_{3}$OH lines (Section ~\ref{sec:radex_nh2}).
In general, $n(\mathrm{H_{\mathrm 2}})$ decreases radially with respect to the continuum center, although some sources appear notably asymmetric and present localised over-densities at large radii.
We use single power-law forms (Equation \ref{eq:rhoradex}) with parameters listed in Table \ref{tab:nprofile_fits}, to describe the radial change of $n(\mathrm{H_{\mathrm 2}})$ maps.
Comparing the observed profile to the fitted single power-law form (Figure \ref{fig:nprofiles}), it can be seen that there are higher density plateaus at the centers of G13 and G31, such that these two sources are better described by piece-wise power-law including a central density profile of slope $\sim$0.
Source G10 harbors even more dense gas at $>$0.1 pc radii as compared to its best-fit single power-law model, which is due to the highly flattened dense gas geometry of an edge-on rotational disk as revealed by gas kinematics (\citealt{Liu17}).
In fact, for G10, the slope of dense gas profile $q_{\mathrm{dense}}$ is an average from two distributions: a geometrically flattened, high-density pseudo-disk with a slope shallower than -0.5, and an envelope of which the gas density decreases sharply (Figure \ref{fig:nmaps}).
Clumps G28, G19 and G08a may also harbor high-density plateaus at the centers although they are not as significantly resolved as those in G10, G13 and G31.

We emphasize that the two set of power-law density slopes derived reflect clump gas in different density regimes: from single-dish continuum the slope represents density distribution of {\it{averaged}} (mass-weighted) bulk gas ($\bar{\rho}$$\sim$10$^{4.5}\,cm^{3}$) where the spherical symmetry might be a rather good approximation of source geometry, while the slope deduced from methanol emission is reflecting the truncated central dense gas portion ($\rho$$\gtrsim$10$^{6.5}\, cm^{-3}$) with an extension of 0.1-0.5 pc.
This dense gas structure is likely highly fragmented and of reduced dimension due to the dominant role of self-gravity at progressingly smaller scales and denser regimes. 

Figure \ref{fig:slopes_lm} reveals the evolutionary trend of the radial density profiles of bulk gas and dense gas structures of the target massive clumps. 
Our results indicate that in the initial stage of massive clump evolution, gas may be less concentrated than the singular isothermal sphere (SIS, \citealt{Shu77}), and even the logatropic model (\citealt{MP97}).
Observations towards low-mass cores at early stages have generally found density profiles with a central plateau (e.g. \citealt{WardThompson99}, \citealt{Bacmann02}). 
This density structure implies low pressure gradients at small radii, and hence in quasi-static models additional support of magnetic fields is required. The magnetic field and subsequent ambipolar diffusion may explain the existence and evolution of the shallower radial density profile of the gas, which is at an initial stage of equilibrium (e.g. \citealt{Mouschovias91}). 
Alternatively, in an isothermal state, the incoherently converging compression wave of shocks propagating outside-in can also disrupt the centrally peaked density profile (\citealt{Whitworth96}).
Finally, hydrostatic models of a self-gravitating gas externally heated (e.g. \citealt{Falgarone85}) can also result in a similar density configuration. 
On the other hand, we can not rule out the possibility that the shallower density profiles are reflecting the underlying fragmentation that is distributed widely within the clump, and that the fragments do not show significant mass segregation (\citealt{ASH19}). Their imprints would result in a rather uniform density profile for the clumps. 

Following definitions in Section \ref{sec:T_rho_profiles}, we denote the power-law slopes for the radial gas density of bulk gas and dense gas as $q_{\mathrm{bulk}}$ and $q_{\mathrm{dense}}$.
In G18 and G28, both $q_{\mathrm{bulk}}$ and $q_{\mathrm{dense}}$ are shallower than $-$1 (Figure \ref{fig:evo_trend}).
This may be due to a combination of the heated gas profiles and a higher level of turbulence, which effectively slow down gravitational collapse at early stage (Figure \ref{fig:vwidthprofiles}; more in Section \ref{sec:velo_width_dis}). 
Although these two sources have L/M $\lesssim$10, and are classified to be younger than or just reaching the ZAMS phase (\citealt{Molinari08}, \citealt{Giannetti17}), it is possible that the enhanced energy release of accretion associated with the massive star formation already introduces an observable temperature gradient (Figure \ref{fig:tprofiles}). 

The more evolved source G08a has a slightly shallower overall distribution of gas ($q_{\mathrm{bulk}}$$\sim$-1.2) than its dense gas component ($q_{\mathrm{dense}}$$\sim$-1.3), although the difference is within the errors. 
Interestingly, source G19 displays an opposite relation between the two slopes ($q_{bulk}$$\sim$-1.4, $q_{\mathrm{dense}}$$\sim$-0.6; Figure \ref{fig:slopes_lm}).  
Comparing the virial state of these two sources (Figure \ref{fig:vwidthprofiles}), it seems G19 is close to a global state of energy balance while G08a has excessive kinetic energy in the central 0.2 pc, possibly caused by energy transfer and induced motions from gravitational collapse (more in Section \ref{sec:velo_width_dis}). 
If a steeper density profile corresponds to a more dominant role of gravitational collapse, then the kinematic differences inferred from line-widths are compatible with the relation between density slopes in local (dense gas regime) and global (bulk gas regime) scales for these two sources. We discuss this in more detail in Section ~\ref{sec:evo}.    

The overall trend seen in the right panel of Figure \ref{fig:evo_trend} can be qualitatively explained by the growing role of self-gravity in the evolution of gas dynamics in massive clumps (\citealt{Lee15}, \citealt{Gomez21}). 
This is illustrated in hydrodynamic simulations of star-forming clouds with continuously driven turbulence (\citealt{Lee15}) that shows self-gravity plays the dominant role for all changes that happen with density structures over the scales up to $>$0.1 pc. 
Their results reveal that the power-law slopes around density peaks change from -1.3 to -1.55 before and after onset of star-formation, of 0.75 and 1.25 times global free-fall timescale ($t_{\mathrm{ff}}$). 
By separating the scales based on dominant mass contributor of gas or (forming) stars, \citet{Murray17} demonstrate that close to the (proto)stars the gas density profile has a slope of -1.5 (an attractor solution, see also \citealt{Coughlin17}) and in the outer envelope the slope ranges from -1.6 to -1.8. 
These values are in quantitative agreement with the evolved sources in our sample.
The three sources having bulk gas density slopes shallower than -1 (left panel of Figure \ref{fig:slopes_lm}) may reflect an initial condition close to a constant-density core when gravitational collapse has not significantly altered the gas density profile, for both the bulk gas and dense gas structures (G18 and G28) or only the dense gas component (G19). 
The tendency of an initial flat inner density profile remaining flat over one $t_{\mathrm{ff}}$ determined by the gas central density is also discussed in \citet{Henriksen97}, which is attributed to the fact that the cloud center has the fastest growing velocity mode. 
The timescale is fleetingly short without further support of turbulence. 

\subsection{The kinematic state of clumps: radial profiles of molecular line-width and virial parameter}\label{sec:velo_width_dis}
\begin{figure*}
\begin{tabular}{p{0.32\linewidth}p{0.32\linewidth}p{0.32\linewidth}}
\hspace{-0.55cm}\includegraphics[scale=0.33]{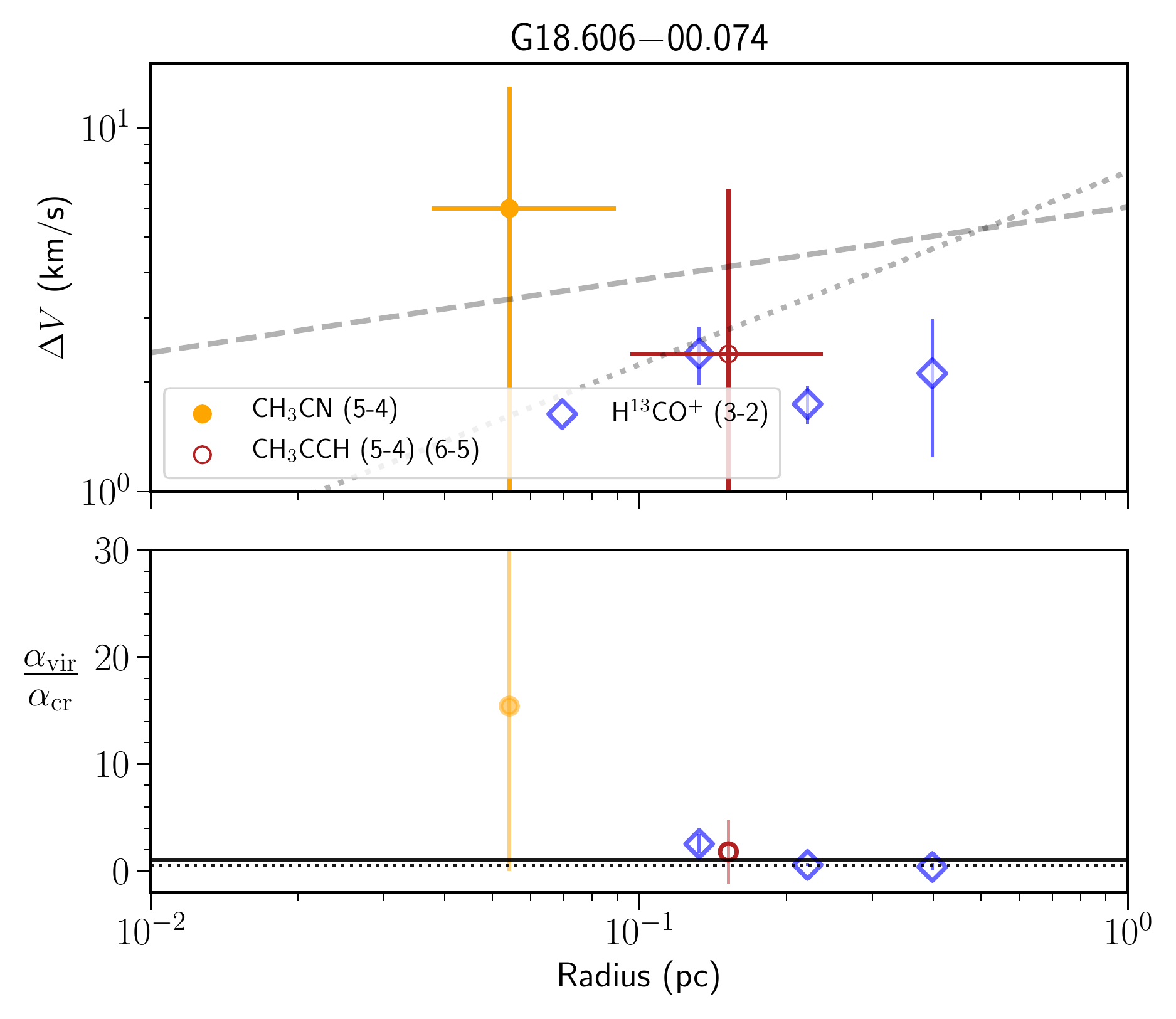}&\hspace{-0.1cm}\includegraphics[scale=0.33]{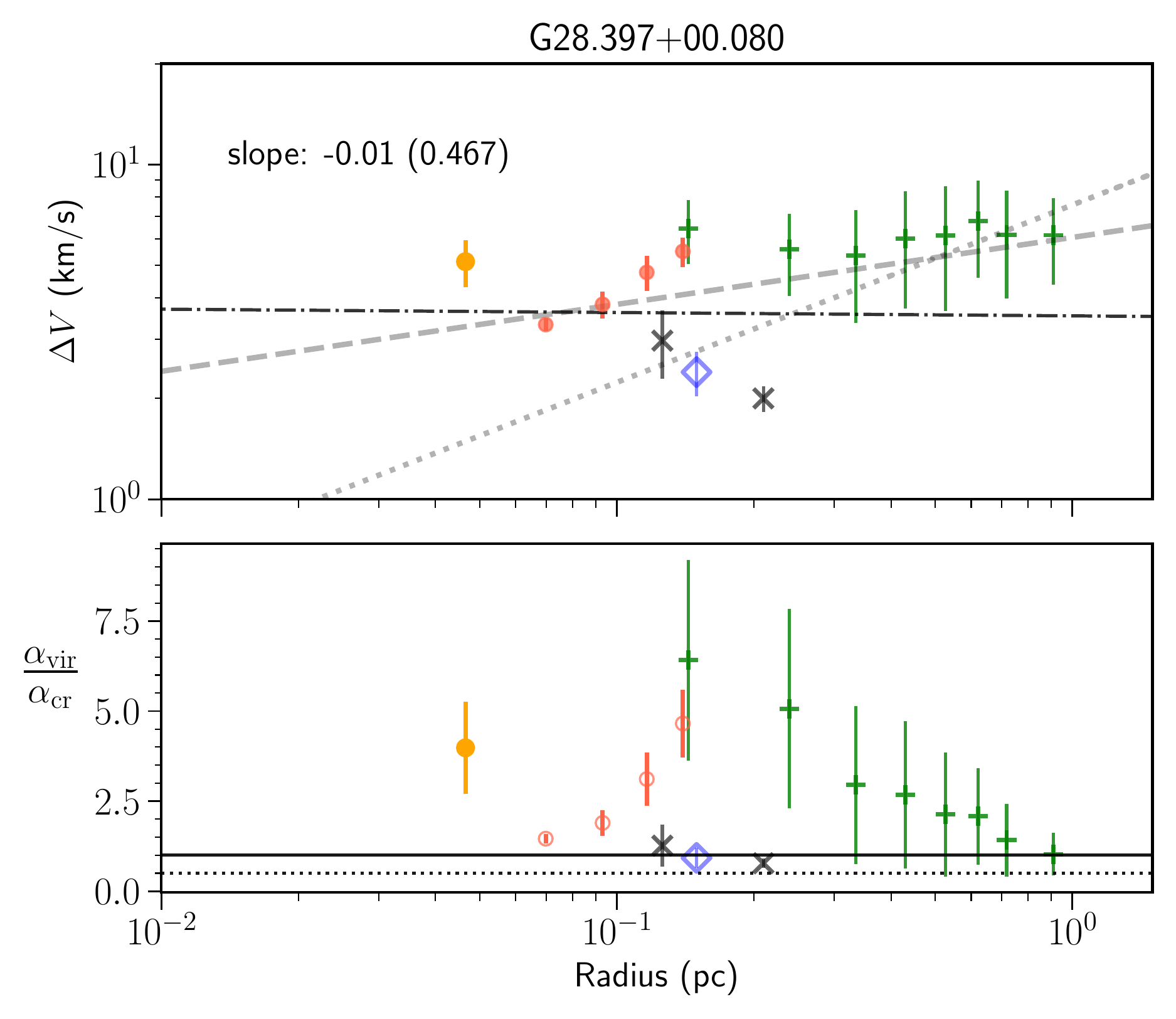}&\hspace{0.1cm}\includegraphics[scale=0.33]{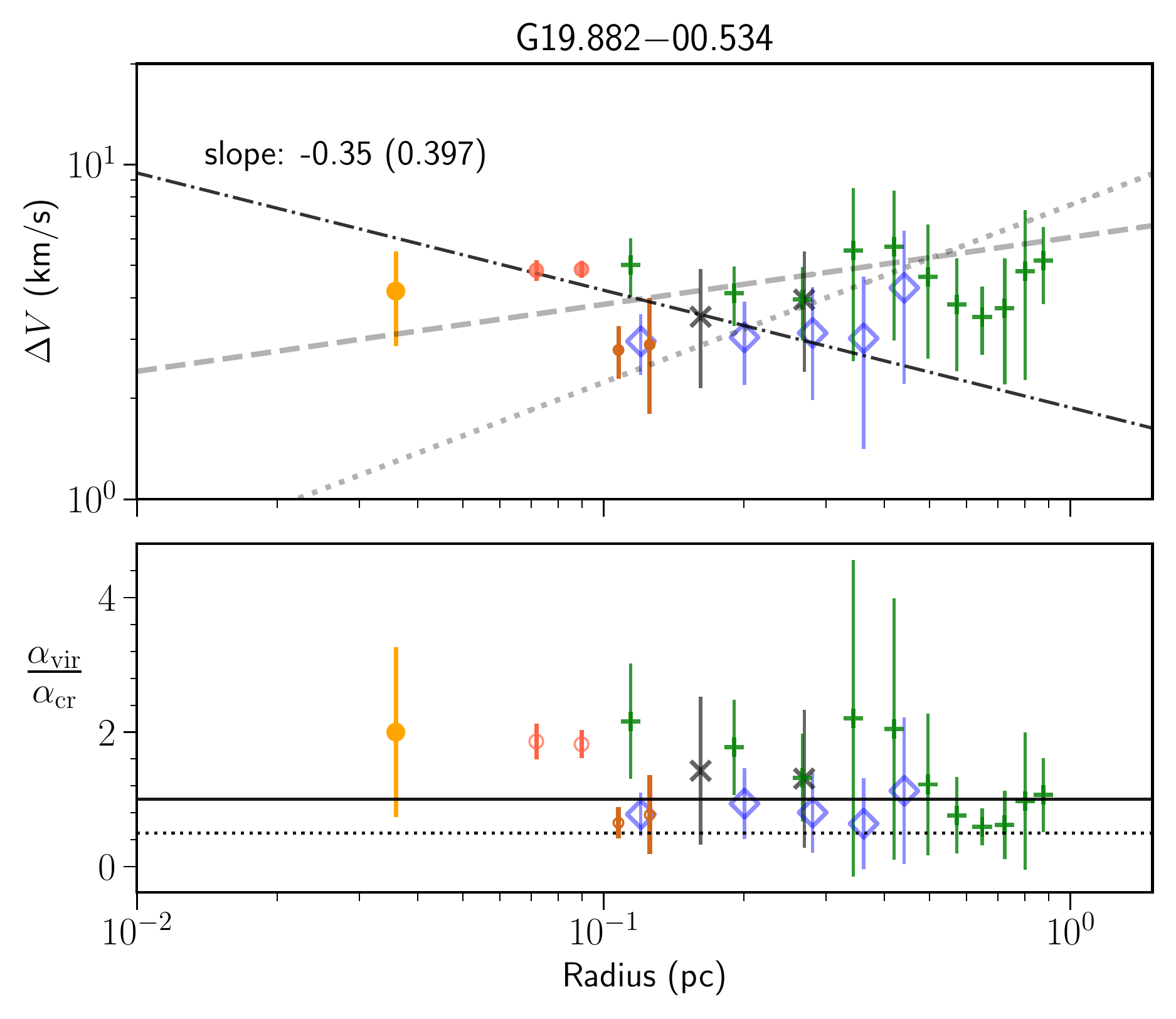}\\
\hspace{-0.55cm}\includegraphics[scale=0.33]{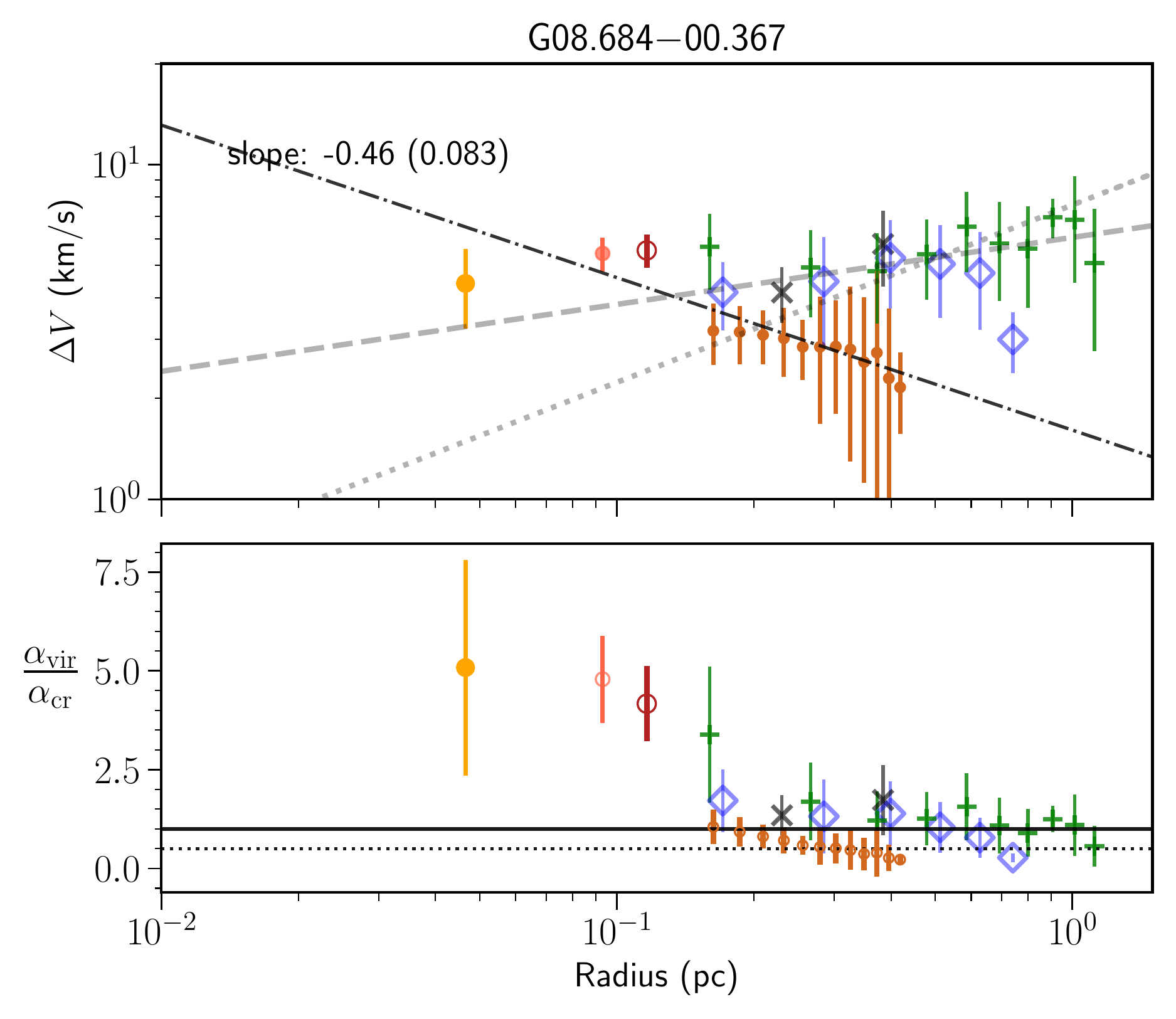}&\hspace{-0.1cm}\includegraphics[scale=0.33]{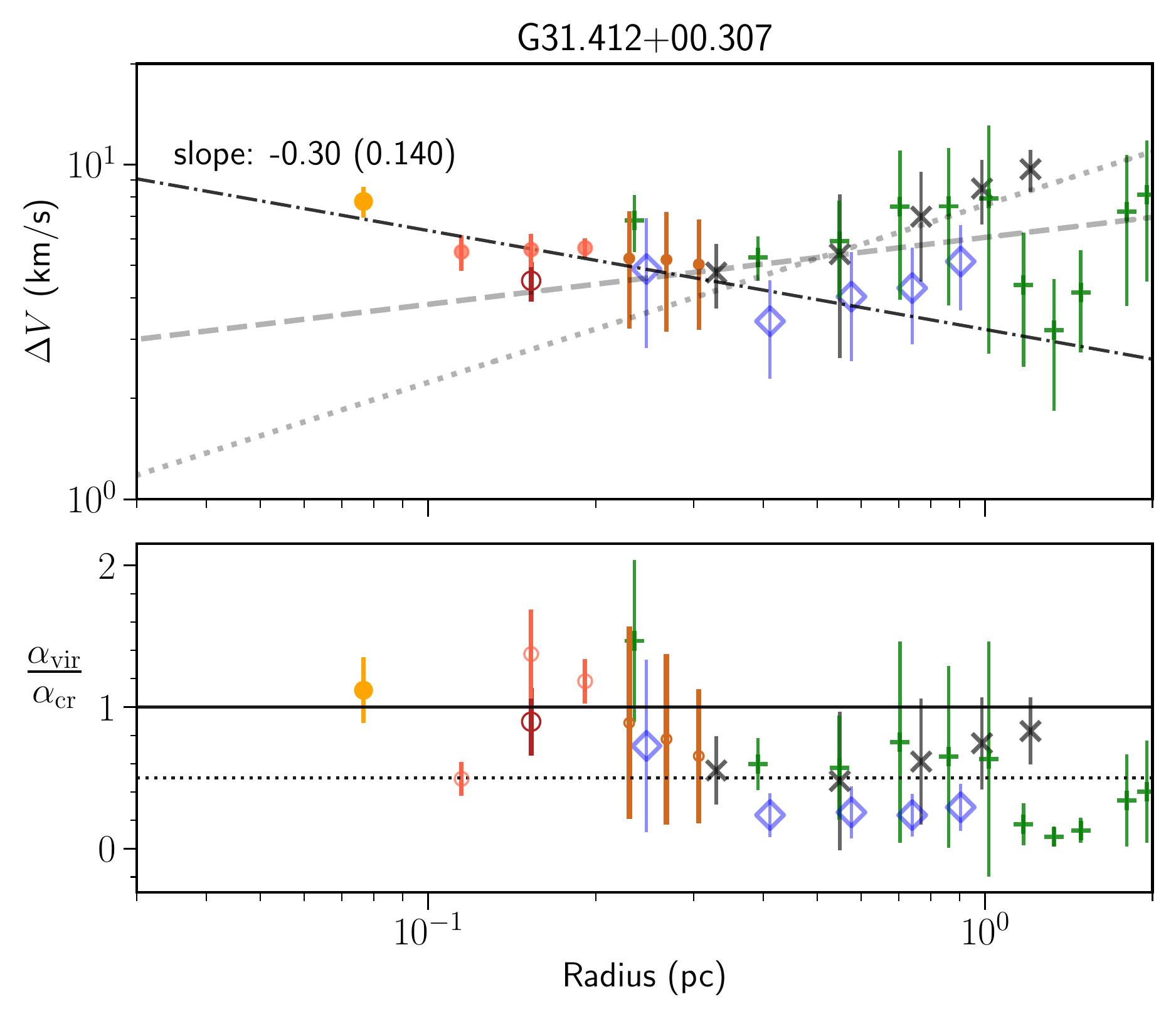}&\hspace{0.1cm}\includegraphics[scale=0.33]{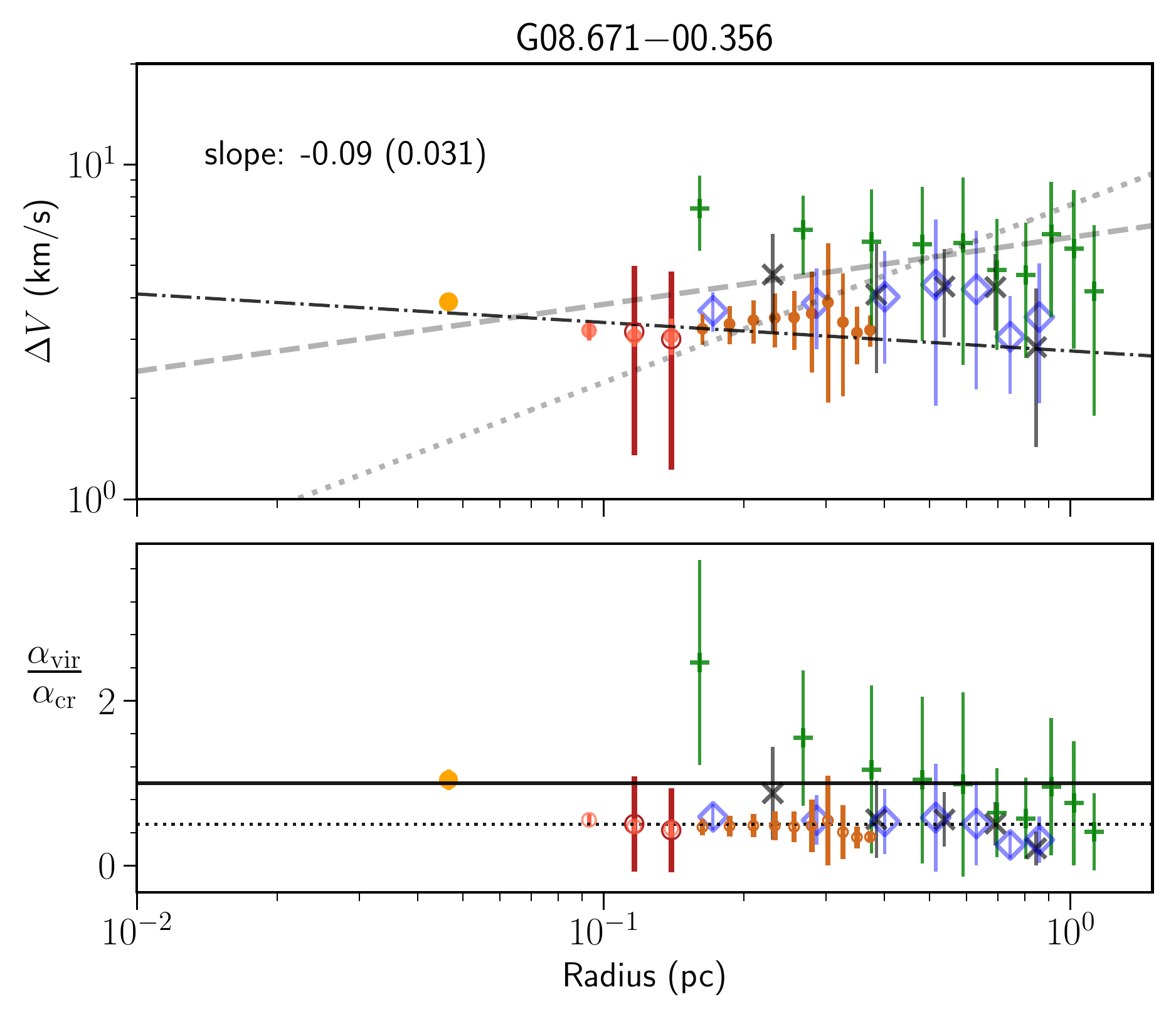}\\
\hspace{-0.45cm}\includegraphics[scale=0.33]{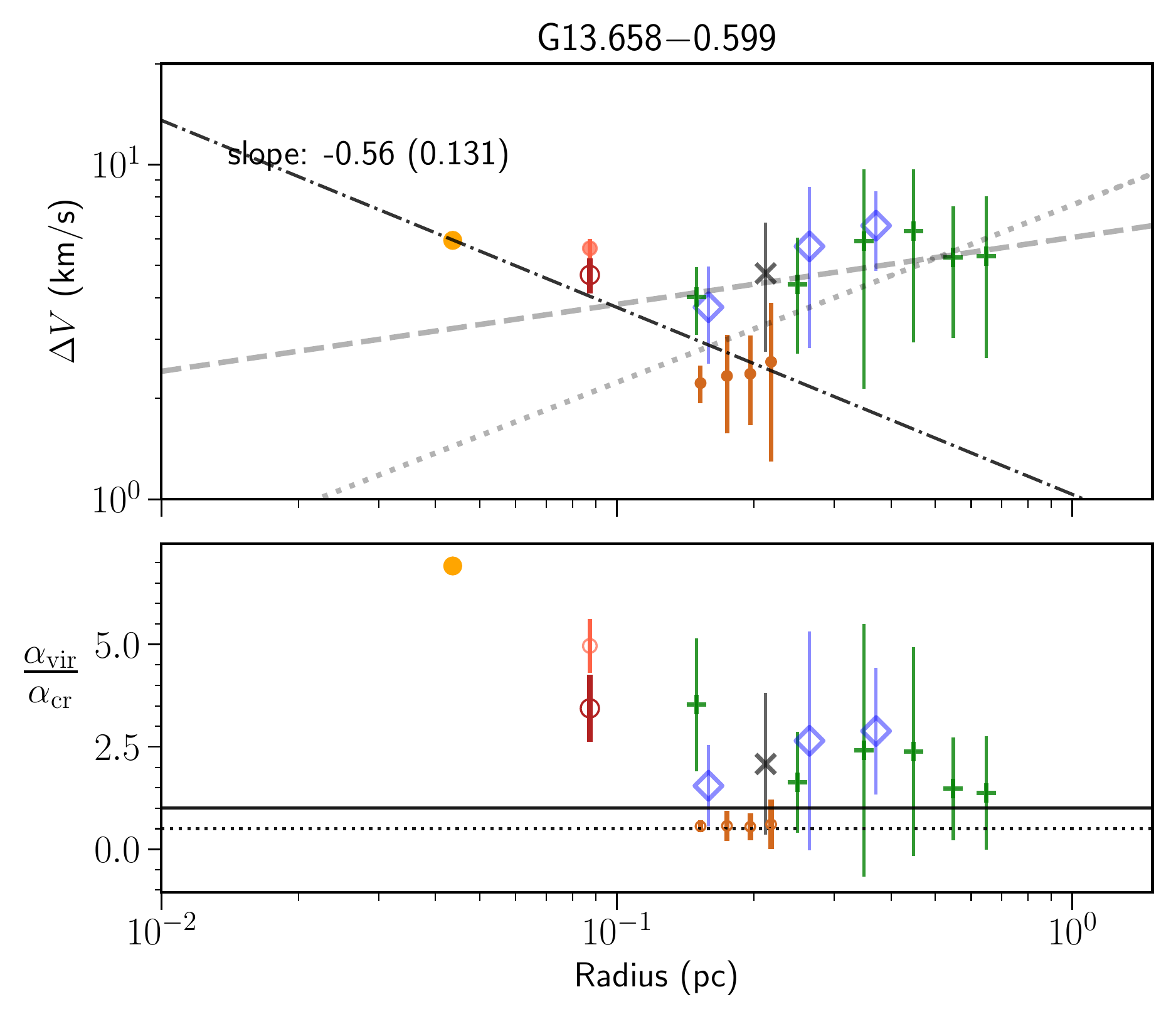}&\hspace{-0.2cm}\includegraphics[scale=0.3]{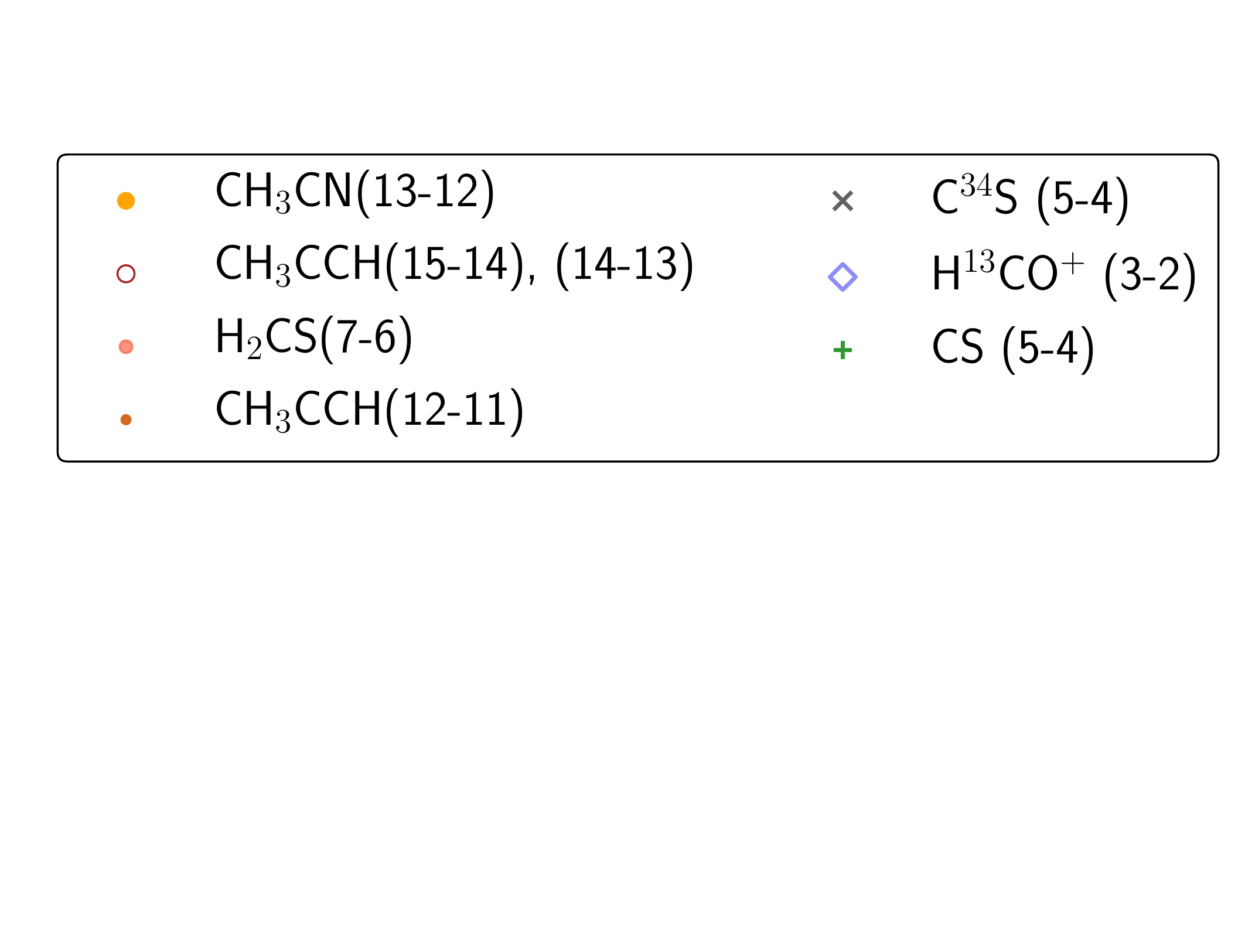}&\hspace{0.1cm}\includegraphics[scale=0.33]{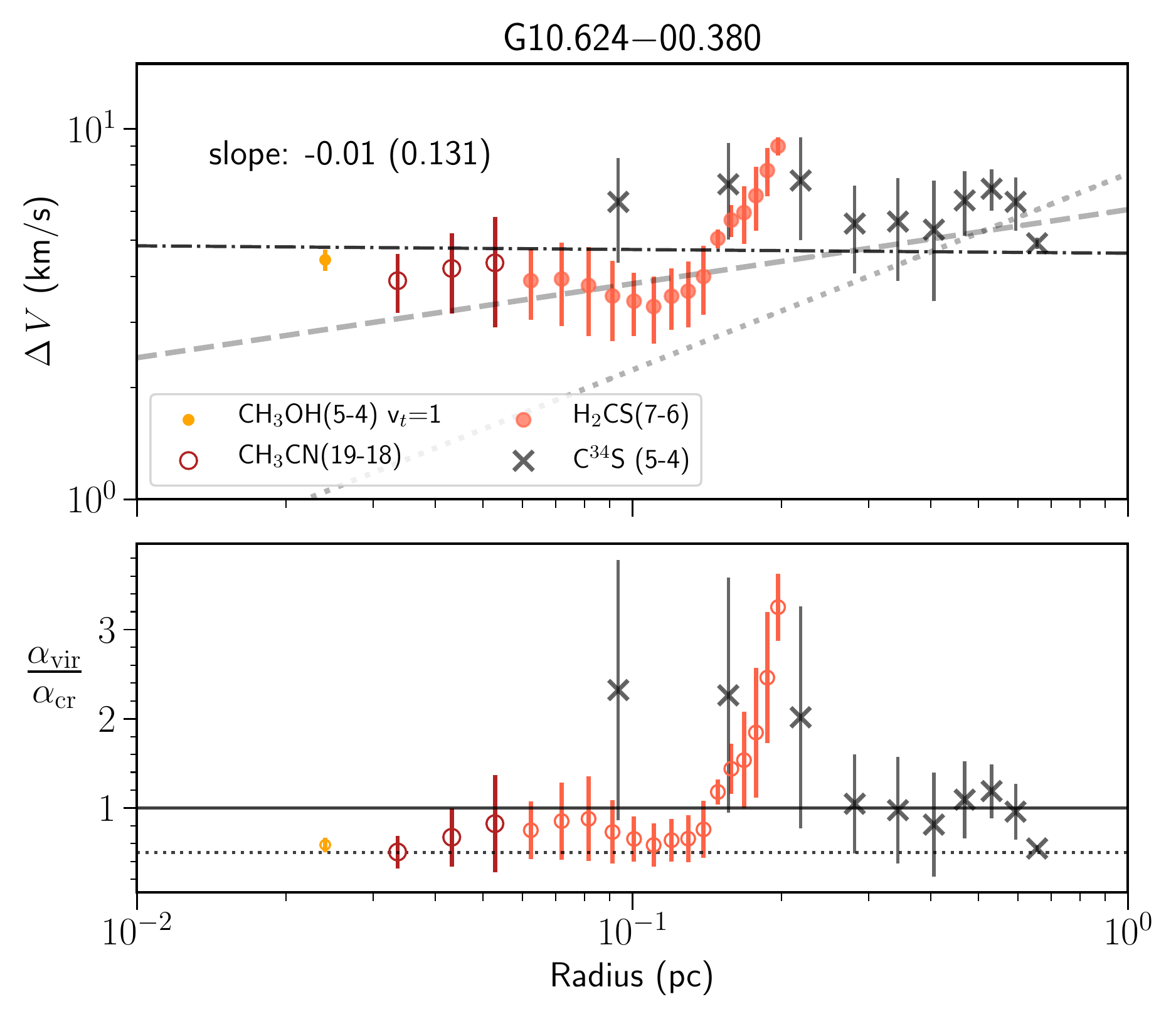}\\
\end{tabular}
\caption{Derived averaged radial linewidth and virial parameter profiles. 
Dashed dotted line in each upper panel indicates the power-law fit to the linewidths from thermometer lines. In each upper panel, gray dotted line indicates relation found by \citet{CM95} of Orion low-mass cores scaled up by a factor 5, which roughly matches the observed linewidth in our case, of $\rm\Delta v/1\,km\,s^{-1}$ = 7.5 (r/1 pc)$^{0.53}$ ; Gray dashed line shows a relation of $\rm\Delta v/1\,km\,s^{-1}$ = 6.0 (r/1 pc)$^{0.2}$. These two reference lines are identical in all plots. In each lower panel, the ratio between $\alpha_{\mathrm{vir}}$ to critical virial parameter $\alpha_{\mathrm{cr}}$ is shown; horizontal solid line and dotted line indicate ratio 1 (equipartition) and 0.5 (virial equilibrium), respectively. All plots except for G10 and G18 share the same legend shown in the middle panel of the last row, in which the color coding for the thermometer lines is the same as in Figure \ref{fig:tprofiles}.}
\label{fig:vwidthprofiles}
\end{figure*}

Based on the calculations in Section \ref{sec:velo_width}, Figure \ref{fig:vwidthprofiles} shows the radial profiles of the molecular linewidths ($\Delta V$ = 2.355 $\sigma_{v}$) for individual species.
We fitted a power-law to the radial linewidth profiles traced by thermometer lines (i.e. excluding the data points from C$^{34}$S and H$^{13}$CO$^+$).
In G08a, G31 and G13, the linewidth traced by these lines decrease radially with power-law indices of -0.40, -0.31, -0.56, respectively. 
Except for the sources G18 and G19, the linewidths on the extended regions traced by C$^{34}$S, H$^{13}$CO$^+$ are systematically larger than that traced by the thermometer lines. Again, for clump G18, we had to rely on pointed observations from IRAM 30m telescope to tentatively mark the radii of the linewidths obtained from thermometer lines, based on a fixed temperature profile (Sect. \ref{sec:xclass}). We therefore note that the analysis for this clump is not based on the same, spatially resolved measurement as the other clumps.

In general, the variation in line widths does not seem to follow a monolithic radial change, as opposed to the $\sigma_{\mathrm{v}}$$\,\propto\,$$r^{1/2}$ universal law observed in giant molecular clouds and low-mass cores (i.e., Larson's first law; \citealt{Larson81}). This is in agreement with the results presented in \citet[][Figure 7]{Izquierdo21a}, where they show that the linewidth-size relationship is not uniform but rather depends on the analysed spatial scale and the physical processes taking place there (see also \citealt{Hacar16}). 

In high-mass star-forming regions, the observed line-widths are systematically larger, and $\sigma_{\mathrm{v}}$ may have a shallower power-law relation with scales (e.g. \citealt{Plume97}, \citealt{CM95}).
In addition, the virial equilibrium naturally infers that $\sigma_{\mathrm{v}}$ has a dependence on gas surface density, such that $\sigma_{\mathrm{v}}\,\propto\,$$(\Sigma R)^{0.5}$ (\citealt{Heyer09}).

Moreover, recent 1D simulations incorporating gravitation-driven turbulence (i.e., adiabatic heating; c.f., \citealt{RG12}) found distinct relations between $\sigma_{\mathrm{v}}$ and spatial scales for star-forming clumps, due to the change of the dominant turbulent driving mechanism from inner to outer regions \citep{Murray15}.
It has been found that $v_{T}$ $\propto$ $r^{-0.5}$ holds within the sphere enclosed by the stellar influence radius, which is defined as the radius where the enclosed gas mass is comparable (e.g., 1-3 times) to the stellar mass (\citealt{Murray15}, \citealt{Murray17}; see also \citealt{Coughlin17}, \citealt{Xu2020} for the analytic derivations).

As a rough estimate of the stellar influence radius, assuming the $L_{\mathrm{bol}}$ of each source is contributed solely from the luminosity of a single ZAMS star, adopting the stellar evolution model of \citet{Choi16}, the stellar mass $M_{\star}$ are estimated to be 12$M_{\odot}$ for G19, 15-16$M_{\odot}$ for G08a, G13 and G28, 30$M_{\odot}$ for G08b. 
\citet{KG09} fit radiative transfer accretion models of YSO to source G19 and G13 by building SED from near-infrared to submm, and their derived stellar masses are consistent with our rough estimates. 
For G10, previous observations suggest a stellar mass of $\sim$200$M_{\odot}$ (\citealt{Liu2010ApJ}). 
We note that inferring stellar mass from luminosity assuming the sole contribution from a single star leads to a lower limit, as these clumps are forming a cluster of stars following the IMF.  With the same total bolometric luminosity, this corresponds to a higher total stellar mass.
Comparing with the central core masses we derived from 1.2mm continuum (Table \ref{tab:sma_cont_direct}), it seems we marginally resolved the influence radius (about half beam FWHM) for sources G13, G08a and G10, but a bit further in the case of G08b. 
This may partially explain why in the inner region of G08b the observed linewidths do not decrease rapidly with radius.
In G10, the observed linewidths have shallower variations with radius.
This may be due to the presence of the $\sim$0.2 pc scale edge-on rotational disc (\citealt{Liu17}), towards which gas settles into coherent rather than random motions.

For the early-stage clumps G18, G19, and G28, the bolometric luminosity is likely dominated by accretion energy. 
This leads to an overestimate of the stellar mass, such that in reality there appears to be a more drastic difference between the embedded stellar mass and the gas mass at the scale probed, than our rough estimates.
Therefore, it is likely that we do not see the decreasing trend of linewidths with radius in their inner regions due to the too coarse resolution to resolve the stellar influence radius.
In addition, assuming turbulent velocity scales as the local infall velocity, \citet{Coughlin17} demonstrates that there is temporal steepening of the radial profile of the rms velocities, due to gravitational field starting to dominate the dynamics of inflowing gas, which changes from $v_{\mathrm T}$ $\propto$ $r^{-0.2}$ to $v_{\mathrm T}$ $\propto$ $r^{-0.5}$. 
Hence, it may also be that early-stage clumps have a shallower inner slope that tend to flatten out, further smeared in coarser resolution measurements.

\begin{table*}
\centering
\begin{threeparttable}
\caption{Properties related to radial linewidth and virial parameter profiles of the target clumps.}
\label{tab:linewidth_virial}
\begin{tabular}{ l c c c c c c c}
\toprule
  
  Source  & 
  $M_{\star}^{a}$&
  Slope$^{b}$ &  
  $\Delta V_{\mathrm{0.1\,pc}}^{b}$ & 
   Radial variations of $\alpha_{\mathrm{vir}}/\alpha_{\mathrm{cr}}$\\
  & ($M_{\odot}$) & 
  & (km/s) &&
  \\
  \midrule
            G18$^{c}$  & 11&-&-&$>$10 in center (<0.1 pc) and $\sim$1-2 at outer envelope (>0.1 pc)\\
            G28  &23&-0.01(0.48)&3.6(0.6)&fluctuate at $\sim$0.5-5 at all radii\\
	        G19&18&-0.35(0.40)&4.2(0.5)&fluctuate at $\sim$0.5-5 at all radii\\
	        G08a&17&-0.46(0.08)&4.6(0.3)&$>$2 in center (<0.2 pc) and $\sim$0.5-2 at outer envelope (>0.2 pc)\\
            G31  &64&-0.30(0.14)&6.3(0.4)&$\sim$1 in center (<0.1 pc) and $\sim$0.5-1 at outer envelope (>0.1 pc)\\
            G08b  &26&-0.09(0.03)&3.3(0.1)&$\lesssim$1 at all radii\\
            G13 & 19&-0.56(0.13)&3.7(0.4)&$>$2 in center (<0.1 pc) and $\sim$1-2 at outer envelope (>0.2 pc)\\
            G10 & 68&-0.01(0.13)&4.7(0.4)&$\sim$0.5 in center (<0.1 pc) and $\sim$1-2 at outer envelope (>0.1 pc)\\
\bottomrule
\end{tabular}
    \begin{tablenotes}
      \small
      \item $^{a}$: Estimated (single) stellar mass based on interpolating the ZAMS models of \citet{Choi16}, from the clump $L_{\mathrm{bol}}$ listed in Table \ref{tab:radmc_para}.  
\item $^{b}$: Fitted slope and linewidth amplitude at 0.1 pc (and errors) of the linewidth radial profile in the inner region. The fitted relation is indicated as the dashed dotted line in the upper panel of Figure
\ref{fig:vwidthprofiles}.  
\end{tablenotes}
  \end{threeparttable}
\end{table*}

Exterior to the stellar influence radius, simulations found that $v_{\mathrm{T}}$ $\propto$ $r^{0.2}$ (see also \citealt{Lee15}). Compared to classical subsonic turbulence following Kolmogorov law ($v_{\mathrm T}$ $\propto$ $r^{1/3}$) or supersonic turbulence ($v_{\mathrm T}$ $\propto$ $r^{1/2}$), the shallower scaling relation between linewidth and spatial scale of the region beyond stellar influence radius, could be due to additional energy converted from (part of) gravitational collapse (\citealt{BP11}, \citealt{Xu2020}) and/or kinetic energy of extended inflow gas transported by radial motions (\citealt{Padoan19}). 
\citet{KH10} demonstrate that the conversion efficiency to internal turbulence depends on the density contrast between the accreting entity and the inflowing gas.
These may explain why in the outer regions of G08a, G08b, G13 and G10, the observed linewidths rise with radius, having a power-law slope of $\gtrsim$0.2-0.3. 

In light of the radial profiles of $\alpha_{\mathrm{vir}}$, it appears that our target clumps can be classified into three types, based on their $\alpha_{\mathrm{vir}}/\alpha_{\mathrm{cr}}$ ratios: sources G08b, G31 and G10 are in an overall virial to sub-virial state; G18, G08a and G13 show super-virial states in the centre and sub-virial states in outer envelope; G28 and G19 have uniform $\alpha_{\mathrm{vir}}$ fluctuating mostly above $\alpha_{\mathrm{cr}}$ (see the detailed description in Table \ref{tab:linewidth_virial}). 
To explain the observed super-virial states in the central regions of some of these clumps, it is required to introduce other line-broadening mechanisms.

The three sources that have large virial parameters in the centers have the lowest central core masses and densities (Table \ref{tab:sma_cont_direct}). 
If central cores are treated as a decoupled structure from the clump envelope, the more massive and denser cores being more sub-virial is consistent with previous observations towards massive star-forming regions (e.g. \citealt{Kauffmann13}, \citealt{Liu15}, \citealt{Traficante18}). However, due to the insufficient angular resolution, the estimate of $\alpha_{\mathrm{vir}}$ in the clump center may suffer from larger uncertainties, e.g. there is possibly higher level of clumpiness with complex velocity structures that confuse the linewidth measurement. 
In addition, the $\alpha_{\mathrm{vir}}$ in the centermost is derived from linewidth of CH$_{3}$CN. CH$_{3}$CN has been shown to trace the interface of convergent flows in high-mass clumps (\citealt{Csengeri11}), which is also enhanced in shocks (\citealt{Bell14}) and associated with hot accretion flows (\citealt{Liu15}).

\citet{Camacho16} and \citet{BP18}) have shown that the observed high $\alpha_{\mathrm{vir}}$ may not be indicative of pressure confining or gas dispersal.
Alternatively, for low column density clumps this picture might be due to highly dynamic externally driven gas accumulation. 
For high column density clumps it might be a result of neglecting the stellar mass or accreting materials outside the cores in contributing to the gravitational energy. 
The latter may qualitatively explain the super-virial state in the center of the three sources G18, G08a and G13, which does not necessarily indicate a halt of collapse. 
On the other hand, the inclusion of infall velocities to the observed linewidth may also cause the source which is undergoing collapse to appear seemingly unbound (e.g. \citealt{Smith09}). 
Infall motions of dense cores that have large $\alpha_{\mathrm{vir}}$$\sim$5-8 are revealed by \citet{Cesaroni19}, inside a massive cluster-forming clump. 

\citet{Giannetti17} suggest that CH$_{3}$CCH is a reliable tracer for the kinematics of bulk gas structures in the massive clumps.
From the $\alpha_{\mathrm{vir}}$ derived based on the CH$_{3}$CCH lines, it appears that all clumps are in sub-virial or virial status.
But at smaller scales in the central region of the clumps, the gas can appear super-virial.
Aside from the three clumps showing large $\alpha_{\mathrm{vir}}$ in the center, other clumps show an evolution of globally (for all radii) decreasing $\alpha_{\mathrm{vir}}$ with increasing $L/M$. 
\citet{BP18} and \citet{Camacho20} investigated the evolution of $\alpha_{\mathrm{vir}}$ in molecular clouds based on numerical simulations.
They showed that the gas structures may be initially assembled due to the large-scale turbulence.
Afterwards, the assembled gas structures evolve from over-virial to sub-virial status, and finally approach energy equipartition, or further become super-virial after one t$_{\mathrm{ff}}$ due to gas expulsion.
Our observations do not show such non-monolithic evolution of $\alpha_{\mathrm{vir}}$ with $L/M$.
This is likely because all target clumps remain in early to intermediate evolutionary stages, which are prior to or in the stage of active star formation and the gas mass dominates the mass budget preceding the gas dispersal stage.

\subsection{Comparison with chemical models: carbon-chain molecules and sulfur-bearing species}\label{sec:ab_dis}
The formation pathways of molecules and chemical rates are tightly influenced by gas temperature and volume density.
In Section \ref{sec:ab_cal}, we used the molecular column densities of CH$_{3}$CCH, H$_{2}$CS, CH$_{3}$CN, ,CH$_{3}$OH, C$^{34}$S, CCH, SO and SO$_{2}$ to derive the molecular abundance distributions and abundance ratios between relevant species.
We discuss comparisons between these results with predictions from chemical models in this section.

\subsubsection{CH$_{3}$OH, CH$_{3}$CCH, CH$_{3}$CN, and CCH}

CH$_{3}$CCH is designated as a `cold molecule' (\citealt{Bisschop07}), since it presents low excitation temperatures compared to other complex organic molecules (COMS) such as CH$_{3}$CN and CH$_{3}$OCH$_{3}$.
The formation pathways of CH$_{3}$CCH include cold gas phase reactions and grain-surface chemistry (\citealt{Calcutt19}). The reactions in gas phase involve thermal desorption which requires low temperatures. \citet{Calcutt19} suggested that CH$_{3}$CCH is a `gateway' molecule which traces the interface between hot cores/corinos and colder envelope, and extends further out in the envelope.
The inset in Figure \ref{fig:ab_profiles} shows measured CH$_{3}$CCH abundances averaged from the outer region ($\gtrsim$0.1 pc) of the clumps with the gas temperature measured at 0.3 pc. 
These properties represent the bulk luke-warm gas. The small variation of CH$_{3}$CCH abundance across our sample indicates that it is not a sensitive molecule to trace the $\gtrsim$100 K thermal desorption of hot core regions. Its emission rather traces the more extended gas and is a good kinematic tracer for $>$0.1 pc gas structures within massive clumps (e.g. \citealt{Giannetti17}). 
There is also not drastic radial change of the abundances of the other two COMs CH$_{3}$OH and CH$_{3}$CN for individual sources (Figure \ref{fig:ab_profiles}). 
This is expected since our angular resolution is not sufficiently high to reveal the $>$100 K region (confined within 0.1 pc) where prominent abundance enhancement of COMs is expected (\citealt{Garrod17}). 

The abundance profiles of CCH exhibit an increase towards the outer clump envelope for most the sources in the sample (Figure \ref{fig:ab_profiles}) and an overall decrease of abundance with clump luminosity.
These results reflect that CCH is abundant at an early stage and then transformed to other molecules in the clump center (\citealt{Beuther02a}). 
The abundance of CCH in outer regions is maintained by reproduction of CO dissociated by an external UV field. Similarly, in early-stage low-mass cores, CCH is also tracing more quiescent gas, e.g. the outer part of a circumbinary envelope (\citealt{vdk95}). 

We compare the abundance ratios between CCH, CH$_{3}$CN and CH$_{3}$CCH with CH$_{3}$OH in Figure \ref{fig:com_ratios}. 
In low-mass star-forming cores, the ratio of [CCH]/[CH$_{3}$OH] is usually used to distinguish less evolved warm carbon-chain cores with hot corinos (e.g. \citealt{Graninger16}). 
The strong anti-correlation of [CCH]/[CH$_{3}$OH] with $L/M$ indicates that in the central region of massive clumps this ratio serves well as an evolutionary indicator.
[CH$_{3}$CN]/[CH$_{3}$OH] has a tight correlation with $L/M$ as well: except the hot massive core G31, the other sources show a clear monotonic increase of [CH$_{3}$CN]/[CH$_{3}$OH] with increasing $L/M$. 
This is consistent with the formation timescale of the two species: CH$_{3}$OH forms at early stages (10 K) on grain surfaces, primarily via successive hydrogenation of CO (e.g. \citealt{Watanabe03}), while the formation of CH$_{3}$CN appears late, through radiative association between HCN and CH$^{+}_{3}$ on grain surface as well as CH$_{3}$ with CN in gas phase (\citealt{Nomura04}). 

\subsubsection{Sulfur-bearing molecules}
Time-dependent theoretical chemical models of sulfur-bearing species find that the evaporation of H$_{2}$S and subsequent chemical reactions in hot cores lead to an increasing abundance ratio of SO$_{2}$/SO (\citealt{Wakelam11}, \citealt{Esplugues14}). 
At a late stage of hot core evolution, H$_{2}$CS and CS are also drastically enhanced; together with SO$_{2}$, they become the most abundant sulphur-bearing species. 
In shocked regions, similar trends along evolution are predicted (\citealt{Wakelam05}, \citealt{Esplugues14}). But observations seem to suggest opposite variations of X(CS)/X(SO) with source evolutionary stages (\citealt{Li15}, \citealt{Gerner14}, \citealt{Herpin09}). 
Assuming same $^{34}$S/S isotopic ratio for our target clumps, we find a similar result which is reflected by the feature in Figure \ref{fig:com_ratios} (left panel, bottom row).  
Similar results of X(H$_{2}$CS)/X(SO) and X(SO$_{2}$)/X(SO) with clump $L/M$ are derived in Figure \ref{fig:com_ratios} (middel and right panel, bottom row): there are no prominent correlations found. 
We note that the temporal evolution of abundance of sulfur-bearing species is highly dependent on the source physical structure (\citealt{Wakelam05}, \citealt{Wakelam11}), which require tailored chemical modeling for individual source to disentangle this impact and the evolutionary time. 

\subsection{Density structure: relation with cloud structure and implications for different dense gas conversions}\label{sec:evo}

In this section, we discuss how the radial density profiles of massive clumps (see Section \ref{sec:dens}) may be linked to statistics of parental cloud structures on large scales ($\gtrsim$10 pc).
We also elaborate on the physical implications of the radial density profiles comparisons between bulk gas and dense gas structures, $\rho_{\mathrm{bulk}}$ and $\rho_{\mathrm{dense}}$.

Gravoturbulence simulations of molecular clouds (e.g., \citealt{Kritsuk11}, \citealt{Lee15}) showed that the cloud volume density probability distribution functions ($\rho$-PDF, $p_{\mathrm{s}}$) have lognormal and power-law forms in the low- and high-density regimes, respectively. 
The lognormal form is attributed to the primordial and maintained supersonic turbulence, while the power-law tail is usually regarded as a sign of gravitational collapse. 
Thus, the study of PDF statistics is a useful method to understand the physical mechanisms at play in molecular clouds.
Gravitational contraction converts a fraction of low-density gas to high-density structures over a timescale comparable to the free-fall timescale ($t_{\mathrm{ff}}$) of the mean density.
With this process, the slope of the power-law tail $s$ changes from s$\approx-$3 to s$\approx-$1.5-$-$1 (e.g. \citealt{Girichidis14}, \citealt{Guz18}) in a mean free-fall time.
When the power-law tail of the $\rho$-PDF is dominated by a single gravitationally bound massive clump with a $\rho\,(r)$ $\propto$ $r^{q}$ density profile, there is a relation $s\,=\,3/q$ (\citealt{Federrath13}).
In other words, $q$ is expected to evolve from $\approx-$1 to $-$3-$-$2 over a mean $t_{\mathrm{ff}}$.  Observational works usually constrain the column density probability distribution function (N-PDF, $p_{\eta}$) rather than $\rho$-PDF.
In this case, the power-law slope of the $N$-PDF ($s_{N}$) can be related to the radial density profile of the molecular clump by $\frac{2}{1+q}$. Column density mapping towards Galactic massive star forming complexes yield power-law slopes for $N$-PDF ranging between $\sim$-4 to -2 (\citealt{Lin17}).
The evolutionary trend of steepening density profiles (the bulk gas density profile) of the massive clumps resolved in this work is in general consistent with the prediction from cloud-scale statistics.


The cloud-scale (10 pc) dense gas fraction can be relatively well described by statistics of $\rho$/$N$-PDFs, which do not factor in features of spatial distribution. Within the $\sim$1 pc scale clump structure, the spatial configuration of dense gas can be critical in determining the properties of final star clusters. 
We discussed the gas radial density profiles probed by dust emission and dense gas tracer (CH$_{3}$OH) separately in Section \ref{sec:dens}.  
Here we note again that dust emission is tracing {\it{averaged}} (mass-weighted) bulk gas structures ($\bar{\rho}$$\sim$10$^{4.5}\,cm^{3}$) while CH$_{3}$OH emission probes dense gas regimes ($\rho$$\gtrsim$10$^{6.5}\, cm^{-3}$, up to several $10^{8}$ cm$^{-3}$) within the massive clumps. 
We note that the gas density difference seen by dust and higher angular resolution CH$_{3}$OH emission can also be partially related to the coarser resolution of the single-dish continuum images, which tend to smear out the gas over-densities at smaller spatial scales. However, the major cause of the difference is that CH$_{3}$OH transitions selectively trace the higher-density gas component, while the dust emission does not critically depend on the volume density of the gas (in the regimes we are interested in), providing a measure of bulk gas properties. 
By relying on two density probes for different gas density regimes, we go beyond the self-similar solution of gravitational collapse by characterising the density configuration of massive clumps with two ``layers'' $\rho_{\mathrm{bulk}}$ and $\rho_{\mathrm{dense}}$, each of which is described by a power-law form. 
The $\rho_{\mathrm{dense}}$ traced by CH$_{3}$OH from RADEX modeling (Section \ref{sec:radex_nh2}) is verified by full radiative transfer non-LTE models described in Appendix \ref{app:lime}, by introducing a volume filling factor $ff_{\mathrm{dens}}$ (Table \ref{tab:lime_radmcpara}) to spherical models. 
The low volume filling factor indicates that the dense gas is highly clumpy and organised into filament- or sheet-like structures, which is a natural outcome of supersonic turbulence and gravitational collapse (\citealt{Smith20}). 
In general there seems to be a higher density contrast (decreasing $ff_{\mathrm{dens}}$) with increasing $L/M$ (as listed sequentially in Table \ref{tab:lime_radmcpara}, last column). 
While $ff_{\mathrm{dens}}$ provides an upper limit of the volume filling factor of high-density gas, we will make here some inferences for the dense gas mass fraction based on intuitive assumptions and references to numerical simulations. The aim is to establish the ratio of the bulk gas and dense gas density profiles, $\rho_{\mathrm{bulk}}$/$\rho_{\mathrm{dense}}$, as a measure of the dense gas fraction, and the radial profile of this ratio as an indicator of different star formation modes (or temporal variations of the SFE) for the massive clumps. 

It is a well-known fact that there is a strong correlation between SFR and dense gas. 
Based on a gravo-turbulent fragmentation scenario, \citet{Padoan12, Padoan17} derived an empirical relation between SFR per free-fall time $\epsilon_{ff}$ (the fraction of a cloud's mass that is converted to stellar mass over a free-fall time scale, \citealt{KM05}) and $\alpha_{vir}$,
\begin{equation}
\epsilon_{\mathrm{ff}} = \dot{M}/(M/t_{\mathrm{ff}})\propto \mathrm{exp}\,(-\alpha_{\mathrm{vir}}^{1/2})\label{eq:padoan1}
\end{equation}.

In a hierarchical description of cloud structure in which gas density increases with decreasing scale, the conservation of $\dot{M}$ (or equivalently SFR) translates into
\begin{equation}
\epsilon_{\mathrm{ff \,,1}} M_{\mathrm{1\,,tot}}/t_{\mathrm{ff\,, \bar{\rho_{1}}}} = \epsilon_{\mathrm{ff\,,2}}M_{\mathrm{2\,, tot}}/t_{\mathrm{ff\,,\bar{\rho_{2}}}}\label{eq:padoan2}
\end{equation}
in which the number indices of 1 and 2 denote two adjacent levels in a hierarchy. 
We assume $\rho_{\mathrm{bulk}}$ represents the gas density of a lower-level structure in the hierarchy from which a higher level structure of gas density $\rho_{\mathrm{dense}}$ originates from.
We further assume that only the dense gas $\rho_{\mathrm{dense}}$ participates in the star formation process. Then with a scale-invariant $\epsilon_{\mathrm{ff}}$ (or equivalently scale-invariant $\alpha_{\mathrm{vir}}$), Equation \ref{eq:padoan2} translates to $M_{\mathrm{dense}}$ = $M_{\mathrm{tot}}$($t_{\mathrm{ff\,,\rho_{\mathrm {dense}}}}/t_{\mathrm{ff\,,{\rho_{bulk}}}}$). 
Substituting the free-fall timescale $t_{\mathrm{ff}}$ $\propto$ $\rho^{-1/2}$ into the equation, we obtain, 
\begin{equation}
M_{\mathrm{dense}} \sim M_{\mathrm{tot}}  (\frac{{\rho_{\mathrm{bulk}}}}{\rho_{\mathrm{dense}}})^{1/2}. \label{eq:dgmf1}
\end{equation}
Naively, this relation means that at a certain scale, the closer the derived dense gas density is to the bulk gas density, the more mass fraction of the clump is occupied by the dense gas. 
We note that even without the assumption of scale-invariant $\epsilon_{\mathrm{ff}}$, considering only that the two set of densities are each the mass-weighted value in their respective density regimes, the mass fraction of dense gas to total clump gas is also largely determined by $\rho_{\mathrm{bulk}}/\rho_{\mathrm{dense}}$, when the ratio between density of bulk gas and gas not seen by CH$_{3}$OH is similar among sources and does not vary with scale.

\begin{figure*}
    \centering
   
   \hspace{-.8cm}\includegraphics[scale=0.38]{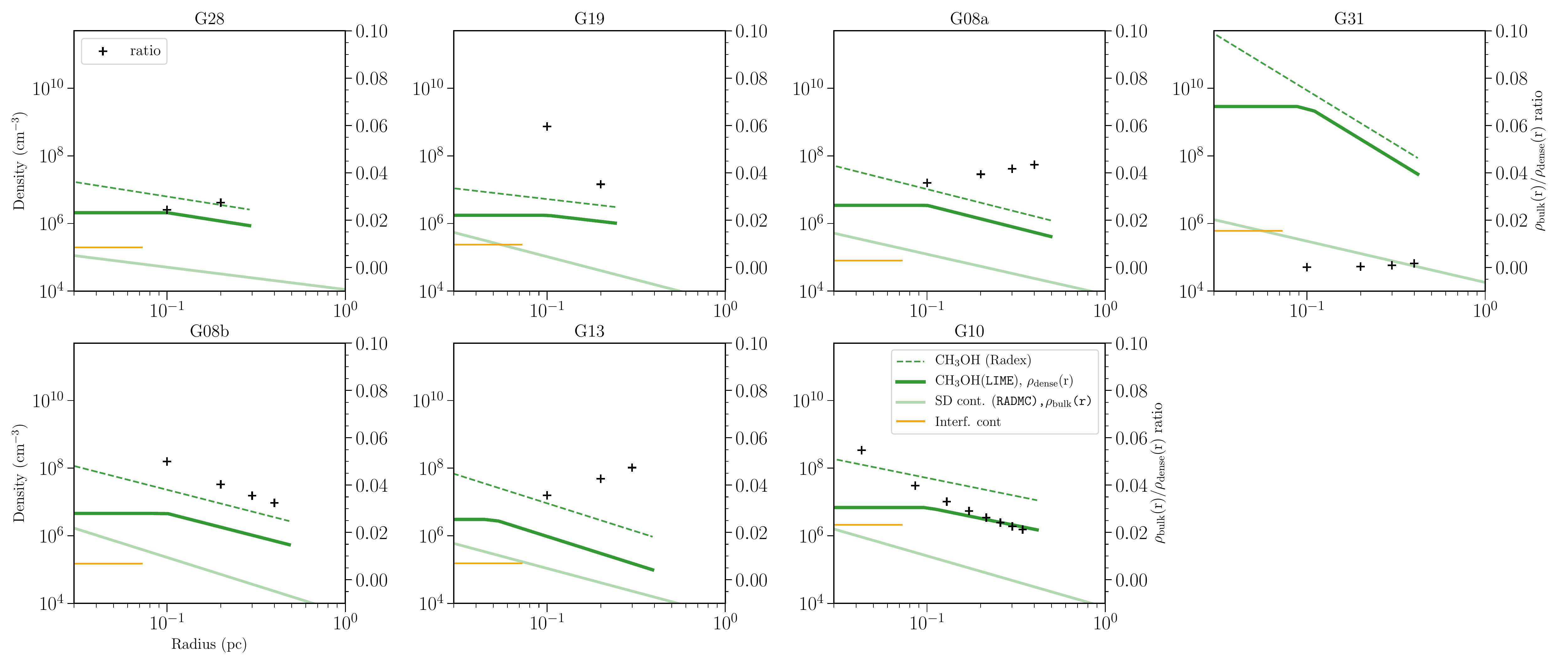}
    \caption{Comparison between gas density profiles derived by modeling continuum and CH$_{3}$OH line emission. The ratio between densities derived by CH$_{3}$OH {\tt{LIME}} modeling ($\rho_{\mathrm{dense}}$) and continuum ($\rho_{\mathrm{bulk}}$) results are shown as pluses (following right y-axis). The density estimated by SMA 1.2 mm continuum observations representing the central core average density is shown as vertical orange line in each plot (Table \ref{tab:sma_cont_direct}).}
    \label{fig:dens_comp}
\end{figure*}

Figure \ref{fig:dens_comp} shows the density profiles $\rho_{\mathrm{dense}}$ and $\rho_{\mathrm{bulk}}$.
This figure also shows the ratios between these two profiles, which provide a measure of dense gas mass fraction (DGMF), i.e. $M_{\mathrm{dense}}$/$M_{\mathrm{tot}}$.
We found that the DGMF in G19, G08b, and G10 decrease with radius, while it increases with radius in G28, G08a, G13 and G31. For G31, since the optical depths of the CH$_{3}$OH lines are high such that the line ratios do not serve as good as a densitometer in the density regimes it associates as the other sources, we omitted discussion on this source while note that the clump indeed holds the highest dense gas densities seen by CH$_{3}$OH, among all target sources. 
Among the other six sources, specifically, G19 shows an decreasing DGMF from 25$\%$ to 19$\%$, G08b from 22$\%$ to 18$\%$ and G10 from 24$\%$ to 15$\%$.  
For G28, DGMF increases from 15\% to $\sim$17\% at 0.1-0.2 pc.
In G08a and G13, DGMF is $\sim$19\% at their centers and achieve $\sim$22\% at 0.2-0.3 pc radii.

Given these radial variations of DGMF, it might be indicative that among these 6 sources, G19 and G10 may be most efficiently converting gas into the dense gas regime, having a focused DGMF towards clump center. 
G08b exhibits a similarly high DGMF that changes slightly with scale, while early-phase clump G08a and the least massive G13 have DGMF peaking around the intermediate scales relative to the center. 
Another early-stage source G28 has the least DGMF overall and also shows an relatively invariant DGMF over scales.
This trend is only partially reflected in the density profiles slopes of $\rho_{\mathrm{bulk}}$(r), which is generally regarded as a measure of dense gas concentration that directly inflates the SFR (e.g. \citealt{Parmentier19}). 
It is noteworthy that radial variations of the DGMF for individual sources is compatible with the radial dependence of the virial parameter shown in Section \ref{sec:velo_width_dis} (Figure \ref{fig:vwidthprofiles}). 
These two measures are both indicative of the capability of the source in converting gas into the denser regime at different positions relative to the bottom of the clump's gravitational potential.  
It is straightforward to see that the three classes of increasing/flattened/decreasing DGMF as a function of radius, for a particular source, correspond to a shallower/similar/steeper dense gas profile ($q_{\mathrm{dense}}$) compared to that of its bulk gas ($q_{\mathrm{bulk}}$) on the basis of Eq.\ref{eq:rhobulk}, \ref{eq:rhoradex} and \ref{eq:dgmf1}. In the non-homologously spherical collapse framework, the physical condition of an outer shell can be regarded as an earlier stage preceding the inner shell (e.g. \citealt{VS19}), then a radially-decreasing (increasing) DGMF corresponds to an accelerating (retarding) dense gas conversion and temporal increase (decrease) of SFE for a particular source.

The localised behavior of dense gas conversion may be causing the chaotic and scattered SFR vs. gas density relation, and such behavior goes beyond the self-similar solutions for collapsing clumps. 
The non-self-similar behavior is suggested to be relevant even for the spherical collapse as shown theoretically by \citet{Coughlin17}, dependent on the initial conditions. 
The spatial variations of the DGMF may also reflect two competing gravitational collapses within these massive clumps: collapse towards the global potential center, and collapse of dense regions into ambient filaments, and the latter process is mostly induced by turbulence (\citealt{Girichidis11}). The dominance of one or another process depends on the initial density profile and turbulence driving modes (\citealt{Girichidis11}, \citealt{Lomax15}). 

We can also establish $\rho_{\mathrm{bulk}}/\rho_{\mathrm{dense}}$ as a measure of DGMF from another perspective, with a less strong assumption.
In the framework of turbulent convergent flows, the gas density enhancement after a compressive shock has an inverse length scale relation following,
\begin{equation}
\rho_{\mathrm{post}}/\rho_{0} \propto L_{0}/l_{\mathrm{post}}\sim\mathcal{M}_{\mathrm s}^{2}\label{eq:shockrho}
\end{equation}
in which $L_{0}$ and $l_{\mathrm{post}}$ denote the length scales of the pre-shock and post-shock gas, and $\rho_{0}$ and $\rho_{\mathrm{post}}$ the gas densities, respectively. $M_{\mathrm s}$ denotes the sonic Mach number, $M_{\mathrm s}$ = $\frac{\sigma_{\mathrm{rms}}}{c_{\mathrm s}}$. This shock jump condition is at the origin of the lognormal distribution of PDF. 
As Equation \ref{eq:shockrho} links the gas length scale to density with an inverse relation, then the mass ratio of post-shock dense gas versus the pre-shock gas, may also be represented by a power-law form following $(\rho_{0}$/$\rho_{\mathrm{post}})^{-s}$ with slope $s$ dependent on the assumed geometry. 
With a cylindrical geometry describing infinite filaments, the gas mass is $\propto$ $\rho l^{2}$ with $l$ denotes thickness (radius), with slope $s$ = 1.

If the enhancement of gas density ($\rho_{\mathrm{dense}}$) as traced by CH$_{3}$OH is regarded as the result of compressive turbulence on the pre-shock gas that has a density represented by $\rho_{\mathrm{bulk}}$, the ratio $(\rho_{0}$/$\rho_{\mathrm{post}})$ still holds as a DGMF measure. 
Compared to Eq \ref{eq:dgmf1} ($s$ = 1/2), the different scaling of $s$ = 1 does not change the general trend of the radial change of this measure, but only to increase the contrast between outer layer and inner region. 

In any case, the dense gas probed by CH$_{3}$OH does have contributions from shock entrainment, as suggested by locally increased $M_{\mathrm{s}}$ and the fact that CH$_{3}$OH is likely enhanced in shocked regions. 
On the other hand, we have already seen in Section \ref{sec:velo_width_dis} that gravitational collapse has altered the general scaling relation of $v_{\mathrm T}$ (hence $\mathcal{M}_{\mathrm s}$), which makes the slope $s$ vary.  
A more stringent comparison of radial change of dense gas mass fraction to indicate the dense gas conversion efficiency would benefit from properly separating the part of dense gas associated with transient gas substructures with virialised cores and coherent flows, which is achievable with finer (both spectral and spatial) resolution observations.    
 
\section{Conclusions}\label{sec:conclusion}
The gas thermal properties are critical to the star-formation process. For massive star formation, the density structure of the nursery gas clumps may not be well predicted by previous hydrostatic equilibrium core models which are generally good representative of low-mass cores, due to a higher level of turbulence and gravitational collapse set at larger scales.   
We conducted a SMA and APEX line survey towards a sample of eight massive star-forming clumps, in order to understand the evolution of temperature and density structures. The major findings are as follows:

\begin{enumerate}
    \item Transitions of multiple molecular species (CH$_{3}$CN, CH$_{3}$CCH, H$_{2}$CS) of distinct critical densities, together with dust emission, provide a reasonably good sampling of different gas temperature regimes ($\gtrsim$200-20 K) over the full clump scale.
    There is not a single power-law relation dependent only on source luminosity that could describe all the radial temperature variations in our sample. The elevated temperature of a less luminous source may be related to the intermittency of accretion and shock-related activities.
    \item CH$_{3}$OH line series are good density probes for massive clumps, selectively tracing density regimes of $\gtrsim$10$^{6}$ cm$^{-3}$. Systematic steepening of density profiles along clump evolution, indicated by $L/M$, is revealed among the sample, from both continuum (bulk gas) and CH$_{3}$OH line (dense gas) modeling. The density slopes change from $>$-1 to $\sim$-1.5. The dense gas proportion becomes denser with increasing L/M, and the volume filling factor of the dense gas decreases, which corresponds to a higher density contrast along clump evolution.
    \item The radial line-width profile traced by multiple lines displays a scale-dependent relation. Several sources, which have comparable stellar mass and gas content at the scale probed, have a central decreasing trend of $v_{\mathrm T}$ $\propto$ $r^{-0.4}$ and  $v_{\mathrm T}$ $\propto$ $r^{0.2}$ in the outer envelope, which may be tentatively related to conversion of gravitational energy to turbulence.
    On larger scales (0.2-0.5 pc) all clumps are close to virial state ($\alpha_{\mathrm{vir}}/\alpha_{\mathrm{cri}}$$\sim$1), as indicated by CH$_{3}$CCH, H$^{13}$CO$+$ lines. Small scale ($\lesssim$0.1 pc) virial parameters traced by CH$_{3}$CN and H$_{2}$CS can exceed equilibrium values for some of the sources, the origin of which is hindered by our limited resolution. Overall, we observe that the clump gas evolves from super-virial to sub-virial state with increasing L/M.  
    \item The abundance of CH$_{3}$OH, CH$_{3}$CN, H$_{2}$CS shows a better correlation with source central temperature than with luminosity. Abundance ratios of [CCH]/[CH$_{3}$OH] and [CH$_{3}$CN]/[CH$_{3}$OH] are in good correlation with clump $L/M$, and can be used as indicators of evolutionary stages of massive star-forming clumps.    
    \item The evolutionary trend of clump density profiles is compatible with cloud-scale diagnosis that frequently reveal the time-varying power-law tail of PDFs. In a hierarchical view, the radial variation of the dense gas mass fraction, which can be approximated by the density ratio between that representing the averaged bulk gas (dust continuum, no spatial filtering) and that probed by a high density tracer (CH$_{3}$OH) may be indicative of the efficiency of the source in dense gas conversion and dense gas focusing to the global gravitational center. The bulk gas density profile is a less distinct measure due to the fact that self-similar approximation is not adequately describing the clumpy gas structures within massive clumps.      
    
\end{enumerate}
In this work, we attempt to understand the role of self-gravity in shaping the gas structure as massive clumps evolve over time. 
Our results are based on $\lesssim$0.1 pc observations, although a dense gas filling factor (probed by CH$_{3}$OH) indicates on smaller scales a clumpy gas environment of higher density contrast along the evolutionary track. Detailed fragmentation properties, cores and filaments and their kinematics, remain unresolved. In any case, deduction of line-of-sight cloud geometry is difficult. With the unknown cloud thickness, the mass scale of the high-density gas regime remains highly uncertain. 

Existing observations have broadly revealed transonic to subsonic line-widths associated with localised substructures in massive star-forming regions, the length scale of which is well above sonic scale, indicating an efficient turbulence dissipation process. More robust velocity line-width profiles require better resolved observations to disentangle multiple velocity components, and confusion from stellar feedback (outflows), etc. In addition, while the scale-dependent line-width vs. radius relation is more indirect evidence (based on assumption that it mirrors radial infall velocity) of changing role of gravitational collapse in dominating gas dynamics, more direct evidence would be measurement of infall velocities, or accretion rates as a function of radius. The latter is indispensable to distinguish different collapse models (e.g. \citealt{Padoan19}). This calls for observations of multiple ``infall tracers" sensitive to different gas density regimes, along with a proper description of the temperature distribution.

\begin{acknowledgements} %
This work was partly funded by the Deutsche Forschungsgemeinschaft (DFG, German Research Foundation) Collaborative Research Council 956, sub-project A6.
Y.L. is a member of the International Max-Planck Research School (IMPRS) for Astronomy and Astrophysics at the Universities of Bonn and Cologne.
H.B.L. is supported by the Ministry of Science and
Technology (MoST) of Taiwan (Grant Nos. 108-2112-M-001-002-MY3 and 110-2112-M-001-069-). 
The Submillimeter Array is a joint project between the Smithsonian Astrophysical Observatory and the Academia Sinica Institute of Astronomy and Astrophysics and is funded by the Smithsonian Institution and the Academia Sinica.
\end{acknowledgements}


\bibliography{sma_aa}

\begin{appendix}
\section{Target sources}\label{app:target_sources}

\subsection{G19.882-00.534}
 G19.882-00.534 (IRAS 18264–1152), is classified as an extended green object (EGO, \citealt{Cyganowski08}), located at 3.7 kpc with luminosity of $>$10$^{4}$ $L_{\odot}$. Most prominent features of this source are the high-velocity outflow (\citealt{Qiu07}, \citealt{Leurini14}) and the high level of maser activity (among a sample of 56 high-mass star-forming regions, \citealt{Rodriguez17}). The outflow is oriented in a east-west direction, showing enhanced H$_{2}$ near-infrared emission as well (\citealt{Varricatt10}). Sensitive 7 mm and 1.3 cm observations resolved the source into a triple system, consisting of two optically thin H\textsc{ii} or dust emitting sources, and a thermal jet or a partially optically thick H\textsc{ii} region (\citealt{Zapata06}). 

\subsection{G08.684-00.366 and G08.671-00.356}
These two massive star-forming clumps are part of the IRAS 18032-2137 star-forming cloud located at 4.8 kpc (\citealt{Purcell06}). Multiple water, methanol and hydroxide masers are detected towards both sources (\citealt{HC96}, \citealt{Caswell98}, \citealt{Gomez10}). 
G08.671-00.356 ($\sim$ 9000 $L_{\odot}$) is a UCH\textsc{ii} region (\citealt{Wood90}), while G08.684-00.366 ($\sim$ 3000 $L_{\odot}$) is a less evolved relatively infrared weak source, lying 1 $'$ offset in the north-eastern direction. Strong SiO emission towards G08.684-00.366 indicates that star formation already takes place in this source (\citealt{Harju98}).  Previous SMA observations resolve the source into 3 dense cores accompanied by extended outflow components traced by CO (2-1) emission; the core masses are around several to ten solar masses, a small fraction of the total clump mass (\citealt{Longmore11}). 

\subsection{G10.624-00.380}
G10.624-00.380 is an extremely luminous and massive ($>$10$^{5}$ $L_{\odot}$, $\sim$5000 $M_{\odot}$) OB cluster forming clump in the galaxy, located at 4.95 kpc (\citealt{Sanna14}). Previous high angular resolution submm and centimeter observations revealed that the clump in the central 0.6 pc is a flattened rotating system (\citealt{Keto87, Keto88}) where multiple UCH\textsc{ii} regions are deeply embedded (\citealt{Ho86}, \citealt{Sollins05}); the clump is fed by the converging flows from ambient filamentary clouds (\citealt{Liu2011ApJ, Liu2012ApJ}). Within the H\textsc{ii} region, absorption lines indicate that the gas accretion continues despite the ionising and radiative pressure (\citealt{Keto90}). Overall, the clumps seems to be in an global collapsing state with a rotating Toomre-unstable disk-like structure in the center (\citealt{Liu17}).

\subsection{G13.658-00.599}
G13.658-00.599 (IRAS 18144-1723) is a molecular clump with a luminosity of $>$10$^{4}$ $L_{\odot}$ at 3.7 kpc. It is associated with multiple water and methanol masers (\citealt{Gomez-Ruiz16}). H$_{2}$ emission displays a bowshock feature 18$''$ offset in the west from the center of the IRAS emission, which seems to originate from an extended K-band continuum source associated with the IRAS source (\citealt{Varricatt10}). The bowshock feature is surrounded by multiple 44 GHz methanol masers (\citealt{Gomez-Ruiz16}), indicating it is likely caused by outflow activity. Deep mid-infrared imaging reveals that the central source hosts two YSOs separated by $\sim$ 10000 au, at different evolutionary stages; outflow traced by CO(3-2) line coincides well with the H$_{2}$ emission, and is likely caused by the younger source in formation (\citealt{Varricatt18}).

\subsection{G31.412+00.307}
G31.412$+$00.307 is a well-known hot massive core (HMC) that has been extensively studied by both single-dish and interferometry observations (e.g. \citealt{Cesaroni94}, \citealt{Beltran04,Beltran18}, \citealt{Rivilla17}), a source showing great chemical richness. 
It is located at a kinematic distance of 7.9 kpc (\citealt{Churchwell90}), and has a luminosity of $\gtrsim$10$^{5}$ $L_{\odot}$. 
The central hot core structure is massive and compact ($\sim$500 $M_{\odot}$ of $\sim$8000 au in size, \citealt{Cesaroni11}, \citealt{Beltran04}). There is an ultra-compact H\textsc{ii} region 5$''$ away from the core (\citealt{Churchwell90}). The kinematic features consistently support a rotating core experiencing infall motions (\citealt{Cesaroni11}, \citealt{Wyrowski12}). Recent ALMA observations of higher angular ($\sim$1700 au) resolution suggest that both the rotation and infall velocities increase towards the center, and that the core is composed of a main core of size $\sim$5300 au and a satellite core of much smaller mass (\citealt{Beltran18}). The overall monolithic feature of source G31 makes it an ideal source for understanding the high-mass star formation scenario in light of gas mass origin and evolution.  

\subsection{G18.606-00.074}
G18.606-00.074, is a massive infrared-dark clump located in a $\sim$4 pc long filamentary cloud associated with IRAS18223 (\citealt{Carey00}, \citealt{Beuther02b}, \citealt{Sridharan05}). The parental molecular cloud is part of an even larger ($>$50 pc) molecular gas filament (e.g. \citealt{Kainulainen11}, \citealt{Ragan14}), which undergoes star formation activities of various evolutionary stages (e.g., \citealt{Beuther02b, Beuther07, Beuther10}, \citealt{Fallscheer09}, \citealt{Ragan12}). G18.606-00.074 (also named as IRDC18223-3, or core 11 in \citealt{Beuther15}), compared to its ambient core and clump structures, appears to have a larger mass reservoir, showing larger linewidth of N$_{2}$H$^+$ line (\citealt{Beuther15}).

\subsection{G28.397+00.080}
G28.397+00.080 (named as P2 in \citealt{Wang08}) is a molecular clump located in a massive ($>$10$^{4}\,M_{\odot}$) filamentary infrared dark cloud, G28.34+0.06 (\citealt{Carey00}, \citealt{Pillai06}, \citealt{Lin17}). The central region of the clump is not associated with any near-infrared compact source counterpart while appears bright at 24/70 $\mu$m wavelength (Fig. \ref{fig:rgb}, \citealt{Wang08}). It also shows line emission of COMs such as CH$_{3}$OCH$_{3}$, CH$_{3}$CHO (\citealt{Zhang09}, \citealt{Vas14}), likely hosting massive protostar(s) which heats the gas up to 45 K, as measured from NH$_{3}$ observations (\citealt{Zhang09}).

\section{Subtraction of free-free emission from SMA 1.2 mm continuum}\label{app:cont_cont}
We use centimeter radio continuum data collected from the NRAO data archive\footnote{http://www.aoc.nrao.edu/~vlbacald/ArchIndex.shtml} for Very Large Array (VLA) for sources G08b, G31 and G10, to estimate the free-free emission contribution to the 1.2 mm flux. Given the time variability of centimeter emission, we select the most recent data products, whose basic information is listed in Table \ref{tab:vla_data}. 

Assuming that the dust and free-free emission are both optically thin, and that the dust emission follows a spectral index $\beta$ = 1.0, then the relations $S_{\nu}$ $\propto$ $\nu^{2+\beta}$ and $S_{\nu}$ $\propto$ $\nu^{-0.1}$ describe the variation of the two emission components, respectively, as a function of frequency. The free-free emission at 1.2 mm can then be solved using simultaneous equations of total flux, at two considered frequencies. The centimeter data was smoothed and regridded to match the 1.2 mm data and subtraction was done in a pixel-by-pixel basis. The total flux contributed from free-free emission to the 1.2 mm continuum is also listed in Table \ref{tab:vla_data}.
\begin{table*}
\centering
\begin{threeparttable}
\caption{Information of the centimeter continuum data collected from NRAO archive.}
 \label{tab:vla_data}
 \begin{tabular}{llllll}
\toprule

Source  &  Freq.& Resolution & Obs. Date & Integrated flux & Extrapolated total flux of free-free emission at 1.2 mm\\
   &(GHz)&($''$)&&(Jy)&(Jy)\\
\midrule
             G31  &43.3 &1.62&2001-11-26&0.42&0.40\\
             G08b  & 8.5&0.85&2005-02-28&0.99&0.70\\
             G10 &23.8&0.09&2002-02-01&3.55&2.80\\
\bottomrule
\end{tabular}
  \end{threeparttable}
\end{table*}

\section{RADEX modeling of CH$_{3}$OH lines: the MCMC procedure}\label{app:radex_mcmc_interp}
\begin{figure}[htb]
\begin{tabular}{p{0.95\linewidth}}
\includegraphics[scale=0.35]{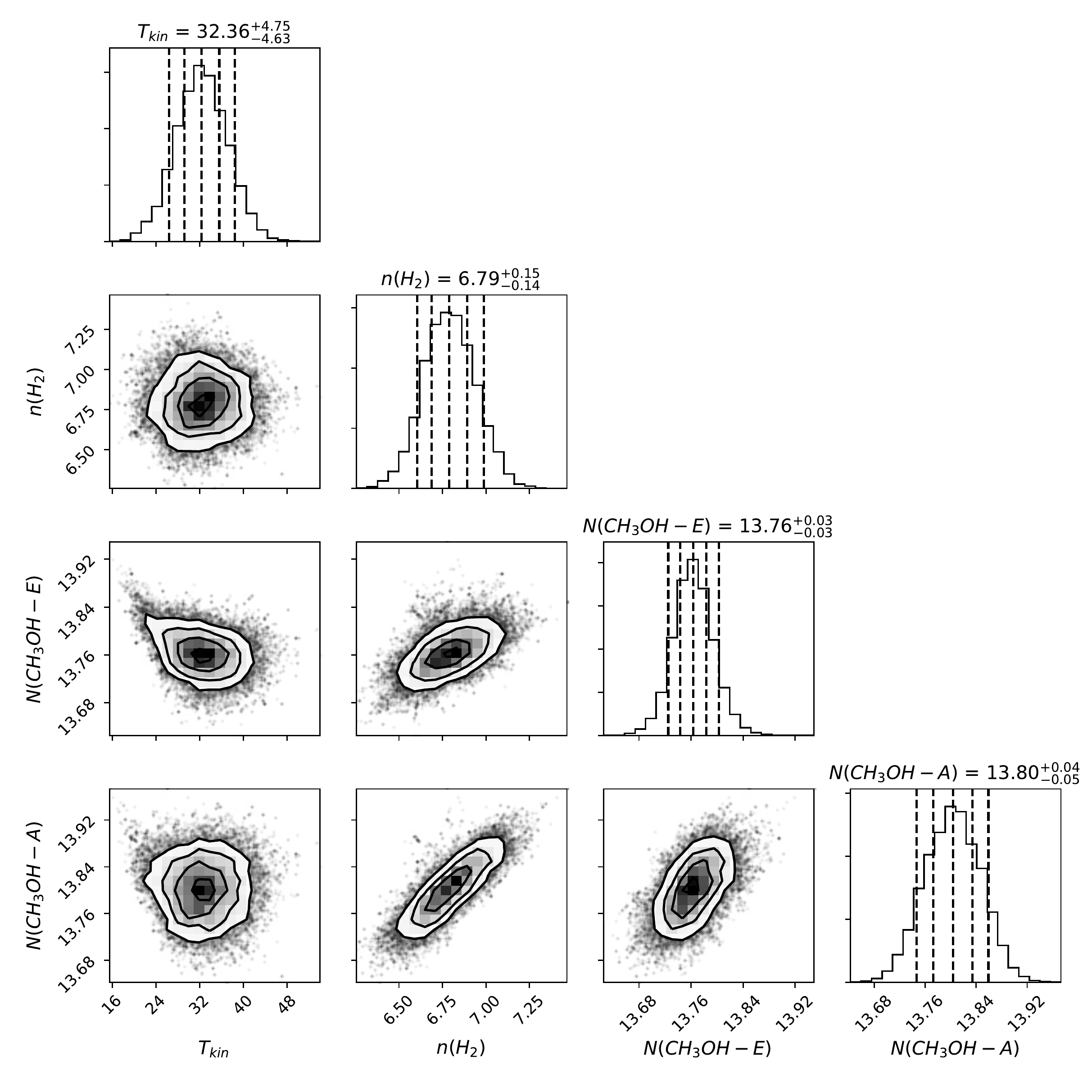}\\
\end{tabular}
\centering\caption{Posterior distribution of parameters from MCMC RADEX fitting of CH$_{3}$OH (5-4) line series. The column densities of $A/E$ type are velocity averaged values (log cm$^{-2}$/km s$^{-1}$). The vertical dashed lines in the 1d histograms are showing the quantiles of 10$\%$, 25$\%$, 50$\%$, 75$\%$, 90$\%$. The contour levels in the 2D histograms indicate 0.5$\sigma$, 1$\sigma$, 1.5$\sigma$ and 2$\sigma$, respectively. The figure shows an example of the fitted parameters of observed lines in one pixel of clump G08b.
}
\label{fig:radex_corner}
\end{figure}
\begin{figure*}
\begin{tabular}{p{0.3\linewidth}p{0.33\linewidth}p{0.33\linewidth}p{0.33\linewidth}}
\hspace{-1.1cm}\includegraphics[scale=0.38]{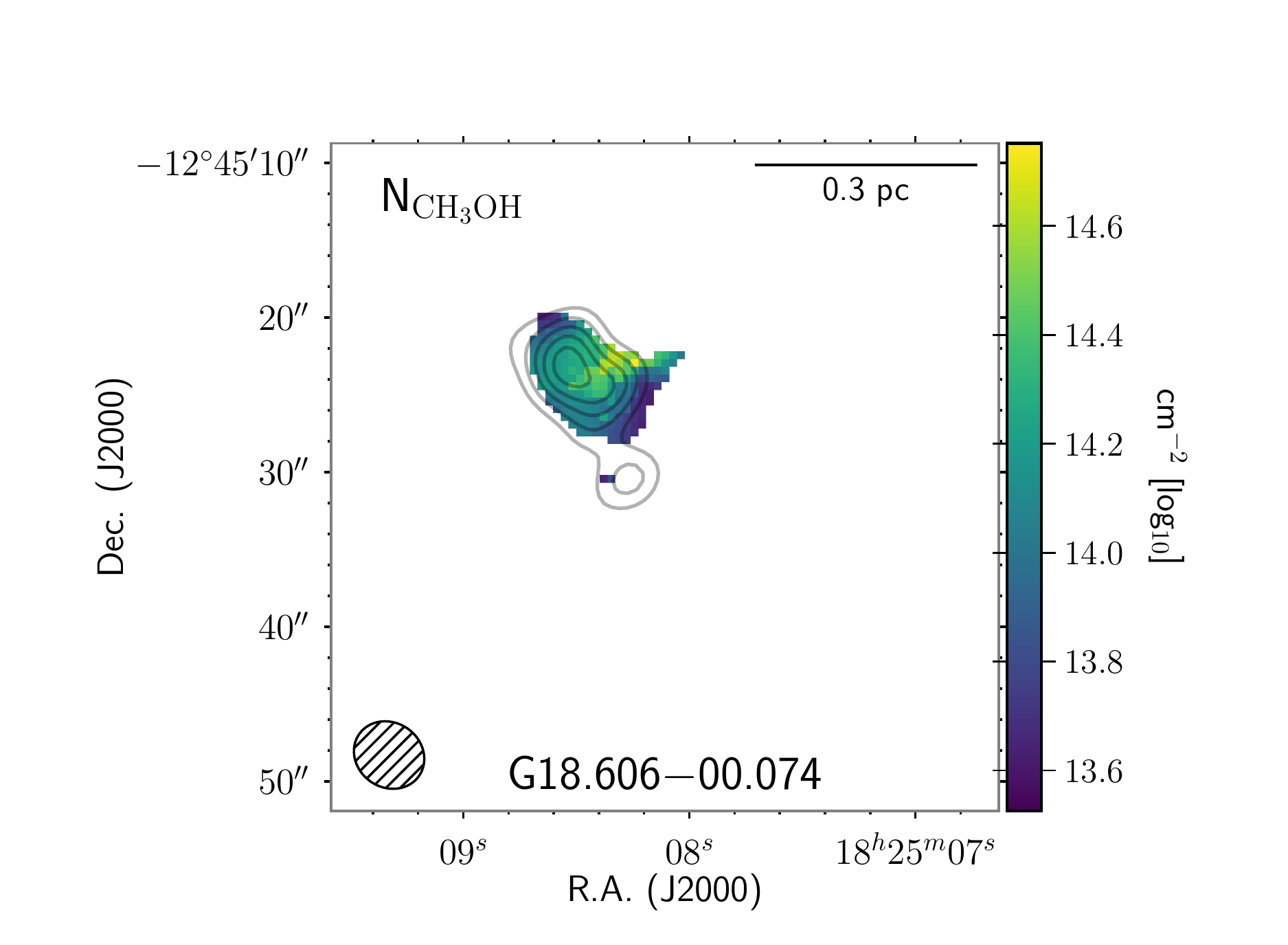}&\hspace{-0.12cm}\includegraphics[scale=0.38]{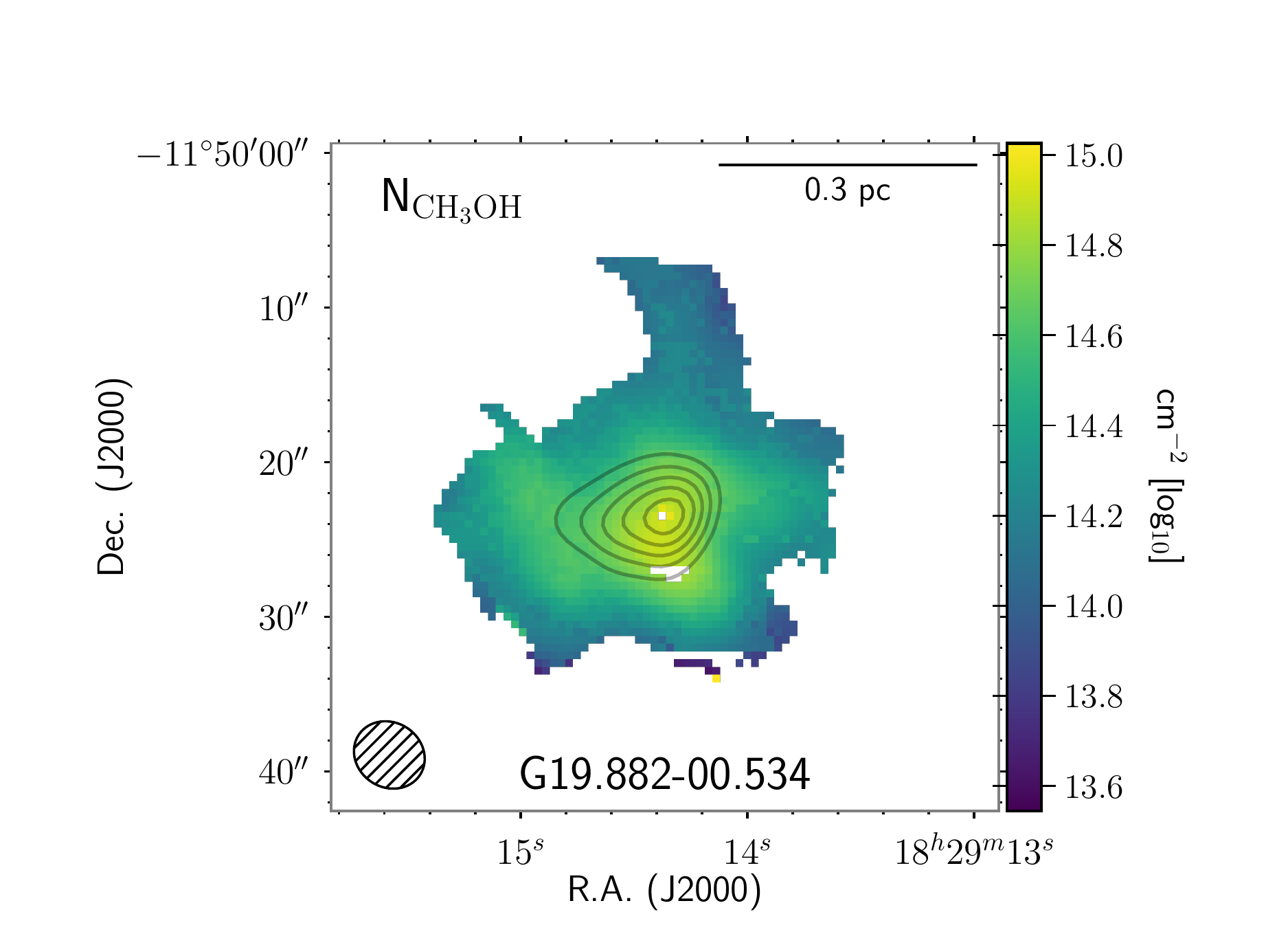}&\hspace{0.25cm}\includegraphics[scale=0.38]{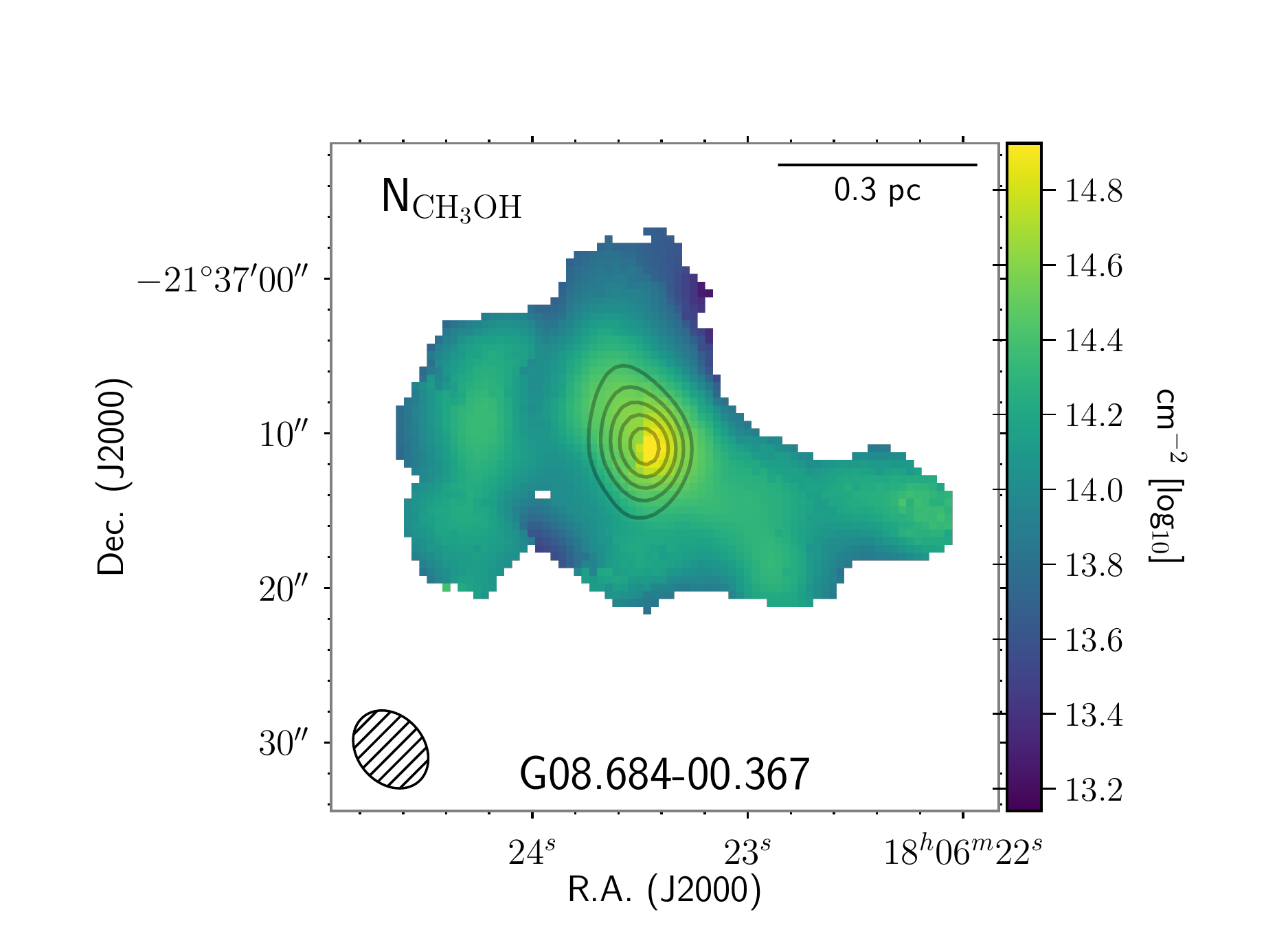}\\
\hspace{-1.1cm}\includegraphics[scale=0.38]{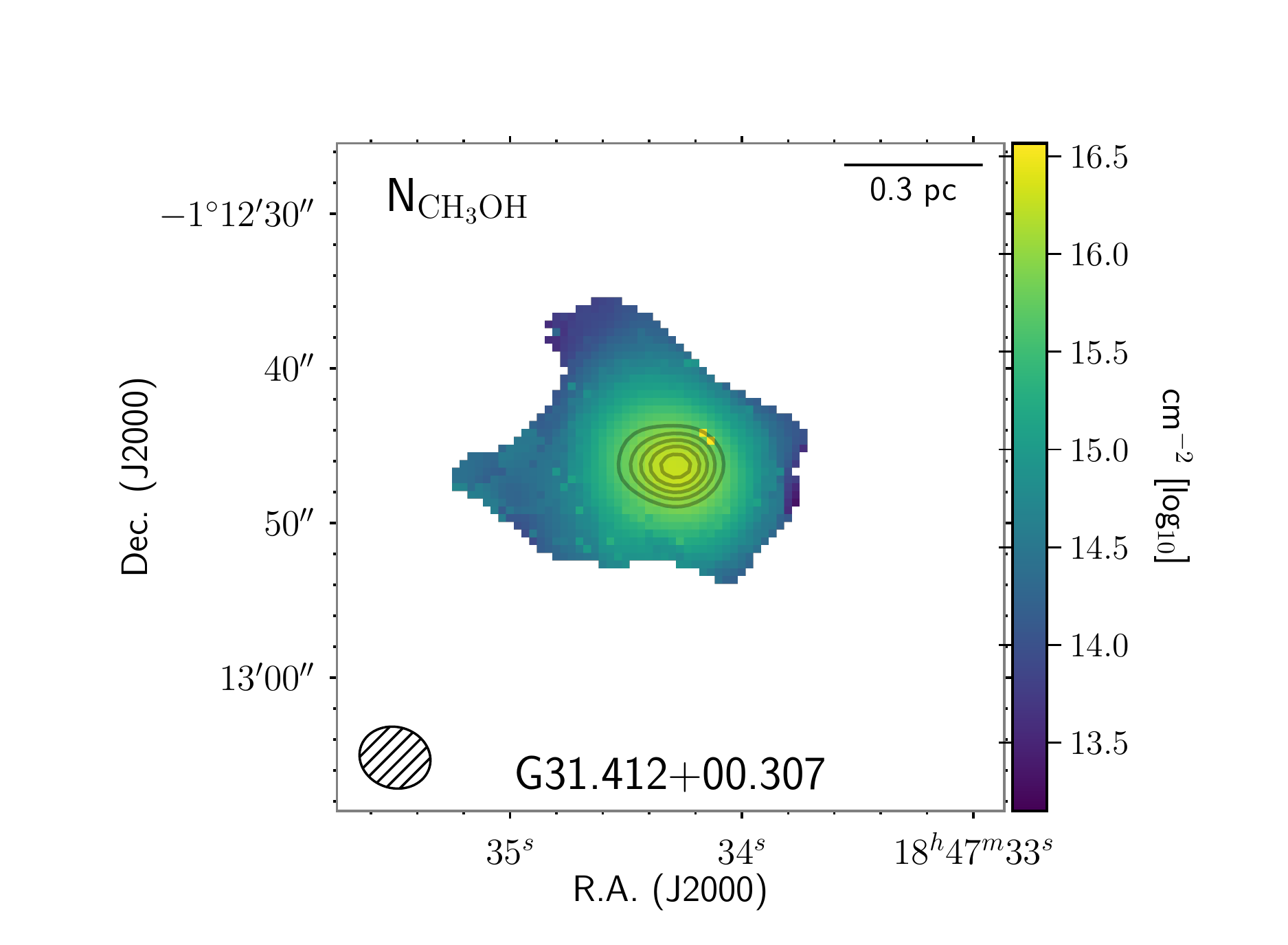}&\hspace{-0.12cm}\includegraphics[scale=0.38]{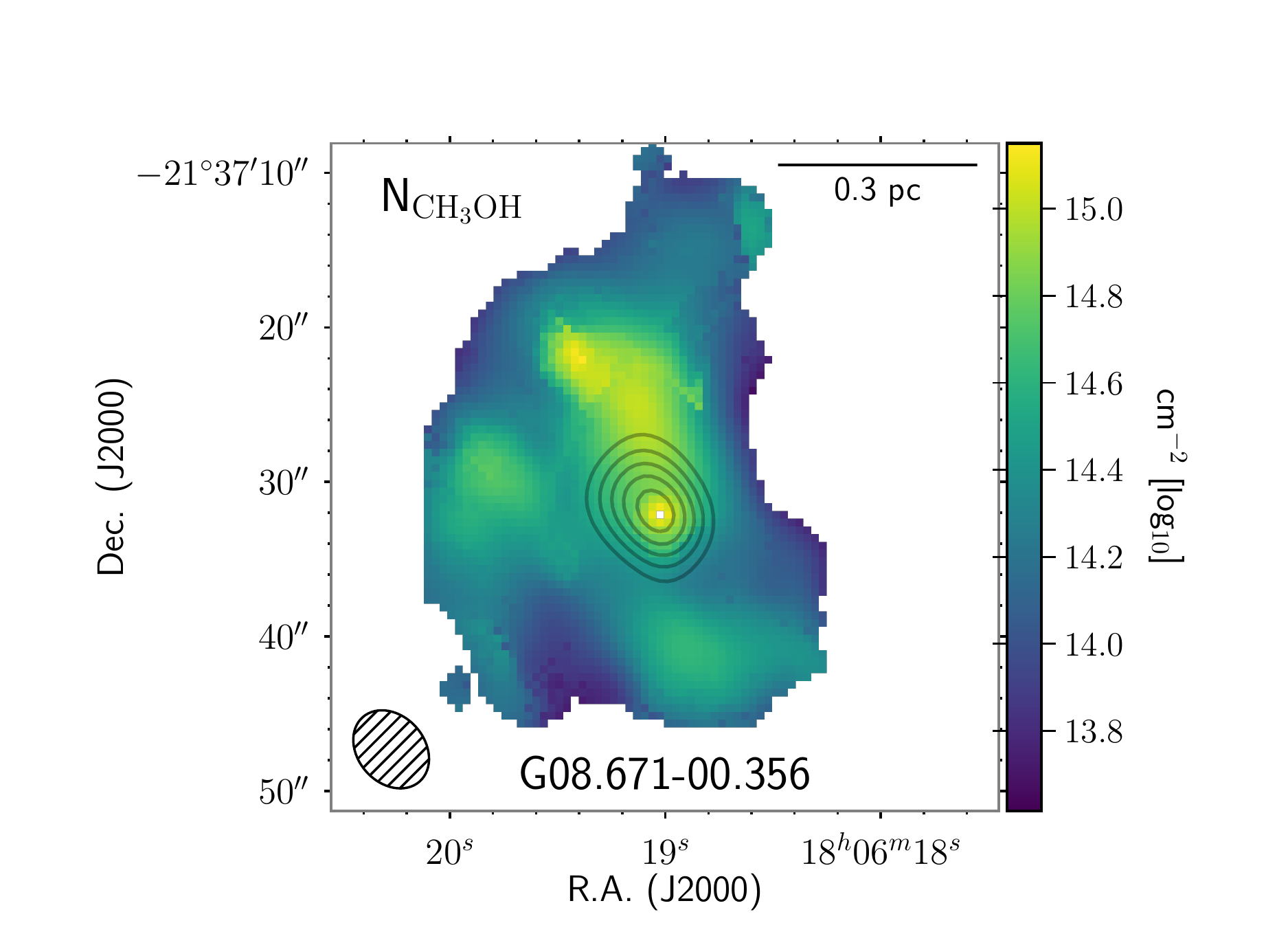}&\hspace{0.25cm}\includegraphics[scale=0.38]{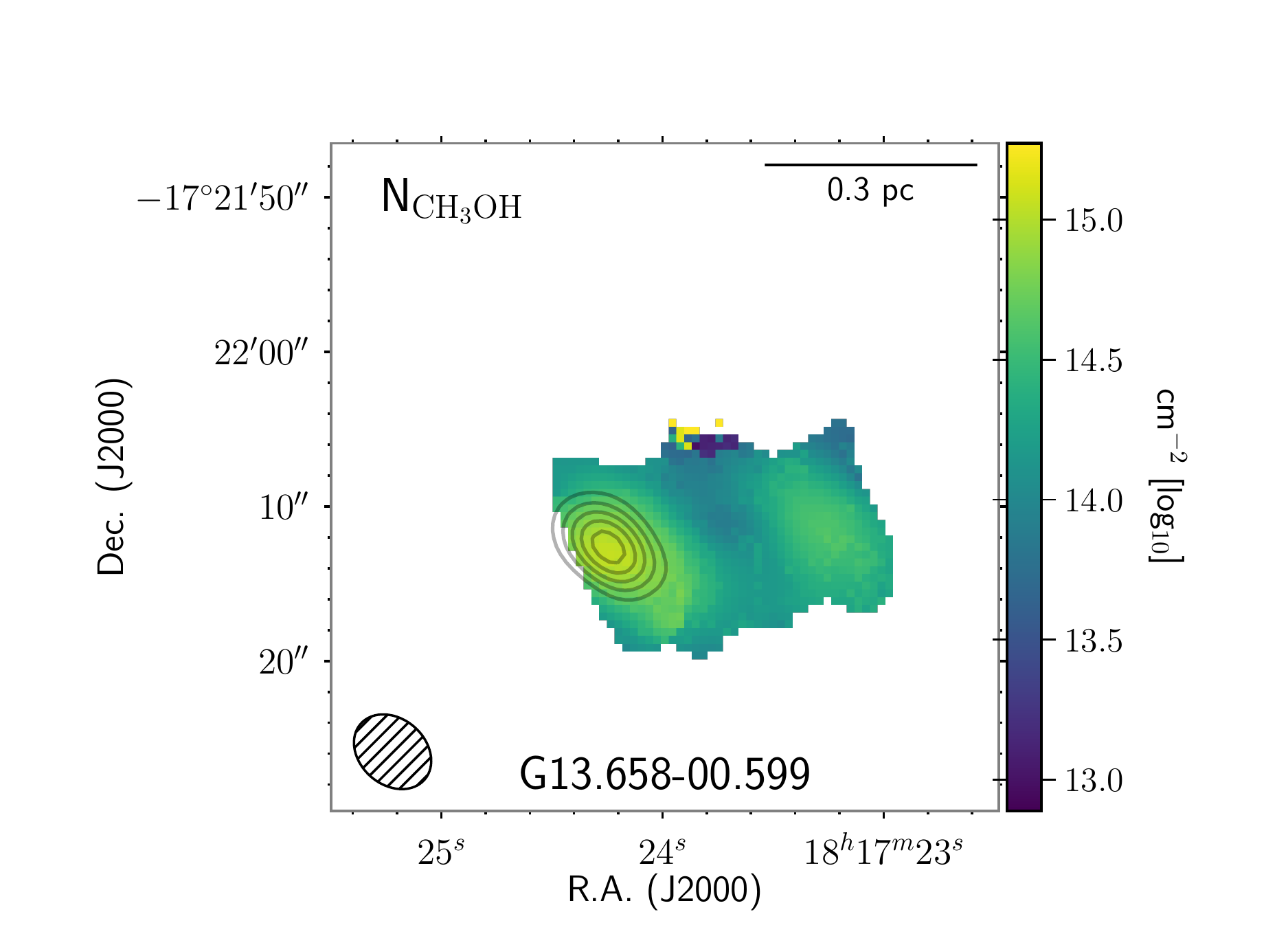}\\
\hspace{-1.1cm}\includegraphics[scale=0.38]{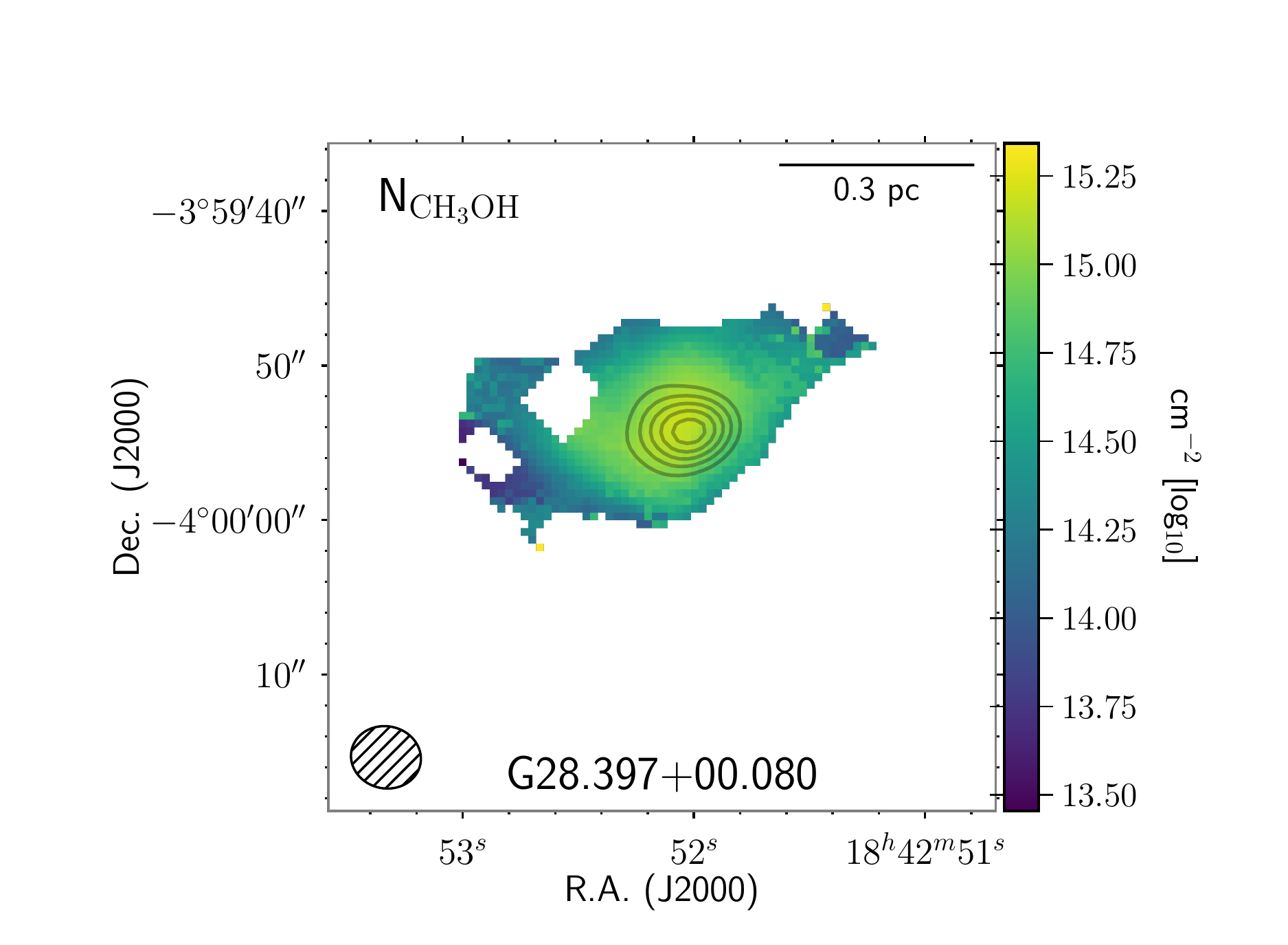}&\hspace{-0.1cm}\includegraphics[scale=0.38]{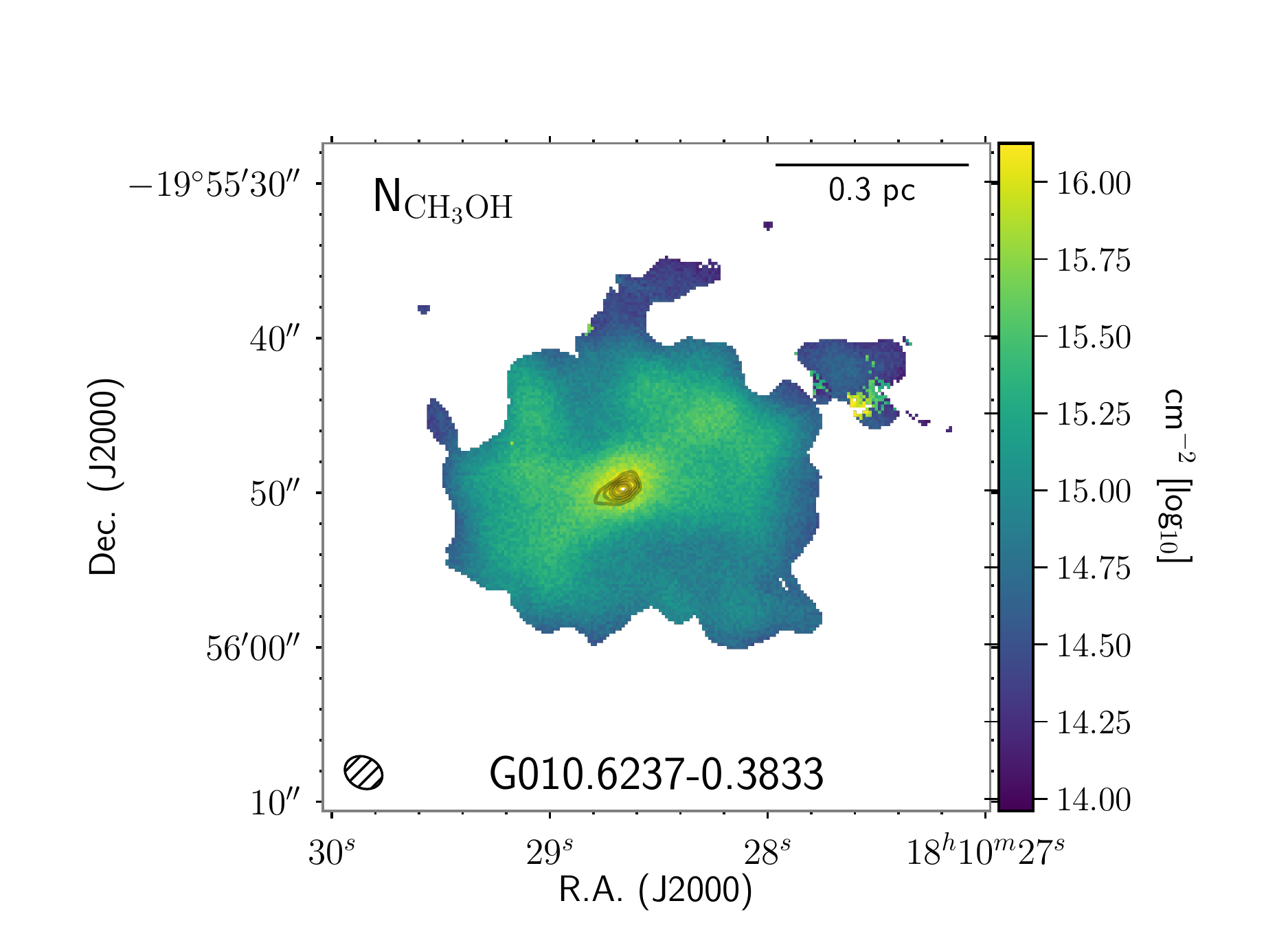}\\
\end{tabular}
\centering\caption{CH$_{3}$OH column density maps ($N_{\mathrm{CH_{3}OH}}$) derived from RADEX modeling of all target sources. The beam of CH$_{3}$OH 5$_{-1}$-4$_{-1}$ $E$ line is indicated in the bottom left corner. Gray contours indicates the SMA 1.2 mm continuum level from 0.1 to 0.9$\times$peak flux represented by 5 levels of uniform interval.}
\label{fig:cdmaps}
\end{figure*}

For both $E$-type and $A$-type CH$_{3}$OH, we generated RADEX model grids in the column density (N/$\Delta v$) range from 10$^{12}$ to 10$^{18}\,$cm$^{-2}$ (with 60 logarithmically spaced uniform intervals), in the density range from 10$^{4}$ to 10$^{9}\,$cm$^{-3}$ (with 100 logarithmically spaced uniform intervals), and in the kinetic temperature range between 10$-$200 K (with 80 uniform intervals).
In the RADEX models, the linewidth is taken as 1 km/s for all the grids of parameters, while the total column densities (shown in Fig. \ref{fig:cdmaps}) are obtained by multiply the fitted N/$\Delta v$ with the linewidth obtained from Gaussian fits (Sect. \ref{sec:radex_nh2}).
The external radiation field was taken to be the cosmic background at 3 K.  
We used a linear interpolator to estimate the line intensities for parameters in between the intervals to better constrain the parameters and to allow for a continuous examination of parameter space.  
Nevertheless, the accuracy of the best-fit parameters remains limited by the resolution of the grid.

We fit the {\it{A}-} and {\it{E}}$-$CH$_{3}$OH (5-4) ($\nu_{t}$ = 0) line profiles with Gaussian models pixel-by-pixel for all the sources.  
We assumed that the line width is identical for all the $K$ components and for $A$ and $E$ types.
We also took into consideration that the HNCO 11$_{0,11}$$\,-\,$10$_{0,10}$ line is blended with the CH$_{3}$OH 5$_{-1}$$\,-\,$4$_{-1}$ $E$ line ($\delta V$ $\sim$ 8.44 km s$^{-1}$) and removed the contribution from this HNCO line by fitting an additional Gaussian component of which the line-width and amplitude are free parameters.


We employ the Markov Chains Monte Carlo (MCMCs) method with an affine invariant sampling algorithm\footnote{A detailed description of the method can be found in the emcee documentation.} \citep[emcee,][]{Foreman13} to perform the pixel-by-pixel fitting.
Our prior assumption for each pixel adopted the temperature $T_{\mathrm{rot}}$ predicted from the temperature profiles in Section \ref{sec:xclass}, which means that for $T_{\mathrm{kin}}$ we assumed a normal distribution centering at $T_{\mathrm{rot}}$ and with a standard deviation of 5 K.
For the other parameters, the priors were assumed to be uniform distributions.
We use a likelihood distribution function which takes into account observational thresholds; the formulas follow, 
\begin{equation}
P\propto\underset{i}\Pi p_{i}\underset{j}\Pi p_{j}
\end{equation}
where $p_{i}$ stands for probabilities of the $i$th data that is a robust detection and $p_{j}$ the $j$th data that gives a constrain by an upper limit; we adopt the normal distribution as likelihood function,
\begin{equation}
p_{i} \propto \mathrm{exp}\,[-\frac{1}{2}(\frac{f^{\mathrm{obs}}_{i}-f^{\mathrm{model}}_{i}}{\sigma^{\mathrm{obs}}_{i}})^{2}]\Delta f_{i}
\end{equation}
\begin{equation}
p_{j} \propto \int _{-\infty}^{f^{\mathrm{obs}}_{lim,j}} \mathrm{exp}\,[-\frac{1}{2}(\frac{f_{j}-f^{\mathrm{model}}_{j}}{\sigma_{j}})^{2}]df_{j}, 
\end{equation}
in which $f^{\mathrm{obs}}_{i}$ (or $f_{j}$) stands for the observed intensity (or intensity upper limit) obtained from Gaussian fit, $f^{\mathrm{model}}_{i}$ (or $f^{\mathrm{obs}}_{i}$) the model intensity, $\sigma^{\mathrm{obs}}_{i}$ is the standard error of the observed intensity which was adopted as the fitted 1$\sigma$ error of the Gaussian fit, $\Delta f_{i}$ being the data offset from the true value of $f_{i}$, and $d f_{j}$ the integrated flux probability to the detection threshold $f^{\mathrm{obs}}_{lim,j}$.

The starting points (initialization) for the chains were chosen to be the parameter set corresponding to a global $\chi^{2}$ minimum calculated between the grid models and the observed values.  
We also employed the "burn-in" phase in the MCMC chains and several resets of the starting-points to ensure the final chains are reasonably stable around the maximum of the probability density. An example posterior distribution of the fitted parameters is shown in Figure \ref{fig:radex_corner}. The obtained $n(\mathrm{H_{\mathrm 2}})$ maps and column density maps of CH$_{3}$OH for all sources are shown in Figure \ref{fig:nmaps} and Figure \ref{fig:cdmaps}, respectively.

After the first run of generating parameter maps, we find that the $n(\mathrm{H_{\mathrm 2}})$ of clump G31 are truncated to $\sim$10$^{9}$ cm$^{-3}$, the parameter boundary of our conducted RADEX models. Therefore we additionally ran a larger grid with $n(\mathrm{H_{\mathrm 2}})$ up to $\sim$10$^{11}$ cm$^{-3}$ with the same range of $T_{\mathrm{kin}}$ and $N_{\mathrm{mol}}$ as the first grid, and re-derive the parameter maps for this clump.

\section{LTE analysis for other lines}\label{app:otherlines}
\subsection{CS/C$^{34}$S (5-4)}
CS transitions are commonly adopted tracers for dense gas. 
Our spectral setup covered the $J$=5-4 transition of CS and its rare isotopologue, C$^{34}$S, for the 7 sources. 
For G10, the data we utilized only covered the C$^{34}$S (5-4) line. 
In the observed sources, the CS line and the 1.2 mm continuum emission trace similar structures.
Additionally, the CS line images revealed some spatially more extended structures.

Assuming that CS and C$^{34}$S trace gas of identical temperature and linewidths, the optical depth of the C$^{34}$S line can be estimated by
\begin{equation}
\frac{W_{\mathrm{CS}}}{W_{\mathrm{C^{34}S}}} = \frac{[1-\mathrm{exp}\,(-R\tau_{\mathrm{C^{34}S}})]}{[1-\mathrm{exp}\,(-\tau_{\mathrm{C^{34}S}})]},
\end{equation}
where $W_{\mathrm{CS}}$ and $W_{\mathrm{C^{34}S}}$ are the integrated line intensities, and we assumed a $^{32}$S/$^{34}$S abundance ratio of $R\,=\,22.6$, which is consistent with terrestrial value, e.g., \citet{Lodders03}, although there are indications that the ratio varies as a function of Galactocentric distance (e.g. \citealt{Humire20}).
The CS line, as expected, is broader than the C$^{34}$S line, and shows self-absorption profiles in most of the cases. 
These are two effects that can mitigate each other's effect on biasing the $\tau_{\mathrm{C^{34}S}}$ estimate as our derivation was based on the integrated intensity ratios.
We found that $\tau_{\mathrm{C^{34}S}}$ ranges from less than 0.1 to 0.25 in most of the sources, while it reaches 0.3-0.45 in G08b and values as high as 0.9 in G31.

Assuming a beam filling factor $f$ = 1, a lower limit of the excitation temperature ($T_{\mathrm{ex}}$) can be estimated from the brightness temperature of C$^{34}$S,
\begin{equation}
    J(T_{\mathrm b}) = \left[J(T_{\mathrm{ex}}) - J(T_{\mathrm{bg}})\right]\left[1-\mathrm{exp}\,(-\tau_{\mathrm{C^{34}S}})\right].
\end{equation}
Since the $n_{\mathrm{crit}}$ of CS is higher than the thermometers except CH$_{3}$CN used in Sect.~\ref{sec:xclass}, and because of the fact that there is more extended cold envelope gas contributing to the emission of CS line, as expected, the derived $T_{\mathrm{ex}}$ is in general lower than the $T_{\mathrm{rot}}$ profiles derived from the observations of multiple transitions of the other molecular thermometers (see Sect.~\ref{sec:xclass}).
The derived $T_{\mathrm{ex}}$ is $<$10 K in G18, in the range of 20-30 K in most of the other sources, and reaches $\sim$45 K at the emission peak of G31.

Under LTE assumption, the overall column densities of the C$^{34}$S ($N_{\mathrm{C^{34}S}}$) can then be estimated by evaluating the partition functions, adopting $T_{\mathrm{ex}}$, with,
\begin{equation}
    N_{\mathrm{tot}} = \frac{8\pi\nu^{3}}{c^{3}A_{\mathrm{ul}}}\frac{Q_{\mathrm{rot}}}{g_{u}}\mathrm{exp}(\frac{E_{\mathrm{up}}}{kT_{\mathrm{ex}}})[\mathrm{exp}(\frac{h\nu}{kT_{\mathrm{ex}}})-1]^{-1}\int\tau_{\mathrm{C^{34}}S}dv,
\end{equation}
in which $g_{\mathrm{u}}$ stands for the level degenaracy and $Q_{\mathrm{rot}}$ the partition function at $T_{\mathrm{ex}}$.
The derived C$^{34}$S column densities are in the range of 4.4$\times$10$^{12}$ - 5.0$\times$10$^{14}$ cm$^{-2}$. 
 
 In an alternative approach, we assumed that the $T_{\mathrm{ex}}$ of C$^{34}$S is the same with the $T_{\mathrm{rot}}$ measured from the other molecular thermometers, and then directly solved for $\tau_{\mathrm{C^{34}S}}$ based on an LTE assumption.
 The $N_{\mathrm{C^{34}S}}$ derived this way is 1.1-1.5 times larger for warmer sources except for G18. 
 In G18, the $N_{\mathrm{C^{34}S}}$ derived with this approach is $\sim$10 times smaller than the derivation based on the CS/C$^{34}$S intensity ratio.
 This is due to the dependence between $N_{\mathrm{C^{34}S}}$ and $T_{\mathrm{ex}}$ under LTE assumption: $N_{\mathrm{C^{34}S}}$ drops significantly below $\sim$30 K and slowly increases for larger values of $T_{\mathrm{ex}}$.

\subsection{CCH (3-2)}
The hyperfine line components of CCH resulting from electron-nucleus interactions allow for an direct measure of the CCH line optical depth. 
We followed fitting procedure similar to that described in \citet{Estalella17}, adopting relative line intensities from LTE predictions. 
From hyperfine line fitting, we derived excitation temperature and optical depth in a pixel-by-pixel basis, and then used these to estimate the CCH column densities. 
We found that the optical depth of the main line, $\tau_{\mathrm m}$$\sim$0.1 in extended regions, but can become optically thick at high column densities in localized regions. 
In most of the sources, $\tau_{\mathrm m}$ reaches up to $\sim$3. 
But in G18, $\tau_{\mathrm m}$ reaches $\sim$10 at the center, which is close to the upper limit in our fitting procedure.
The column density of CCH ranges between 1.2$\times$10$^{14}$ - 6$\times$10$^{15}$ cm$^{-2}$ across the emission region for all the sources. 
In general, the CCH column density distribution shows porosity over extended regions ($\sim$0.2-0.3 pc) except G18, and is systematically larger in outer regions, resembling a ring-like structure, especially for clump G08b and G31. 
The central region of the hot massive core G31 is almost completely devoid of CCH emission. 
\citet{Jiang15} presented high-angular resolution CCH observations towards clump G10 together with other 3 more evolved high-mass clumps than our sample, which also exhibit ring-like distributions. 

\subsection{SO and SO$_{2}$}
Our spectral setup covered three SO lines (Table \ref{tab:lines_more}) which have similar upper level energies $E_{\mathrm{up}}$. 
These three lines were detected in all clumps except G18.
The resolved SO intensities were spatially compact and were confined within where 1.2 mm continuum emission was detected.

The SO$_{2}$ 14$_{0,14}$-13$_{1,13}$ ($E_{\mathrm{up}}$$\sim$94 K) line was detected in the all clumps except G18 and G28; in G28, a lower excitation transition SO$_{2}$ 3$_{2,2}$-2$_{1,1}$ ($E_{\mathrm{up}}$$\sim$15 K) was marginally detected.


These SO and SO$_{2}$ transitions are likely thermalized given that the emission closely follows the extension of the central core, and are of relatively high critical densities (Table \ref{tab:lines_more}). 
We cannot directly derive the excitation temperatures of the SO and SO$_{2}$ molecules owing to not covering multiple transitions, or owing to that the covered transitions have similar $E_{\mathrm{up}}$.
Therefore, we adopted the gas temperature derived from the thermometer lines (Section \ref{sec:xclass}) when estimating their column densities, assuming LTE condition. 


\section{Radiative transfer modeling of multi-wavelength continuum}\label{app:radmc}

We divided the 870 $\mu$m (using APEX/LABOCA), 450 $\mu$m (using JCMT/SCUBA-2), or 350 $\mu$m (using CSO/SHARCII or APEX/SABOCA) images (whenever available) of each clump into annuli which have intervals equal to half of the beam FWHM, and then (projected) radially averaged the intensities in the annuli.
When the intensity distribution is largely asymmetric with respect to the clump center (e.g., when there is a bright adjacent core/clump or a bright external gas filament), we trimmed the contribution from those asymmetric (sub)structures before making the averages.
In practice, at each radius we first derived the mean and standard deviation ($I_{\mathrm{std}}$) of the intensities and then masked the pixels at which the sum of the radial intensities deviates from the mean by more than 0.9 times the sum of radial $I_{\mathrm{std}}$.

We used radiative transfer models to invert these derived radial intensity profiles to radial density profiles.
We employed the publicly available Monte Carlo radiative transfer code {\tt{RADMC-3D}} (\citealt{radmc3d}).
To find the simulated intensity profiles which match the observed ones, our modeling ergodically visited the parameter spaces of the assumed functional form for density (Equation \ref{eq:rhobulk}, Sect.~\ref{sec:T_rho_profiles}). This means that the resultant models, when the number of which is sufficiently large (as in our case we adopt 10,000 models), can represent the average statistical properties of the models constructed from the full parameter space.

To benchmark and refine the radial temperature profile, we conducted aperture photometry with multi-wavelength dust continuum data. We used the tools in Python package {\tt{photutils}} (\citealt{larry_bradley_2021_5525286}). The total fluxes at each wavelength of 24, 70, 160, 250, 350, and 500 $\mu$m from {\it{Spitzer}/MIPS} (\citealt{Carey09}), {\it{Herschel}}/PACS, and {\it{Herschel}}/SPIRE images are measured and they provide an observed SED profile.
We then let {\tt{RADMC-3D}} generate the SEDs ranging from mid-IR to mm wavelength and ensure that the best-fit model can reproduce an SED profile which is broadly consistent with the observed SED.

We assumed that dust opacity does not have spatial variation and quoted the opacity model from \citet{OH94}. 
Specifically, we quoted the column evaluated for the thin ice mantle coated dust which was coagulated for 10$^{5}$ years in an environment with 10$^{6}$ cm$^{-3}$ gas density (hereafter OH5 model). This model has been successfully applied in previous studies to explain the radial profiles of dust emission around high-mass embedded protostars (\citealt{Mueller02}; \citealt{Rolffs11}).
The maximum grain size in the OH5 opacity model is under the Rayleigh limit such that the scattering opacity can be neglected in the modeling.


The hydrogen gas density profile ($\rho_{\mathrm{bulk}}(r)$) is described by the functional form of Equation \ref{eq:rhobulk}.
Using Monte Carlo radiative transfer to evaluate the temperature distributions (e.g., based on assumptions of heating sources and interstellar radiation field) is subject to a very large degree of freedom.
Instead, we adopted the radial gas temperature profile probed by the thermometers in Section ~\ref{sec:xclass}, following the form of Equation \ref{eq:tformula}.

 The overall mass of individual clumps derived by single-component SED fitting ($M_{\mathrm{sed}}$ as in Table \ref{tab:sources}) is used as a prior for our subsequent modeling.
Given that we only need to fit two free parameters ($\bar{\rho}$, $q$, c.f. Equation \ref{eq:rhobulk}), the Markov chain Monte Carlo (MCMC) method has no clear advantage.
Instead, for each observed molecular clump, we drew 10,000 samplers from the parameter space of $q$ = 0.0-2.5, $\bar{\rho}$=10$^{3}$-10$^{6}$ cm$^{-3}$ with a random process.
The likelihood function of the clump total mass follows the truncated normal distribution (e.g. \citealt{Lynch2007}): 
\begin{equation}
    ln \mathcal{L} = \left\{
\begin{array}{rcl}
-0.5(\frac{M-M_{\mathrm{sed}}}{3M_{\mathrm{sed}}})^{2} & & {(0.2M_{\mathrm{sed}}\leq M\leq 5M_{\mathrm{sed}})}\\
-inf & & {(M < 0.2M_{\mathrm{sed}}\,\,\, or\,\,\, M > 5M_{\mathrm{sed}})}\\
\end{array} \right.
\end{equation}

We then ran \texttt{RADMC-3D} for each sampler and convolved the derived images with the corresponding Gaussian beams to compare with the multi-wavelengths observations.
When summing the $\chi^{2}$, we assumed that the observational data can have nominal $\sim$20\% errors, and adopted one standard deviation in the radial profile calculation as the uncertainty of the observational data. 
The $\chi^{2}$ calculation follows,
\begin{equation}
    \chi^{2} = \sum_{\lambda}\sum_{r}\frac{(f_{\mathrm{mod}}-I_{\mathrm{obs}})^{2}}{I_{\mathrm{std}}^{2}+(0.2I_{\mathrm{obs}})^{2}}
\end{equation}
where $\sum\limits_{\lambda}$ and $\sum\limits_{r}$ denote summing over all wavelengths and sampled radii, respectively. $f_{\mathrm{mod}}$ stands for the model intensity at a certain radius.

\begin{figure*}
\begin{tabular}{p{0.245\linewidth}p{0.245\linewidth}p{0.245\linewidth}p{0.245\linewidth}}
\hspace{-.4cm} \includegraphics[scale=0.25]{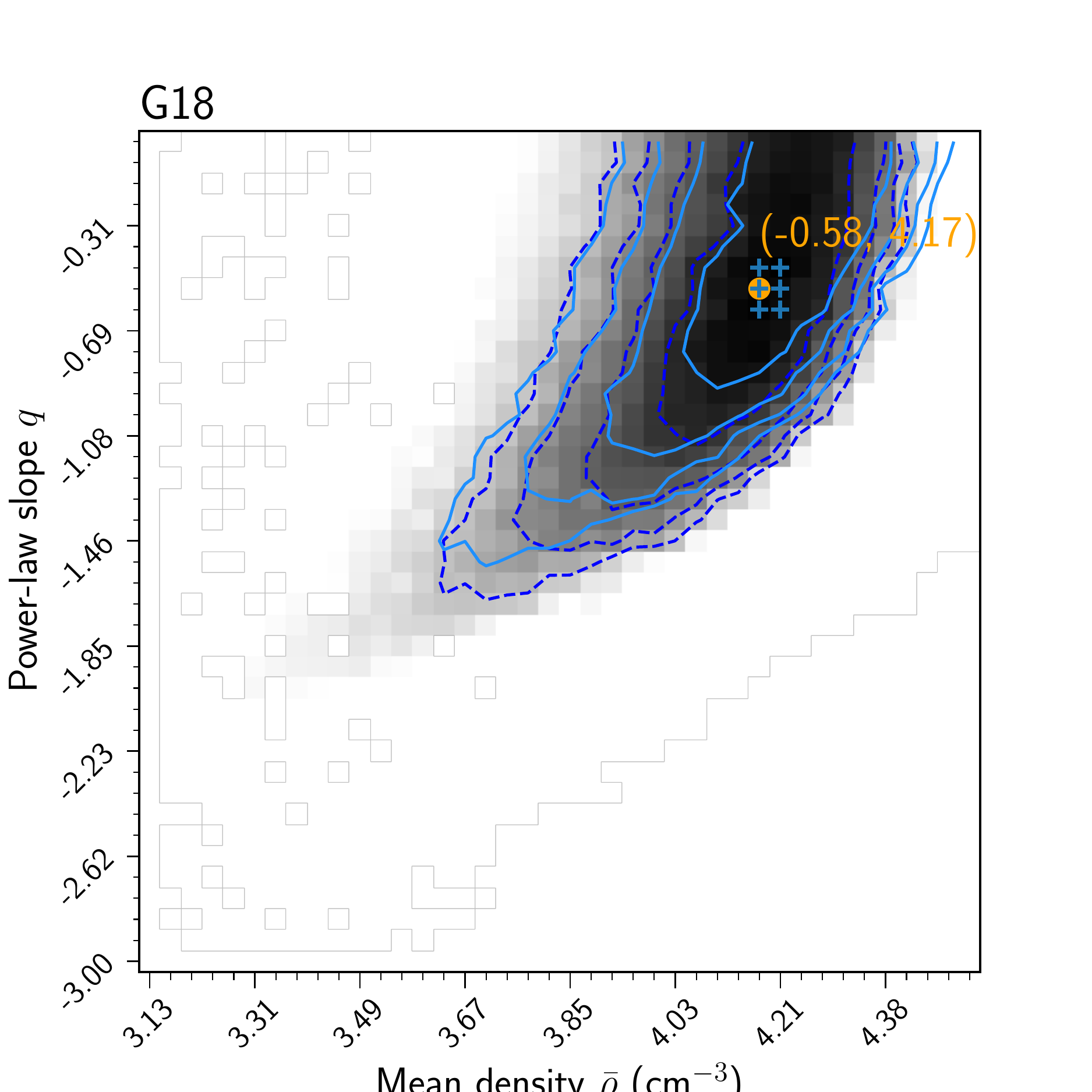}&  \hspace{-.3cm}\includegraphics[scale=0.23]{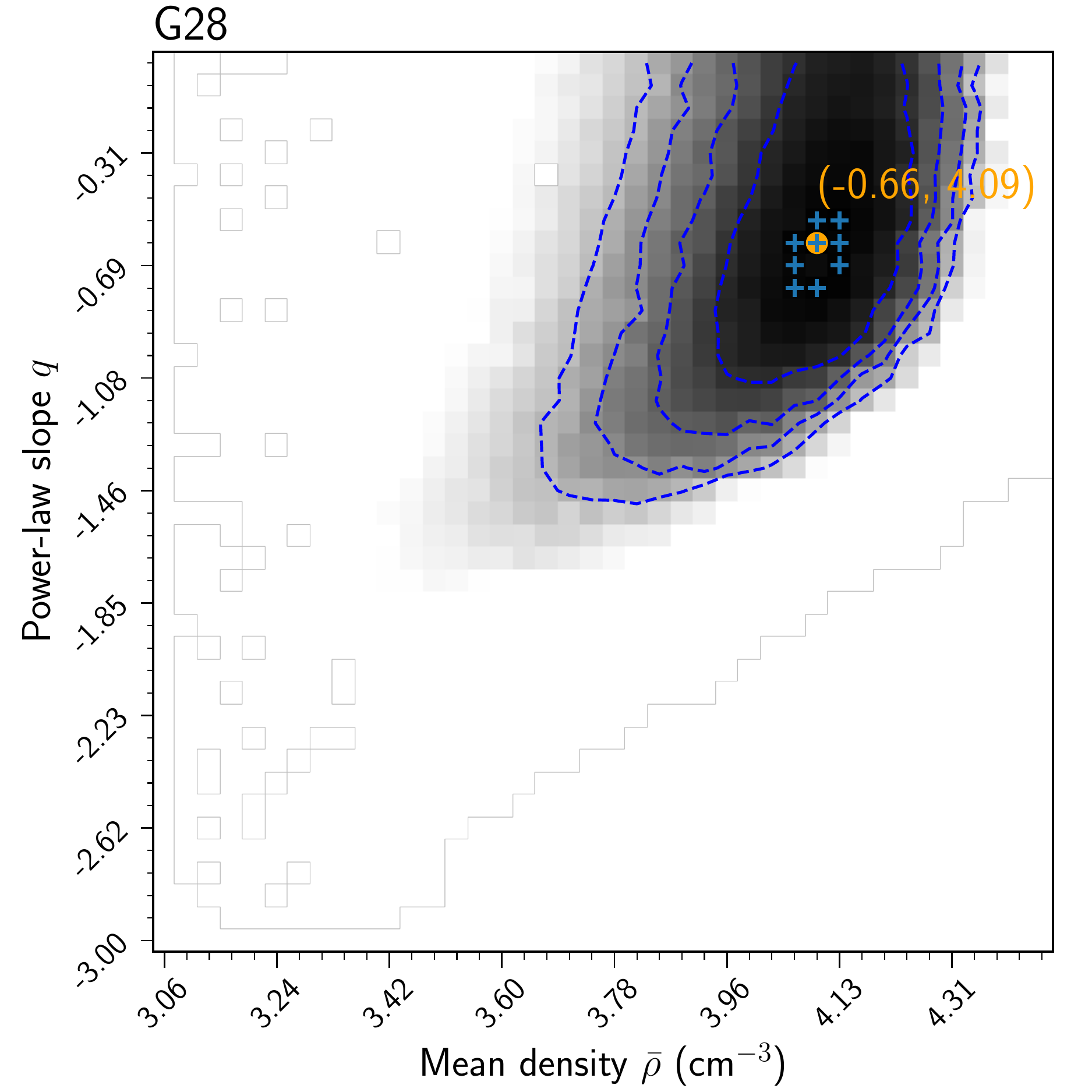}&
\hspace{-.4cm}\includegraphics[scale=0.25]{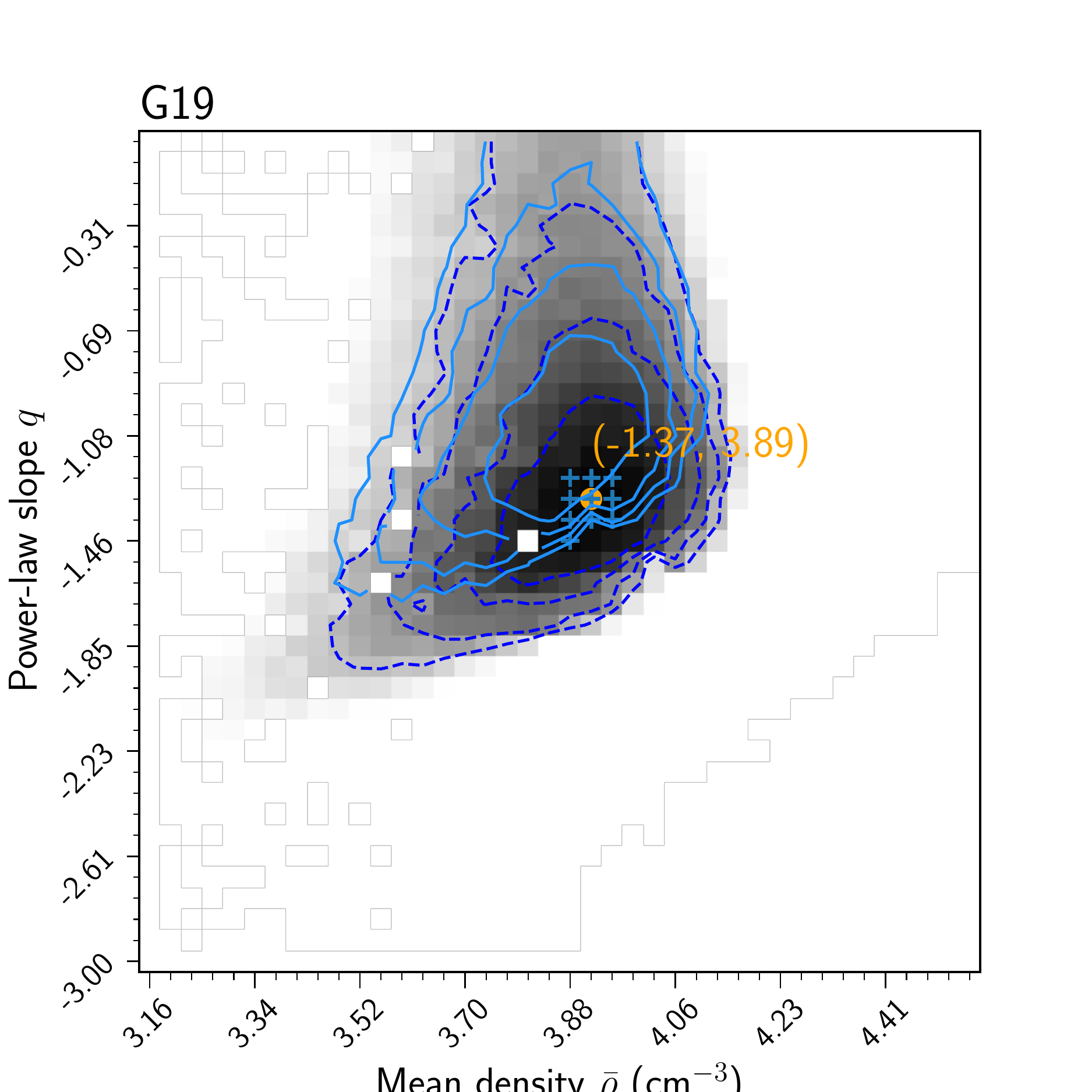}&
\hspace{-.3cm}\includegraphics[scale=0.25]{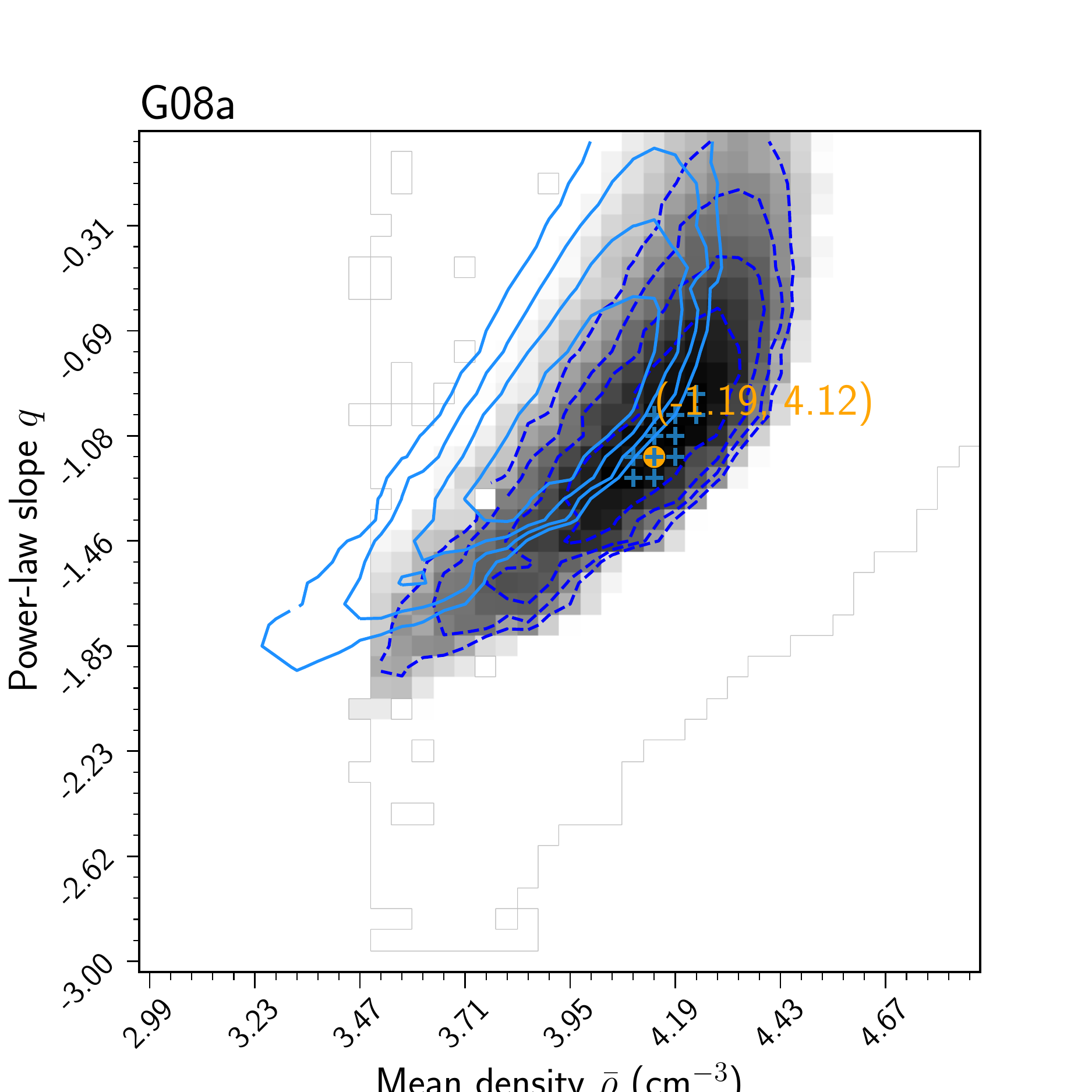}\\
\hspace{-.4cm}\includegraphics[scale=0.25]{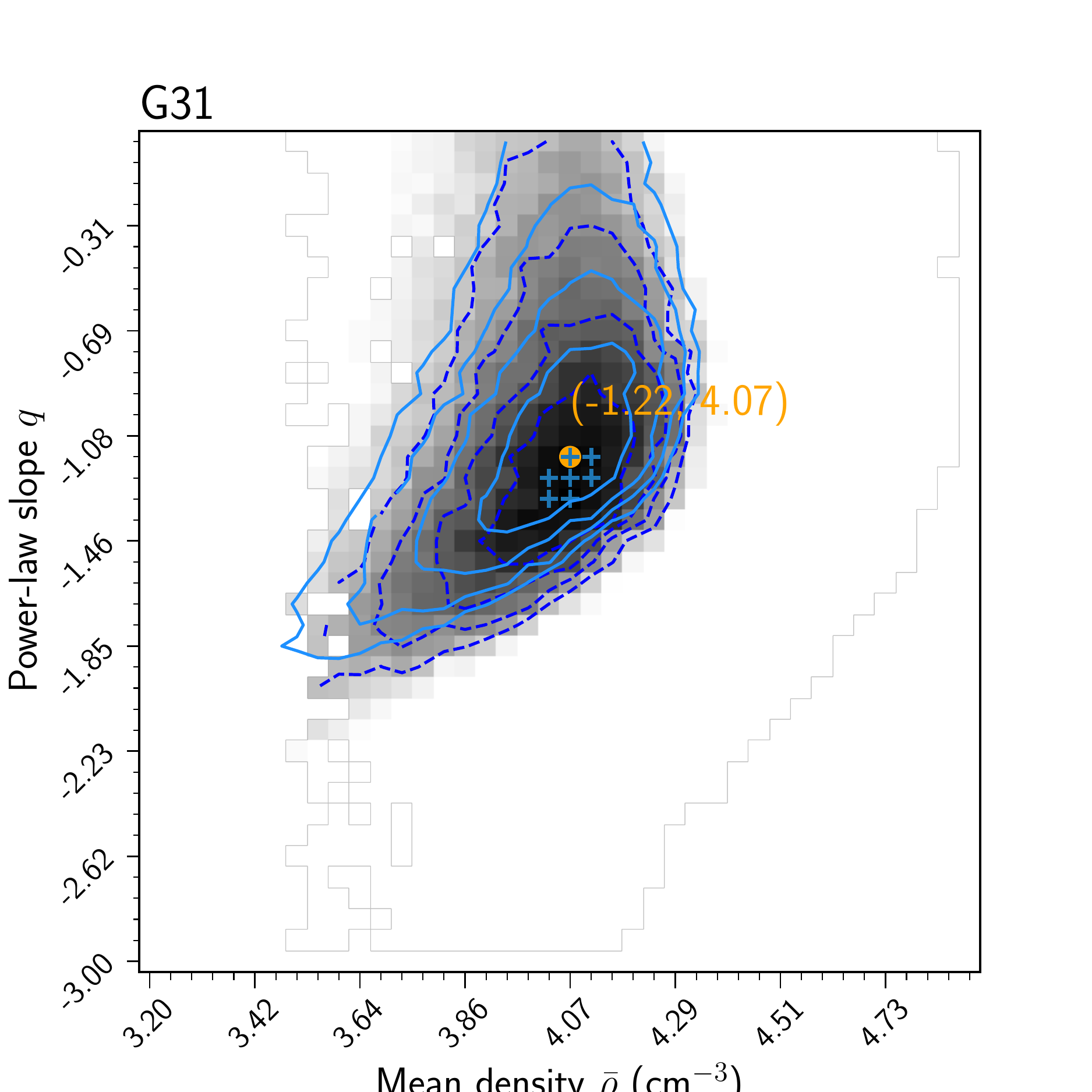}&
\hspace{-.3cm}\includegraphics[scale=0.23]{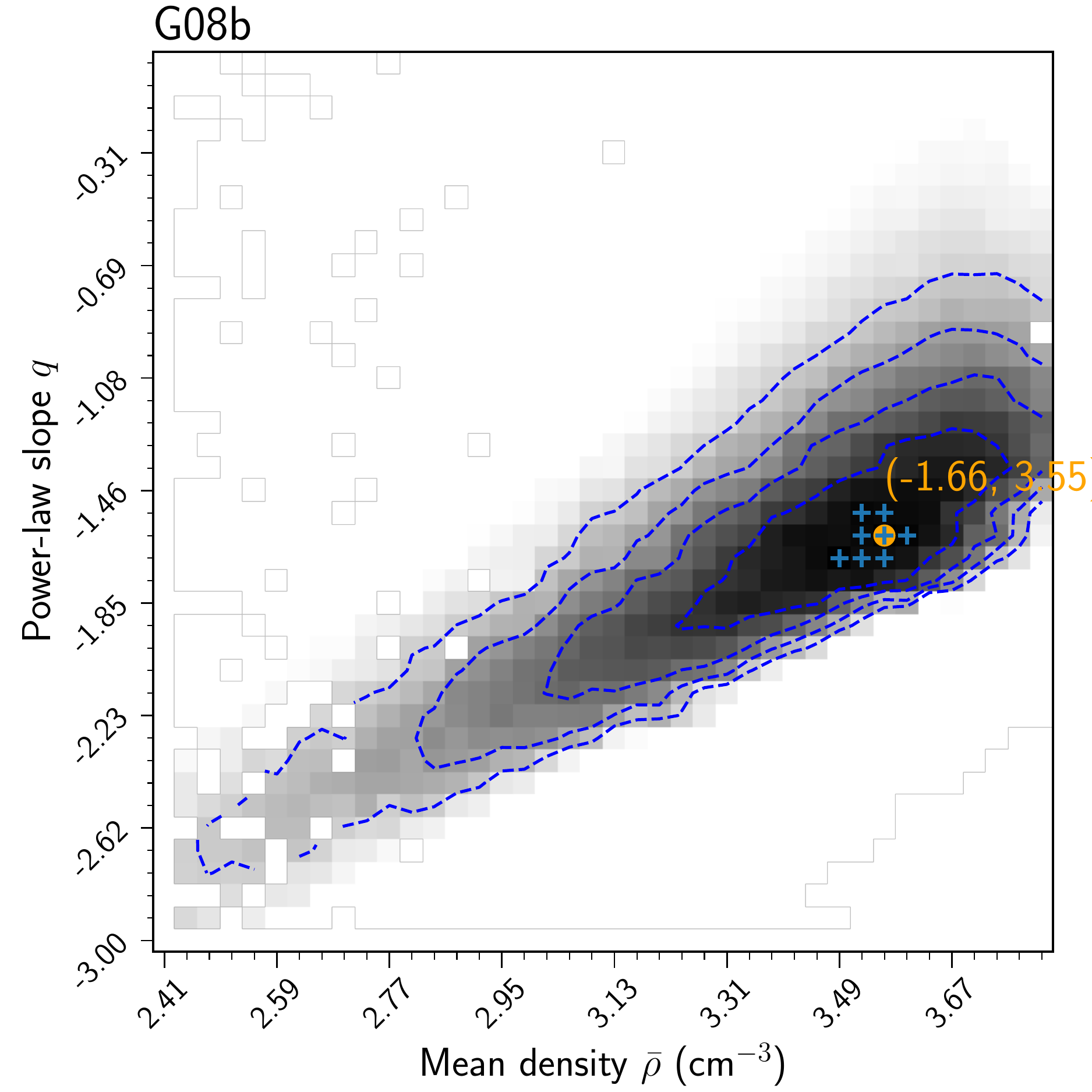}&
\hspace{-.4cm}\includegraphics[scale=0.25]{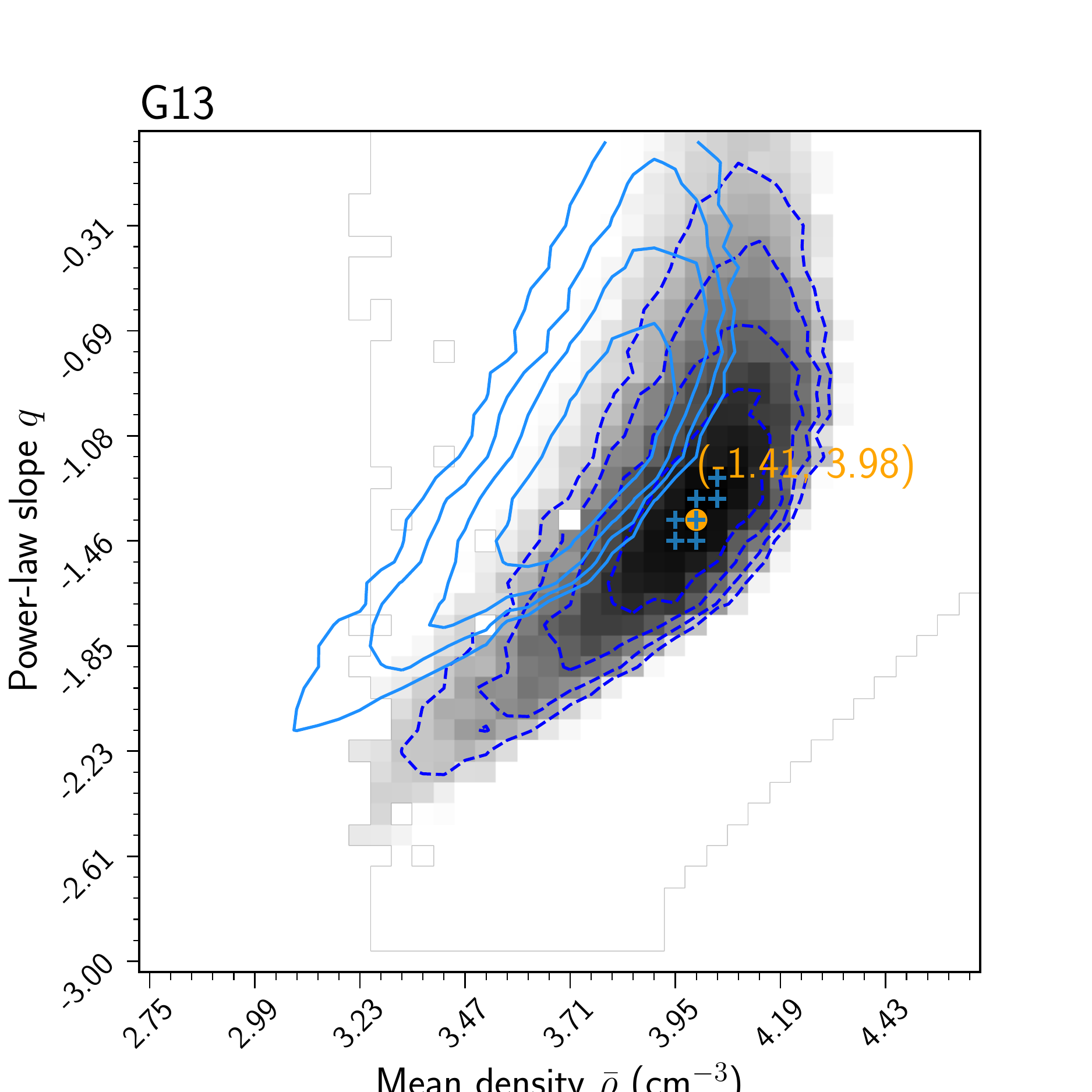}&
 \hspace{-.3cm}\includegraphics[scale=0.23]{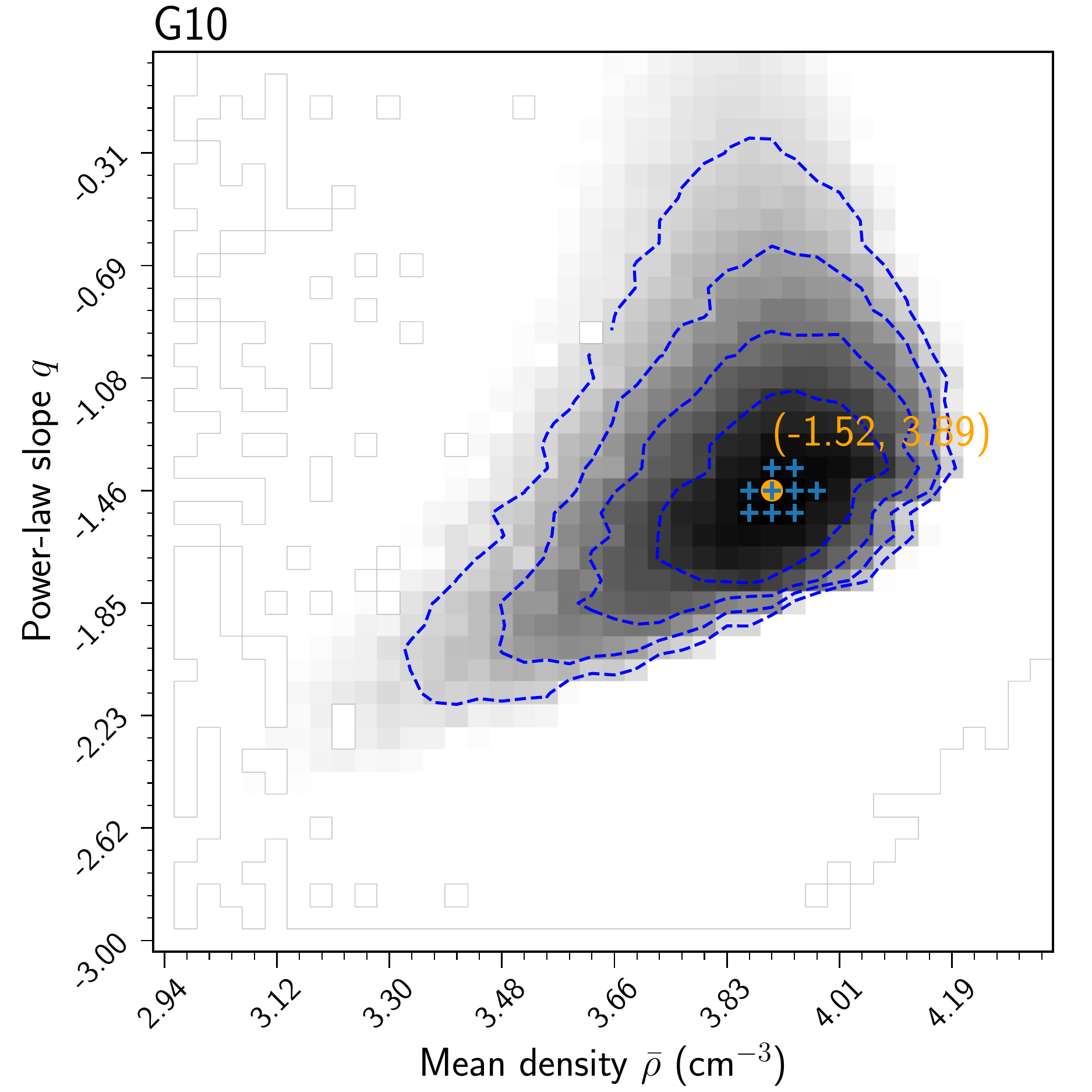}\\
\end{tabular}
\caption {$\chi^{2}$ converted probability distribution of the 10 000 parameter set of {\tt{RADMC-3D}} models for all clumps. Orange point indicates the best-fit model. Blue crosses mark the positions of the 30 best-fit models, so there could be overlaps between different parameter set due to the binning. For sources G31, G13, G08a and G19, results before re-adjusting $T(r)$ based on SED are shown as light blue contours.}
\label{fig:radmc_corner}
\end{figure*}

Figure \ref{fig:radmc1} compares the observed radial intensity profiles with the best-fit models.
The fitted values of $q$ and $\bar{\rho}$, the (pre-determined) clump radius, overall clump mass, and the characteristic density determined from the best-fit model at 0.1 pc radius, are summarized in Table \ref{tab:radmc_para}. 
The posterior probability distributions of $\bar{\rho}$-$q$ are shown in Figure \ref{fig:radmc_corner}, with best-fit parameter set marked. 
In Figure \ref{fig:sed_radmc} the observed multi-wavelength flux densities are compared with the SED generated from best-fit {\tt{RADMC-3D}} models. 

It can be seen that the observed SEDs of sources G18, G19, G08a, G13 and G31 show large deviation from that of the best-fit models which were produced based on the assumed $T(r)$ profiles. 
For sources G18, G19, G08a and G13, the deviations are partially expected as these three sources exhibit the smallest densities in the sample, from the clump center to outer region. 
It is likely that $T_{\mathrm{rot}}$ estimated from aforementioned thermometers is biased to the small proportion of the dense gas at each radius (clumpiness), while the temperature for the bulk gas at each layer, or the {\it{average}} gas temperature is smaller. 
In addition, for the intermediate-scale gas having densities $\lesssim$10$^{5}$ cm$^{-3}$, which are mostly probed by CH$_{3}$CCH (12-11) and H$_{2}$CS lines in terms of gas temperature, the gas temperature being higher than dust temperature may also have an origin from turbulent heating, as under such gas densities thermal coupling between dust and gas is weaker (\citealt{PanPadoan09}).
For G31, which is much denser in terms of bulk gas radially but shows the most prominent monolithic core in the center, the over-estimation of temperature likely originates from optical depth effect. 

To refine $T(r)$ as guided by the observed SED shape, we retain the parametric form (Equation \ref{eq:tformula}) as elaborated in Section \ref{sec:T_rho_profiles}, and only adjust the parameter $r_{\mathrm{in}}$ (Equation \ref{eq:tformula}), scaled by a factor $<$1. Based on the adjustment, we regenerate SED profiles from {\tt{RADMC}} modeling and find the best-fit rescaled $r_{\mathrm{in}}$, which is listed in Table \ref{tab:radmc_para} for the 5 sources. 
For these 5 sources, we then iterate the fitting of radial density profiles from {\tt{RADMC-3D}} calculations, based on the adjusted temperature profiles $T(r)$. From Figure \ref{fig:radmc_corner}, the best-fit parameter set before and after adjusting $T(r)$ are shown together. Decreasing $r_{\mathrm{in}}$ in the temperature form, is equivalent to reduce the steepness and absolute value of temperature radial profile, which results in an increment of mean gas density and density profile slope in the radial intensity profile fits, as expected.   
As a further benchmark, we use {\tt{RADMC-3D}} to self-consistently calculate the dust temperature. To convert the clump luminosity to a central stellar source, we use the stellar evolution models of solar metallicity from \citet{Choi16} which give relations between luminosity, mass, radius and temperature for ZAMS to estimate the stellar $T_{\mathrm{eff}}$. The re-iterated best-fit density model is used to describe the envelope structure. The resultant SED is also shown in Figure \ref{fig:sed_radmc}. 


\section{Radiative transfer modeling of CH$_{3}$OH and CH$_{3}$CCH lines: benchmarking the results from one-component non-LTE/LTE models}\label{app:lime}

The RADEX analysis of CH$_{3}$OH (5-4) lines in Section \ref{sec:radex_nh2}, with one-component non-LTE assumption, can be biased due to the mixed contribution along the LOS.
On the other hand, the {\tt RADMC-3D} modelings for dust continuum in Appendix \ref{app:radmc} better represents the enclosed masses within certain radii by constraining $\rho_{\mathrm{bulk}}$ while they cannot provide constraint on the higher volume density internal structures.
To refine our estimates of physical parameters, we build on these two efforts to conduct 3D radiation transfer forward modelings of CH$_{3}$OH (5-4) and CH$_{3}$CCH (12-11) lines.
Specifically, we performed the non-LTE modeling for the excitation conditions using the {\tt LIME} code (\citealt{Brinch10}) and then compared the synthetic spectra with the observed spectral cubes.
The spatial distributions of CH$_{3}$OH (5-4) and CH$_{3}$CCH (12-11) line emission are rather extended such that they better characterize the majority of dense gas in the clumps.
We focus on benchmarking the radial density profile of the dense gas ($\rho_{\mathrm{dense}}$).
We fixed the gas temperature profiles ($T(r)$; Section \ref{sec:T_rho_profiles}) according to the profiles obtained in Section \ref{sec:xclass} (and refined in Appendix \ref{app:radmc}), when constructing the input models. 
As $T(r)$ has been verified by full radiative transfer of dust continuuum by building SEDs, the CH$_{3}$CCH (12-11) modeling conducted here is further used as a sanity check that the assumed temperature profile can reproduce the line emission of this thermometer.

The gas density radial profiles were initially specified to be the best fits from the modeling continuum intensity profiles, $\rho_{\mathrm{bulk}}$ (Equation \ref{eq:rhobulk}, see details in Appendix \ref{app:radmc}).
Based on this density model and temperature profile $T(r)$, we then experimented various assumptions of the abundance spatial variations of CH$_{3}$OH and CH$_{3}$CCH, motivated by the results of chemical network from \citet{Belloche17} and \citet{Calcutt19}.

\begin{figure*}
    \centering
    \hspace{-0.35cm}\includegraphics[scale=0.6]{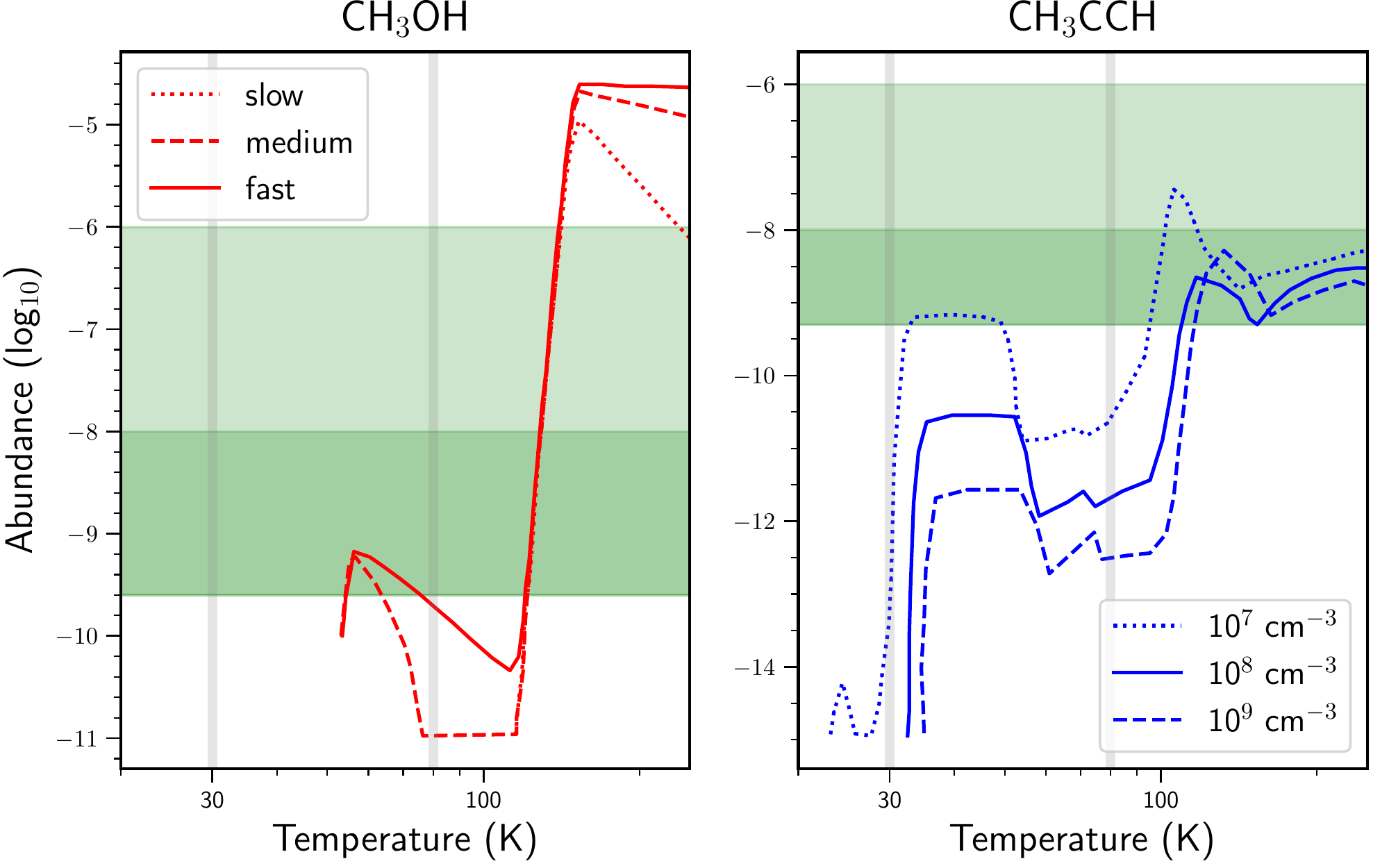}
    \caption{The abundance variations of CH$_{3}$OH and CH$_{3}$CCH from chemical models (lines) and the explored parameter space (filled) for the {\tt{LIME}} models based on $\rho_{\mathrm{bulk}}$ density model (column A of Table 7). The results from chemical models are shown in lines (solid, dashed and dotted): for CH$_{3}$OH the abundance profiles for different warm-up timescales are shown (\citealt{Garrod17}); for CH$_{3}$CCH, abundance profiles of different final collapse densities are shown (\citealt{Calcutt19}). The dark green filled region indicates the lower abundance range explored for $X_{\mathrm{out}}$ by the {\tt{LIME}} models, and the light green filled region the upper range explored for $X_{\mathrm{in}}$. The vertical gray lines indicate the jump temperature of 30 and 80 K.}
    \label{fig:ab_profiles_chem}
\end{figure*}

Based on the chemical network of \citet{Belloche17}, \citet{Calcutt19} calculated the variation of gas-phase abundance of CH$_{3}$CCH as a function of warm-up time at different densities which range from 10$^{7}$ to 10$^{10}$ cm$^{-3}$. 
In these calculations, a two-stage physical evolution is assumed: a cold collapse stage is followed by a static warm-up stage which reaches a gas temperature of 400 K.
The CH$_{3}$CCH abundance is enhanced when the gas temperature reaches 30-40 K due to desorption of CH$_{4}$ to form CH$_{3}$CCH (i.e., dissociative recombination of larger hydrocarbons).
The CH$_{3}$CCH abundance is significantly enhanced again when the gas temperature reaches 80-100 K, due to the direct desorption of CH$_{3}$CCH from grain surfaces.
Finally, the models with higher final gas density present systematically lower CH$_{3}$CCH abundance since in these high-density models CH$_{4}$ desorbs at slightly higher temperature (lower panel of Figure \ref{fig:ab_profiles_chem}). 
Similarly, according to the chemical model of \citet{Garrod17}, the abundance of CH$_{3}$OH experiences two significant enhancements at 30-40 K and 80-100 K gas temperatures (upper panel of Figure \ref{fig:ab_profiles_chem}).  The chemical modeling include three warm-up models, depending on the timescale for the system to increase from 10-200 K, over 5$\times$10$^{4}$ (fast), 2$\times$10$^{5}$ (medium) and 1$\times$10$^{6}$(slow) yr (\citealt{GH06}). 

To mimic the abundance enhancement in the warm or luckwarm regions as described by these chemical models, we parameterized the CH$_{3}$OH and CH$_{3}$CCH abundance profiles as Equation \ref{eq:xmol}.
Again, we assumed that the $A$ and $E$-type CH$_{3}$OH have the same abundance.
The parameter ranges of $X_{\mathrm{in}}$ and $X_{\mathrm{out}}$ were chosen by referencing to $N_{\mathrm{mol}}$ for CH$_{3}$CCH and CH$_{3}$OH from XCLASS/RADEX modeling results (Section , \ref{sec:xclass}-\ref{sec:radex_nh2}) and the aforementioned chemical models.
Figure \ref{fig:ab_profiles_chem} shows a comparison between the parameter space we explored and the chemical modeling results.

Presently, only the collisional coefficients of CH$_{3}$OH with para-H$_{2}$ are available, although it seems that only in hot shocked gas there is a significant difference between the thermal rate coefficients of collisions with ortho- and para-H$_{2}$ (\citealt{Flower10}).
Our non-LTE models were not affected by the uncertain collisional coefficients because we assumed a low ortho- to para-H$_{2}$ ratio (OPR), such that the collisions with ortho-H$_{2}$ is negligible.
Chemical models and observations towards early stage dense cores indeed indicate an OPR value of $\sim$10$^{-3}$-10$^{-2}$ (\citealt{Flower06}, cf. \citealt{Troscompt09}), which is well below the equilibrium value of 3. 
In the post-shock gas, the OPR may remain low since the short timescale does not allow significant conversion from para-H$_{2}$ to ortho-H$_{2}$ (\citealt{Leurini16}).

When performing non-LTE modeling of CH$_{3}$CCH, the collisional rates of CH$_{3}$CN (\citealt{Green86}) were substituted by those of CH$_{3}$CCH.
This is a common approach since CH$_{3}$CCH and CH$_{3}$CN have similar molecular weights and configuration, while there is not yet published collisional rates for CH$_{3}$CCH.
The collisional rates of CH$_{3}$OH were quoted from \citet{Rabli10} (c.f., Section \ref{sec:radex_nh2}).

To create the distributions of physical properties we adopted the {\tt sf3dmodels} package (\citealt{Izquierdo18}) to generate homogeneous grids in Cartesian coordinates, which are then interpolated onto the {\tt LIME} input format of randomly generated set of points.
We used the linear resolution of $\sim$0.015-0.03 pc (depending on the source radius $R_{\mathrm{clump}}$, Table \ref{tab:radmc_para}) as grid size which corresponds to better than 1/5 beam size for each source. 
On each grid, we specified a mean gas velocity using a random process to mimic the turbulent velocity field: the direction was uniformly sampled from the $4\pi$ solid angle while the magnitude of the velocity was drawn from a Gaussian distribution with $\sigma=$3.5 km\,s$^{-1}$. 
We additionally adopted a uniform Doppler broadening of 0.4 km\,s$^{-1}$ ($\sigma_{\mathrm{turb}}$ = 0.4 km\,s$^{-1}$) to accommodate the unresolved (micro-)turbulence velocity. 
Therefore the intrinsic line-width for each model is $\sigma_{\mathrm{1D}}$ = $\sqrt{\sigma_{\mathrm{turb}}^{2}+\sigma_{\mathrm{thermal}}^{2}}$, where $\sigma_{\mathrm{thermal}}$ is determined by the assumed gas temperature. 
This yields linewidths comparable to the observed values ($\Delta V$ = 3.5$\pm$1.2 km\,s$^{-1}$).

We post-processed the output of {\tt LIME} to match the angular and velocity resolutions of our observations, and then compared the annularly and beam averaged synthetic spectra with those from observed spectral cubes.
For each observed clump, the best-fit model was taken as the one with the lowest $\chi^{2}$.
The parameters of these best-fit chemical models are summarized in Table \ref{tab:lime_radmcpara}, column A.

\begin{figure*}[htb]
\hspace{-.5cm}\includegraphics[scale=0.4]{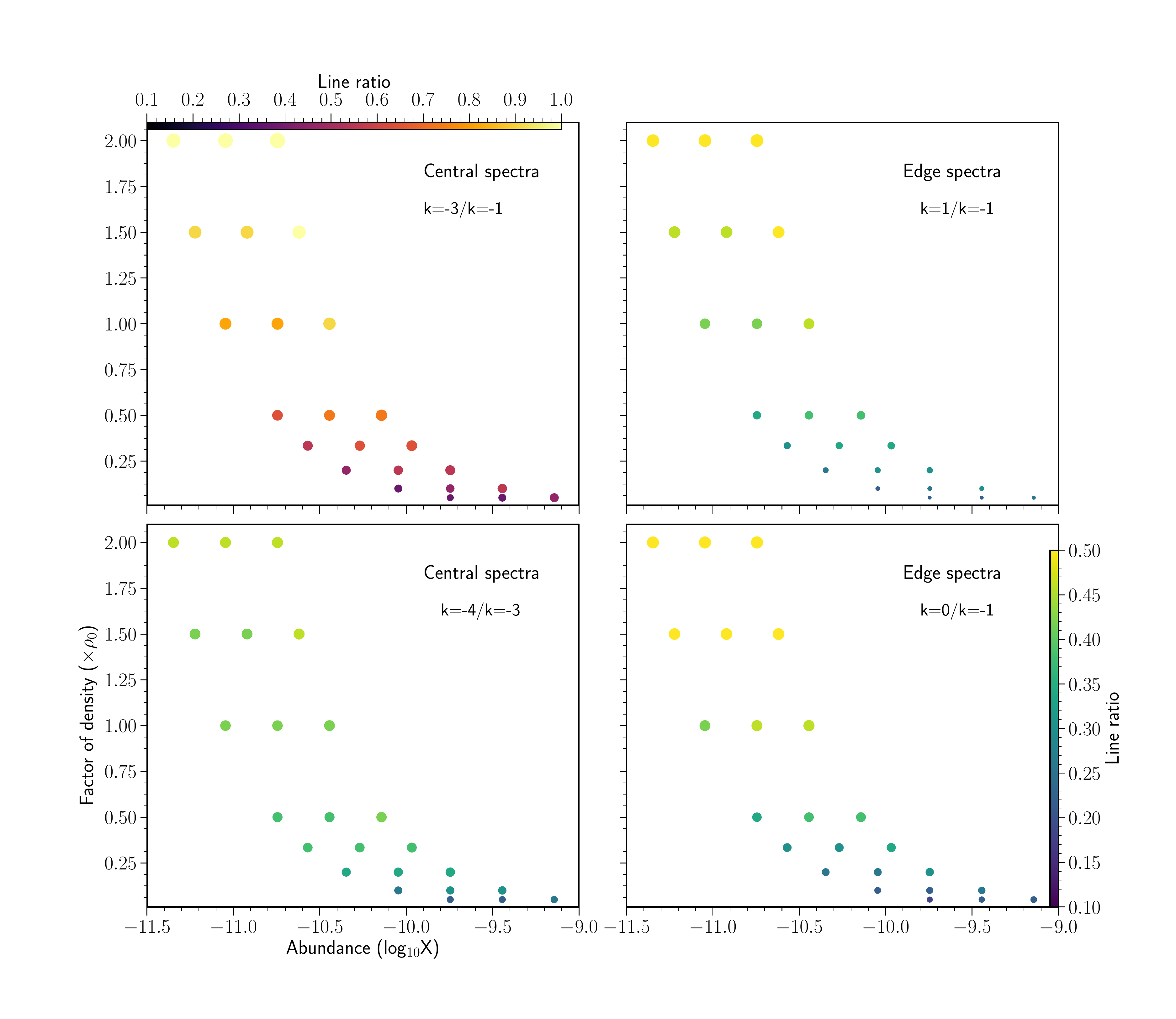}
\centering\caption{Line ratios of CH$_{3}$OH (5-4) $K$ components from {\tt{LIME}} models based on density profile of G08b ($\rho_{0}$ corresponds to the adjusted reference density at 0.1 pc for this source, as in Equation \ref{eq:limeB}). Abundance value corresponds to the outer abundance ($X_{\mathrm{out}}$) as in Table \ref{tab:lime_radmcpara}. The central marker at horizontal line of y = 1 corresponds to the best-fit model listed in Table. \ref{tab:lime_radmcpara} (column B). The comparison of observations with model spectra is presented in Figure \ref{fig:lime_radmc_dens_radexadj}, top panel.}
\label{fig:lime_dens_ab_degeneracy}
\end{figure*}

We found that these initial models systematically underestimated the intensities of the higher $K$ components of CH$_{3}$OH lines in the inner regions for all sources (Figure \ref{fig:lime_radmc_dens}).
There are some sources in which the intensities of the higher $K$ components were underestimated also at outer radii.
This implies that, in general, the density profiles $\rho_{\mathrm{bulk}}$ derived from dust continuum modeling (Appendix \ref{app:radmc}) were not high enough to collisionally excite the high $K$ levels of CH$_{3}$OH (e.g., gas may be concentrated in substructures of higher volume density).
This was expected, as was revealed by the comparison to the $\rho_{\mathrm{dense}}$ derived by RADEX modeling: gas densities are 50-200 times larger than that of $\rho_{\mathrm{bulk}}$ derived by single-dish dust continuum.
Therefore, we updated the radial density profile in the models according to the RADEX results $\rho_{\mathrm{dense}}$ (Section ~\ref{sec:radex_nh2}), following Equation \ref{eq:limeB}.
We manually adjusted the flexible parameters, which are the density scaling factor $f_{\mathrm{n}}$ (Equation \ref{eq:limeB}) and $f_{\mathrm{inc}}$ to quantify the abundance profile of $X_{\mathrm{mol}}(r)$ in Equation \ref{eq:xmol}, in a trial-and-error manner to seek for better fits to the observational data.
Figure \ref{fig:lime_dens_ab_degeneracy} demonstrates how the line ratios of CH$_{3}$OH $K$ components vary with gas density and molecular abundance while keeping a fixed overall molecular column density, as an example using the density model of G08b.

Our best-fit model parameters for all the clumps are summarized in Table \ref{tab:lime_radmcpara}, column B.
In these results, $f_{\mathrm{r}}$ ranges from 1.5 to 5 in all the sources except G31, indicating RADEX results only moderately overestimated the gas densities.
As mentioned in Section \ref{sec:comp_rough}, the rather large $f_{\mathrm r}$ in G31 is due to the very high optical depth of its lower CH$_{3}$OH $K$ components in the central region.
In this case the CH$_{3}$OH (5-4) lines do not provide meaningful constraints for the RADEX modeling which was based on the assumption of moderate optical depth.

When spatially integrating Equation \ref{eq:limeB} with a spherical symmetric assumption, the resulting overall molecular gas masses ($M_{\mathrm{mod}}$) considerably exceed those derived based on modeling dust continuum emission, by integrating $\rho_{\mathrm{bulk}}$ (see Appendix \ref{app:radmc}).
This implies that the dense gas structures traced by CH$_{3}$OH do not have spherically symmetric distributions.
Instead, they are local gas concentrations that have small volume filling factors. 
To reconcile the mass difference, we defined $ff_{\mathrm{dens}}\equiv$ $M_{\mathrm{enc}}$/$M_{\mathrm{mod}}$, where $M_{\mathrm{enc}}$ designates the enclosed molecular gas mass within 0.5 pc radius (a scale encompassing the CH$_{3}$OH emission entirely for all sources) derived from the dust continuum models following $\rho_{\mathrm{bulk}}$ (Equation \ref{eq:rhobulk}, Appendix \ref{app:radmc}). 
$ff_{\mathrm{dens}}$ can be regarded roughly as an upper limit of volume filling factor of the dense gas traced by CH$_{3}$OH. 
From the radiative transfer modeling point of view, to fit the observed line profiles, in the optically thin limit, varying the dense gas volume does not affect line intensity ratios, while the values of $X_{\mathrm{mol}}(r)$ can be adjusted accordingly such that the overall molecular column densities are not altered.
We discuss the implications from the inferred {\it{volume filling}} factor of dense gas, and the dense gas mass fraction based on comparing $\rho_{\mathrm{bulk}}$ and $\rho_{\mathrm{dense}}$ in Section \ref{sec:evo}.

\section{Comparison between observed spectra and modeling results: other targets}\label{app:other_sps}
\label{app:other_sps}
\begin{figure*}
\begin{tabular}{p{0.95\linewidth}}

\hspace{-2.75cm}\includegraphics[scale=0.32]{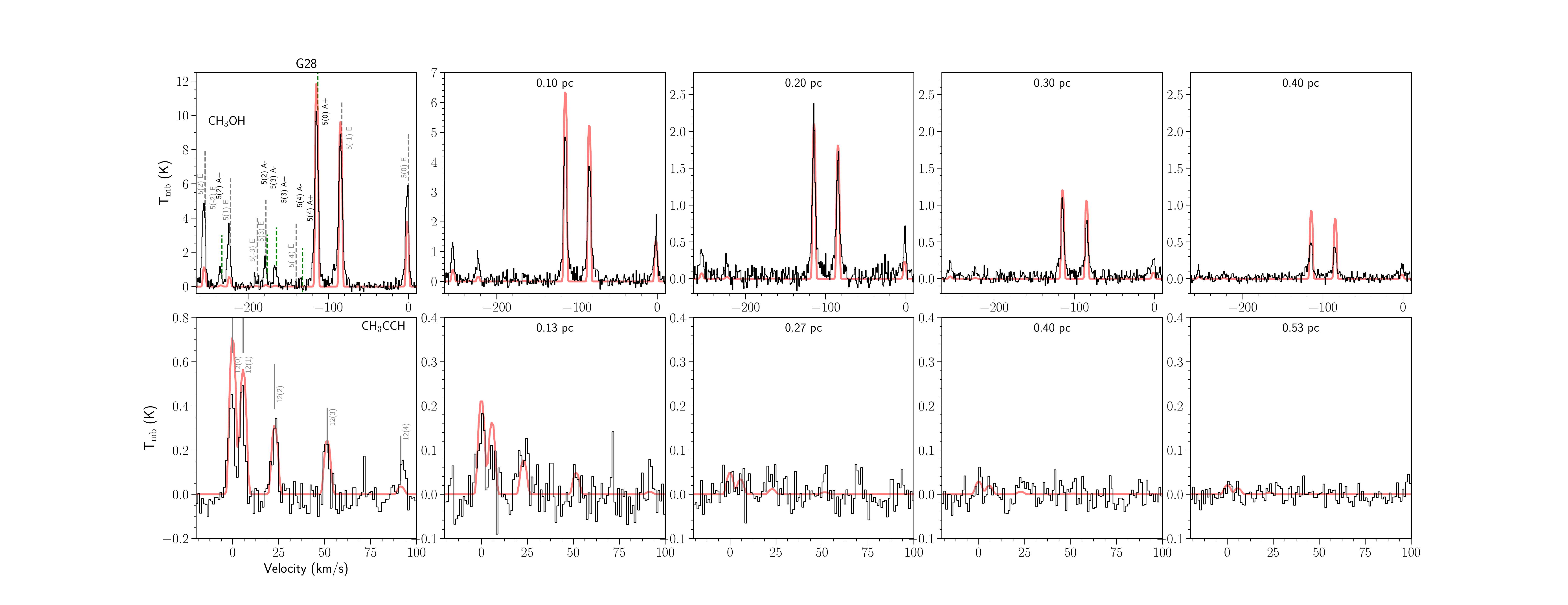}\\
\hspace{-2.75cm}\includegraphics[scale=0.32]{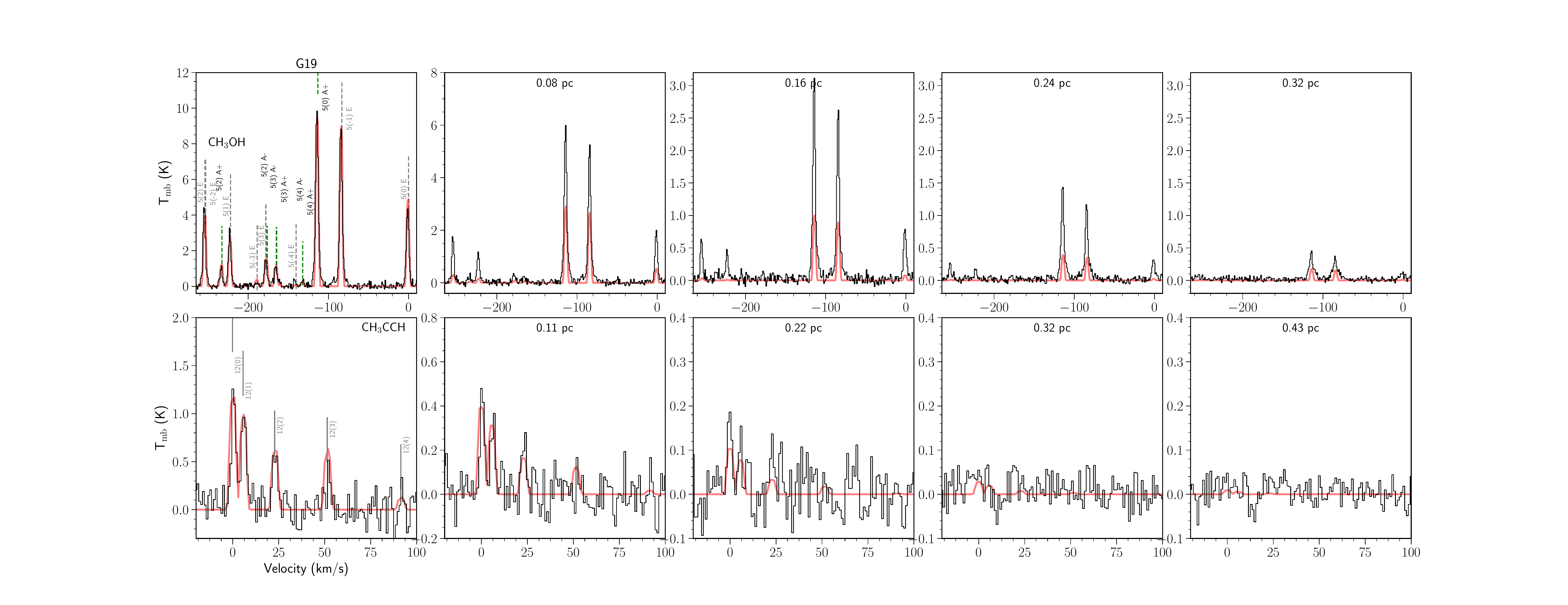}\\
\end{tabular}

\centering\caption{Same as Figure \ref{fig:lime_radmc_dens}
, for other target sources. \label{fig:lime_radmc_dens1}}
\end{figure*}

\begin{figure*}[htb]
\begin{tabular}{p{0.95\linewidth}}
\hspace{-2.75cm}\includegraphics[scale=0.32]{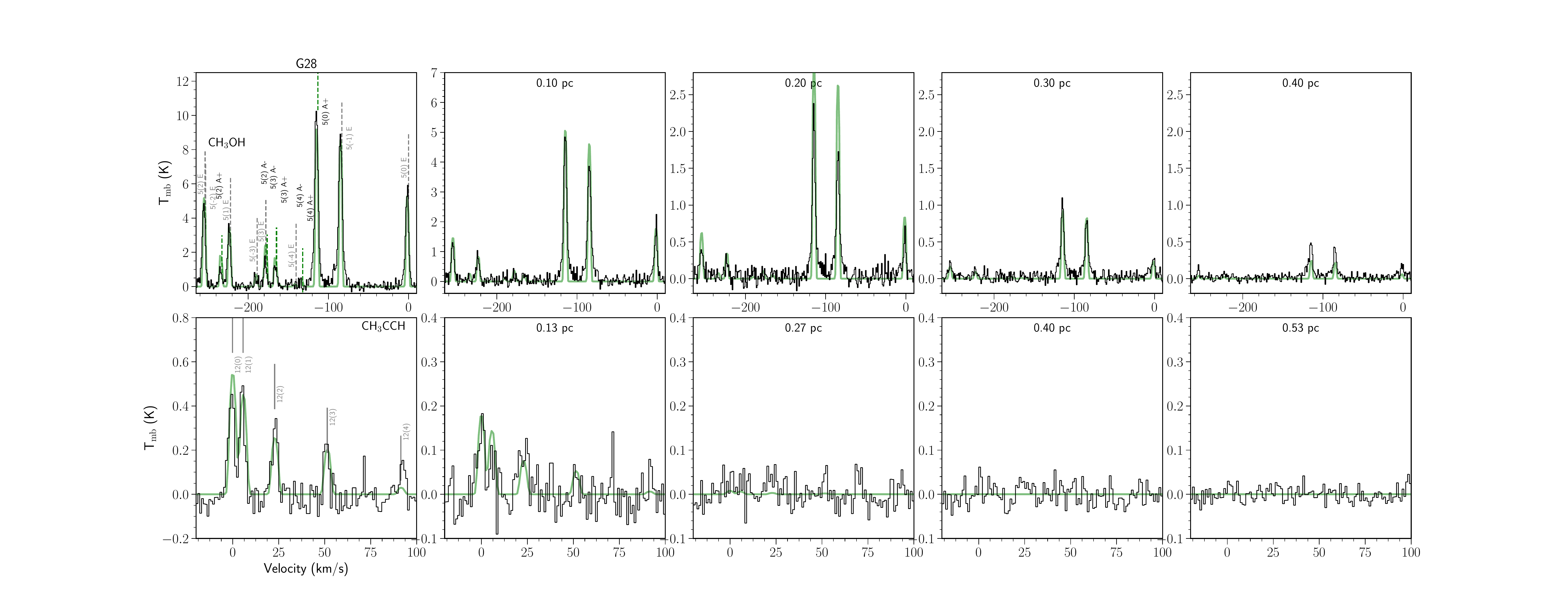}\\
\hspace{-2.75cm}\includegraphics[scale=0.32]{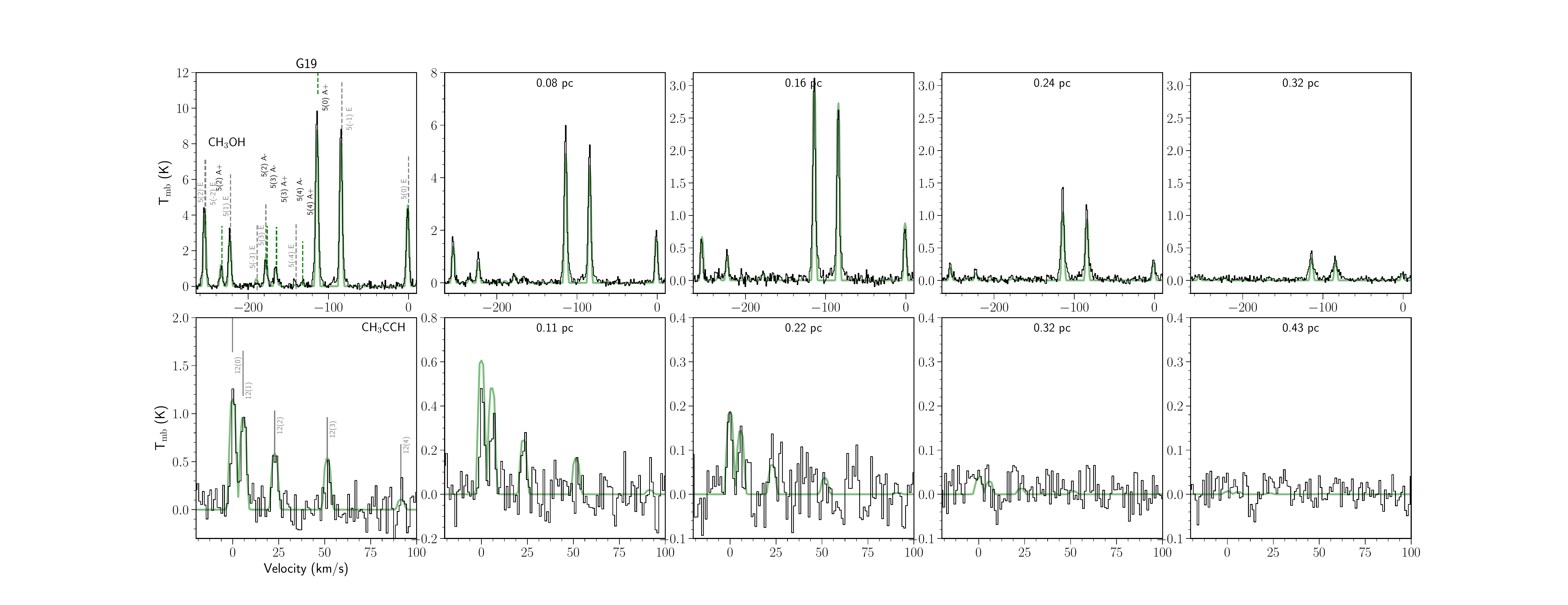}\\
\end{tabular}
\centering\caption{Same as Figure \ref{fig:lime_radmc_dens_radexadj}
, for other target sources. }
\label{fig:lime_radmc_dens_radexadj1}
\end{figure*}

\begin{figure*}[htb]
\begin{tabular}{p{0.95\linewidth}}
\hspace{-2.75cm}\includegraphics[scale=0.32]{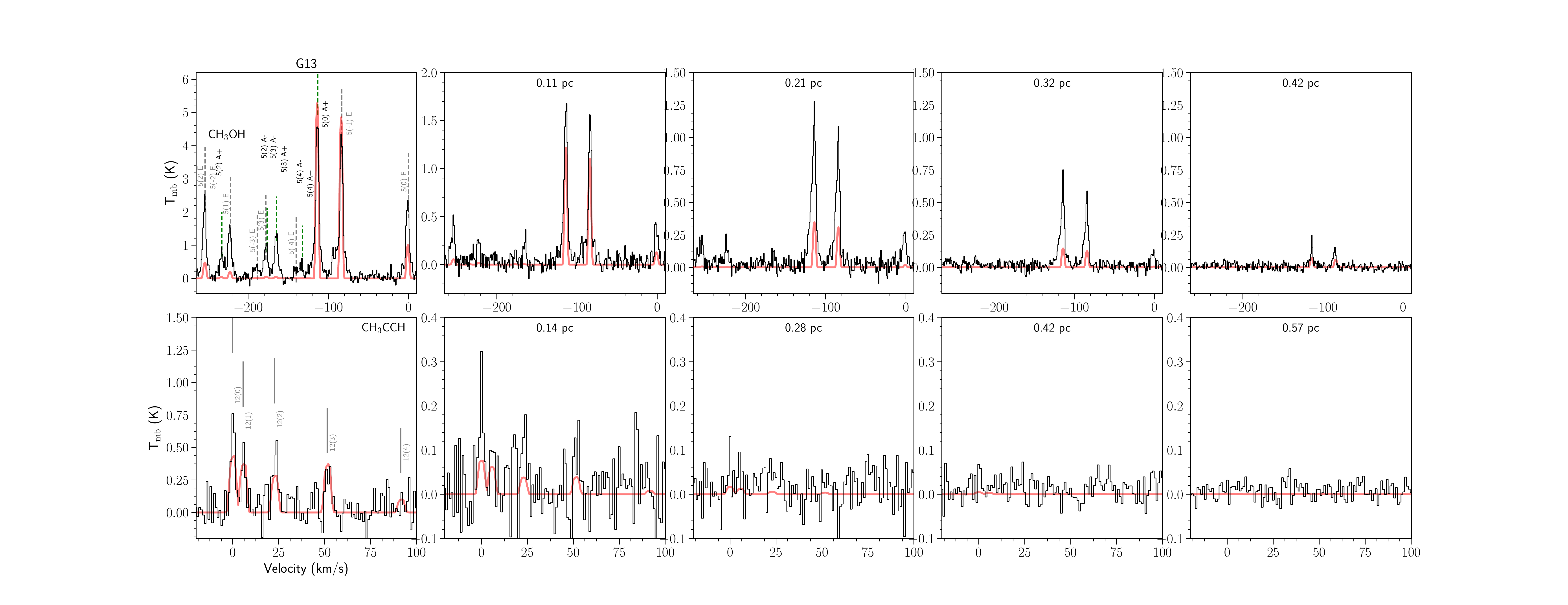}\\
\hspace{-2.75cm}\includegraphics[scale=0.32]{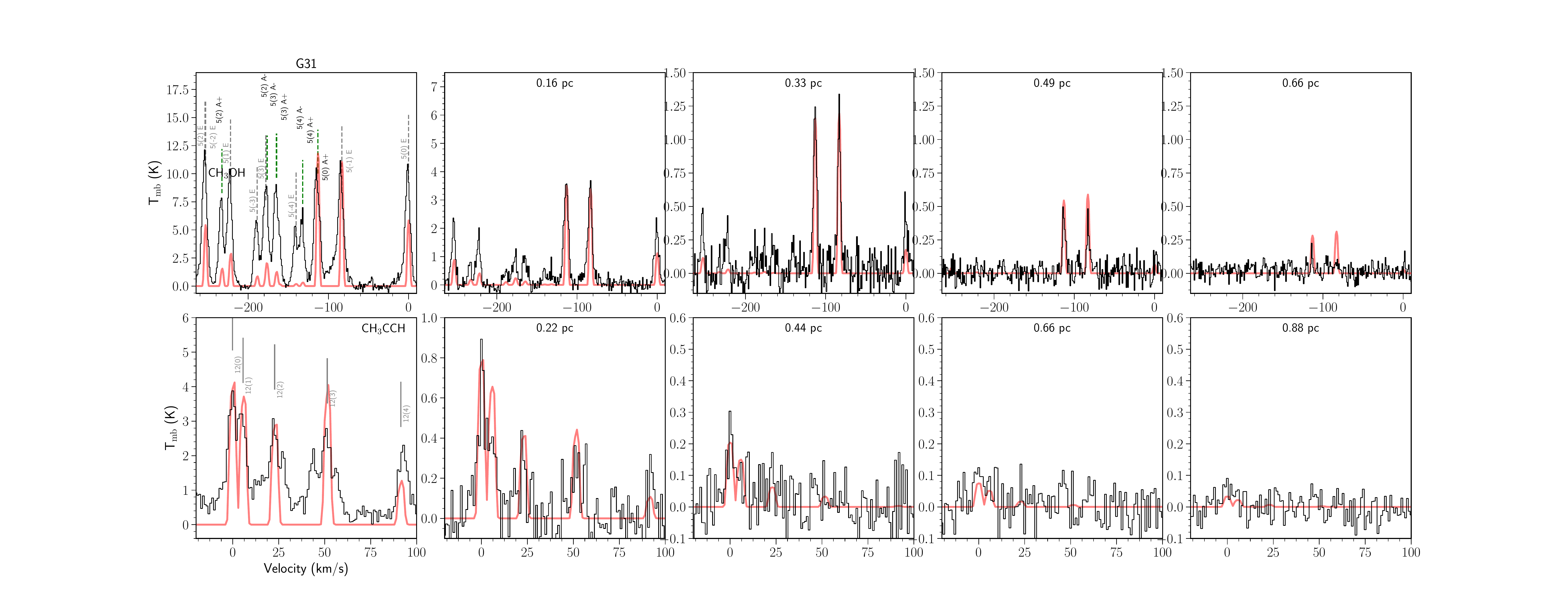}\\
\end{tabular}
\centering\caption{Same as Figure \ref{fig:lime_radmc_dens}
, for other target sources. \label{fig:lime_radmc_dens2}}
\end{figure*}

\begin{figure*}[htb]
\begin{tabular}{p{0.95\linewidth}}
\hspace{-2.75cm}\includegraphics[scale=0.32]{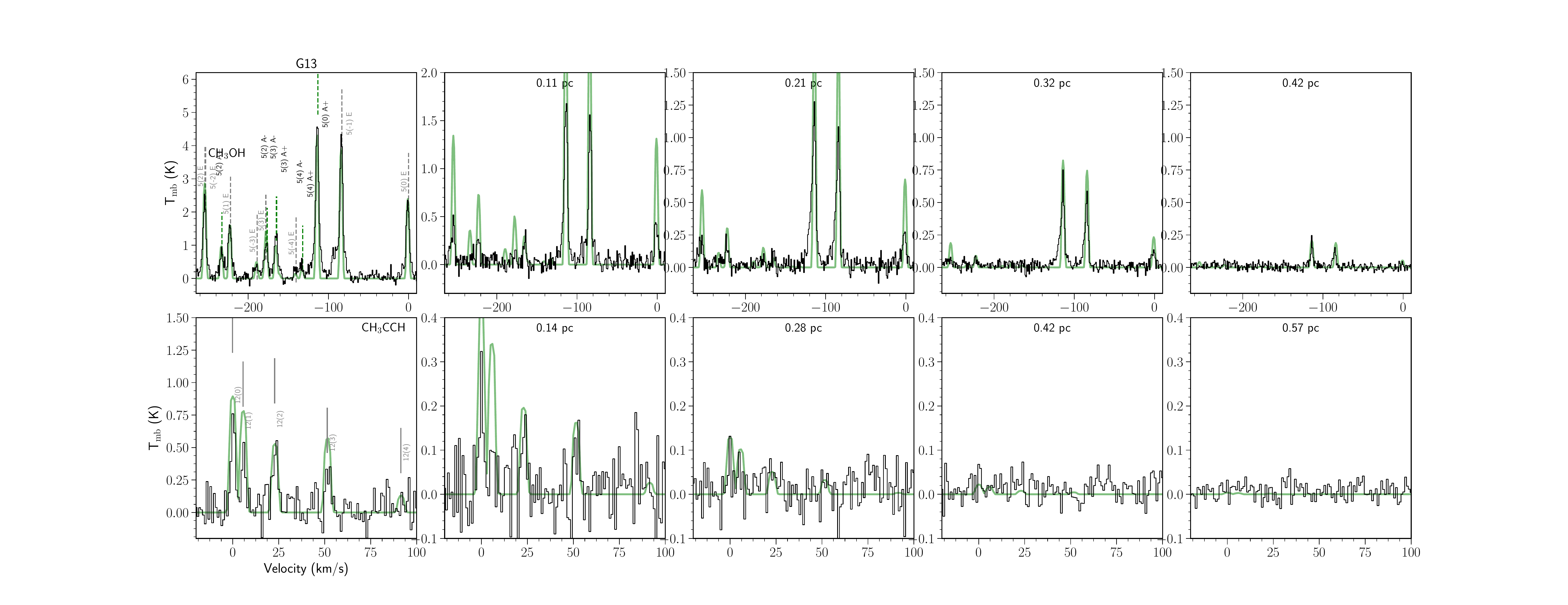}\\
\hspace{-2.75cm}\includegraphics[scale=0.32]{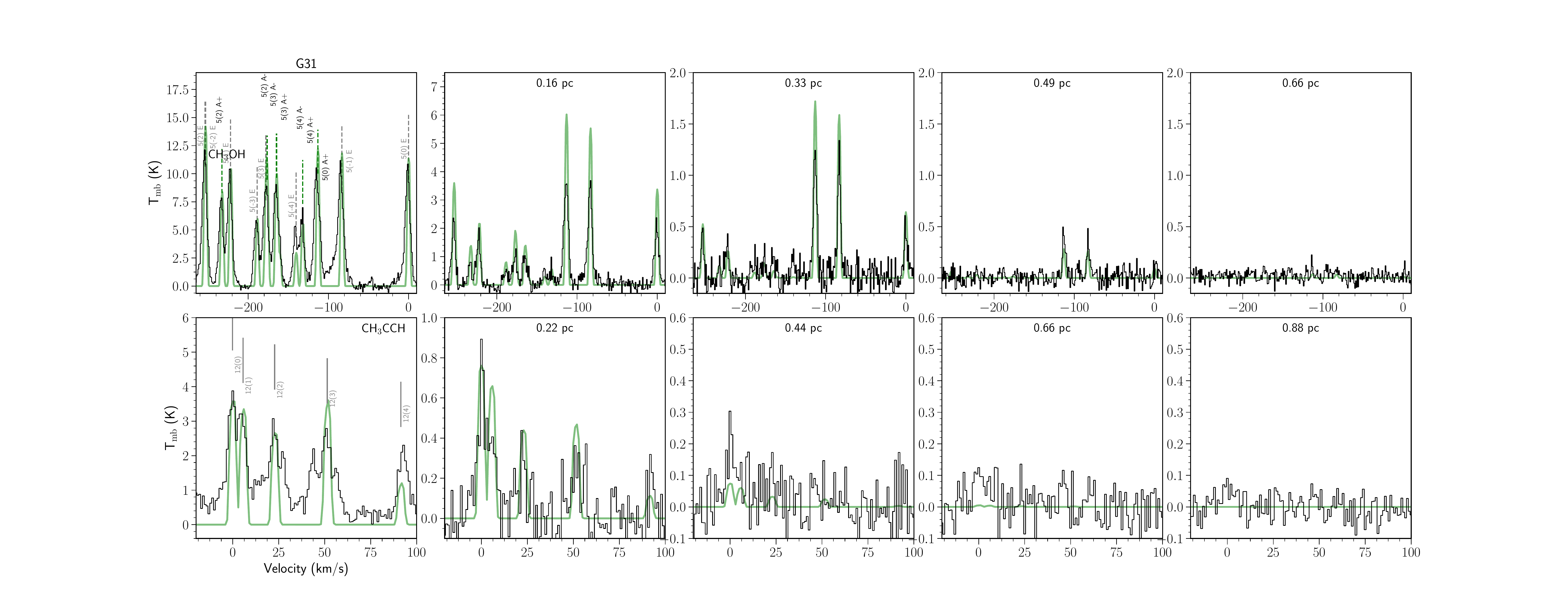}\\
\end{tabular}
\centering\caption{Same as Figure \ref{fig:lime_radmc_dens_radexadj}
, for other target sources. }
\label{fig:lime_radmc_dens_radexadj2}
\end{figure*}

\begin{figure*}[htb]
\begin{tabular}{p{0.95\linewidth}}
\hspace{-.9cm}\includegraphics[scale=0.165]{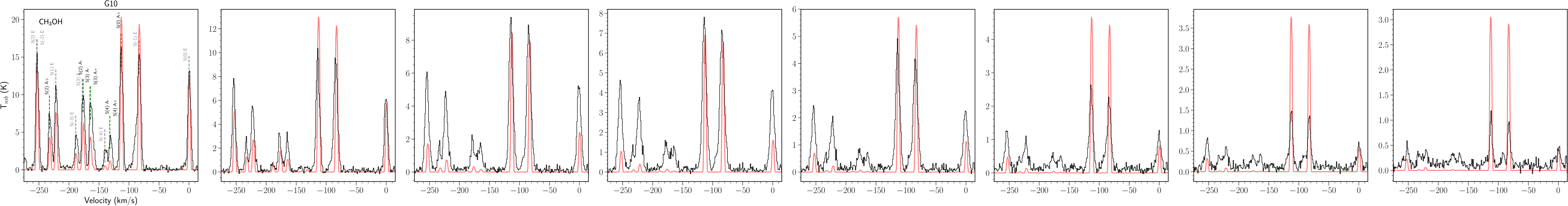}\\
\end{tabular}
\centering\caption{Same as Figure \ref{fig:lime_radmc_dens}
, for other target sources. \label{fig:lime_radmc_dens3}}
\end{figure*}

\begin{figure*}[htb]
\begin{tabular}{p{0.95\linewidth}}
\hspace{-.4cm}\includegraphics[scale=0.16]{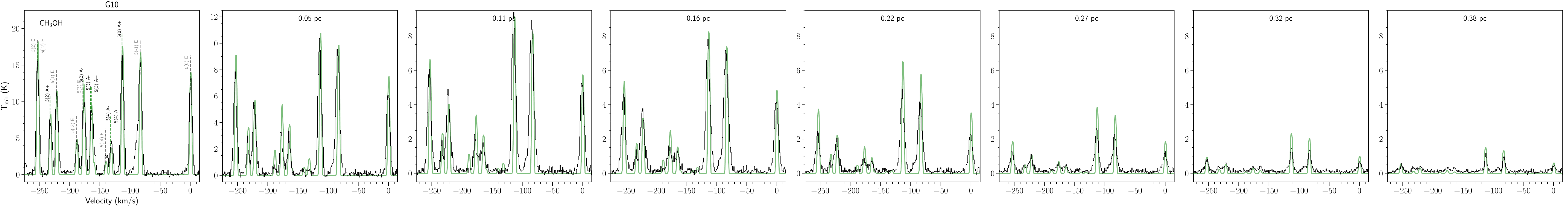}\\
\end{tabular}
\centering\caption{Same as Figure \ref{fig:lime_radmc_dens_radexadj}
, for other target sources. }
\label{fig:lime_radmc_dens_radexadj3}
\end{figure*}

\begin{figure*}
\includegraphics[scale=0.3]{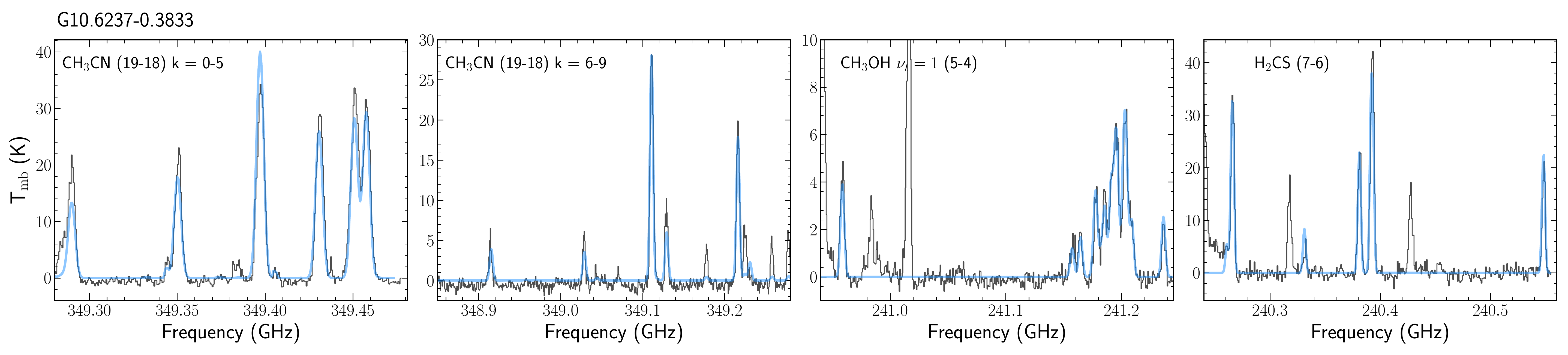}
\centering\caption{Example spectra (archival SMA data, Sect.~\ref{sec:observation}) of thermometer lines and the XCLASS fits of G10.}
\end{figure*}


\end{appendix}

\end{document}